\documentclass[11pt]{article}

\usepackage[margin=1in]{geometry}
\usepackage{setspace}
\usepackage{amsmath,amssymb,amsthm,amsopn,amsbsy,amsgen,amstext}
\usepackage{bm}
\usepackage{mathtools}
\usepackage{mathrsfs}
\usepackage{dsfont}
\RequirePackage[numbers]{natbib}
\RequirePackage[colorlinks,citecolor=blue,urlcolor=blue]{hyperref}%% uncomment this for coloring bibliography citations and linked URLs
\RequirePackage{graphicx}%% uncomment this for including figures
\usepackage{float}
\usepackage{xcolor}
\usepackage[linesnumbered,ruled,vlined]{algorithm2e}
\usepackage{authblk}
\usepackage{subfiles}
\usepackage{titletoc}
\theoremstyle{definition}
\newtheorem{Theorem}{Theorem}

\newtheorem{Proposition}{Proposition}
\newtheorem{condition}{Condition}

\newtheorem{Lemma}{Lemma}

\newtheorem{Remark}{Remark}
\newcommand{\RR}{\mathbb{R}}
\newcommand{\PP}{\mathbb{P}}
\newcommand{\QQ}{\mathbb{Q}}
\newcommand{\TT}{\mathbb{T}}
\newcommand{\E}{\mathbb{E}}
\newcommand{\GG}{\mathcal{G}}
\newcommand{\A}{\mathcal{A}}
\newcommand{\B}{\mathcal{B}}
\newcommand{\Dc}{\mathcal{D}}
\newcommand{\U}{\mathcal{U}}
\newcommand{\Ec}{\mathcal{E}}
\newcommand{\Oc}{\mathcal{O}}
\newcommand{\Nc}{\mathcal{N}}
\newcommand{\argmin}{\mathop{\arg\min}}
\newcommand{\argmax}{\mathop{\arg\max}}
\newcommand{\thetainit}{\widehat{\theta}}  
\newcommand{\X}[1]{X^{(#1)}}
\newcommand{\Y}[1]{Y^{(#1)}}
\newcommand{\f}[1]{f^{(#1)}}
\newcommand{\w}[1]{\omega^{(#1)}}
\newcommand{\hf}[1]{\widehat{f}^{(#1)}}

\newcommand{\hgammam}{{\widehat{\gamma}^{[m]}}}
\newcommand{\hw}[1]{\widehat{\omega}^{(#1)}}
\newcommand{\XQ}{X^\QQ}
\newcommand{\bmmu}{\U}
\newcommand{\bmmum}{\widehat{\bmmu}^{[m]}}
\newcommand{\err}{{\rm err}_n}
\newcommand{\CGDRO}{CG-DRO}

\title{Statistical Analysis of Conditional Group Distributionally Robust Optimization with Cross-Entropy Loss}
\author[1]{Zijian Guo}
\author[2]{Zhenyu Wang}
\author[2]{Yifan Hu}
\author[3]{Francis Bach}
\affil[1]{Center for Data Science, Zhejiang University, China}
\affil[2]{Department of Statistics, Rutgers University, USA}
\affil[3]{Inria, Ecole Normale Supérieure, PSL Research University, France}
\begin{document}
\maketitle
\begin{abstract}
In multi-source learning with discrete labels, distributional heterogeneity across domains poses a central challenge to developing predictive models that transfer reliably to unseen domains. We study multi-source unsupervised domain adaptation, where labeled data are available from multiple source domains and only unlabeled data are observed from the target domain. To address potential distribution shifts, we propose a novel {\bf C}onditional {\bf G}roup {\bf D}istributionally {\bf R}obust {\bf O}ptimization (CG-DRO) framework that learns a classifier by minimizing the worst-case cross-entropy loss over the convex combinations of the conditional outcome distributions from sources domains. We develop an efficient Mirror Prox algorithm for solving the minimax problem and employ a double machine learning procedure to estimate the risk function, ensuring that errors in nuisance estimation contribute only at higher-order rates.
 We establish fast statistical convergence rates for the empirical CG-DRO estimator by constructing two surrogate minimax optimization problems that serve as theoretical bridges. A distinguishing challenge for CG-DRO is the emergence of nonstandard asymptotics: the empirical CG-DRO estimator may fail to converge to a standard limiting distribution due to boundary effects and system instability. To address this, we introduce a perturbation-based inference procedure that enables uniformly valid inference, including confidence interval construction and hypothesis testing.
\end{abstract}

% \begin{keyword}[class=MSC]
% \kwd[Primary ]{62G09}
% \kwd[; secondary ]{62E10}
% \end{keyword}

% \begin{keyword}
% \kwd{Convergence Rate for Minimax Optimization}
% \kwd{Uncertainty Quantification for Minimax Optimization}
% \kwd{Generalization Bound}
% \kwd{Non-standard Inference}
% \kwd{Mirror Prox}
% \end{keyword}
% \end{frontmatter}
%%%%%%%%%%%%%%%%%%%%%%%%%%%%%%%%%%%%%%%%%%%%%%
%% Please use \tableofcontents for articles %%
%% with 50 pages and more                   %%
%%%%%%%%%%%%%%%%%%%%%%%%%%%%%%%%%%%%%%%%%%%%%%
%\tableofcontents

%%%%%%%%%%%%%%%%%%%%%%%%%%%%%%%%%%%%%%%%%%%%%%
%%%% Main text entry area:

\section{Introduction}

In modern applications, data are frequently collected from multiple sources, such as hospitals, geographic locations, time periods, or experimental batches. These multi-source datasets often exhibit a certain degree of distributional heterogeneity, with each source characterized by potentially different but related distributions. This heterogeneity poses challenges for prediction models trained under the empirical risk minimization (ERM) framework, which tends to perform well on average but may fail to generalize when there are distributional shifts between the source and target domains \cite{koh2021wilds, malinin2021shifts}. The central objective of multi-source data analysis is to learn a generalizable prediction model that performs reliably not only within the observed domains but also under potential distributional shifts in future or previously unseen domains.

\begin{figure}[t]
\centering
\includegraphics[width=0.85\linewidth]{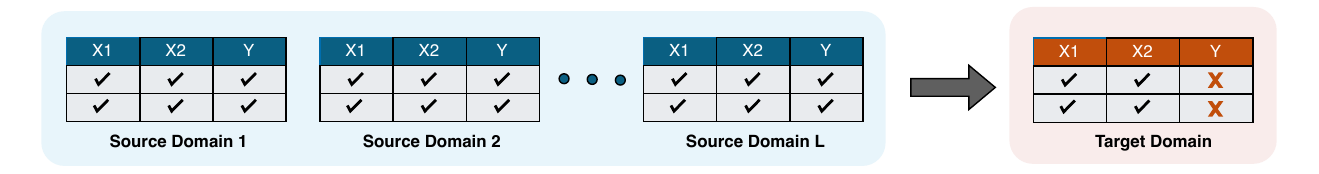}
\caption{Illustration of Multi-source Unsupervised Domain Adaptation. The source domains have labeled data, while the target domain only has unlabeled data.}
\label{fig:illus_MSDA}
\end{figure}

We focus on the multi-source unsupervised domain adaptation regime for multi-category classification: multiple source domains have labeled data, while the target domain only has unlabeled data \cite{duan2012domain, sun2015survey, perone2019unsupervised}. 
We illustrate this regime in Figure \ref{fig:illus_MSDA}, and the term ``unsupervised'' refers to the absence of target labels. Such a setting is prevalent in practice, as acquiring outcome labels in the target domain is often costly or time-consuming. Our primary goal is to learn information shared across multiple source domains and transfer such generalizable knowledge to the unlabeled target domain. To achieve this goal, we propose a {\bf C}onditional {\bf G}roup {\bf D}istributionally {\bf R}obust {\bf O}ptimization (\CGDRO)  model that optimizes the worst-case risk over the collection of all possible mixture conditional outcome distributions, {given covariate data from the target domain.}

Classification tasks play a central role in modern applications, with loss functions such as cross-entropy loss. Despite their widespread use, the theoretical understanding of \CGDRO\ under cross-entropy loss remains limited. In particular, fundamental questions concerning statistical convergence rates and uncertainty quantification remain largely unaddressed. Existing theoretical results have predominantly focused on regression problems involving the squared loss \cite{meinshausen2015maximin,guo2022statistical}, which fail to capture the distinctive challenges of classification under distributional shifts. This paper addresses these gaps by developing a unified theoretical framework for \CGDRO\ under cross-entropy loss, encompassing logistic loss as a special case. We solve the empirical \CGDRO\ problem via optimization algorithms, derive sharp statistical convergence rates for the \CGDRO\ estimator, defined as the optimal solution of the empirical \CGDRO\ problem. We further propose a method for constructing uniformly valid confidence intervals for the population-level \CGDRO\ model.

\subsection{Our Results and Contributions}

We study the multi-source unsupervised domain adaptation setting illustrated in Figure \ref{fig:illus_MSDA} and propose a novel \CGDRO\ framework to extract a generalizable predictive structure shared across sources. Since we do not have access to samples of the conditional outcome distribution from the target domain, we consider an uncertainty set as mixtures of conditional outcome distributions from the source domains. Thus, our goal is to minimize the worst-case cross-entropy risk with respect to the uncertainty set of the conditional outcome distributions. 

%\Zhenyu{We study the multi-source unsupervised domain adaptation setting illustrated in Figure \ref{fig:illus_MSDA} and propose a novel \CGDRO\ framework to extract predictive structure shared across sources that generalizes well to the target domain. Note that we have no access to the target labels, meaning the target's conditional outcome distribution is unobserved. Thus, we define an uncertainty set for the plausible target conditional outcome distribution, consisting of convex mixtures of the source ones, and minimize the worst-case cross-entropy risk over this set.}
Unlike the classical Group DRO~\cite{hashimoto2018fairness, sagawa2019distributionally, lahoti2020fairness}, which considers uncertainty classes over joint distributions, our formulation focuses on conditional outcome distributions, motivating the term \textit{Conditional} Group DRO, by leveraging the feature information from the target domain. 
Our model admits a minimax formulation and encourages identifying {stable} features that are consistently predictive across all source domains, thereby reducing sensitivity to domain-specific spurious correlations. %

%The \CGDRO\ involves a minimax optimization problem, where the maximum is taken to calculate the worst-case loss over the uncertainty set, and the minimization is to search for a model minimizing the worst-case loss.  
We adopt the Mirror Prox algorithm~\cite[e.g.]{juditsky2011solving,bubeck2015convex} to solve the minimax optimization problem for \CGDRO, where we leverage optimistic gradient updates to efficiently handle the minimax optimization problem. To address the covariate shift between the source and target domains, we further develop {\bf D}ouble {\bf M}achine {\bf L}earning (DML) or double robustness estimator \cite[e.g.]{chernozhukov2018double, vansteelandt2022assumption, wang2024debiased} for a key component of the empirical risk. This estimator is integrated into the Mirror Prox algorithm to ensure that the error arising from the machine learning estimation of nuisance models enters only at a higher-order rate. 
%\Zhenyu{Currently, I feel like the role of DML estimators is not clarified. For me, I think DML are used to construct the empirical \CGDRO\ problem (from the population $\theta^*$ to empirical $\widehat{\theta}$), and Mirror Prox is applied to solve the problem (from the empirical $\widehat{\theta}$ to iterated $\widehat{\theta}_T$). Plz see my following rewriting of this paragraph.}

%\Zhenyu{In the finite-sample setting, we construct the empirical \CGDRO\ problem. To accurately approximate the population objective, instead of using simple plug-in estimators, we develop Double Machine Learning (DML) estimators \cite[e.g.]{chernozhukov2018double, vansteelandt2022assumption, wang2024debiased} for the key component in the objective. By construction, errors from nuisance model estimation using machine learning contribute only at higher-order rates, ensuring that the empirical \CGDRO\ closely approximate its the population counterpart. We then adopt the Mirror Prox algorithm~\cite[e.g.]{juditsky2011solving, bubeck2015convex}, enhanced with optimistic gradient updates, to solve the minimax problem globally and achieve faster optimization convergence.}

%\Zhenyu{If the paragraph reads good, we can remove the first sentence related to Mirror Prox, and probably need to modify the abstract as well.}

While it is well established that the Mirror Prox algorithm converges to the optimal solution of the empirical \CGDRO\ problem, there is a significant lack of theoretical understanding for this optimal solution's statistical behavior, 
particularly its statistical convergence rate and limiting distribution. We close this gap by developing a unified theoretical framework that characterizes the statistical properties of the empirical \CGDRO\ estimator, defined as the solution to the empirical problem. In particular, we derive sharp finite-sample convergence rates for this estimator relative to its population counterpart. Standard M-estimation theory does not directly yield fast parametric rates in this setting, as the maximization taken after the expectation introduces non-smoothness into the primal objective function. To overcome this challenge, we construct two carefully designed surrogate minimax problems that approximate the empirical and population \CGDRO\ problems. These surrogates serve as analytical intermediaries, enabling us to derive fast statistical convergence rates for the empirical estimator. As a key technical contribution, we characterize the discrepancy between the solutions of the surrogate and original minimax problems in Theorem \ref{thm: quad convergence}. These theoretical tools may be of independent interest beyond the current setting.

Statistical inference for the CG-DRO model is an important yet challenging task. Our proposed CG-DRO model is interpreted as a generalizable model capturing stable predictive patterns across diverse environments, making it a robust and meaningful choice for deployment in new target domains. This makes testing whether a certain element of the CG-DRO model is significant an important task, i.e., whether the corresponding coordinate is zero. However, as illustrated in Figure~\ref{fig: Inference Challenge}, the empirical \CGDRO\ estimator typically does not admit an asymptotically normal limiting distribution, complicating standard inferential approaches. This nonstandard behavior arises in at least two common scenarios: (i) when the source domains exhibit highly similar conditional outcome distributions, leading to instability in solving the empirical maximization problem; and (ii) when the shifts between conditional outcome distributions are sparse but pronounced, resulting in one domain receiving a dominating weight and inducing boundary effects that complicate inference. To address these challenges, we propose a novel perturbation-based inferential procedure for quantifying uncertainty in the empirical \CGDRO\ estimator. We show that the resulting confidence intervals achieve desired coverage and parametric length, even in settings where the estimator fails to exhibit an asymptotically normal distribution. This provides a principled and flexible framework for inference in minimax optimization problems.

To summarize, the main contributions of this paper are threefold:

\begin{enumerate}
\item We propose the \CGDRO\ framework that extracts stable predictive structure across source domains under cross-entropy loss, and solve the resulting minimax problem via a Mirror Prox algorithm enhanced with a DML estimator of the risk function. %\Zhenyu{We propose the \CGDRO\ framework for extracting stable predictive structure from multiple sources that generalizes to the target domain under the cross-entropy loss. In the finite-sample setting, we construct its empirical version using Double Machine Learning (DML) techniques, and solve the resulting minimax problem globally via the Mirror Prox algorithm.}
\item We establish a fast statistical convergence rate for the empirical \CGDRO\ estimator by constructing surrogate minimax problems that serve as analytical bridges.
\item We develop a perturbation-based inference method that yields valid confidence intervals with parametric length for the \CGDRO\ model, even when its empirical estimator does not admit a standard normal limiting distribution. 
\end{enumerate}

\subsection{Additional Related Literature} 
\label{subsec: literature}
Beyond the works discussed above, we provide further relevant literature, organized by topic. Further literature related to minimax optimization, algorithms, and generalization is provided in the Supplement Section \ref{sec: add literature}.

\vspace{1mm}
\noindent\textbf{Maximin Effect with Least-Square Type Loss.}
This paper is closely related to research on the maximin effect in the no-covariate-shift setting \cite{meinshausen2015maximin, buhlmann2015magging, rothenhausler2016confidence} and the covariate-shift setting \cite{guo2022statistical, wang2023distributionally}, 
%\Yifan{What does it mean by \CGDRO\ framework under covariate shift?} \Zijian{we write this to contradict the maximin effect without covariate shift.} \Yifan{Do they also study CG-DRO or a similar but different model?} 
and minimax optimization with regret loss \cite{mo2024minimax, zhang2024minimax}. These works define minimax optimization by minimizing the worst-case squared losses {or its variants}, and a key aspect of their analysis is that the optimal solution of these maximin problems attains a closed-form expressions.
In contrast, our paper focuses on classification tasks with cross-entropy loss, where such closed-form expressions are unavailable. This means existing analytical and inferential methods do not directly apply. To overcome these challenges, we construct surrogate minimax optimization problems to facilitate theoretical analysis and build perturbation-based confidence intervals.%

\vspace{1mm}
\noindent\textbf{Group DRO.}
Group Distributionally Robust Optimization approaches build generalizable prediction models using source domain data without target domain information \cite{sagawa2019distributionally, lahoti2020fairness}. Recently, in the context of online data, \cite{zhang2024optimal} proposed a novel algorithm that achieves optimal sample complexity for solving Group DRO. Our work, however, focuses on unsupervised domain adaptation, leveraging available target covariates, as illustrated in Figure \ref{fig:illus_MSDA}. This extra information allows us to design a {conditional} distributionally robust model specifically tailored for the unlabeled target domain. A more detailed comparison is in Section \ref{sec: group dro}.

\vspace{1mm}
\noindent\textbf{Similarity-based Multi-source Learning.} Apart from Group DRO, a broader body of work in multi-source learning aims to identify generalizable patterns from heterogeneous data sources, by assuming certain types of shared structure across domains. Notable examples include transfer learning \cite{li2022transfer,tian2023transfer,xiong2023distributionally}, meta-analysis \cite{sun2024optimal,guo2025robust} and multi-task learning \cite{duan2023adaptive}. 
 In contrast, our proposed CG-DRO framework does not require such similarity conditions and offers a distinct approach to learning generalizable models.

\vspace{1mm}
\noindent\textbf{Nonstandard Inference.} Nonstandard (or nonregular) inference frequently arises in modern data science, where estimators do not admit standard limiting distributions (e.g., asymptotically normal), and hence classical inference tools based on computing the normal distribution's quantiles fail to apply \cite{wasserman2020universal,xie2024irregular,kuchibhotla2021hulc, guo2022statistical, guo2025robust}. We demonstrate in Section~\ref{sec: challenge} that the empirical \CGDRO\ estimator exhibits nonstandard behavior due to boundary effects and/or system singularity. While classical subsampling and bootstrap methods fail in such nonstandard settings \cite{andrews1999estimation}, we propose a novel perturbation-based inference procedure that guarantees valid coverage even when the estimator deviates from asymptotic normality.

\vspace{1mm}
\noindent\textbf{Double Machine Learning (DML).} DML provides a powerful framework for incorporating nonparametric estimation of nuisance components while still allowing valid inference for parameters of interest; see \cite{chernozhukov2018double, wang2024debiased, vansteelandt2022assumption} and references therein. In this work, we extend the DML framework beyond its traditional focus on low-dimensional target parameters. Specifically, we use DML to construct an unbiased estimator of the risk function in settings with covariate shift. This shift in perspective, from estimating a finite-dimensional parameter to debiasing a full risk function, enables us to analyze a novel empirical minimax optimization problem based on this debiased risk. Our approach highlights how DML ideas can be leveraged not only for inference but also as a tool for robust optimization under distributional shift.

\vspace{1mm}

\subsection{Notations}
For a positive integer $m$, we define $[m] = \{1,2, ...,m\}$.  For a real value $x$, we denote by $x_m$ the vector with all entries equal to $x$, and let $\mathbf{I}_m$ be the
$m\times m$ identity matrix. For a matrix $A\in \RR^{m\times n}$, we denote its singular values by $\sigma_1(A)\geq...\geq \sigma_p(A)\geq 0$ with $p=\min\{m,n\}$, its spectral norm by $\|A\|_2$, and its Frobenius norm by $\|A\|_F$. For a symmetric matrix $A$, we use $\lambda_{\min}(A)$ and $\lambda_{\max}(A)$ to denote its smallest and largest eigenvalues, respectively. We write $A \preceq B$ (or $B\succeq A$) if $B - A$ is positive semidefinite, for two symmetric matrices $A,B$.
For a vector $x\in \RR^d$, its $\ell_p$-norm ($p\geq 1$) is given by $\|x\|_p = (\sum_{j=1}^d |x_j|^p)^{1/p}$. 
For a sequence $x_n$, we write $x_n\xlongrightarrow{p} x$ and $x_n\xlongrightarrow{d}x$ to denote convergence in probability and in distribution, respectively. For positive sequences $a(n)$ and $b(n)$, we write $a(n)\lesssim b(n)$ or $a(n) = \mathcal{O}(b(n))$ if there exists a constant $C > 0$ such that $a(n) \leq C b(n)$ for all 
$n\geq 1$, and write $a(n) \asymp b(n)$ if both $a(n)\lesssim b(n)$ and $b(n)\lesssim a(n)$. We use $a(n)\ll b(n)$, $a(n)=o(b(n))$ or $b(n)\gg a(n)$ when $\limsup_{n\rightarrow\infty} (a(n)/b(n)) = 0$.
We use $c_1,c_2,c_3,C$ to denote generic positive constants that may vary from place to place.

\section{Conditional Group Distributionally Robust Optimization}
\label{sec: CGDRO}
In this section, we introduce the data structure of  the multi-source unsupervised domain adaptation and propose the \CGDRO\ model to learn the generalizable information shared across domains. For the $l$-th source dataset with $1\leq l\leq L$, we have access to independent and identically distributed (i.i.d.) observations $\{X_i^{(l)}, Y_i^{(l)}\}_{1\leq i\leq n_l}$ sampled from the joint distribution $\PP^{(l)} =(\mathbb{P}_{X}^{(l)}, \mathbb{P}_{Y|X}^{(l)})$, where $X_i^{(l)} \in \RR^d$ and $\Y{l}_i\in \RR$ denote the covariates and the outcome, respectively. In the target domain, we observe covariates $\{X^{\mathbb{Q}}_i\}_{1\leq i\leq N}$ that are i.i.d. generated from $\mathbb{Q}_X$, while the corresponding outcome variables remain unobserved. We focus on the regime with $N\gg \max_{1\leq l\leq L}n_l$, that is, the target domain has a much larger amount of unlabeled data compared to the amount of labeled data in the source domains. Such a regime is common in the unsupervised domain adaptation as the unlabeled data are easier to access in practice. We capture distributional differences across domains by allowing both covariate shift and conditional distributional shift between the source and target domains.
Specifically, the target covariate distribution $\mathbb{Q}_X$ may differ from any of $\{\mathbb{P}_X^{(l)}\}_{1 \leq l \leq L}$, and similarly, the conditional distribution $\mathbb{Q}_{Y|X}$ may differ from any of $\{\mathbb{P}_{Y|X}^{(l)}\}_{1 \leq l \leq L}$.
We refer to the special case where $\mathbb{Q}_X = \mathbb{P}_X^{(l)}$ for all $1 \leq l \leq L$ as the {no covariate shift} setting. 

To extract transferable knowledge shared across source domains and adapt it to the target domain, we propose in the following the \CGDRO\ model that guarantees robust performance over a family of plausible target distributions. While \(\mathbb{Q}_X\) is identifiable through observed target covariates, the conditional distribution \(\mathbb{Q}_{Y|X}\) remains non-identifiable without additional structural assumptions. Instead of making identifiability assumptions, we construct an uncertainty class encapsulating potential target distributions:
\begin{equation}
\mathcal{C} = \left\{ (\mathbb{Q}_X, \mathbb{T}_{Y|X}) : \mathbb{T}_{Y|X} = \sum_{l=1}^L \gamma_l \mathbb{P}^{(l)}_{Y|X}, \; \text{with}\;\gamma \in \Delta^L \right\},
\label{eq:uncertainty_class}
\end{equation}
where \(\Delta^L\) denotes the \((L-1)\)-dimensional simplex. This class \(\mathcal{C}\) contains the true target distribution \((\mathbb{Q}_X, \mathbb{Q}_{Y|X})\) if \(\mathbb{Q}_{Y|X}\) admits a mixture representation of the source conditional distributions. %\Yifan{I remember that Fanny Young has a paper about group DRO is optimal if no further information is not provided. Let me ask her student which paper.} 
Given the distributional uncertainty, we define the worst-case risk:
\begin{equation}
\max_{\mathbb{T} \in \mathcal{C}}\  \mathbb{E}_{(X, Y)\sim \mathbb{T}}  \ \ell(X, Y, \theta),
\label{eq: worst-case loss}
\end{equation}
{where $\ell(X,Y,\theta)$ is the loss of a prediction model parameterized by $\theta$, and the expectation is taken under the joint distribution $\mathbb{T}$.} We define the \CGDRO\ model $\theta^*$ as the solution to the following minimax optimization:
\begin{equation}
\theta^* = \argmin_{\theta } \ \left[\max_{\mathbb{T} \in \mathcal{C}} \ \mathbb{E}_{(X, Y)\sim \mathbb{T}} \ \ell(X, Y, \theta)\right].
\label{eq:general_minimax}
\end{equation}
When $\ell(X, Y, \theta)$ is strictly convex in $\theta$, the worst-case risk $\max_{\mathbb{T} \in \mathcal{C}} \ \mathbb{E}_{(X, Y)\sim \mathbb{T}} \ \ell(X, Y, \theta)$ inherits its  strict convexity, ensuring a unique solution of $\theta^*$ in \eqref{eq:general_minimax}. 
Importantly, we interpret $\theta^*$ as a transferable prediction model since it checks through all possible target distributions belonging to $\mathcal{C}$ and optimizes the worst-case risk function. 

Leveraging the definition of the uncertainty set about the mixture distribution used in \eqref{eq:uncertainty_class}, we reformulate \eqref{eq:general_minimax} as:
\begin{equation}
\theta^* = \argmin_{\theta } \ \left[\max_{\gamma \in \Delta^L} \sum_{l=1}^L \gamma_l \cdot \mathbb{E}_{X \sim \mathbb{Q}_X} \mathbb{E}_{Y\sim \mathbb{P}^{(l)}_{Y|X}} \ell(X, Y, \theta)\right].
\label{eq:h_loss_opt}
\end{equation}

\subsection{Comparison to Group DRO and ERM} 
\label{sec: group dro}
We now review the related Group DRO works \cite{sagawa2019distributionally, lahoti2020fairness} and further compare our \CGDRO\ model to Group DRO and ERM. Our proposed model is different from Group DRO, which applies to settings with data only from multiple sources, but not any target data. Specifically, Group DRO defines the uncertainty class comprising all mixtures of the joint distributions $(X,Y)$ from the sources,
\begin{equation*}
    \mathcal{C}_{\rm GDRO}=\left\{\TT: \TT=\sum_{l=1}^{L} \gamma_l\cdot\PP^{(l)},\; \gamma\in \Delta^{L}\right\}.
\end{equation*}
With the above uncertainty class,
Group DRO solves the following robust prediction model:
\begin{equation}
\begin{aligned}
\theta^*_{\rm GDRO}&=\argmin_{\theta}\  \max_{\TT \in \mathcal{C}_{\rm GDRO}}\  \E_{(X, Y)\sim \TT} \ell(X,Y,\theta). %
\end{aligned}
\label{eq: general minimax DRO}
\end{equation}
The uncertainty class $\mathcal{C}_{\rm GDRO}$ considers all possible mixtures of the full joint distributions across sources, whereas our uncertainty class $\mathcal{C}$ in \eqref{eq:uncertainty_class} focuses on mixtures of the conditional outcome distributions, holding the target covariate distribution $\QQ_X$ fixed.
When the covariate distributions align, i.e., $\PP_X^{(l)} = \QQ_X$ for all $1 \leq l \leq L$, the two uncertainty classes coincide. In this case, the distributionally robust prediction model in \eqref{eq: general minimax DRO} reduces to our formulation in \eqref{eq:general_minimax}.
However, under general covariate shift, the two models typically yield different solutions.

We utilize a numerical example to illustrate the distinctions of our proposed \CGDRO\ compared to both the Group DRO method in \eqref{eq: general minimax DRO} and the ERM approach that fits a model on the pooled source data. All methods are evaluated based on their worst-case risk \eqref{eq: worst-case loss}. Since the worst-case risk provides a summarized measure of performance over a class of plausible target conditional outcome distributions, 
it is well-suited for the unsupervised domain adaptation setting, where the target domain provides only covariate samples.
We consider a numerical example with $L = 2$ source domains and one target domain under covariate shift, i.e. $\QQ_X \neq \PP^{(l)}_X$ for $l=1,2$; see Supplement Section \ref{appendix: setups} for details. We vary the source mixture by adjusting the proportion of samples from the first source across $\{0.1, 0.2, \dots, 0.9\}$. %\Zhenyu{For completeness, additional simulations under the no covariate shift regime are provided in Supplement Section \ref{appendix: groupdro}.}

\begin{figure}[ht!]
    \centering
    \includegraphics[width=0.65\linewidth]{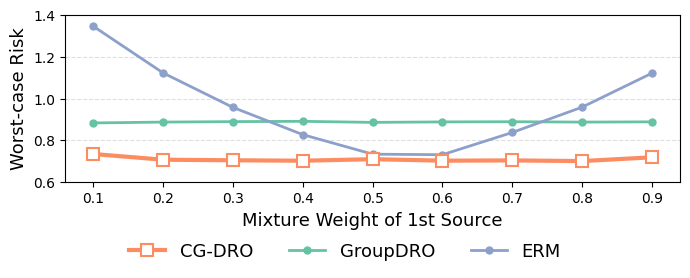}
    \caption{Comparison of the worst-case risk defined in \eqref{eq: worst-case loss} for CG-DRO, Group DRO, and ERM.} %
    \label{fig:worst-case}
\end{figure}
As shown in Figure~\ref{fig:worst-case}, the proposed \CGDRO\ method consistently achieves lower worst-case risk than both Group DRO and ERM, demonstrating its robustness to distributional shifts between the source and target domains. While the ERM approach is sensitive to the source mixture, both Group DRO and our method remain stable across different proportions. Notably,
\CGDRO\ outperforms Group DRO, as it is specifically designed to ensure robustness on the target domain by leveraging the unlabeled target data during training. 

% As the benchmarks, Figure~\ref{fig:worst-case} also includes the non-reducible losses \Zijian{again, I have questions about non-reducible losses here.}\Zhenyu{discuss.} for the two source domains, each computed under its own distribution using the corresponding true conditional distributions \(\PP^{(l)}_{Y| X}\) for $l = 1, 2$; see the exact formula in Section \ref{appendix: setups} of the supplements. %

\subsection{CG-DRO with cross-entropy loss}
\label{sec:CE}
In this subsection, we focus on the \CGDRO\ model with  commonly used cross-entropy loss function for categorical outcome variables. We consider the categorical outcome variable $Y$ taking values from $\{0,1,\dots,K\}$ with $K\geq 1$, with the binary classification setting corresponding to the special case with $K=1$. 
We define the cross-entropy loss function as, 
\begin{equation*}
\ell(X,Y,\theta)=-\sum_{c=0}^{K} {\bf 1}(Y=c) \log \left[\frac{\exp(\theta_c^{\intercal}X)}{1+\sum_{k=1}^{K}\exp(\theta_k^{\intercal}X)}\right],
\end{equation*}
with each class-specific parameter vector $\theta_c\in \RR^d$ for $c\in \{0,1,...,K\}$. 
{For identifiability, we set class $c = 0$ as the reference category and fix $\theta_0$ as a zero vector.} We then stack the remaining parameters into a single vector $\theta = (\theta_1^{\intercal}, \theta_2^{\intercal}, \dots, \theta_K^{\intercal})^{\intercal} \in \mathbb{R}^{dK}$, which represents the parameters of the predictive model.
In the context of cross-entropy loss, we further simplify the \CGDRO\ model in \eqref{eq:h_loss_opt} as follows, 
\begin{equation}
\theta^*= \argmin_{\theta\in \RR^{dK}} \ \max_{\gamma\in \Delta^L}\  \phi(\theta,\gamma) \quad \text{with}\quad \phi(\theta,\gamma) \coloneqq
\sum_{l=1}^{L} \gamma_{l} \cdot \theta^{\intercal}\mu^{(l)}+S(\theta),
\label{eq: minimax model}
\end{equation}
where $\mu^{(l)}=\left([\mu^{(l)}_1]^{\intercal},[\mu^{(l)}_2]^{\intercal},\cdots,[\mu^{(l)}_K]^{\intercal}\right)^{\intercal}\in \mathbb{R}^{dK}$ with $\{\mu^{(l)}_{c}\}_{1\leq c\leq K}$ defined as, 
\begin{equation}
\mu^{(l)}_c=-\mathbb{E}_{X \sim \mathbb{Q}_X} \mathbb{E}_{Y\sim \mathbb{P}^{(l)}_{Y|X}} \left[{\bf 1}(Y=c) \cdot X\right], \;\; \;\; S(\theta)=\E_{X\sim \QQ_{X}} \log\left[{1+\sum_{c=1}^{K}\exp(\theta_c^{\intercal}X)}\right].
\label{eq:mu def}
\end{equation} 
We establish in Lemma \ref{lem: convexity of H} of supplements that  $\phi(\theta,\gamma)$ is strictly convex towards $\theta$ for any~$\gamma$, implying that $\theta^*$ in \eqref{eq: minimax model} is uniquely defined.
In the above definition, $\mu^{(l)}_c$ measures the association between the one-hot encoding ${\bf 1}(Y=c)$ and $X\in \RR^{d}$. The vector $\mu^{(l)}$ is constructed by stacking the class-specific vectors $\{\mu^{(l)}_c\}_{1\leq c\leq K}$ and does not depend on $\theta$. We separate the log sum exponential function $S(\theta)$ as it does not depend on the source index $l$.

In Supplement Section \ref{appedix: prior}, we also discuss how to integrate prior information about sourcing mixture into constructing \CGDRO\ model.

%It is essential to note that our analysis does not rely on a correctly specified conditional mean model for each source distribution. \Zijian{We do not assume the data generating process follows the multi-logistic distribution. Instead, our target is to find the optimal solution $\theta^*$ of the optimization problem in (6).} \Yifan{Maybe I am in lack of knowledge. What does this mean?} \Zijian{this means that we do not assume that the true codntional mean model in each source is following multi-logistic model. We use machine learning to fit and then correct it. That is what I meant to say.} Instead, we define the population parameter of interest $\theta^*$ as the solution to the optimization problem in \eqref{eq: minimax model}. 

\section{Mirror Prox Algorithm with Double Machine Learning} %
\label{sec: refined analysis}
%\Zhenyu{plz double check this section, while we decide to move all mirror prox algorithms to appendix.}

In this section, we formulate the empirical \CGDRO\ problem utilizing the observed data and solve the empirical problem.
We first substitute the population-level quantities $S(\theta)$ and $\{\mu^{(l)}\}_{1\leq l\leq L}$ in \eqref{eq: minimax model} with their data-dependent estimators $\widehat{S}(\theta)$ and $\{\widehat{\mu}^{(l)}\}_{1\leq l\leq L}$, respectively. We then construct the empirical \CGDRO\ {estimator} $\thetainit$ as the solution to the following optimization problem:
\begin{equation}
\widehat{\theta}=\argmin_{\theta\in \RR^{dK}} \ \max_{\gamma\in \Delta^{L}} \ \widehat{\phi}(\theta,\gamma) \quad \text{with}\quad \widehat{\phi}(\theta,\gamma)=\sum_{l=1}^{L}\gamma_l\cdot \theta^{\intercal}\widehat{\mu}^{(l)}+ \widehat{S}(\theta).
\label{eq: sample optimizers}
\end{equation}
Here, $\widehat{S}(\theta)$ is computed over the unlabeled target data $\{\XQ_i\}_{1\leq i\leq N}$ as follows:
\begin{equation}
    \widehat{S}(\theta)=\frac{1}{N}\sum_{i=1}^{N} \log\left({1+\sum_{k=1}^{K}\exp(\theta_k^{\intercal}\XQ_i)}\right).
    \label{eq: hatS def}
\end{equation} 
In Sections \ref{sec: Mirror Prox} and \ref{sec: double robustness}, we detail the construction of $\{\widehat{\mu}^{(l)}\}_{1\leq l\leq L}$, where its form differs depending on whether covariate shift is present. 

Given the expressions of $\widehat{\mu}^{(l)}$ and $\widehat{S}(\theta)$,  we evaluate the objective $\widehat{\phi}(\theta,\gamma)$ defined in \eqref{eq: sample optimizers}, which is strictly convex in $\theta$ and linear (and thus concave) in $\gamma$. We then solve the empirical \CGDRO\ problem to obtain $\widehat{\theta}$ using a variant of the gradient descent-ascent method~\cite{lin2020gradient} {called the Mirror Prox algorithm. It performs gradient descent on $\theta$ and mirror ascent~\cite{nemirovski2009robust} on $\gamma$ due to the simplex constraint. In addition, Mirror Prox first performs such a descent-ascent update to find an intermediate point and then performs another descent-ascent update from the original point using gradients computed at the intermediate point. The Mirror Prox algorithm  effectively reduces oscillations near saddle points and achieves faster optimization convergence rates \cite{nemirovski2004prox, bubeck2015convex, mokhtari2020unified}. Details of the algorithm and its convergence guarantees are provided in Supplement Section~\ref{subsec: MP}. 

In the remaining of this section, we consider the regime without and with covariate shift and present the corresponding estimator $\widehat{\mu}^{(l)}$ of $\mu^{(l)}$. 
%\Zhenyu{I suggest removing the second sentence. It replicates the last paragraph.}
%\Yifan{Add: A key component of Mirror Prox algorithm is access to unbiased gradient estimators of the form [add equations], which requires us to estimate... } \Zijian{Zhenyu, can you help fix this?}

%\Zijian{do we want to include more details here as the section title is called "mirror prox with..."} \Yifan{See what I added.}

% \Zhenyu{In Section XXX of the supplement,} we present the Mirror Prox algorithm for solving the empirical problem in \eqref{eq: sample optimizers} along with the discussions regarding its statistical convergence. We highlight that the literature has shown that such an algorithm effectively solves a minimax optimization problem by reducing oscillations near saddle points and achieves faster optimization convergence rates \cite{nemirovski2004prox, bubeck2015convex, mokhtari2020unified}.

% This section is organized as follows. In Section \ref{sec: Mirror Prox}, we derive explicit expressions for $\{\widehat{\mu}^{(l)}\}_{1\leq l\leq L}$ for the setting without covariate shift in \eqref{eq: concat mu def}. Section \ref{sec: double robustness} involves the general covariate shift regime, for which we construct DML estimators $\{\widehat{\mu}^{(l)}\}_{1\leq l\leq L}$ in \eqref{eq: DR component}. 
% and introduce the optimistic-gradient Mirror Prox algorithm to solve the empirical problem \eqref{eq: sample optimizers}. 

% and develop a corresponding Mirror Prox algorithm with key components estimated by DML.

\subsection{No Covariate Shift Regime}
\label{sec: Mirror Prox}

In the no covariate shift setting, with $\PP^{(l)}_X = \QQ_X$ for each source $1\leq l\leq L$, we have access to $\{X_i^{(l)}, Y_i^{(l)}\}_{1\leq i\leq n_l} \stackrel{i.i.d.}{\sim} \left(\PP^{(l)}_X, \PP^{(l)}_{Y|X}\right)$, whose distribution is the same as $\left(\QQ_{X}, \PP^{(l)}_{Y|X}\right)$. By the definitions of $\mu^{(l)}_{c}$ in \eqref{eq:mu def}, we construct the following unbiased estimator of $\mu^{(l)}_{c}$, %\Yifan{where is unbiased estimator for $\hat S(\theta)$? What is the purpose to do this? Mention we need to know the gradient with respective to $\theta$ and $\gamma$: To implement Mirror Prox, unbiased estimators of ....} \Zijian{I have answered the first question. For the remaining question, have we agreed that we shall define the empircal optimization as in (8) and then we implement mirror prox? I think we are clear here.}
\begin{equation}
\widehat{\mu}^{(l)}_c=-\frac{1}{n_l} \sum_{i=1}^{n_l} {\bf 1}(Y^{(l)}_i=c) \cdot X^{(l)}_i \quad \text{for}\quad 1\leq c\leq K.
\label{eq: no shift est}
\end{equation}
% Given the expressions of $\widehat{\mu}^{(l)}$ in \eqref{eq: no shift est} and $\widehat{S}(\theta)$ in \eqref{eq: hatS def},  we evaluate the objective $\widehat{\phi}(\theta,\gamma)$ defined in \eqref{eq: sample optimizers}, \Zhenyu{and solve the empirical \CGDRO\ $\widehat{\theta}$ using the Mirror Prox algorithm presented in Section XXX of the supplement.}
% which is strictly convex in $\theta$ and linear (and thus concave) in $\gamma$. 

\subsection{Covariate-shift Regime: DML Estimator of $\mu^{(l)}$}
\label{sec: double robustness}
When there is covariate shift between the target and source domains, unlike Section \ref{sec: Mirror Prox}, the paired data $\{X_i^{(l)}, Y_i^{(l)}\}_{1\leq i\leq n_l} \sim \left(\PP^{(l)}_{X}, \PP^{(l)}_{Y|X}\right)$ are biased samples of the distribution $\left(\QQ_{X}, \PP^{(l)}_{Y|X}\right)$ due to the covariate shift. To address this, we propose a DML estimator $\widehat{\mu}^{(l)}$ of $\mu^{(l)}$ even when we do not directly have access to data sampled from $\left(\QQ_{X}, \PP^{(l)}_{Y|X}\right).$ 

The main idea of DML is to estimate the nuisance models used to define $\mu^{(l)}$ via flexible machine learning methods, and then correct the bias by simply plugging in these estimated nuisance models. The key advantage of DML is that the nuisance estimation errors enter only at a higher-order rate, ensuring that $\widehat{\mu}^{(l)}$ remains an accurate estimator of $\mu^{(l)}$, even under covariate shift. Consequently, the empirical \CGDRO\ estimator $\widehat{\theta}$ also remains close to its population counterpart $\theta^*$, where $\widehat{\theta}$ and $\theta^*$ are defined through $\widehat{\mu}^{(l)}$ and $\mu^{(l)}$ in \eqref{eq: sample optimizers} and \eqref{eq: minimax model}, respectively. Since this idea originates from the double robustness or DML literature \cite{robins1994estimation,bang2005doubly,chernozhukov2018double}, we refer to $\widehat{\mu}^{(l)}$ as a DML estimator.
%of $\mu^{(l)}$ using the cross-fitting technique . 
% \Zhenyu{We then solve the empirical \CGDRO\ model $\widehat{\theta}$ \eqref{eq: sample optimizers} with this refined estimator $\widehat{\mu}^{(l)}$.}
% , by applying the \CGDRO\
% implement the Mirror Prox algorithm in Algorithm~\ref{algo: mp} with this refined estimator $\widehat{\mu}^{(l)}$. 

%The construction of $\widehat{\mu}^{(l)}$ involves estimating two nuisance components, which will be specified in the following discussion. 

The construction of the DML estimator leverages the following equivalent definition of $\mu^{(l)}_c$ in \eqref{eq:mu def}: for each category $c=1,...,K$, 
\begin{equation}
\mu^{(l)}_c=-\E_{X_i\sim \QQ_{X}} \left[f^{(l)}_c(X_i)\cdot X_i\right]\quad \text{with}\quad f^{(l)}_c(X_i)\coloneqq \PP^{(l)}(Y_i=c\mid X_i). 
    \label{eq: concat mu def}
\end{equation}
A naive approach is to estimate $f^{(l)}_c(\cdot)$ with a Machine Learning (ML) approach, and then plug in this estimated $\widehat{f}^{(l)}_c(\cdot)$ to construct an initial estimator. However, such a plug-in estimator suffers from a dominating bias component resulting from the error $\widehat{f}^{(l)}_c(\cdot)-f^{(l)}_c(\cdot)$. 
To mitigate this issue, we estimate and correct the leading bias component. This bias correction step leads to a refined estimator of ${\mu}^{(l)}$ by reducing the estimation error $\widehat{f}^{(l)}_c(\cdot)-f^{(l)}_c(\cdot)$ to a higher-order term. We now provide more details about our two-step procedure. %

\vspace{2mm}
\noindent{\bf Step 1: Plug-in ML estimator.} We start with estimating 
$\mu^{(l)}_c$ by plugging in a ML estimator of $f^{(l)}_c$ into \eqref{eq: concat mu def}. Particularly, for each source domain $1\leq l\leq L$, we use the labeled source data $\{\X{l}_i, \Y{l}_i\}_{1\leq i\leq n_l}$ to fit the conditional probability model $\f{l}$ using a preferred ML classification method (e.g., multinomial logistic regression, random forest classifier, neural networks). The resulting fitted model is denoted by $\hf{l}(x) = \left(\hf{l}_1(x),..., \hf{l}_K(x)\right)$, where each $\hf{l}_c(x)$ is the predicted probability that $x$ is classified in the category $c\in \{1,...,K\}$. The plug-in estimator of $\mu_c^{(l)}$ is $-\frac{1}{N}\sum_{j=1}^N \widehat{f}^{(l)}_c(X^\QQ_j)X^\QQ_j$, using the unlabeled target data $\{\XQ_j\}_{1\leq j\leq N}$. We decompose the estimation error $-\frac{1}{N}\sum_{j=1}^N\widehat{f}^{(l)}_c(X^\QQ_j)  X^\QQ_j - \mu^{(l)}_c(x) $ as

\begin{equation}
    \begin{aligned}
    \frac{1}{N}\sum_{j=1}^N \left[-\widehat{f}^{(l)}_c(X^\QQ_j)  X^\QQ_j\right] - \E_{\QQ_X}\left[-\widehat{f}^{(l)}_c(X) X\right] + \E_{\QQ_X}\left[\left(f^{(l)}_c(X) -\widehat{f}^{(l)}_c(X)\right) X\right].
\end{aligned}
\label{eq: error decompose}
\end{equation}
In the above decomposition, the term $\frac{1}{N}\sum_{j=1}^N \left[-\widehat{f}^{(l)}_c(X^\QQ_j) \cdot X^\QQ_j\right] - \E_{\QQ_X}\left[-\widehat{f}^{(l)}_c(X) \cdot X\right]$ measures the discrepancy between the finite-sample average and the corresponding population mean. Its $\ell_2$ norm admits the convergence rate of $\sqrt{d/N}$ and cannot be further improved. However, the term $\E_{\QQ_X}\left[\left(f^{(l)}_c(X) -\widehat{f}^{(l)}_c(X)\right) \cdot X\right]$ arises from the estimation error in the ML prediction model. We refer to this term as the bias component, and we shall estimate the bias component in the second step and further correct the plug-in estimator of $\mu^{(l)}$.  %

\vspace{2mm}
\noindent {\bf Step 2: Correct the bias of the plug-in estimator.} We now explain how to obtain an accurate estimator of the last term in \eqref{eq: error decompose}.
Suppose that $\QQ_X$ is absolutely continuous with respect to $\PP^{(l)}_X$ for $l\in [L]$. We define the density ratio $\omega^{(l)}(x) = d\QQ_X(x)/d\PP^{(l)}_X(x)$. For $(X,Y)$ drawn from the distribution $\PP^{(l)} = (\PP^{(l)}_X, \PP_{Y|X}^{(l)})$ and independent of the fitted $\hf{l}(\cdot)$, we obtain the following by taking the conditional expectation of the outcome variable, 
\[
\begin{aligned}
    &\E_{(X,Y)\sim (\PP^{(l)}_X, \PP^{(l)}_{Y|X})}\left[\omega^{(l)}(X)\cdot ({\bf 1}(Y=c) - \hf{l}_c(X))\cdot X\right] \\
    &= \E_{X\sim\PP^{(l)}_X}\left[\omega^{(l)}(X)\cdot (f^{(l)}_c(X) - \hf{l}_c(X))\cdot X\right]=\E_{\QQ_X}\left[(f^{(l)}_c(X) - \hf{l}_c(X))\cdot X\right],
\end{aligned}
\] 
where the last step follows from the definition of the density ratio $\w{l}$. 
Since the $l$-th source data $\{\X{l}_i, \Y{l}_i\}_{1\leq i\leq n_l}$  are i.i.d.~drawn from the distribution $(\PP^{(l)}_X, \PP^{(l)}_{Y|X})$, we leverage the above expression to  construct the following estimator of the last term in \eqref{eq: error decompose},
\[
\frac{1}{n_l}\sum_{i=1}^{n_l}\left[\hw{l}(\X{l}_i)\cdot ({\bf 1}(\Y{l}_i = c) - \hf{l}_c(\X{l}_i))\cdot \X{l}_i\right],
\]
where $\hw{l}(\cdot)$ denotes a density ratio estimator. Consequently, we leverage the above expression to correct the bias of the plug-in estimator $-\frac{1}{N}\sum_{j=1}^N \widehat{f}^{(l)}_c(X^\QQ_j) \cdot X^\QQ_j$ and construct the DML estimators for each category $c\in \{1,...,K\}$ as:
\begin{equation}
    \widehat{\mu}_c^{(l)} = -\frac{1}{N}\sum_{j=1}^N \widehat{f}^{(l)}_c(X^\QQ_j) \cdot X^\QQ_j - \frac{1}{n_l} \sum_{i=1}^{n_l}\widehat{\omega}^{(l)}(X_i^{(l)})\cdot \left({\bf 1}(Y_i^{(l)}=c)-\widehat{f}^{(l)}_c(X_i^{(l)})\right) \cdot X_i^{(l)}.
    \label{eq: DR component}
\end{equation}
In Supplement Section \ref{sec: doubly robust}, we provide the full details on the construction of the conditional probability estimator $\hf{l}(\cdot)$ and the density ratio estimator $\hw{l}(\cdot)$ using the cross-fitting idea.
{For the special case with no covariate shift, we can simply set $\hw{l}\equiv 1$ in the above estimator, which shares the same theoretical property with the estimator in \eqref{eq: no shift est}.}
We then stack the above estimators and construct $\widehat{\mu}^{(l)}=\left([\widehat{\mu}^{(l)}_1]^{\intercal},[\widehat{\mu}^{(l)}_2]^{\intercal},\cdots,[\widehat{\mu}^{(l)}_K]^{\intercal}\right)^{\intercal}\in \RR^{dK}$. 
Lastly, we apply this DML estimator $\widehat{\mu}^{(l)}$ in \eqref{eq: DR component} to the empirical \CGDRO\ problem \eqref{eq: sample optimizers} and solve it with Mirror Prox algorithm. The full procedure is presented in Supplement Section \ref{sec: doubly robust}.

% \Zhenyu{In the following Algorithm \ref{algo:overall}, we apply this updated estimator of $\mu^{(l)}$ to the empirical \CGDRO\ model in \eqref{eq: sample optimizers}, and solve it by Mirror Prox algorithm presented in Supplement Section XXX.}
% Since $\widehat{\mu}^{(l)}$ is constructed as the DML estimator, we refer to our proposal as the Mirror Prox Algorithm with DML. \Zijian{We can also move this to the supplement.}

\section{Statistical Convergence Rate}
\label{sec: conv rate}

In this section, we establish the convergence rate of $\|\widehat{\theta}-\theta^*\|_2$ and defer the construction of confidence intervals for any coordinate of $\theta^*$ to  Section \ref{sec: UQ}.
To facilitate theoretical analysis, we use $n=\min_{1\leq l\leq L} n_l$ to denote the minimum source sample size. We impose the following conditions:

\begin{condition}
    \label{cond:A1}
   There exist constants $c_0, c_1>0$ and a sequence $\delta_n\to0$ as $n\to\infty$ such that the estimators $\{\widehat{\mu}^{(l)}\}_{1\leq l\leq L}$ satisfy: for any $t \geq c_0$,
    \begin{equation}
    \mathbb{P}\left(\max_{1\leq l\leq L}\|\widehat{\mu}^{(l)}-\mu^{(l)}\|_2\geq t\sqrt{d/n}\right)\leq \exp(-c_1 t^2)+\delta_n.
    \label{eq: rate condition}
    \end{equation}
\end{condition}
This condition requires that, with high probability, $\widehat{\mu}^{(l)}$ is an accurate estimator of $\mu^{(l)}$ in the order of $\sqrt{d/n}$.  In Supplement Section \ref{subsec: properties of hatmu}, we verify that our proposed estimators $\{\widehat{\mu}^{(l)}\}_{1\leq l\leq L}$, given in \eqref{eq: no shift est} or \eqref{eq: DR component} of Section \ref{sec: refined analysis}, satisfy \eqref{eq: rate condition} under regimes without and with covariate shift. This holds under mild regularity assumptions, along with the following standard condition from the DML literature \cite{chernozhukov2018double}: with high probability $1-\delta_n$,
\begin{equation}
\|\widehat{\omega}^{(l)}/\omega^{(l)}-1\|_{\ell_2(\PP^{(l)})}\cdot\|\widehat{f}^{(l)}-f^{(l)}\|_{\ell_2(\PP^{(l)})} \ll n^{-1/2},
    \label{eq: DML cond}
\end{equation}
where $\|g\|_{\ell_2(\mathbb{T})} = \left(\E_{X\sim \mathbb{T}}\left[\|g(X)\|_2^2\right]\right)^{1/2}$ denotes the $\ell_2$ norm of a function $g(\cdot)$ evaluated on the independent data drawn from the distribution $\mathbb{T}$.  In both \eqref{eq: rate condition} and \eqref{eq: DML cond}, $\delta_n$ represents a small probability that the ML estimators $\widehat{\omega}^{(l)}(\cdot)$ and $\widehat{f}^{(l)}(\cdot)$ do not converge at a sufficiently fast speed.
In \eqref{eq: rate condition}, we require $t \geq c_0$ for some positive constant $c_0$ to account for the scale of $\E[\|\widehat{\mu}^{(l)}-\mu^{(l)}\|_2]$, which itself is of order $\sqrt{d/n}$. 
\begin{condition}
    The target covariate $\XQ_i$ is almost surely bounded for $1\leq i\leq N$. There exist constants $\kappa_1,\kappa_2>0$ such that 
    $ \kappa_1 \leq \lambda_{\min}\left( \E[\XQ_i X^{\QQ\intercal}_i]\right)\leq \lambda_{\max}\left( \E[\XQ_i X^{\QQ\intercal}_i]\right)\leq \kappa_2$. The population \CGDRO\ model $\theta^*$ in \eqref{eq: minimax model} exists and satisfies $\|\theta^*\|_1\leq C$ for some constant $C>0$. The sample sizes $n,N$ and the dimension $d$ satisfy $n\gg d^{3}$ and $N\gg n \log N.$
    \label{cond:A2}
\end{condition}
The boundedness condition of the target covariate $\XQ_i$ and $\|\theta^*\|_1$ is mainly imposed to ensure that the conditional probabilities $\frac{\exp([\theta^{*}_c]^\intercal X_i^{\QQ})}{1+\sum_{k=1}^K \exp([\theta^{*}_k]^\intercal X_i^{\QQ})}$ are bounded away from $0$ and $1$ for each category $c\in \{1,...,K\}$. Such boundedness conditions were adopted to facilitate the theoretical analysis of the binary and multi-class classification models \cite{van2014asymptotically, athey2018approximate, guo2021inference}. Moreover, the boundedness of the eigenvalues for $\E[\XQ_i X^{\QQ\intercal}_i]$ guarantees the strong convexity and smoothness of the function $S(\theta)$ defined in \eqref{eq:mu def} over a compact neighborhood near $\theta^*$. 
Lastly, we consider the growing-dimensional regime and requires $d\ll n^{1/3}$. We assume $N\gg n\log N$, that is, the amount of unlabeled data in the target domain is much larger than the amount of labeled data in the source domains.  %\Yifan{Is this more like a setting that should be specified earlier?} \Zijian{I put a sentence in the second paragraph.} 

In the following proposition, we establish an initial convergence rate of $\widehat{\theta}$, the solution of the empirical \CGDRO\ problem. In Sections~\ref{sec: approx} and \ref{sec: refined theory}, we derive a faster convergence rate using more sophisticated analytical tools. 
\begin{Proposition} 
Suppose that {\rm Conditions \ref{cond:A1} and \ref{cond:A2}} hold. Then there exist positive constants $c_0,c_1>0$ such that with probability at least $1-N^{-c_1 d}-\exp(-c_1 t^2)-\delta_n,$ 
\begin{equation}
\|\widehat{\theta}-\theta^*\|_2\lesssim \sqrt{t}\cdot (d/n)^{1/4},
\label{eq: initial bound}
\end{equation}
where $t$ is a positive value satisfying $c_0\leq t\lesssim \sqrt{n/d^3}$ 
and both $c_0$ and the vanishing sequence $\delta_n \to 0$ are specified in {\rm Condition~\ref{cond:A1}}.
\label{prop: initial bound}
\end{Proposition}

The parameter $t>0$ is introduced to control the tail probability term $\exp(-c_1 t^2)$, which we aim to make small by choosing a sufficiently large $t$. Throughout the paper, \( t \) satisfies
\begin{equation}
c_0 \leq t \lesssim \sqrt{n/d^3} \quad \text{with } c_0 \text{ specified in Condition \ref{cond:A1}.}
\label{eq: t condition}
\end{equation}
We have explained the reason for including the lower bound $t\geq c_0$ after stating Condition ~\ref{cond:A1} while the upper bound $t \lesssim\sqrt{n/d^3}$ is introduced  to guarantee that the $\ell_1$-error $\|\widehat{\theta} - \theta^*\|_1$ remains bounded with a high probability. Under the assumption $n \gg d^3$ in Condition~\ref{cond:A2}, this range for $t$ is sufficiently broad to make $\exp(-c_1 t^2)$ small. {Importantly, $t$ can be chosen as a large fixed positive constant or allowed to grow slowly with $n$ while still satisfying \eqref{eq: t condition}; for example, we may choose $t=\sqrt{\log n}$ if $n\gtrsim d^3\log n$.}

%We shall mention that the positive value $t>0$ is involved with the tail probability $\exp(-c_1 t^2)$ and the range $c_0\leq t\lesssim \sqrt{n/d^3}$ allows us to choose a wide range of $t$ values. We have explained the reason for requiring \( t \geq c_0 \) after stating Condition \ref{cond:A1}. The reason for requiring \( t \lesssim \sqrt{n/d^3} \) is to ensure that {the $\ell_1$ norm error} \( \|\widehat{\theta} - \theta^*\|_1 \) remains bounded with a high probability. Importantly, {under the assumption \( n \gg d^3 \) of Condition \ref{cond:A2},} the positive value \( t \) can still take a wide range of values. {For instance, one may choose $t$ as a large positive constant or allow it to grow slowly with $n$, such as $t = \sqrt{\log n}$, both of which satisfy \eqref{eq: t condition}. 

The above proposition establishes a slow convergence rate for $\widehat{\theta}$ of order $n^{-1/4}$. This suboptimal rate arises because the inner maximization problem within \eqref{eq: minimax model} and \eqref{eq: sample optimizers} is concave but not strongly concave, allowing the existence of multiple maximizers. As a result, quantifying the distance between the sets of maximizers for empirical and population problems leads to a slow rate. A similar phenomenon is observed in~\cite{zhang2024generalization}. In the following Theorem \ref{thm: final rate}, we establish a faster rate $\sqrt{d/n}$ through a more refined analysis. The key idea is to construct an approximate optimization problem whose dual problem is strongly concave in $\gamma$. 

%\Yifan{The connection between this paragraph to the previous one is not very clear. Maybe say such a fast convergence rate of order $n^{-1/2}$ cannot be readily obtained by applying standard M-estimation techniques. (Also need to mention how we get the slow rate).} \Zijian{Let us discuss this.}
\begin{Remark}
We highlight that existing M-estimation techniques cannot be applied to achieve the fast convergence rate $\sqrt{d/n}$ in our setting.
For M-estimation, one typically replaces a population mean with its empirical counterpart. In contrast, we take an extra maximization over the weight after replacing the population quantity with the corresponding sample average, which introduces additional challenges in controlling the statistical error. Moreover, as the maximization is taken after the expectation, standard M-estimation techniques for non-smooth loss functions, such as least absolute deviation (LAD) minimization discussed in Chapter~5 of \cite{vandervaart1998asymptotic}, cannot be directly applied in our setting.
\end{Remark}

\subsection{Main Idea: Faster Convergence Rate via Risk Approximation}
\label{sec: approx}
The essential idea of establishing fast convergence is to identify an approximate optimization problem that serves as a surrogate for the original minimax optimization problem. We utilize this approximate optimization problem as a bridge to establish the fast convergence rate of $\|\widehat{\theta}-\theta^*\|_2$. We now introduce the approximate optimization problems and describe the main proof idea. 

For the original population problem in \eqref{eq: minimax model}, we approximate $S(\theta)$ with its second-order Taylor expansion near $\widehat{\theta}$, defined as 
\begin{equation}
Q(\theta)=S(\thetainit)+\langle \nabla S(\thetainit),\theta-\thetainit\rangle+\frac{1}{2} (\theta-\thetainit)^{\intercal} H(\thetainit) (\theta-\thetainit),
\label{eq: approximation}
\end{equation}
where $\nabla S(\theta)$ and $H(\theta)$ respectively denote the gradient and Hessian of $S(\theta)$ in \eqref{eq:mu def}. We then approximate the original population problem in \eqref{eq: minimax model} by replacing $S(\theta)$ with $Q(\theta)$ and define ${\theta}^*_{\rm ap}$ as the solution to the following approximate population problem:
\begin{equation}
{\theta}^*_{\rm ap}=\argmin_{\theta\in \RR^{dK}}\max_{\gamma\in \Delta^L}\phi_{\rm ap}(\theta,\gamma) \quad \text{with}\quad \phi_{\rm ap}(\theta,\gamma) \coloneqq
\sum_{l=1}^{L} \gamma_{l} \cdot\theta^{\intercal}\mu^{(l)}+Q(\theta).
\label{eq: approximate minimax}
\end{equation}
The objective function $\phi_{\rm ap}(\theta,\gamma)$ can be viewed as a quadratic approximation of $\phi(\theta,\gamma)$ near $\widehat{\theta}$, and the optimizer ${\theta}^*_{\rm ap}$ serves as a surrogate of $\theta^*$ for the theoretical analysis. %\Yifan{I changed everything to approximate population optimization problem to distinguish from approximate empirical problem.}

Similarly, we define the approximate empirical problem for the original one in \eqref{eq: sample optimizers}:
\begin{equation}
\widehat{\theta}_{\rm ap}=\argmin_{\theta\in \RR^{d}} \max_{\gamma\in \Delta^{L}} \widehat{\phi}_{\rm ap}(\theta,\gamma) \quad \text{with}\quad \widehat{\phi}_{\rm ap}(\theta,\gamma)\coloneqq \sum_{l=1}^{L} {\gamma}_l\cdot \theta^{\intercal}\widehat{\mu}^{(l)}+ \widehat{Q}(\theta),
\label{eq: inter estimator}
\end{equation}
where $\widehat{Q}(\theta)\coloneqq \widehat{S}(\thetainit)+\langle \nabla \widehat{S}(\thetainit),\theta-\thetainit\rangle+\frac{1}{2} (\theta-\thetainit)^{\intercal} \widehat{H}(\thetainit) (\theta-\thetainit)$ and $\nabla \widehat{S}(\theta)$ and $\widehat{H}(\theta)$ denote the gradient and Hessian of $\widehat{S}(\theta)$ in \eqref{eq: hatS def}. In \eqref{eq: inter estimator}, we substitute the objective $\widehat{\phi}(\theta,\gamma)$ in the original empirical problem \eqref{eq: sample optimizers} by $\widehat{\phi}_{\rm ap}(\theta,\gamma)$, with $\widehat{Q}(\theta)$ serving as a quadratic approximation of $\widehat{S}(\theta).$ As a remark, ${\theta}^*_{\rm ap}$ and $\widehat{\theta}_{\rm ap}$ are mainly theoretical intermediate quantities used for analysis; we do not implement the approximate problems defined in \eqref{eq: approximate minimax} and \eqref{eq: inter estimator}. 

%\Zijian{I have switched the order following Yifan's suggestion.} 
With the definition of the approximate population and empirical problems, we demonstrate our proof idea for establishing the fast rate of $\|\widehat{\theta}-\theta^*\|_2$, via the following triangle inequality: 
\begin{equation*}
\|\widehat{\theta}-\theta^*\|_2\leq \|\widehat{\theta}-\widehat{\theta}_{\rm ap}\|_2+\|\thetainit_{\rm ap}-{\theta}^*_{\rm ap}\|_2+\|{\theta}^*_{\rm ap}-\theta^*\|_2,
\label{eq: tria}
\end{equation*}
where $\widehat{\theta}_{\rm ap}$ and $\theta^*_{\rm ap}$ are the solutions for two approximate problems linking $\widehat{\theta}$ and $\theta^*$, as defined in \eqref{eq: inter estimator} and \eqref{eq: approximate minimax}, respectively. As illustrated in Figure \ref{fig:illus_est}, the following
Theorem \ref{thm: quad convergence} establishes that the first and third terms on the right-hand side converge at a fast rate, while the following Theorem \ref{thm: refined rate} further establishes that $\|\widehat{\theta}_{\rm ap}-\theta^*_{\rm ap}\|_2$ also converges at a fast rate.

\begin{figure}[ht!]
    \centering
    \includegraphics[width=0.75\linewidth]{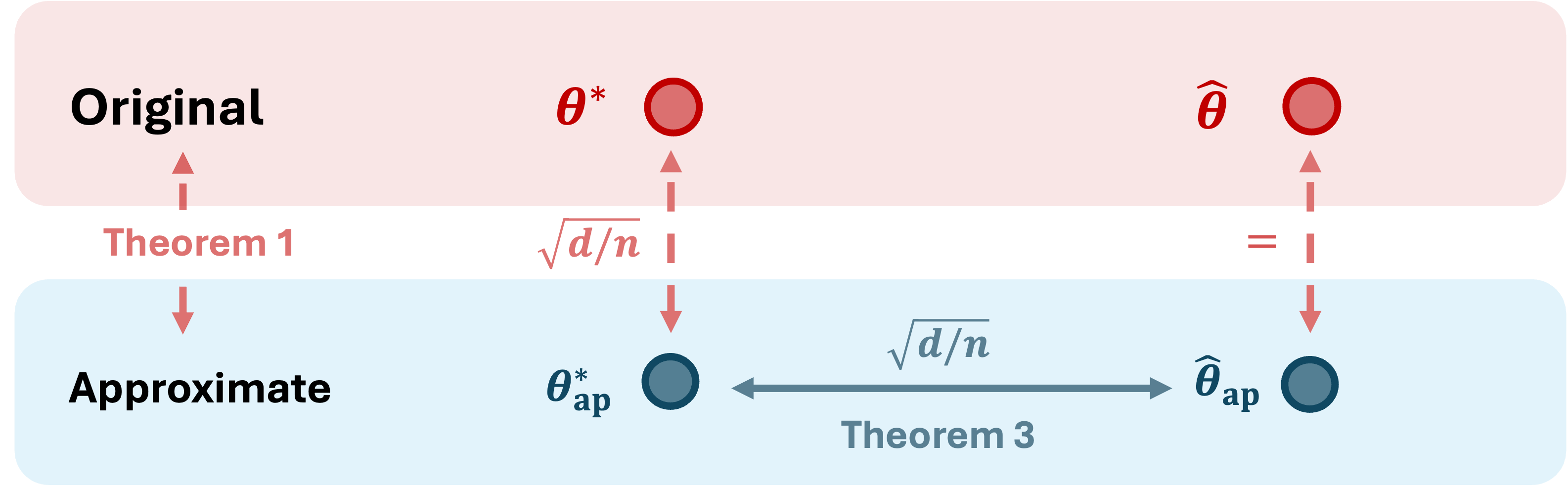}
    \caption{Illustration of the proof strategy for establishing the fast convergence rate of $\|\widehat{\theta}-\theta^*\|_2$.  The following Theorem \ref{thm: quad convergence} establishes the approximation errors of $\|\theta^*_{\rm ap} - \theta^*\|_2$ and $\|\widehat{\theta}_{\rm ap} - \widehat{\theta}\|_2$, while the following Theorem \ref{thm: refined rate} establishes the convergence rate of $\|\widehat{\theta}_{\rm ap} - \theta^*_{\rm ap}\|_2$.}
        \label{fig:illus_est}
\end{figure}

The following theorem quantifies the error between the solutions of the approximate and original minimax optimization problems. This approximation analysis plays a key role in justifying our surrogate formulation and may be of independent theoretical interest. The full proof is provided in Supplement Section \ref{sec: quad convergence}.

\begin{Theorem}
Suppose that {\rm Conditions \ref{cond:A1} and \ref{cond:A2}} hold. Then there exists a positive constant $c_1>0$ such that with probability at least $1-N^{-c_1d}-\exp(-c_1t^2)-\delta_n$, 
 \begin{equation}
\|{\theta}^*_{\rm ap}-\theta^*\|_2\lesssim  \|\thetainit-\theta^*\|_2^2 \lesssim t\sqrt{{d}/{n}},
\label{eq: quad convergence}
\end{equation}
\begin{equation}
\thetainit_{\rm ap}=\thetainit,
\label{eq: stable optimizer}
\end{equation}
where $t>0$ satisfies \eqref{eq: t condition} and the sequence $\delta_n \to 0$ is specified in {\rm Condition~\ref{cond:A1}}. 
\label{thm: quad convergence}
\end{Theorem}

%\Yifan{Just for curiosity, why the probability losses so strange?} \Zhenyu{Because the Condition (A1) is satisfies with such a probability. Do we need to spell it out? which also occurs in Proposition 1.} \Zijian{I am not sure that I fully understand this but such a high probability is common in stat.}
In the above theorem, the approximation error of ${\theta}^*_{\rm ap}$ satisfies $\|{\theta}^*_{\rm ap} - \theta^*\|_2 \lesssim \|\widehat{\theta} - \theta^*\|_2^2$, indicating that the approximation error $\|{\theta}^*_{\rm ap} - \theta^*\|_2$ scales quadratically with the estimation error  $\|\widehat{\theta} - \theta^*\|_2^2$. 
Together with Proposition \ref{prop: initial bound}, this quadratic dependence implies a fast parametric convergence rate for the approximation error $\|{\theta}^*_{\rm ap}-\theta^*\|_2$. Importantly, we show that $\widehat{\theta}_{\rm ap} = \widehat{\theta}$, indicating that the approximate empirical problem in \eqref{eq: inter estimator} shares the same solution as the original empirical problem in \eqref{eq: sample optimizers}. %

%\Yifan{I would suggest to demonstrate the proof idea for the fast convergence rate first, then state Theorem 1, which should be more natural.}

\subsection{Convergence Rate of the Approximate Minimax Problems}
\label{sec: refined theory}
To derive a fast convergence rate for $
\|\thetainit_{\rm ap} - \theta^*_{\rm ap}\|_2$, we derive in the following alternative representations for both $\theta^*_{\rm ap}$ and 
$\thetainit_{\rm ap}$ by solving dual problems of the approximate minimax problems.  
%Specifically, we compute the associated optimal weights and then express $\theta^*_{\rm ap}$ and $\thetainit_{\rm ap}$ as solutions to standard minimization problems parameterized by these optimal weights. 
We primarily focus on the approximate population minimax optimization problem  in~\eqref{eq: approximate minimax}, while similar arguments extend to the empirical version in~\eqref{eq: inter estimator}. To provide the intuition of deriving the dual form, we interchange the max and min operations in~\eqref{eq: approximate minimax}, which yields the following reformulation:
\begin{equation*}
\min_{\theta\in \RR^{dK}}\left[\max_{\gamma\in \Delta^L}\phi_{\rm ap}(\theta,\gamma)\right]=\max_{\gamma\in \Delta^L} \left[\min_{\theta\in \RR^{dK}}\phi_{\rm ap}(\theta,\gamma)\right].
\end{equation*}
The right-hand side of the above equation provides a dual form of identifying $\theta^*_{\rm ap}$:
\begin{itemize}
\item {\bf Step 1.} We solve the outer maximization problem with respect to $\gamma$ and compute the optimal weight as
\begin{equation}
\gamma^*_{\rm ap}=\argmax_{\gamma\in \Delta^L} F(\gamma)  \quad \text{with}\quad F(\gamma) \coloneqq \min_{\theta\in \RR^{dK}}\phi_{\rm ap}(\theta,\gamma).
\label{eq: step 1}
\end{equation}

\item {\bf Step 2.} With the optimal weight $\gamma^*_{\rm ap},$ we then solve the inner minimization problem, 
\begin{equation}
\theta^*_{\rm ap}\coloneqq\argmin_{\theta\in \RR^{dK}}\ \phi_{\rm ap}(\theta,\gamma^*_{\rm ap}). 
\label{eq: step 2}
\end{equation}
\end{itemize}
%The benefit of \eqref{eq: step 1} and \eqref{eq: step 2} is that both objective functions $F(\gamma)$ and $\phi_{\rm ap}(\theta,\gamma^*_{\rm ap})$ admit an explicit quadratic form, which facilitates analytical tractability and enables sharper theoretical characterization. 
%\Zhenyu{Zijian, I feel the previous sentence is hard to follow, as even after reading Theorem 2, readers still don't know the expressions of both $F(\gamma)$ and $\phi_{\rm ap}(\theta,\gamma^*_{\rm ap}$. See my writings below, which omit these details.} 
%\Yifan{The next sentence is not clear to me. What does it mean by identifying? We will not compute it but only for theoretical analysis. I suggest that we could delete the next sentence.}
%We provide a rigorous justification for identifying $\theta^*_{\rm ap}$ in theory using  \eqref{eq: step 1} and \eqref{eq: step 2} in the following Theorem \ref{thm: equi expressions}. 

{We show in the following Theorem \ref{thm: equi expressions} that, after such a reformulation, $\theta^*_{\rm ap}$, originally defined through minimax optimization in \eqref{eq: approximate minimax}, now admits a closed form expression given $\gamma^*_{\rm ap}$, while $\gamma^*_{\rm ap}$ itself can be characterized as the optimizer of the quadratic optimization problem $F(\gamma)$ defined in \eqref{eq: step 1}. This closed-form expression enables sharper theoretical analysis between $\theta^*_{\rm ap}$  and $\widehat{\theta}_{\rm ap}$, where the latter one also has a similar closed-form expression.}

To facilitate the discussion, we define the matrices 
\begin{equation}
\U=(\mu^{(1)}, \mu^{(2)},..., \mu^{(L)})\in \RR^{dK\times L} \quad \text{and}\quad \widehat{\U} =\begin{pmatrix} \widehat{\mu}^{(1)},\widehat{\mu}^{(2)},...,\widehat{\mu}^{(L)}\end{pmatrix}\in \RR^{dK\times L},
\label{eq: U definition}
\end{equation}
where each $\mu^{(l)}\in \RR^{dK}$ is defined in \eqref{eq:mu def} and $\widehat{\mu}^{(l)}\in \RR^{dK}$ is the estimator of $\mu^{(l)}$. Our analysis requires $\{\mu^{(l)}\}_{1\leq l\leq L}$ to be linearly independent, stated in the following assumption.
\begin{condition}
    The $L$-th largest singular value of $\U$ is positive, i.e., $\sigma_L(\U)>0$.
    \label{cond:A3}
\end{condition}
The above condition holds as long as the vectors $\{\mu^{(l)}\}_{1\leq l\leq L}$ are not collinear. When there exist distributional shifts across distributions, the linear independence assumption among $\{\mu^{(l)}\}_{1\leq l\leq L}$ tends to hold plausibly. However, when two source distributions are identical, their corresponding $\mu^{(l)}$
  also coincide, violating Condition \ref{cond:A3}. This highlights that some degree of heterogeneity across environments is necessary for Condition \ref{cond:A3} to hold. Such heterogeneity is common in practice, supporting the applicability of our theoretical analysis. We also emphasize that our theoretical analysis does not assume $\sigma_L(\U)$ to be a fixed constant; instead, we allow it to decay to zero with the sample size $n$. %reflecting scenarios with increasing collinearity as more data are collected.

%\Zhenyu{Zijian, plz check the statement.}
The following theorem establishes equivalent closed-form representations for ${\theta}^*_{\rm ap}$ and $\widehat{\theta}_{\rm ap}$.
\begin{Theorem}
Suppose that {\rm Conditions \ref{cond:A1}--\ref{cond:A3}} hold. Then
${\theta}^*_{\rm ap}$ defined in \eqref{eq: approximate minimax} can be expressed as
\begin{equation}
{\theta}^*_{\rm ap}=\thetainit-[H(\thetainit)]^{-1}\left(\U\gamma^*_{\rm ap}+\nabla S(\thetainit)\right) \quad \text{with}\quad {\gamma}^*_{\rm ap}= \argmax_{\gamma\in \Delta^{L}} F(\gamma),
\label{eq: closed form theta}
\end{equation}
where $F(\gamma)=\min_{\theta\in \RR^{dK}}\phi_{\rm ap}(\theta,\gamma)$ is a quadratic function of $\gamma$, with the explicit expression provided in the following equation \eqref{eq: resampled gamma}. Similarly, there exists a positive constant $c_1>0$ such that with probability larger than $1-N^{-c_1d} -e^{-c_1t^2}-\delta_n$,
\begin{equation}
\widehat{\theta}_{\rm ap}=\thetainit-[\widehat{H}(\thetainit)]^{-1}\left(\widehat{\U}\widehat{\gamma}_{\rm ap}+\nabla \widehat{S}(\thetainit)\right) \quad \text{with}\quad \widehat{\gamma}_{\rm ap}=\argmax_{\gamma\in \Delta^{L}} \widehat{F}(\gamma),
\label{eq: inter theta}
\end{equation}
where $\widehat{F}(\gamma)=\min_{\theta\in\RR^{dK}}\widehat{\phi}_{\rm ap}(\theta,\gamma)$ is a quadratic function of $\gamma$, with the explicit expression provided in \eqref{eq: expression of F(gamma) hat} of Supplement Section \ref{proof of thm: equi expressions}. Here, $t>0$ satisfies \eqref{eq: t condition}, and the sequence $\delta_n \to 0$ is specified in {\rm Condition~\ref{cond:A1}}.
\label{thm: equi expressions}
\end{Theorem}

We elaborate on the key insights of the expression \eqref{eq: closed form theta}, which naturally extend to \eqref{eq: inter theta}. First, the expression in \eqref{eq: closed form theta} provides an explicit solution for ${\theta}^*_{\rm ap}$ given the optimal weight  ${\gamma}^*_{\rm ap}$, which arises because the unconstrained optimization problem in \eqref{eq: step 2} admits a closed-form minimizer. 
Second, we emphasize the definitions of $\phi_{\rm ap}(\theta,\gamma)$ and $F(\gamma)$ depend on the data-driven estimator $\widehat{\theta}$, and consequently, so do the optimizers $({\theta}^*_{\rm ap}, \gamma^*_{\rm ap})$. Therefore, despite the use of a star in their notations, ${\theta}^*_{\rm ap}$ and $\gamma^*_{\rm ap}$ are not population-level quantities, but are instead data-dependent intermediate quantities to facilitate the theoretical analysis.

% Second, following from the definition of $F(\gamma)$ in \eqref{eq: step 1}, we also obtain an explicit expression of $F(\gamma)$ used to compute ${\gamma}^*_{\rm ap}$,
% \begin{equation*}
% \begin{aligned}
% F(\gamma)=-
% \frac{1}{2}\left(\U \gamma+\nabla S(\thetainit)\right)^{\intercal}[H(\thetainit)]^{-1}\left(\U \gamma+\nabla S(\thetainit)\right)+ \widehat{\theta}^\intercal \U \gamma.
% \end{aligned}
% \label{eq: closed form gamma}
% \end{equation*} 
% We emphasize that the definitions of $\phi_{\rm ap}(\theta,\gamma)$ and $F(\gamma)$ depend on the data-driven estimator $\widehat{\theta}$, and consequently, so do the optimizers $({\theta}^*_{\rm ap}, \gamma^*_{\rm ap})$. Therefore, despite the use of a star in their notations, ${\theta}^*_{\rm ap}$ and $\gamma^*_{\rm ap}$ are not population-level quantities, but are instead data-dependent intermediate quantities to facilitate the theoretical analysis.

We now explain why Theorem \ref{thm: equi expressions} leads to a fast convergence rate for $\|\widehat{\theta}_{\rm ap}-{\theta}^*_{\rm ap}\|_2$. The main idea is that both $\widehat{F}(\gamma)$ and ${F}(\gamma)$ are smooth quadratic functions of $\gamma$. When these two functions are close, their minimizers, $\widehat{\gamma}_{\rm ap}$ and ${\gamma}^*_{\rm ap}$, are also close. Given the explicit formulas for $\widehat{\theta}_{\rm ap}$ and ${\theta}^*_{\rm ap}$ in Theorem \ref{thm: equi expressions}, the closeness of their corresponding optimal weights directly implies the closeness of the estimators, yielding the fast convergence rate for $\|\widehat{\theta}_{\rm ap}-{\theta}^*_{\rm ap}\|_2$, which is stated in the following theorem. 

\begin{Theorem}
Suppose that {\rm Conditions \ref{cond:A1}--\ref{cond:A3}} hold. Then there exists a positive constant $c_1>0$ such that with probability at least $1-N^{-c_1 d}-\exp(-c_1 t^2)-\delta_n,$
\begin{equation}
\|\thetainit_{\rm ap} - \theta_{\rm ap}^*\|_2 \lesssim \frac{t}{\sigma_L^2(\U)} \left(1+ \|\nabla S(\theta^*)\|_2 + \|\U\|_2 + \|\U\|_2^2\right)\sqrt{d/n} ,
\label{eq: general rate}
\end{equation}
where $t>0$ satisfies \eqref{eq: t condition} and the sequence $\delta_n \to 0$ is specified in {\rm Condition~\ref{cond:A1}}.
\label{thm: refined rate}
\end{Theorem}

As illustrated in Figure \ref{fig:illus_est}, we combine the above theorem and Theorem \ref{thm: quad convergence} to establish the fast convergence rate of $\|\thetainit - \theta^*\|_2.$ We impose the following condition to further simplify the final upper bound.

\begin{condition}
    \label{cond:A4}
There exists $C_0>0$ such that
$\max\{\|\U\|_2, \|\nabla S(\theta^*)\|_2\}\leq C_0$.
\end{condition}

\begin{Theorem}
    Suppose that {\rm Conditions \ref{cond:A1}--\ref{cond:A4}} hold. Then there exist  positive constants $c_1,C>0$ such that with probability at least $1-N^{-c_1 d}-\exp(-c_1t^2)-\delta_n,$
    \begin{equation}
    \|\widehat{\theta}-\theta^*\|_2\leq \tau\sqrt{d/n} \quad \text{with}\quad \tau= C \left(1+\frac{1}{\sigma_L^2(\U)}\right) t,
    \label{eq: final rate}
    \end{equation}
where $t>0$ satisfies \eqref{eq: t condition} and the sequence $\delta_n \to 0$ is specified in {\rm Condition~\ref{cond:A1}}.
    \label{thm: final rate}
\end{Theorem}
This theorem establishes an upper bound on $\|\widehat{\theta} - \theta^*\|_2$, showing that $\widehat{\theta}$ converges to $\theta^*$ at the rate $\sqrt{d/n}$, up to a multiplicative factor $\tau$. The factor $\tau$ remains of constant order if $\sigma_L(\U)$ is bounded away from zero. Compared to the slower rate established in Proposition~\ref{prop: initial bound}, our refined analysis, based on constructing two approximate optimization problems, yielding a sharper statistical convergence rate for $\|\widehat{\theta}-\theta^*\|_2$. %Notably, this rate matches the optimal rate for estimating regression parameters in the single source setting \cite{wainwright2019high}.

\section{Uncertainty Quantification for Minimax Optimization}
\label{sec: UQ}
In this section, we focus on statistical inference for the \CGDRO\ model $\theta^*$ in \eqref{eq:general_minimax}, since we interpret $\theta^*$ as a generalizable model that captures stable predictive patterns across diverse domains. Our goal is to construct the confidence interval for each component $\theta^*_j$ with $1\leq j\leq dK$, or to test hypotheses of the form $H_j: \theta^*_j = 0$. When $\theta^*_j \neq 0$, we say that the $j$-th element is distributionally robust significant, meaning that it is likely to exert a consistent and non-negligible effect even under distribution shifts. For clarity, we focus on the inference for $\theta^*_1$, though the proposed methodology naturally extends to any $\theta^*_j$ with $1\leq j\leq dK$.

In Section~\ref{sec: challenge}, we demonstrate the challenge of inference for $\theta^*_1$, that is, the empirical distribution of $\widehat{\theta}_1$ may be nonstandard. Throughout this paper, we use the terms \textit{nonnormal} and \textit{nonstandard} interchangeably to indicate that the limiting distribution of the estimator is not Gaussian. The standard inference approach based on taking quantiles of the Gaussian distribution cannot be applied to such a nonstandard inference problem. In Sections~\ref{sec: resampling} and \ref{sec: procedure}, we propose novel inference procedures to address this challenge.

\subsection{Nonnormal Limiting Distribution of the Data-driven Estimator}
\label{sec: challenge}
We demonstrate in the following that the empirical distribution of $\widehat{\theta}_1$ is not asymptotically normal. Consider a simplified setting with $L = 2$ sources and $K = 1$ (binary classification), where each source contains $n_l = 400$ samples with the covariate dimension  $d = 20$. The number of unlabeled target samples is fixed at $N = 4000$; {see Supplement Section \ref{appendix: setups} for the detailed specification. We perform $500$ independent simulations, where in each run the empirical \CGDRO\ problem is solved according to the procedure described in Section \ref{sec: refined analysis}. Figure~\ref{fig: Inference Challenge} displays the resulting empirical distributions of the first coefficient $\widehat{\theta}_1$ and, as a companion result, the weight assigned to the first source $\widehat{\gamma}_1$; see Supplement Section \ref{subsec: DML MP} for implementation details.}
% , where $\widehat{\theta}$ and $\widehat{\gamma}$ are computed using Algorithm~\ref{algo:overall}. 
As shown in the left and middle panels of Figure~\ref{fig: Inference Challenge}, the empirical distribution of $\widehat{\theta}_1$ deviates from normality and is not centered at the true value $\theta^*_1$.

\begin{figure}[!ht]
    \centering
    \includegraphics[width=0.85\linewidth]{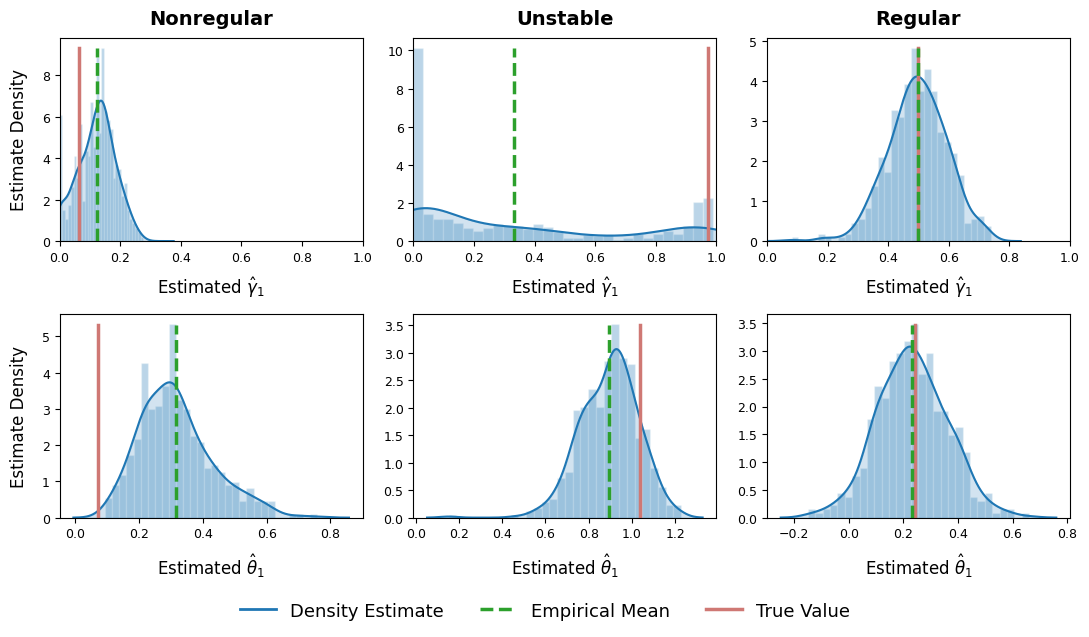}
    \caption{Empirical distributions of $\widehat{\gamma}_1$ and $\widehat{\theta}_1$ in the  nonregular, unstable, and regular settings. The top row corresponds to the first source's estimated weight $\widehat{\gamma}_1$, while the bottom row corresponds to the estimated $\widehat{\theta}_1$. Vertical red lines indicate true parameter values, while dashed green lines show the empirical average across $500$ simulations. Blue histograms with overlaid kernel density estimates depict the empirical distributions of estimates across 500 simulations. The exact simulation settings are reported in Supplement Section \ref{appendix: setups}.
 }
    \label{fig: Inference Challenge}
\end{figure}

% We provide insight into the emergence of the nonnormal
% limiting distribution reported in Figure~\ref{fig: Inference Challenge}.  To highlight the core idea, we
% informally assume that the minimax problem allows for the exchange of the max and
% min operations, and write the empirical problem
% \eqref{eq: sample optimizers} as
% \begin{equation}
% \begin{aligned}
% \widehat{\theta} = \argmin_{\theta\in \RR^{dK}} \widehat{\phi}(\theta, \widehat{\gamma}) 
% \quad \text{with} \quad \widehat{\gamma} \in \argmax_{\gamma \in \Delta^{L}} 
% \left[ \min_{\theta \in \RR^{dK}} \widehat{\phi}(\theta, \gamma) \right].
% \end{aligned}
% \label{eq: simplification}
% \end{equation}
% The above expression describes a dual method of solving the empirical \CGDRO\ problem, where we first solve for the optimal weight $\widehat{\gamma}$ by exchanging the max
% and min operations, and then minimize the objective 
% $\widehat{\phi}(\theta, \widehat{\gamma})$ with respect to $\theta$. Under
% regularity conditions, for a fixed $\gamma \in \Delta^{L}$, the standard
% M-estimation theory implies that
% $\widehat{\theta}(\gamma) \coloneqq \argmin_{\theta} \widehat{\phi}(\theta, \gamma)$
% is asymptotically normal. However, as demonstrated in Figure~\ref{fig: Inference Challenge}, the data-driven
% weight $\widehat{\gamma}$ in \eqref{eq: simplification} may follow a nonstandard limiting distribution and further induce a nonnormal limiting distribution for $\widehat{\theta}$.
We now explain the  insight behind the emergence of the nonnormal
limiting distribution reported in Figure~\ref{fig: Inference Challenge}. Note that, for a fixed $\gamma \in \Delta^{L}$, the standard
M-estimation theory implies that
$\widehat{\theta}(\gamma) \coloneqq \argmin_{\theta} \widehat{\phi}(\theta, \gamma)$
is asymptotically normal under
regularity conditions. However, the computation of $\widehat{\theta}$ involves a data-driven estimator $\widehat{\gamma}$, which possibly leads to nonnormal limiting distribution of $\widehat{\theta}$. In particular, we assume that the minimax problem allows for the exchange of the max and
min operations, and write the empirical problem
\eqref{eq: sample optimizers} as
\begin{equation}
\begin{aligned}
\widehat{\theta} = \argmin_{\theta\in \RR^{dK}} \widehat{\phi}(\theta, \widehat{\gamma}) 
\quad \text{with} \quad \widehat{\gamma} \in \argmax_{\gamma \in \Delta^{L}} 
\left[ \min_{\theta \in \RR^{dK}} \widehat{\phi}(\theta, \gamma) \right].
\end{aligned}
\label{eq: simplification}
\end{equation}
The above expression describes a dual method of solving the empirical \CGDRO\ estimator, where we first solve for the optimal weight $\widehat{\gamma}$, and then minimize the objective 
$\widehat{\phi}(\theta, \widehat{\gamma})$ with respect to $\theta$. The data-driven
weight $\widehat{\gamma}$ in \eqref{eq: simplification} may follow a nonstandard limiting distribution due to nonsmoothness and further induce a nonnormal limiting distribution for $\widehat{\theta}$. Figure~\ref{fig: Inference Challenge} validates such an observation.

In fact, there are two common data-generating mechanisms leading to nonnormal limiting distributions, as shown in Figure \ref{fig: Inference Challenge}. We demonstrate these two mechanisms by specifying the source-specific conditional 
probabilities as 
$
\PP^{(l)}(Y=1|X) = 
\left[1+\exp(-X^\intercal \beta^{(l)})\right]^{-1}$  for $1\leq l\leq 2$ with $\beta^{(1)},\beta^{(2)}\in\RR^d.$
\begin{itemize}
\item \textbf{Scenario 1: sparse shifts between $\beta^{(1)}$ and $\beta^{(2)}$.} When 
$\beta^{(1)}$ differs significantly from $\beta^{(2)}$ only in a few entries, \CGDRO\  tends to put more weight on a source domain with a smaller magnitude of $\|\beta^{(l)}\|_2.$ Consequently, most of  $\widehat{\gamma}_1$ out of 
500 simulations lie near the boundary $0$, as reported in the left panel of Figure 
\ref{fig: Inference Challenge}. The non-negative constraint 
induced by the simplex $\Delta^L$ over $\gamma$ leads to a mixture distribution of 
$\widehat{\gamma}_1$ (with a positive mass at $0$) and further induces both substantial bias in 
$\widehat{\theta}_1$ and a nonnormal distribution of $\widehat{\theta}_1$.
\item \textbf{Scenario 2: similar $\beta^{(1)}$ and $\beta^{(2)}$.} 
In this regime, all entries of $\beta^{(1)}$ are similar to the 
corresponding entries of $\beta^{(2)}$, indicating the existence of a 
weak shift. This causes similarity among $\{{\mu}^{(l)}\}_{1\leq l\leq L}$,
resulting in the instability in calculating the data-driven weight 
$\widehat{\gamma}$. As reported in the middle panel of Figure 
\ref{fig: Inference Challenge}, $\widehat{\gamma}_1$ fluctuates widely 
between $0$ and $1$, leading again to a nonnormal empirical distribution 
for $\widehat{\theta}_1$.
\end{itemize}
The above two scenarios showcase the difficulty of identifying a limiting distribution of $\widehat{\theta}_1$ when there exist certain similarity patterns between the two source sites. 
More generally, {Scenario 1} is a primary example of a 
\textbf{nonregular} regime, which refers to estimators having a 
nonnormal distribution due to the boundary constraint \cite{self1987asymptotic,andrews1999estimation,drton2009likelihood}, i.e., the simplex 
constraint in our problem. {Scenario 2} is an example of an 
\textbf{unstable} regime, referring to estimators having a nonnormal 
limiting distribution due to the instability of the system, i.e., the 
similarity among $\{{\mu}^{(l)}\}_{1\leq l\leq L}$. Moreover, there are scenarios where the limiting distribution of
$\widehat{\theta}$ is normal, and we refer to this regime as the 
\textbf{regular} regime. An example of the regular regime is that
$\beta^{(1)}$ and $\beta^{(2)}$ are very different, as shown in the rightmost panel of Figure~\ref{fig: Inference Challenge}.

\subsection{Main Idea of Perturbation and Filtering}
\label{sec: resampling}
We present the main idea of constructing a uniformly valid confidence interval for $\theta^*_1$, ensuring coverage even when the estimator $ \widehat{\theta}_1$ does not follow an asymptotic normal distribution. 
The intuition comes from the dual identification of $\theta^*$ through a minimization problem. Instead of treating $\theta^*$ as the optimal solution of a minimax optimization problem in \eqref{eq: minimax model}, suppose that we are granted access to a specific optimal population weight $\gamma^* \in \Delta^L$ (though in practice unknown), defined later in \eqref{eq: one saddle} and we are able to express $\theta^*$ as the solution to the following minimization problem,
\begin{equation}
\label{eq: main idea theta_star}
\theta^*=\argmin_{\theta \in \RR^{dK}} \ \phi(\theta,\gamma^*);
\end{equation}
see the following Proposition~\ref{prop: local theta} for the justification of 
% the equivalent expression 
\eqref{eq: main idea theta_star}. With the expression \eqref{eq: main idea theta_star}, we may apply the standard M-estimation theory to establish the asymptotic normality of the semi-oracle global minimizer defined as $\widehat{\theta}(\gamma^*) \coloneqq \argmin_{\theta} \widehat{\phi}(\theta, \gamma^*).$ We then construct the confidence interval for $\theta^*$ by computing the quantiles associated with the limiting  distributions of semi-oracle quantity $\widehat{\theta}(\gamma^*)$. 

Since $ \gamma^*$ is unknown in practice, we generate a collection of perturbed weight vectors and show that our particular perturbation scheme ensures that at least one of the perturbed weights lies close to $\gamma^* $.  These perturbed weights are then used to construct a valid confidence interval, as one of them effectively mimics the true but unknown $\gamma^*$. Intuitively, this procedure separates the overall uncertainty into two components: the uncertainty arising from estimating \( \gamma^* \), and the uncertainty in estimating \( \theta^* \) conditional on knowing \( \gamma^* \). Instead of directly recovering $\gamma^*$, we aim to recover  $\gamma^*_{\rm ap}$, an approximation of $\gamma^*$, defined in \eqref{eq: step 1}. 
We show in the following Theorem \ref{thm: gamma true} that $\gamma^*_{\rm ap}$ is sufficiently close to the targeted optimal weight $\gamma^*$. Consequently, one of the perturbed weight vectors that recovers $\gamma^*_{\rm ap}$ also nearly recovers $\gamma^*$. Throughout the paper, we say that one point is sufficiently close to or nearly recovers another if their distance is of order $\ll n^{-1/2}$.
We illustrate this overall idea in Figure~\ref{fig:illu_resample}.

\begin{figure}[ht!]
\centering
\includegraphics[width=0.9\linewidth]{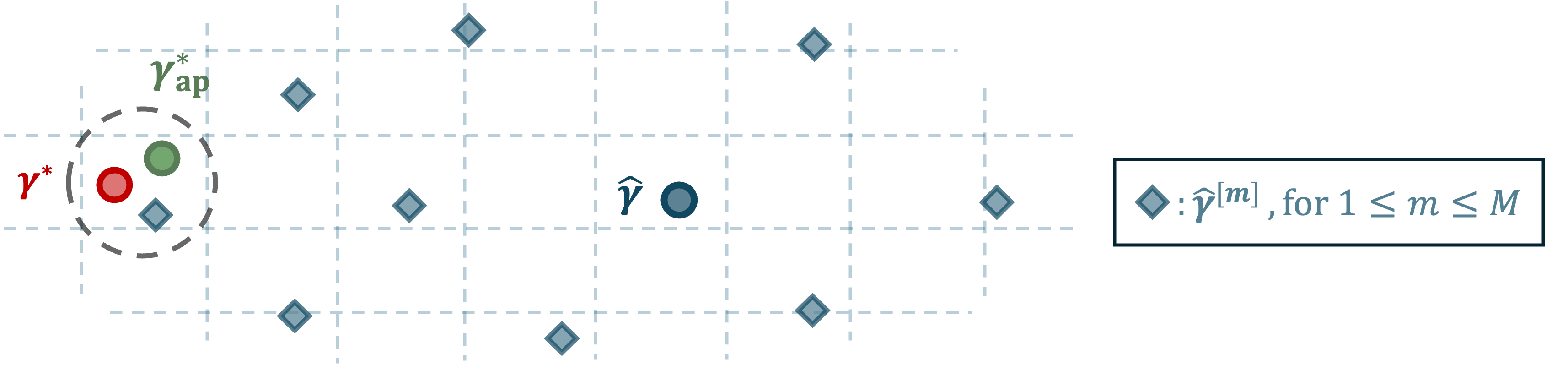}
\caption{Demonstration of the perturbation idea. We generate a collection of $\widehat{\gamma}^{[m]}$ for $1\leq m\leq M$ and show that there exists one of $\{\widehat{\gamma}^{[m]}\}_{1\leq m\leq M}$ that  is nearly equal to $\gamma^*_{\rm ap},$ which is close to $\gamma^*$ by itself.}
\label{fig:illu_resample}
\end{figure}

The main reason for aiming at $\gamma^*_{\rm ap}$ instead of $\gamma^*$ is that  $\gamma^*_{\rm ap}$ is much easier to recover using the perturbation technique proposed below. Particularly, we recall the definition of $\gamma^*_{\rm ap}$ as defined in \eqref{eq: closed form theta}: 
\begin{equation}
\gamma^*_{\rm ap}=\argmax_{\gamma\in \Delta^L} F(\gamma),\;\;  F(\gamma)=-
\frac{1}{2}\left(\U \gamma+\nabla S(\thetainit)\right)^{\intercal}[H(\thetainit)]^{-1}\left(\U \gamma+\nabla S(\thetainit)\right)+ \widehat{\theta}^\intercal \U \gamma,
\label{eq: closed form gamma}
\end{equation}
where $\gamma^*_{\rm ap}$ is expressed as the optimizer of the quadratic-form objective function $F(\gamma)$.

To generate $\widehat{\gamma}^{[m]}$ as an approximation of $\gamma^*_{\rm ap}$, we shall generate a collection of the objective functions $\{\widehat{F}^{[m]}(\cdot)\}_{1\leq m\leq M}$ approximating the quadratic function $F(\cdot)$ and approximate $\gamma^*_{\rm ap}$ by solving this collection of perturbed optimization problems.} In the following, we leverage the limiting distribution of $\widehat{\U} - \U$ to generate the collection of perturbations. Specifically, for the $l$-th column of $\widehat{\U} - \U$, we establish in Supplement Section \ref{subsec: properties of hatmu} that
\begin{equation}
\left[\widehat{\bf V}_{\mu}^{(l)}\right]^{-1/2}\left(\widehat{\mu}^{(l)}-\mu^{(l)}\right)\xlongrightarrow{d} \mathcal{N}(0,{\bf I}_{dK}),\quad \textrm{for $1\leq l\leq L$,}
\label{eq: limit mu}
\end{equation}
where $\widehat{\bf V}_{\mu}^{(l)}$ is an estimated covariance matrix for $\widehat{\mu}^{(l)}$ as defined in Supplement  Section \ref{subsec: properties of hatmu}. We then condition on the observed data and generate $M$ independent $\{\widehat{\U}^{[m]}\}_{1\leq m\leq M}$:
\begin{equation*}
\widehat{\U}^{[m]} = \begin{pmatrix}\widehat{\mu}^{(1,m)},\dots, \widehat{\mu}^{(L,m)}\end{pmatrix} \quad \text{with}\quad \widehat{\mu}^{(l,m)} \sim \Nc\left(\widehat{\mu}^{(l)}, \widehat{\bf V}^{(l)}_{\mu} +\frac{1}{n_l}{\bf I}_{dK}\right) \quad \text{for } 1\leq l\leq L,
\end{equation*}
{where we add a small inflation term $\tfrac{1}{n_l}{\bf I}_{dK}$ to the covariance to avoid degeneracy. }
For each matrix $\widehat{\U}^{[m]}$, we define the perturbed objective function:
\begin{equation*}
\widehat{F}^{[m]}(\gamma)=-\frac{1}{2}\left(\widehat{\U}^{[m]}\gamma+\nabla \widehat{S}(\thetainit)\right)^{\intercal}[\widehat{H}(\widehat{\theta})]^{-1}\left(\widehat{\U}^{[m]}\gamma+\nabla \widehat{S}(\thetainit)\right) +\widehat{\theta}^\intercal \widehat{\U}^{[m]}\gamma,
\end{equation*}
and obtain the perturbed weight vector $\widehat{\gamma}^{[m]}$ by solving
\begin{equation}
\label{eq: resampled gamma}
\widehat{\gamma}^{[m]}=\argmax_{\gamma\in \Delta^{L}} \widehat{F}^{[m]}(\gamma).
\end{equation}

We now provide more intuition for designing the above perturbed optimization problem. The accuracy of approximating $\gamma^*_{\rm ap}$ depends on how well we approximate the objective $F(\gamma)$, which involves three population-level components: $\U, \nabla S(\cdot)$, and $H(\cdot)$. Our goal is to estimate all these components at a rate much smaller than the parametric rate $\sqrt{1/n}$, so that the resulting maximizer remains sufficiently close to $\gamma^{*}_{\rm ap}$. Among these three terms, controlling the estimation error of $\mathcal{U}$ is most crucial. Indeed, one may directly average over the unlabeled target samples $\{\XQ_i\}_{1\leq i\leq N}$ to estimate $\nabla S(\cdot)$ and $H(\cdot)$,  the gradient and Hessian of $S(\cdot)$ defined in \eqref{eq:mu def}, yielding estimators $\nabla \widehat{S}(\cdot)$ and $\widehat{H}(\cdot)$ with an error at the order of $\sqrt{1/N}$. Since we focus on the unsupervised domain adaptation regime with $N$ being much larger than $n$, the estimation errors of $\nabla \widehat{S}(\cdot)$ and $\widehat{H}(\cdot)$ are negligible compared to the parametric rate $\sqrt{1/n}.$  
%However, the optimal rate of estimating $\U$ is of the order $\sqrt{1/n}$, which is achieved by our proposed estimator $\widehat{\U}$ defined in \eqref{eq: U definition}. 
Therefore, once we obtain some perturbed $\widehat{\U}^{[m]}$ lying sufficiently close to the true $\U$, its corresponding perturbed objective \( \widehat{F}^{[m]}(\gamma) \) and perturbed optimizer \( \widehat{\gamma}^{[m]} \)  are respectively sufficiently close to the population-level objective \( F(\gamma) \) and \( \gamma^*_{\rm ap} \).

To formalize the above intuition, we establish in Theorem~\ref{thm: min_gamma_m} that, 
for a sufficiently large resampling size  $M$, there exists an index \( m^* \) such that
\begin{equation}
\|\widehat{\gamma}^{[m^*]} - \gamma^*_{\rm ap}\|_2 \lesssim \frac{{\rm err}(M)}{\sqrt{n}} 
+ \sqrt{\frac{d \log N}{N}}, \quad \text{with} \quad \lim_{M \to \infty} {\rm err}_n(M) = 0.
\label{eq: illu gamma m_star}
\end{equation}
This means that with sufficiently many resampling, there exists $\widehat{\gamma}^{[m^*]}$ closely recovering $\gamma^*_{\rm ap}$, which is itself nearly identical to the population weight $\gamma^*$ as the intuition shown in 
Figure~\ref{fig:illu_resample}. As a result, if we knew this particular index $m^*$ (but we do not know it when implementing the method), it allows us to perform inference as if we knew $\gamma^*$.

%\vspace{1mm}
%\noindent {\bf Filtering.} 

Instead of using all perturbed weights $\{\widehat{\gamma}^{[m]}\}_{1\leq m\leq M}$ defined in \eqref{eq: resampled gamma}, we construct a filtered index set $\mathcal{M}\subseteq[M]$ to exclude those extreme {perturbations}. Note that our goal is to obtain a resampled $\widehat{\U}^{[m]}$ that is sufficiently close to the true $\U$. Based on the limiting distribution in \eqref{eq: limit mu}, an ideal resample $\widehat{\U}^{[m]}$ should not deviate significantly from $\widehat{\U}$. This motivates filtering out extreme resamples that fall in the tails of the $\widehat{\U}$'s distribution. In particular, for a prespecified significance level  $\alpha_0\in (0,0.01]$, we define the filtered index set $\mathcal{M}$ as follows:
\begin{equation}
    \mathcal{M} = \left\{1\leq m\leq M: \max_{1\leq l\leq L}\max_{1\leq j\leq dK}\frac{\left|\widehat{\mu}^{(l,m)}_j - \widehat{\mu}^{(l)}_j\right|}{\sqrt{[\widehat{\bf V}^{(l)}_\mu]_{j,j}+ 1/n_l}} \leq (1+\eta_0)\cdot z_{\alpha_0/(dKL)}\right\},
    \label{eq: filtered set}
\end{equation}
where $\eta_0>0$ is a small positive constant to account for finite-sample variability (with a default value of $0.1$). {The construction of $\mathcal{M}$ achieves two purposes simultaneously: (i) The index set excludes extreme resamples far away from $\widehat{\U}$. (ii) $\mathcal{M}$ contains at least one index $m^*$ ensuring that the corresponding $\widehat{\gamma}^{[m^*]}$ lies sufficiently close to $\gamma^*$,} even though the exact index $m^*$ is unknown. 

\subsection{Inferential Procedure}
\label{sec: procedure}

We are now ready to  construct the confidence interval (CI) for $\theta^*_1$ using $\mathcal{M}$.
%As implied by the expression of $\theta^*$ in \eqref{eq: main idea theta_star}, the semi-oracle estimator $\widehat{\theta}(\gamma^*)\coloneqq \argmin_{\theta}\widehat{\phi}(\theta,\gamma^{*})$ is asymptotically normal and centered on $\theta^* = \argmin_{\theta} \phi(\theta, \gamma^*)$.
As discussed in Section \ref{sec: resampling}, we know that, with a high probability, there exists $m^*\in \mathcal{M}$ such that $\widehat{\gamma}^{[m^*]}\approx \gamma^*$. This implies that the limiting distribution of $\widehat{\theta}(\widehat{\gamma}^{[m^*]})\coloneqq \argmin_{\theta} \widehat{\phi}(\theta, \widehat{\gamma}^{[m^*]})$ is asymptotically normal and centered on $\theta^* = \argmin_{\theta} \phi(\theta, \gamma^*)$, enabling us to construct CI for $\theta^*_1$ using the quantiles of the limiting distribution of $\widehat{\theta}_1(\widehat{\gamma}^{[m^*]})$. However, since the exact identity of the index $m^*$ is unknown in practice, we construct the individual interval for each $m \in \mathcal{M}$ and take their union to form the final CI.

For each $\widehat{\gamma}^{[m]}$ constructed via \eqref{eq: resampled gamma}, we construct a perturbed estimator of $\theta^*$ as
\begin{equation}
\widehat{\theta}^{[m]}= \argmin_{\theta\in \RR^{dK}}\widehat{\phi}(\theta,\widehat{\gamma}^{[m]})=\argmin_{\theta\in \RR^{dK}}\sum_{l=1}^{L}\widehat{\gamma}^{[m]}_l\cdot \theta^{\intercal}\widehat{\mu}^{(l)}+\widehat{S}(\theta).
\label{eq: m-opt}
\end{equation}
When $N \gg \max\{n_1,\cdots,n_L\}$, the error of $\widehat{\theta}^{[m]}$ approximately decomposes as:
\begin{equation}
\begin{aligned}
\widehat{\theta}^{[m]}-\theta^*\approx
-\left[H(\theta^*)\right]^{-1}\sum_{l=1}^{L}\left(\widehat{\gamma}^{[m]}_l-{\gamma}^*_l\right){\mu}^{(l)}-\left[H(\theta^*)\right]^{-1}\sum_{l=1}^{L}{\gamma}^*_l\left(\widehat{\mu}^{(l)}-{\mu}^{(l)}\right),
\end{aligned}
\label{eq: limit dist decomp}
\end{equation}
where $H(\theta^*)$ denotes the Hessian matrix of $S(\theta^*)$; see the exact statement in Supplement Proposition \ref{prop: simplify theta_hat_m}. In the above decomposition, the first term on the right-hand side represents the uncertainty of estimating $\gamma^*$ with a perturbed weight vector  $\widehat{\gamma}^{[m]}$, while the second term captures the uncertainty of estimating $\mu^{(l)}$. As illustrated in Figure~\ref{fig:illu_resample}, our perturbation strategy ensures that there exists at least one $\widehat{\gamma}^{[m]}$ that is nearly the same as $\gamma^*$. Consequently, 
we mainly quantify the uncertainty arising from $-\left[H(\theta^*)\right]^{-1}\sum_{l=1}^{L}{\gamma}^*_l\left(\widehat{\mu}^{(l)}-{\mu}^{(l)}\right)$, leveraging the asymptotic normality of $\widehat{\mu}^{(l)}$.
Given a pre-specified significance level $\alpha$ with $\alpha>\alpha_0$, we define $\alpha' = \alpha-\alpha_0$ and construct {
\begin{equation}
{\rm Int}^{[m]}=\left(\widehat{\theta}^{[m]}_1-z_{\alpha'/2} {\widehat{\rm se}^{[m]}},\widehat{\theta}^{[m]}_1+z_{\alpha'/2}{\widehat{\rm se}^{[m]}}\right),
\label{uq: ci_m}
\end{equation}
where $z_{\alpha'/2}$ denotes the upper $\alpha'/2$ quantile of the standard normal distribution, and ${\widehat{\rm se}^{[m]}}$
represents the standard error to the first entry of $-\left[H(\theta^*)\right]^{-1}\sum_{l=1}^{L}{\gamma}^*_l\left(\widehat{\mu}^{(l)}-{\mu}^{(l)}\right)$ with its explicit expression provided in Supplement Section \ref{subsec: hat_V_m}.}

\begin{figure}[ht!]
    \centering
    \includegraphics[width=0.95\linewidth]{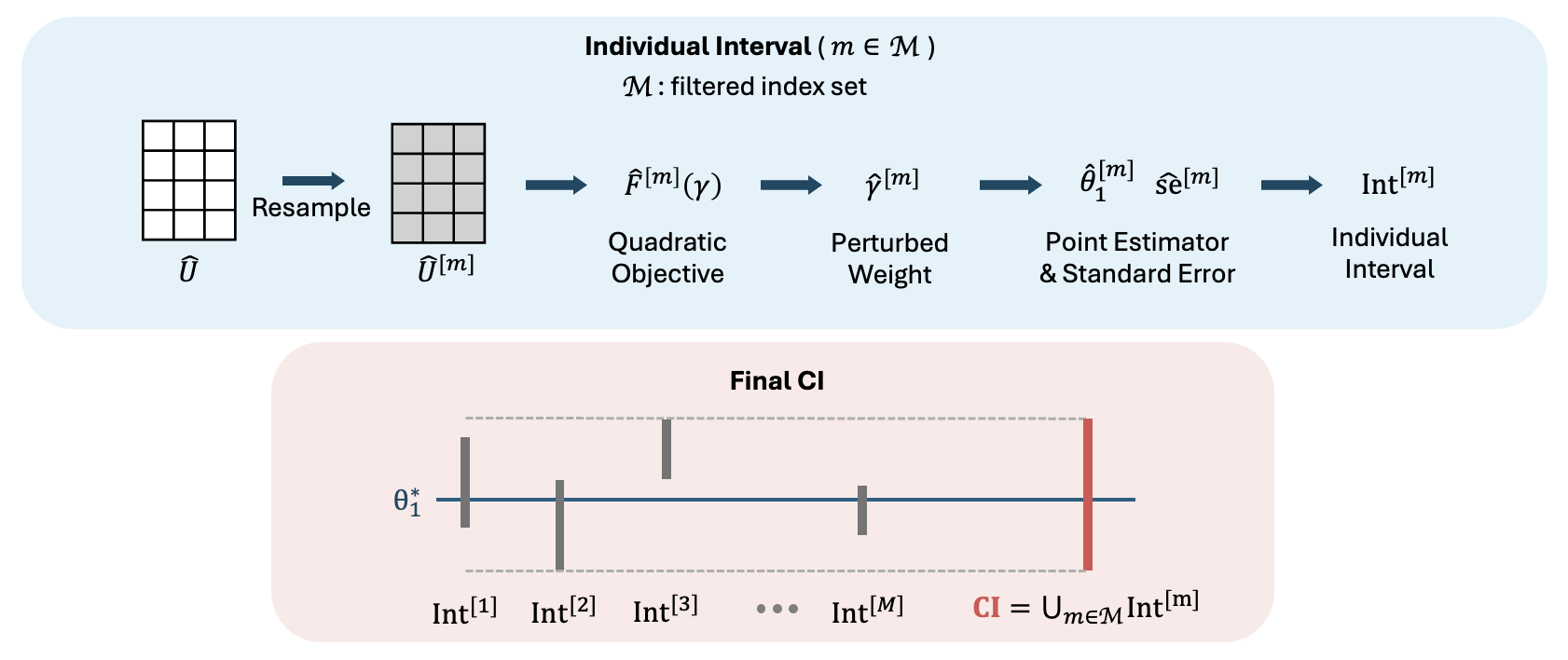}
    \caption{The top panel illustrates the construction of ${\rm Int}^{[m]}$, 
    and the bottom panel illustrates the construction of our final confidence interval by taking a union of ${\rm Int}^{[m]}$ over $m\in \mathcal{M}$.}
    \label{fig:procedure}
 \end{figure}

Since we do not know the index $m^*$, we take a union of the {perturbed} intervals for $m\in \mathcal{M}$,
\begin{equation}
{\rm CI}= \bigcup_{m\in \mathcal{M}}{\rm Int}^{[m]}.
\label{uq:union ci}
\end{equation}
Regarding a level-$\alpha$ test of  $H_0: \theta^{*}_1=0$, we reject $H_0$ if $0\not \in {\rm CI}$. We illustrate the proposed perturbation-based inferential approach in Figure \ref{fig:procedure}.

The validity of CI in \eqref{uq:union ci} requires $M$ to be sufficiently large. In practice, we typically choose $M = 500$ or $1000$, which achieves uniform coverage in the experiments presented in Section~\ref{sec: simus}. 
Even though we are taking a union to construct the final CI, we have shown that our proposed CI attains a parametric length in the following Theorem \ref{thm: inference-length}. Moreover, we observe that the length of the resulting CI does not get much longer with an increasing $M$; see Appendix \ref{appendix: varyM} for the empirical results. %\Zijian{Zhenyu, provide the location of the observation.}

Regarding the computational cost of the proposed perturbation-based inferential procedure, for each index $m\in \mathcal{M}$, one solves a  strongly convex problem in \eqref{eq: resampled gamma} to obtain the weight vector $\widehat{\gamma}^{[m]}$. Given $\widehat{\gamma}^{[m]}$, we then compute the estimator $\widehat{\theta}^{[m]}$ {by solving another convex problem} in \eqref{eq: m-opt}. Thus, the computation of the perturbed estimator $\widehat{\theta}^{[m]}$ requires solving only two convex problems, both of which can be solved computationally efficiently. Moreover, in practice, we further accelerate the entire procedure by parallelizing the computations across $M$ independent {perturbations}. For example, in our simulation experiments under setting \textbf{(S4)} in Section~\ref{sec: simus}, the inferential procedure takes approximately 20 seconds in total with $M=500$. Compared to the methods that require repeated sampling and retraining based on the original data, our method avoids repeatedly solving machine learning models when constructing the DML estimators of ${\U}=(\mu^{(1)}, \mu^{(2)},..., \mu^{(L)})$. 
This difference results in improvements in both computational and memory efficiency.

\begin{Remark}[Inference for Minimax Problem] \cite{guo2022statistical} employed perturbation techniques to address nonregular inference in minimax optimization. However, their method relied on a closed-form solution under squared loss in regression settings, which is hard to generalize to other loss functions. In contrast, we tackle the more challenging cross-entropy loss without a closed-form solution and devise a computationally efficient {perturbation-based} method by carefully designing a surrogate problem. 
We address the nonstandard inference through decoupling the origins of the randomness into two components: the nonstandard limiting distribution of the weight estimator $\widehat{\gamma}$ and the standard limiting distribution of $\widehat{\theta}(\gamma^*)=\argmin_{\theta}\widehat  \phi(\theta,\gamma^*)$. Thus, we introduce a novel perturbation approach to handle the nonstandard component $\widehat{\gamma}$, while using the Gaussian quantile to track the standard component $\widehat{\theta}(\gamma^*)$. 
Such an inferential tool addresses the challenging nonstandard inference for a general class of minimax problems and could be of independent interest for other problems. %
\label{remark: resampling}
\end{Remark}

\section{Theoretical Justification for Uncertainty Quantification}
\label{sec: theory UQ}
In this section, we provide a theoretical justification for the perturbation-based inference method introduced in Section \ref{sec: resampling}, and analyze the confidence interval constructed in \eqref{uq:union ci}.
In Section~\ref{subsec: gamma_star}, we formally define the population weight vector $\gamma^*$, and show that the approximate weight $\gamma^*_{\rm ap}$, defined in \eqref{eq: step 1}, lies sufficiently close to $
\gamma^*$. In Section~\ref{subsec: justification CI}, we establish theoretical guarantees for the proposed confidence interval, including its coverage and length results.

\subsection{Definition of the Population Optimal Weight $\gamma^*$}
\label{subsec: gamma_star}
As discussed in Section \ref{subsec: gamma_star}, the validity of our perturbation-based inference method relies on the existence of a population weight vector $\gamma^*$ with two properties: (i) the population \CGDRO\ model $\theta^*$ can be characterized as the minimizer of $\phi(\theta,\gamma^*)$ such that $\theta^* = \argmin_{\theta\in \RR^{dK}}\phi(\theta,\gamma^*)$, and (ii) the approximation weight $\gamma^*_{\rm ap}$ lies sufficiently close to $\gamma^*$. 

To construct such a $\gamma^*$, we first define a local neighborhood around $\theta^*$ as follows:
\begin{equation}
\Theta_{\rm loc}=\left\{\theta\in \RR^{dK}: \|\theta-\theta^*\|_2\leq 2\tau \sqrt{d/n}\right\} \quad \text{with} \; \tau \; \text{defined in} \; \eqref{eq: final rate}.
\label{eq: Theta local}
\end{equation}
By Theorem~\ref{thm: final rate}, the empirical estimator $\widehat{\theta}$ lies in $\Theta_{\rm loc}$ with high probability. Upon this local region, we then define $\gamma^*\in \Delta^L$ as the solution to the following optimization problem:
\begin{equation}
\gamma^*\in \argmax_{\gamma\in \Delta^{L}}F_{\rm loc}^*(\gamma) \quad\text{with}\quad F_{\rm loc}^*(\gamma)\coloneq\min_{\theta\in \Theta_{\rm loc}}\phi(\theta,\gamma).
\label{eq: one saddle}
\end{equation}

The following proposition confirms that $\theta^*=\argmin_{\theta\in \RR^{dK}}\phi(\theta, \gamma^*)$. This implies that if $\gamma^*$ were known, the population \CGDRO\ model $\theta^*$,  which was originally defined through the minimax problem in \eqref{eq: minimax model}, reduces to a standard minimization problem.
\begin{Proposition}
\label{prop: local theta}
For any $\gamma^*$ defined in \eqref{eq: one saddle}, the population \CGDRO\ model $\theta^*$ in \eqref{eq: minimax model} satisfies $\theta^* = \argmin_{\theta \in \Theta_{\rm loc}} \ \phi(\theta,\gamma^*) = \argmin_{\theta \in \RR^{dK}} \ \phi(\theta,\gamma^*).$
\end{Proposition}

We now explain why restricting to the local set $\Theta_{\rm loc}$ ensures that the constructed $\gamma^*$ lies close to the approximate weight $\gamma^*_{\rm ap}$. By their definitions in \eqref{eq: step 1} and \eqref{eq: one saddle}, $\gamma^*_{\rm ap}$ and $\gamma^*$ are solutions to dual problems with the objective functions $\phi_{\rm ap}(\theta,\gamma)$ and $\phi(\theta,\gamma)$, respectively. Since $\phi_{\rm ap}(\theta,\gamma)$ provides a second-order approximation to $\phi(\theta,\gamma)$ around $\widehat{\theta}$ as defined in \eqref{eq: approximate minimax}, the two objectives are uniformly close to each other for all $\theta\in \Theta_{\rm loc}$. Consequently, their corresponding solutions $\gamma^*_{\rm ap}$ and $\gamma^*$, must also be close.

The following theorem establishes that $\gamma^*_{\rm ap}$ is sufficiently close to $\gamma^*$.
\begin{Theorem} 
\label{thm: gamma true}
Suppose that {\rm Conditions \ref{cond:A1}--\ref{cond:A4}} hold. Then there exists a positive constant $c_1>0$ such that with probability at least $1-n^{-c_1 d}-e^{-c_1 t^2}-\delta_n$,
\begin{equation*}
\|{\gamma}^*_{\rm ap}-\gamma^*\|_2 \lesssim \tau^2(d/n)^{3/4},\quad \textrm{with}\quad \tau= C\left(1 + \frac{1}{\sigma_L^2(\U)}\right) t.
\end{equation*}
where the value $t$ satisfies $c_0\leq t\ll \sigma_L^2(\U) (n/d)^{1/4}$ and the constant $c_0>0$ and the sequence $\delta_n \to 0$ specified in {\rm Condition~\ref{cond:A1}}.
\end{Theorem}

While Theorem \ref{thm: quad convergence} already establishes the closeness between the primal variable $\theta$ of the original and approximate minimax problems, Theorem \ref{thm: gamma true} quantifies the approximation error associated with the dual variable $\gamma$. We refer to the analysis in Theorem~\ref{thm: gamma true} as the localization technique since its proof builds on constructing the local set $\Theta_{\rm loc}$. Beyond its role in our theoretical development, this localization approach may be of independent interest for the analysis of empirical minimax problems involving approximate objective functions.

The condition $c_0\leq t\ll \sigma_L^2(\U)(n/d)^{1/4}$  implies that $\sigma_L(\U)\gg (d/n)^{1/8}$, {suggesting that $\{\mu^{(l)}\}_{l\in [L]}$ cannot be overly collinear, meaning that the data distributions of source domains should not be identical but allow for a certain degree of similarity.} In the special case where $\sigma_L(\U)$ is lower bounded by a constant, the condition on $t$ is satisfied for any $c_0\leq t\ll (n/d)^{1/8}$. Similar to the discussion on the range of $t$ in \eqref{eq: t condition}, one may choose $t$ as a large positive constant or allow it to grow slowly with $n$, such as $t=\sqrt{\log n}$.

\subsection{Justification for Confidence Interval}
\label{subsec: justification CI}
To justify the CI in \eqref{uq:union ci} from a theoretical perspective, the key is to establish the existence of one perturbed weight $\widehat{\gamma}^{[m^*]}$ that nearly recovers $\gamma^*$. %With this, the remaining proof is based on the decomposition \eqref{eq: limit dist decomp} and the standard M-estimation inferential theory. 
For clarity, in this subsection, we focus on the fixed dimension setting, although our analysis can be directly extended to a growing dimension regime at the expense of more involved notation. We define  $\err(M)$ as a decreasing function of the resampling size $M$, 
\begin{equation}
    \err(M)= \left(\frac{1}{2}c^*(\alpha_0,d){\rm Vol}(dK)\right)^{-1/dK} \left(\frac{\log n}{M}\right)^{1/dKL},
    \label{eq: errM}
\end{equation}
where $\alpha_0\in (0, 0.01]$ is the pre-specified constant used to construct the filtered index set $\mathcal{M}$ in \eqref{eq: filtered set}, $c^*(\alpha_0,d)$ is a positive value depending on $\alpha_0$ and $d$ (with its explicit form given in Supplement Section \ref{subsec: expression of c_star}), 
and ${\rm Vol}(dK)$ denotes the volume of the unit ball in $\RR^{dK}$. 
When the dimension $d$ is fixed, both $c^*(\alpha_0,d)$ and ${\rm Vol}(dK)$ become positive constants independent of the sample size $n$ and the resampling size $M$. Consequently, $\err(M)$ scales as $(\log n/ M)^{1/dKL}$, which tends to zero as the resampling size $M$ increases.

We now introduce the following limiting distribution condition of $\widehat{\U}=(\widehat{\mu}^{(1)},...,\widehat{\mu}^{(L)})$.
\begin{condition}
In the fixed dimension regime,
the estimators $\{\widehat{\mu}^{(l)}\}_{1\leq l\leq L}$ satisfy
    \begin{equation}
    \left[{\bf V}_{\mu}^{(l)}\right]^{-1/2}\left(\widehat{\mu}^{(l)}-\mu^{(l)}\right)\xlongrightarrow{d} \mathcal{N}(0,{\bf I}_{dK}),\;\;\textrm{and}\; \;\|n_l(\widehat{\bf V}_\mu^{(l)} - {\bf V}_\mu^{(l)})\|_2\xlongrightarrow{p} 0,
    \label{eq: limiting condition}
    \end{equation}
    where ${\bf V}_\mu^{(l)}$ and $\widehat{\bf V}_\mu^{(l)}$ denote the population and estimated covariance matrices of $\mu^{(l)}$. Additionally, there exist  $c_1, c_2 > 0$ such that the scaled covariance matrix satisfies $\|n_l {\bf V}_\mu^{(l)}\|_2 \leq c_1$, and its diagonal entries are bounded below with $\min_{1 \leq j \leq dK} n_l [{\bf V}_\mu^{(l)}]_{j,j} \geq c_2$.
    \label{cond:A5}
\end{condition}
Compared to Condition \ref{cond:A1}, this condition additionally requires that each $\widehat{\mu}^{(l)}$ admits a valid limiting distribution together with a consistent covariance estimator  $\widehat{\bf V}_\mu^{(l)}$. In Supplement Section \ref{subsec: properties of hatmu}, we provide the explicit expressions of both ${\bf V}_\mu^{(l)}$ and $\widehat{\bf V}_\mu^{(l)}$, and verify that this condition holds for the proposed DML estimators $\widehat{\mu}^{(l)}$ in \eqref{eq: no shift est} and \eqref{eq: DR component}  under covariate shift and non-shift regimes, respectively.

In the following theorem, we control $\min_{m\in \mathcal{M}}\|\hgammam - \gamma^*_{\rm ap}\|_2.$ 
\begin{Theorem} 
Suppose that {\rm Conditions \ref{cond:A1}--\ref{cond:A5}} hold. Then
\begin{equation*}
\begin{aligned}
\liminf_{n\to\infty}\liminf_{M\to\infty} \mathbf{P}\left\{\min_{m\in \mathcal{M}}\|\hgammam - \gamma^*_{\rm ap}\|_2 \lesssim \frac{1}{\sigma_L^2(\U)}\cdot \left(\frac{\err(M)}{\sqrt{n}} + \sqrt{\frac{\log N}{N}}\right)\right\}\geq 1-\alpha_0,
\end{aligned}
\end{equation*}
where $\alpha_0\in (0, 0.01]$ is the pre-specified constant used to construct $\mathcal{M}$ in \eqref{eq: filtered set}.
\label{thm: min_gamma_m}
\end{Theorem}

Note that the error term $\err(M)$ vanishes as the resampling size $M$ increases, indicating that the best resampled discrepancy is of a much smaller order than $\sqrt{1/n}$ for a sufficiently large $M$, {if $\sigma_L^2(\U)\gg \sqrt{n\log N/N}$.}
The pre-specified constant $\alpha_0\in (0,0.01]$ is the small probability that is reserved such that $\widehat{\mu}^{(l)}$ does not lie on the extreme tail of the limiting distribution in \eqref{eq: limiting condition}.
The theorem shows that there exists an index $m^*\in \mathcal{M}$ such that its corresponding $\widehat{\gamma}^{[m^*]}$ is sufficiently close to $\gamma^*_{\rm ap}$, where $m^*$ is defined as the minimizer of $\min_{m\in \mathcal{M}}\|\widehat{\gamma}^{[m]} - \gamma^*_{\rm ap}\|_2$.
Combining Theorems \ref{thm: gamma true} and \ref{thm: min_gamma_m}, we establish  
\begin{equation}
  \min_{m\in \mathcal{M}}\|\hgammam - \gamma^*\|_2 \leq 
  \min_{m\in \mathcal{M}}\|\hgammam - \gamma^*_{\rm ap}\|_2+
  \|\gamma^*_{\rm ap}-\gamma^*\|_2\ll \sqrt{1/n}.
  \label{eq: main idea gamma m_star}
\end{equation}
It means that there exists a perturbed weight $\widehat{\gamma}^{[m^*]}$ that converges to $\gamma^*$ at a rate faster than $\sqrt{1/n}.$ 
The following theorem establishes the coverage validity of the proposed CI in \eqref{uq:union ci}.
\begin{Theorem}
Suppose $\sigma_L^2(\U)\gg \max\{\sqrt{n\log N/N}, \; \sqrt{\log n/n^{1/4}}\}$ and {\rm Conditions \ref{cond:A1}--\ref{cond:A5}} hold. Then the ${\rm CI}$ proposed in \eqref{uq:union ci} satisfies
\begin{equation}
\liminf_{n\to\infty}\ \liminf_{M\to\infty}\  \mathbf{P}(\theta_1^* \in {\rm CI})\geq 1-\alpha,
\label{eq: resampling coverage}
\end{equation}
where $\alpha\in (0,1/2)$ is the pre-specified significance level.
\label{thm: inference-coverage}
\end{Theorem}
The above theorem suggests that the proposed CI achieves valid coverage uniformly, even when $\widehat{\theta}_1$ exhibits a nonnormal limiting distribution, as illustrated in Figure \ref{fig: Inference Challenge}. The imposed condition $\sigma_L^2(\U)\gg \max\{\sqrt{n\log N/N}, \; \sqrt{\log n/n^{1/4}}\}$ requires that $\{\mu^{(l)}\}_{1\leq l\leq L}$ are not overly collinear. %\Yifan{Does it mean that Scenarios 2 do not appear?}
When this condition holds, we show in \eqref{eq: main idea gamma m_star} that there exists one perturbed weight vector $\widehat{\gamma}^{[m^*]}$ that accurately approximates $\gamma^*$ with $\|\widehat{\gamma}^{[m^*]}-\gamma^*\|_2\ll \sqrt{1/n}$. 

Next, we analyze the length of the proposed confidence interval in \eqref{uq:union ci}.
\begin{Theorem}
Suppose that {\rm Conditions} {\rm \ref{cond:A1}}--{\rm \ref{cond:A5}} hold. Then there exists a positive constant $c_1>0$ such that the {\rm CI}'s length ${\bf Leng}({\rm CI})$ satisfies
\begin{equation}
    \liminf_{n\to\infty}\ \liminf_{M\to\infty} \ \mathbf{P}\left({\bf Leng}({\rm CI})\lesssim \left(t+\frac{z_{\alpha_0}}{\sigma_L^2(\U)}\right)\frac{1}{\sqrt{n}}\right)\geq 1-\alpha_0-e^{-c_1 t^2},
    \label{eq: resampling length}
\end{equation}
where $t$ satisfies $c_0\leq t \ll \sigma_L^2(\U) n^{1/8}$ with the constant $c_0>0$ from {\rm Condition \ref{cond:A1}}, $\alpha_0\in (0,0.01]$ is the pre-specified parameter used to construct $\mathcal{M}$ in \eqref{eq: filtered set}, and $z_{\alpha_0}$ denotes the upper $\alpha_0$ quantile of the standard normal distribution.
\label{thm: inference-length}
\end{Theorem}
A few remarks are in order for Theorem \ref{thm: inference-length}. {Firstly, although the CI in \eqref{uq:union ci} is defined as the union of $|\mathcal{M}|$ individual intervals, its length still achieves the parametric rate $\sqrt{1/n}$ with high probability, if $\sigma_L(\U)$ is of a constant order. Secondly, the control of the CI length consists of two parts: the deviation of perturbed intervals'  centers $\max_{m\in \mathcal{M}}|\widehat{\theta}^{[m]}_1 - \theta^*_1|$ and  perturbed intervals with a radius of order $\sqrt{1/n}$. We bound the deviation of the centers $\max_{m\in \mathcal{M}}|\widehat{\theta}^{[m]}_1 - \theta^*_1|$ by $\max_{l}\|\widehat{\mu}^{(l)} - \mu^{(l)}\|_2$, which in turn is controlled by $t/\sqrt{n}$ with a probability larger than $1-\exp(-c_1t^2)$, as specified in Condition~\ref{cond:A1}.} Third, the condition $c_0\leq t\ll \sigma_L^2(\U)n^{1/8}$ implies that $\sigma_L(\U)\gg n^{-1/16}$, indicating that $\{\mu^{(l)}\}_{l\in [L]}$ cannot be too collinear. In the special case, when $\sigma_L(\U)$ is bounded below by a positive constant, the requirement on $t$ is satisfied for any $c_0\leq t\ll n^{1/8}$.
Lastly, we note that both the coverage and length properties can be extended directly to the settings for any linear transformation of $\theta^*$,  enabling inference on general linear contrasts of the parameter vector.

\section{Simulations}
\label{sec: simus}

We start with the description of the simulated data setup. %
For each source domain $1\leq l\leq L$, we generate $\{\X{l}_i,\Y{l}_i\}_{1\leq i\leq n_l}$ in an i.i.d. fashion, where $\X{l}_i\sim \Nc(0_d, {\bf I}_d)$ and $\Y{l}_i$ are sampled from a multinomial distribution following
\begin{equation}
    \PP^{(l)}(Y=c|X) = \frac{\exp(\phi^{(l)}_c(X))}{\sum_{k=0}^K\exp(\phi_{k}^{(l)}(X))}, \quad \textrm{for $c=0,1,\dots, K$.}
    \label{eq: phi_cl}
\end{equation} 
The functions $\{\phi_c^{(l)}(\cdot)\}_{l,c}$ could be linear and nonlinear that vary across different settings. We set $n_l = n$ for all $1\leq l\leq L$ and $N=10,000$, and vary $n, d, L, K, \{\phi^{(l)}_c\}_{l,c}$ and the target covariate distribution $\QQ_X$ across different setups. To ensure consistency across simulation rounds, all relevant parameters are generated once and held fixed throughout the 500 independent replications. All reported results below take averages over these 500 simulations. Detailed specifications of each setting in this section are deferred to Supplement Section \ref{sec: setup simus}.
%For the target domain, the unlabeled data $\{X^\QQ_j\}_{j\in [N]}$ are i.i.d.~drawn from $\QQ_X$, which will be specified for each simulation regime. 

\vspace{1mm}
\noindent {\bf Estimation Error.} We empirically evaluate the convergence behavior of the estimator $\widehat{\theta}$, under two different settings \textbf{(S1)} and \textbf{(S2)}.
The difference between the two settings lies in the complexity of the classification task. While \textbf{(S1)} is a linear binary classification problem, \textbf{(S2)} introduces a highly nonlinear and multi-class classification task. We utilize both settings to highlight the effectiveness of our method under both parametric and nonparametric regimes.

As discussed in Section \ref{sec: double robustness}, solving $\widehat{\theta}$ involves the construction of DML estimators $\{\widehat{\mu}^{(l)}\}_{1\leq l\leq L}$, which requires estimating the density ratio $\widehat{\omega}^{(l)}$ and the conditional probability $\widehat{f}^{(l)}$ for each source $1\leq l\leq L$. For \textbf{(S1)}, we employ logistic regression, but we also implement the XGBoost algorithm and observe similar results. For \textbf{(S2)}, we apply XGBoost with cross-validation to handle this more complex task.

\begin{figure}[ht!]
    \centering
    \includegraphics[width=0.8\linewidth]{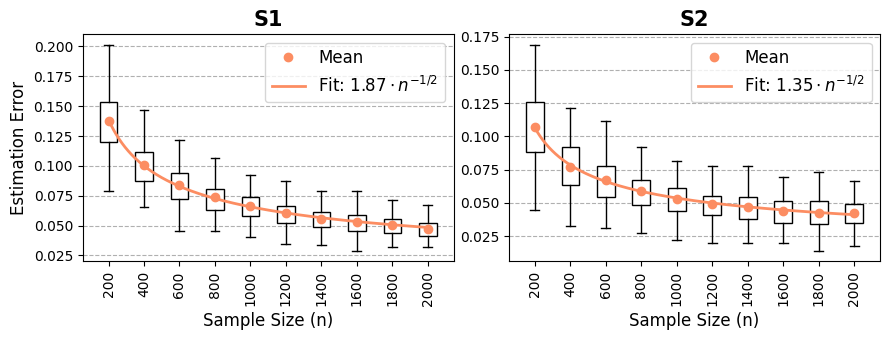}
    \caption{The estimation error $\|\widehat{\theta} - \theta^*\|_2/\sqrt{d}$ in settings \textbf{(S1)} and \textbf{(S2)} with the source sample size $n$ varied across the range of $[200, 2000]$. The results are summarized from 500 simulations, and the orange curves are fitted on the empirical means of 500 estimation errors for each sample size.}
    \label{fig:estimation}
\end{figure}

In Figure \ref{fig:estimation}, for each source sample size $n\in \{200,400,..., 2000\}$, we report the box plot of the estimation errors $\|\widehat{\theta}-\theta^*\|_2/\sqrt{d}$ over 500 simulations, where the normalization by $\sqrt{d}$ ensures the comparative scale across settings. The orange dots on each box correspond to the empirical mean estimation error at the given sample size. We then fit the orange curve to these empirical means across all considered sample sizes to visualize the convergence rate. It is evident that the estimation error of $\widehat{\theta}$ decays at the rate of $\sqrt{1/n}$, aligning with our theoretical result established in Theorem \ref{thm: final rate}.

\vspace{1mm}
\noindent {\bf Inference Validation.} In this section, we assess the effectiveness of our inference procedure for constructing the confidence interval for the target parameter $\theta_1^*$. %
We introduce two settings \textbf{(S3)} \textbf{(S4)} involving parametric modeling, and provide additional simulations under more complex, highly nonlinear setups in Supplement Section \ref{appendix: infer simus}.
\begin{itemize}
    \item[\textbf{(S3)}] This captures a spectrum of regimes ranging from \textbf{regular} to \textbf{nonregular}, controlled by the sparse perturbation parameter {$\delta\in \{0, 0.5, 1.0, ..., 4.0\}$}. As $\delta$ increases, the discrepancy between the two source grows, increasing the nonregularity of the problem.
    \item[\textbf{(S4)}] This captures a spectrum of regimes ranging from \textbf{regular} to \textbf{unstable}, governed by the instability parameter {$\sigma \in [0.1, 0.5]$}. When $\sigma$ is small, the data distribution across sources are nearly identical, inducing strong instability. As $\sigma$ increases, the variability between sources grows, alleviating the instability and leading to a more regular setting.
\end{itemize}

In addition to our proposed approach as detailed in Section \ref{sec: resampling}, we consider the following benchmark inference procedures for comparison in terms of coverage and interval length: 
\begin{itemize}
    \item \texttt{Normality.} This method assumes  $(\widehat{\theta}_1 - \theta^{*}_1)/\widehat{\rm se}_{\rm ora} \xlongrightarrow{d} \mathcal{N}(0, 1)$, where $\widehat{\rm se}_{\rm ora}$ denotes the oracle standard error of $\widehat{\theta}_1$ calculated from 500 simulations. The corresponding $1-\alpha$ confidence interval is constructed as:
    $(\widehat{\theta}_1-z_{\alpha/2} \widehat{\rm se}_{\rm ora}, \widehat{\theta}_1+z_{\alpha/2} \widehat{\rm se}_{\rm ora})$. %
    \item \texttt{Bootstrap.} 
    For the $b$-th bootstrap with $b\in [B]$, we randomly draw observations with replacement from the source and target datasets, and compute the corresponding estimator $\widehat{\theta}^{b}_1$. Repeating this process $B=500$ times yields a collection of estimators $\{\widehat{\theta}^b_1\}_{b\in [B]}$, from which we compute the empirical standard error $\widehat{\rm se}_{\rm boot}$. We construct the bootstrap confidence interval as $(\widehat{\theta}_1 - z_{\alpha/2} \widehat{\rm se}_{\rm boot}, \widehat{\theta}_1 + z_{\alpha/2} \widehat{\rm se}_{\rm boot})$. %
    \item \texttt{Oracle Bias-awareness (OBA).} 
    This method augments \texttt{Normality} by incorporating oracle knowledge of the bias.
    %It assumes $(\widehat{\theta}_1 - \theta^{*}_1)/\widehat{\rm se}_{\rm ora} \xlongrightarrow{d} \mathcal{N}(b, 1)$, where $b$ represents the oracle bias. 
    Following the work \cite{armstrong2020bias}, it uses the oracle knowledge of the bias $|\mathbb{E} \widehat{\theta}_1-\theta^{*}_1|$ and form the oracle bias-aware CI as
    $(\widehat{\theta}_1-\chi, \widehat{\theta}_1+\chi)$, 
    where $\chi=\widehat{\rm se}_{\rm ora} \cdot \sqrt{\mathrm{cv}_\alpha(|\mathbb{E} \widehat{\theta}_1-\theta^{*}_1|^2 / \widehat{\rm se}_{\rm ora}^2)}$, and $\mathrm{cv}_\alpha\left(B^2\right)$ denotes the $1-\alpha$ quantile of the non-central $\chi^2$ distribution with one degree of freedom and non-centrality parameter $B^2$ .
\end{itemize}

\begin{figure}[ht!]
    \centering
    \includegraphics[width=0.7\linewidth]{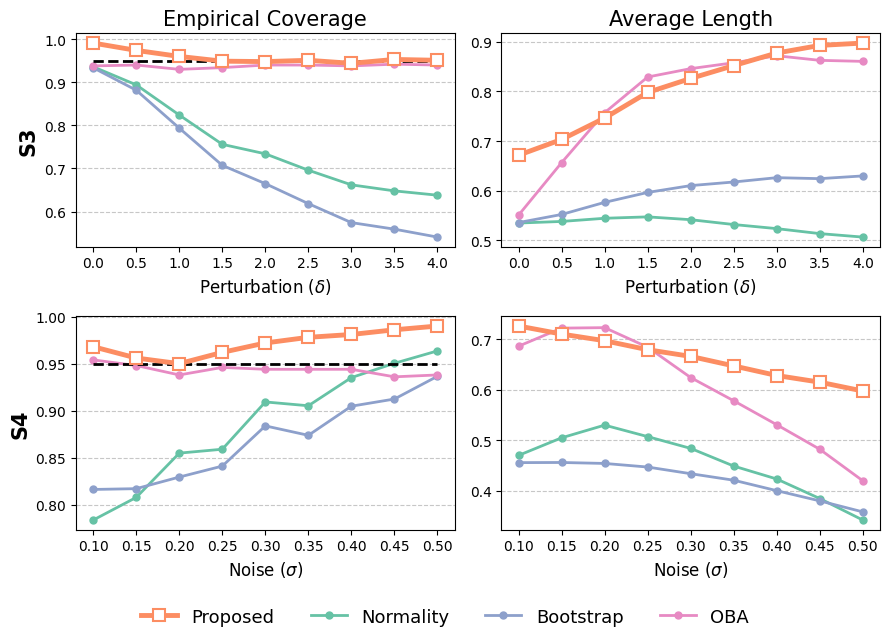}
    \caption{Empirical coverage and average interval lengths of different inference methods under two settings: \textbf{(S3)} (top panels) with sparse perturbation level $\delta \in \{0, 0.5, \ldots, 4.0\}$, and \textbf{(S4)} (bottom panels) with instability level $\sigma \in \{0.1, 0.15, \ldots, 0.50\}$. Each point is averaged over 500 simulation rounds.}
    \label{fig:infer-parametric}
\end{figure}
Note that \texttt{OBA} serves as an oracle benchmark procedure that cannot be implemented in practical problems. Figure~\ref{fig:infer-parametric} reports the empirical coverage and average interval lengths for all the aforementioned inference methods.

We begin with results for \textbf{(S3)}, shown in the top panels, where we vary the sparse perturbation level $\delta\in \{0, 0.5, \dots, 4.0\}$. When $\delta=0$, the setup is regular, and all methods achieve the nominal $95\%$ coverage. However, as $\delta$ increases, the degree of nonregularity becomes more pronounced. Consequently, both the \texttt{Normality} and \texttt{Bootstrap} methods fail to maintain valid coverage, with coverage deteriorating as $\delta$ grows. In contrast, our \texttt{Proposed} method, along with the oracle benchmark method \texttt{OBA}, retains uniformly valid coverage across all values of $\delta$ considered in this experiment. %We emphasize that \texttt{OBA} relies on oracle knowledge of the true parameter $\theta^*_1$ and is thus not implementable in practice. 
For interval lengths, our \texttt{Proposed} method yields interval lengths that are comparable to those of the oracle benchmark \texttt{OBA}.

We now turn to the setup \textbf{(S4)}, with results shown in the bottom panels of Figure~\ref{fig:infer-parametric}. In this experiment, we vary the instability level across $\sigma \in \{0.1, 0.15, \ldots, 0.50\}$. When $\sigma$ is small, the coefficients across source domains are highly similar, inducing an unstable estimation regime. As $\sigma$ increases, the variation among source coefficients becomes more pronounced, and the instability gradually diminishes. The results show that when $\sigma$ is small, both the \texttt{Normality} and \texttt{Bootstrap} methods fail to achieve valid $95\%$ coverage, though their performance improves as $\sigma$ increases. 
In contrast, the \texttt{Proposed} method consistently attains the nominal $95\%$ coverage across all values of $\sigma$, while achieving comparable lengths to those of the oracle \texttt{OBA} method. 

In summary, Figure~\ref{fig:infer-parametric} illustrates that the proposed confidence interval in \eqref{uq:union ci} achieves uniform coverage even when the empirical \CGDRO\ estimator $\widehat{\theta}$ exhibits a nonstandard limiting distribution in the \textbf{nonregular} and \textbf{unstable} settings. Moreover, the confidence interval attains a comparable length to that of the oracle procedure.

%%%%%%%%%%%%%%%%%%%%%%%%%%%%%%%%%%%%%%%%%%%%%%
%% Single Appendix:                         %%
%%%%%%%%%%%%%%%%%%%%%%%%%%%%%%%%%%%%%%%%%%%%%%
%\begin{appendix}
%\section*{???}%% if no title is needed, leave empty \section*{}.
%\end{appendix}
%%%%%%%%%%%%%%%%%%%%%%%%%%%%%%%%%%%%%%%%%%%%%%
%% Multiple Appendixes:                     %%
%%%%%%%%%%%%%%%%%%%%%%%%%%%%%%%%%%%%%%%%%%%%%%
%\begin{appendix}
%\section{???}
%
%\section{???}
%
%\end{appendix}

%%%%%%%%%%%%%%%%%%%%%%%%%%%%%%%%%%%%%%%%%%%%%%
%% Support information, if any,             %%
%% should be provided in the                %%
%% Acknowledgements section.                %%
%%%%%%%%%%%%%%%%%%%%%%%%%%%%%%%%%%%%%%%%%%%%%%
\section*{Acknowledgments}
The authors are grateful to Dr. Qiyang Han for insightful discussions on the proof of Proposition 1. We also thank Ziang Fu and Fan Liu for their contributions to the initial implementation of the methods and assistance with the simulations. %Yifan Hu is also affiliated with the Department of Computer Science, ETH Z\"urich. 
A large proportion of this research was conducted when Z. Guo was the associate professor at Rutgers University and when Y. Hu was a postdoc at EPFL and ETH Zurich. 
During this period, the research of Z. Guo and Z. Wang was partly supported by the NSF grant DMS 2015373 and NIH grants R01GM140463 and R01LM013614. Y. Hu was supported by NCCR Automation, funded by the Swiss National Science Foundation (grant
number 51NF40$\_$225155).
\bibliographystyle{plain}
\bibliography{ref}       % Bibliography file (usually '*.bib')

%% or include bibliography directly:
% \begin{thebibliography}{}
% \bibitem{b1}
% \end{thebibliography}
\appendix
\newpage
\setcounter{page}{1}
\renewcommand{\thesection}{\Alph{section}}
\renewcommand{\thesubsection}{\Alph{section}.\arabic{subsection}}
\setcounter{section}{0} 
{\centering{\large {\bf Supplements of ``Statistical Analysis of Conditional Group Distributionally Robust Optimization with Cross-Entropy Loss''}}}

\vspace{2mm}

This appendix provides additional methodological details, proofs of theoretical results, and experiments that supplement the main text. We present a summary of the appendix as follows.  
\begin{itemize}
\item Section~\ref{sec: proof of approx} contains the proof of two key approximation results: Theorem~\ref{thm: quad convergence} and Theorem~\ref{thm: gamma true}. We also include the proof of Theorem \ref{thm: equi expressions} as it is an integral part of establishing  Theorem~\ref{thm: gamma true}.
\item Section~\ref{appendix: additional methods} mainly introduces additional methods and theoretical results omitted from the main text. 
\begin{itemize}
    \item We first review related literature about minimax optimization in Section~\ref{sec: add literature}. 
    \item Section \ref{sec: doubly robust} presents the complete procedure to solve the empirical \CGDRO\ model, including the detailed construction of DML estimators $\widehat{\mu}^{(l)}$ and the introduction of Mirror Prox algorithm. 
    \item Section \ref{subsec: properties of hatmu} establishes the theoretical properties of DML estimators $\widehat{\mu}^{(l)}$. 
    \item Section~\ref{sec: entire infer} provides further implementation details for the proposed perturbation-based statistical inference procedure. 
    \item Finally, Section~\ref{appedix: prior} discusses how to incorporate prior information on the target distribution into the \CGDRO\ formulation.
\end{itemize}
\item Section~\ref{sec: main proof supp} contains the proofs of the remaining main theoretical results, while Section~\ref{sec: lemma proof} presents the proofs of supporting lemmas.
\item Section~\ref{appendix: simus} describes the experimental details within the main article and presents additional simulation studies.
\end{itemize}

%%%%%%%%%%%%%%%%%%%%%%%%%%%%%%%%%%%%%%%%%%%%%%%%%%%%%%%%%%%%%%%%%%%%%%%%
\section{Proof of Approximation Theorems}
\label{sec: proof of approx}
%%%%%%%%%%%%%%%%%%%%%%%%%%%%%%%%%%%%%%%%%%%%%%%%%%%%%%%%%%%%%%%%%%%%%%%%
In this section, we present the proofs of two key approximation results: Theorem~\ref{thm: quad convergence} and Theorem~\ref{thm: gamma true}. These results are of independent interest, as they rigorously quantify the discrepancy between the solutions of the approximate and original minimax optimization problems. Specifically, Theorem~\ref{thm: quad convergence} establishes convergence on the primal side, while Theorem~\ref{thm: gamma true} addresses the dual convergence. 
We start with introducing some important notations and lemmas in Section \ref{proof prepare} and then prove Theorems~\ref{thm: quad convergence} and \ref{thm: gamma true} in Sections \ref{sec: quad convergence} and \ref{proof of thm: gamma true}, respectively. 

The proof of Theorem~\ref{thm: quad convergence} relies on Proposition \ref{prop: initial bound} and the proof of Theorem~\ref{thm: gamma true} relies on the closed-form expressions established in Theorem \ref{thm: equi expressions}. Hence, we present the proof of Proposition \ref{prop: initial bound} in Section \ref{sec: initial rate} and the proof of Theorem \ref{thm: equi expressions} in Section \ref{proof of thm: equi expressions}.
%%%%%%%%%%%%%%%%%%%%%%%%%%%%%%%%%%%%%%%%%%%%%%%%%%%%%%%%%%%%%%%%%%%%%%%%%%%%%%%%%%%%%%
\subsection{Proof Preparations}
\label{proof prepare}
%%%%%%%%%%%%%%%%%%%%%%%%%%%%%%%%%%%%%%%%%%%%%%%%%%%%%%%%%%%%%%%%%%%%%%%%%%%%%%%%%%%%%%
We begin with the introduction of additional notations and definitions to facilitate the theoretical analysis. Let $\otimes$ denote the Kronecker product. 
Given a covariate vector $X_i\in \RR^d$ and parameter $\theta\in \RR^{dK}$, we define the probability vector $p(X_i,\theta)\in \RR^{K}$ as  
\begin{equation}
p(X_i,\theta)=(p_1(X_i,\theta), \cdots,p_{K}(X_i,\theta))^{\intercal} \quad \text{with}\quad 
p_c(X_i,\theta)=\frac{\exp(\theta_c^{\intercal}X_i)}{1+\sum_{k=1}^{K}\exp(\theta_k^{\intercal}X_i)},
\label{eq: p def}
\end{equation}
where $p_c(X_i,\theta)$ represents the predicted probability for the category $c\in[K]$ under the parameter $\theta$. We define the matrix $$D(X_i,\theta)={\rm diag}(p(X_i,\theta))-p(X_i,\theta)p^{\intercal}(X_i,\theta)\in \RR^{K\times K}.$$ 
We use $\nabla S(\theta)$ and $H(\theta)$ to respectively denote the gradient and Hessian of $S(\theta),$ with the following expression  
$$
\nabla S(\theta)=\E_{X\sim \QQ_{X}}\left(p(X,\theta) \otimes X\right) \quad \text{and}\quad H(\theta)=\E_{X\sim \QQ_{X}}[D(X,\theta)\otimes XX^{\intercal}].$$
For their empirical counterparts, given the unlabeled data $\{\XQ_i\}_{i\in [N]}$ from the target domain,  we also obtain the following gradient and Hessian of $\widehat{S}(\theta)$,
\begin{equation*}
\nabla \widehat{S}(\theta)=\frac{1}{N}\sum_{i=1}^{N} p(X_i^\QQ,\theta)\otimes X_i^\QQ \quad \text{and} \quad \widehat{H}(\theta)=\frac{1}{N}\sum_{i=1}^{N}D(X_i^\QQ,\theta)\otimes X_i^\QQ X_i^{\QQ \intercal}.
\end{equation*}
We introduce the bounded parameter space (in $\ell_1$ norm) centered around $\theta^*$, 
\begin{equation}
\Theta_{B}=\left\{\theta\in \mathbb{R}^{dK}: \|\theta-\theta^*\|_1\leq 1 \right\}.
\label{eq: bound para}
\end{equation}
%which serves as the foundation for our theoretical analysis.

In the following two lemmas, we establish the positiveness and the continuity of $H(\theta)$. The proofs of the following two lemmas are presented in Sections \ref{proof of lem conveixty of H} and \ref{proof of lem key property}.
%The following lemma shows the continuity of $H(\theta)$.

\begin{Lemma}
Suppose that Condition {\rm (A2)} holds. For all $\theta \in \RR^{dK}$, we have
$$\lambda_{\min}(H(\theta)) > 0,$$
which implies that $S(\theta)$ is strictly convex in $\theta$ for all $\theta\in \RR^{dK}$.
Moreover, there exists some constant $c_1\in (0,1/2)$ such that: for all $\theta\in \Theta_{B}$, 
$$c_1^2\kappa_1 \leq \lambda_{\min}(H(\theta))\leq \lambda_{\max}(H(\theta))\leq (1-c_1)\kappa_2,$$
where $\kappa_1,\kappa_2$ are positive constants defined in Condition {\rm (A2)}. 
\label{lem: convexity of H}
\end{Lemma}
Lemma \ref{lem: convexity of H} implies that the function $S(\theta)$ is strictly convex over the entire space $\theta\in \RR^{dK}$, and becomes strongly convex when restricted to the bounded parameter set $\theta\in \Theta_B$. Consequently, the objective function $\phi(\theta)$ defined in \eqref{eq: minimax model} is also strictly convex over $\mathbb{R}^{dK}$ and strongly convex when restricted to $\Theta_B$.

The following lemma establishes the continuity of the Hessian matrix.

\begin{Lemma}
Suppose that Condition {\rm (A2)} holds. There exists a constant $C>0$ such that for any $\theta,\theta' \in \RR^{dK}$, 
\begin{equation}
\|H(\theta')-H(\theta)\|_2\leq C \left(\|\theta'-\theta\|_2+\|\theta'-\theta\|_2^2\right),
\label{eq: Lip H}
\end{equation}
and for any $X\in \RR^d$,
\begin{equation}
\left\|D(X,\theta')-D(X,\theta)\right\|_2\leq 3 \Delta(X,\theta'-\theta)+ \Delta(X,\theta'-\theta)^2,
\label{eq: Lip D}
\end{equation}
with 
$\Delta(X,\theta'-\theta) = \left(\sum_{k=1}^K[X^\intercal(\theta_k'-\theta_k)]^2\right)^{1/2}.$
\label{lem: key property}
\end{Lemma}

Define $n=\min_{1\leq l\leq L}n_l.$We introduce the following events
\begin{equation}
\begin{aligned}
\GG = \GG_0\cap &\GG_1\cap \GG_2 \cap \GG_3,\\
\textrm{with}\quad\GG_0&=\left\{\|\widehat{\U}-\U\|_2\leq t\sqrt{d/n}\right\}\\
\GG_1&=\left\{\max_{\theta\in \Theta_{B}}|\widehat{S}(\theta)-S(\theta)|\leq C\sqrt{d \log N/N} \right\}\\
\GG_2&=\left\{\max_{\theta\in \Theta_{B}}\|\nabla \widehat{S}({\theta})-\nabla {S}({\theta})\|_2\leq C\sqrt{d\log N/N}\right\}\\
\GG_3&=\left\{\max_{\theta\in \Theta_{B}}\|\widehat{H}({\theta})-{H}({\theta})\|_2\leq C\sqrt{d\log N/N}\right\},\\
\end{aligned}
\label{eq: events}
\end{equation}
where $t$ is any value larger than the positive constant $c_0$ specified in Condition {\rm (A1)}, $\Theta_B$ is the bounded set defined in \eqref{eq: bound para}, and $C>0$ denotes some positive constant.
We now briefly explain these events. Event \( \GG_0 \) quantifies the distance between the estimated vectors $\{\widehat{\mu}^{(l)} \}_{1\leq l\leq L}$ and their population counterparts $\{ \mu^{(l)} \}_{1\leq l\leq L}$, which holds with high probability under Condition (A1) of the main article. Events \( \GG_1 \), \( \GG_2 \), and \( \GG_3 \) provide uniform concentration bounds for the empirical risk function $\widehat{S}(\theta)$, its gradient $\nabla\widehat{S}(\theta)$, and its Hessian $\widehat{H}(\theta)$, respectively, over the bounded parameter space $\Theta_B$. We show that $\GG$ occurs with high probability in the following lemma, whose proof is presented in \ref{proof of lem concentration lemma}.
\begin{Lemma} Suppose Conditions {\rm (A1)} and {\rm (A2)} hold. Then there exists a positive constant $c_1>0$ such that
\[
\mathbf{P}\left(\GG_1\cap\GG_2\cap\GG_3\right)\geq 1-N^{-c_1d},
\]
and
\begin{equation}
\mathbf{P}\left(\GG\right)\geq 1-N^{-c_1d}-e^{-c_1t^2}-\delta_n,
\label{eq: ALL concen}
\end{equation}
where $t$ is any value larger than $c_0$, and both $c_0$ and the vanishing sequence $\delta_n\to 0$ are specified in Condition {\rm (A1)}.
\label{lem: concentration lemma}
\end{Lemma}

%Now, on the restricted set $\Theta$, we establish the positive definiteness of $H(\theta)$ and $\widehat{H}(\theta)$ below.
We use the following lemma to establish the positive definiteness of $\widehat{H}(\theta)$, whose proof is presented in \ref{proof of lem: strong convex of hatH}.
\begin{Lemma}
Suppose that Condition {\rm (A2)} holds, and 
$N\gtrsim d \log N/ \kappa_1^2.$ 
%$N >  frac{C}{4\kappa_1^2c_1^4}d \log N$ for some constant $C>0$. 
Then on the event $\GG_3$, for all $\theta\in \Theta_B$,
    \[
    \lambda_{\rm min}\left(\widehat{H}(\theta)\right) \geq \frac{\kappa_1 c_1^2}{2},
    \]
where $c_1>0$ is a constant specified in Lemma \ref{lem: convexity of H} and $\kappa_1>0$ is specified in Condition {\rm (A2)}.
    \label{lem: strong convex of hatH}
\end{Lemma}

%%%%%%%%%%%%%%%%%%%%%%%%%%%%%%%%%%%%%%%%%%%%%%%%%%%%%%%%%%
\subsection{Proof of Proposition \ref{prop: initial bound}}
\label{sec: initial rate}
%%%%%%%%%%%%%%%%%%%%%%%%%%%%%%%%%%%%%%%%%%%%%%%%%%%%%%%%%%
We rewrite the true parameter $\theta^*$ in \eqref{eq: minimax model} and the estimator $\thetainit$ in \eqref{eq: sample optimizers} as
\begin{equation}
    \theta^* = \argmin_{\theta\in \RR^{dK}}\Phi(\theta), \quad \textrm{with}\quad \Phi(\theta) = \max_{\gamma\in \Delta^L} \left[\theta^\intercal \U \gamma + S(\theta)\right].
    \label{eq: theta_star def appendix}
\end{equation}
\begin{equation}
\thetainit = \argmin_{\theta\in \RR^{dK}}\widehat{\Phi}(\theta), \quad \textrm{with}\quad \widehat{\Phi}(\theta) = \max_{\gamma\in \Delta^L}\left[\theta^\intercal \widehat{\U} \gamma + \widehat{S}(\theta)\right].
    \label{eq: theta_init def appendix}
\end{equation}
The definition of $\Theta_{B}$ in \eqref{eq: bound para} implies $
\theta^*=\argmin_{\theta\in \Theta_{B}} {\Phi}(\theta)$.
To facilitate the discussion, we define $\widehat{\theta}_{B}$ as 
\begin{equation}
\widehat{\theta}_{B}\in \argmin_{\theta\in \Theta_{B}} \widehat{\Phi}(\theta),\quad\text{where}\; \Theta_{B} \; \text{is the bounded set defined in}\; \eqref{eq: bound para}.
\label{eq: bound def}
\end{equation}
It follows from Lemma \ref{lem: strong convex of hatH} that, on the event $\GG_3$, the function $\widehat{\Phi}(\theta)$ is strongly convex over $\theta\in \Theta_{B}$ and hence $\widehat{\theta}_B$ is uniquely defined in \eqref{eq: bound def}.

The main idea of the proof is to show that $\widehat{\theta}_{B}$ is the same as $\widehat{\theta}$ with a high probability, and then we establish the convergence rate of $\|\widehat{\theta}-\theta^*\|_2$ by controlling $\|\widehat{\theta}_{B}-\theta^*\|_2$.  
We start with establishing the convergence rate for $\|\widehat{\theta}_{B}-\theta^*\|_2.$ 
% Note that, for any $\theta\in \Theta_{B},$ we have that both $\|\theta\|_1$ and $|X_i^{\intercal}\theta|\leq C$ are bounded, which further implies $\min_{1\leq k\leq K} p_k(X_i,\theta)\geq c_0$ and $\sum_{k=1}^K p_k(X_i,\theta)\leq 1-c_0$, for some constant $c_0$. 
% Recall that $H(\theta)$ is the Hessian matrix of $S(\theta)$, and together with \eqref{eq: positive def}, we have shown that $S(\theta)$ is a strongly convex function with respect to $\theta$. Together with the fact that $g(\theta)$ is a convex function, it follows from \eqref{eq: rewriting optimization} that $\Phi(\theta)$ is strongly convex with parameter $\lambda_0 c_0^2$. 
Since both $\theta^*$ and  $\widehat{\theta}_{B}$ belong to $\Theta_{B}$, we apply the strong convexity of $\Phi(\theta)$ over the parameter space $\Theta_B$, as established in Lemma \ref{lem: strong convex of hatH}, and obtain that
\begin{equation}
\Phi(\widehat{\theta}_{B})-\Phi(\theta^*)\geq \langle \partial \Phi(\theta^*), \widehat{\theta}_{B}-\theta^*\rangle+ \frac{\kappa_1c_1^2}{2}\|\widehat{\theta}_{B}-\theta^*\|_2^2\geq \frac{\kappa_1 c_1^2}{2} \|\widehat{\theta}_{B}-\theta^*\|_2^2,
\label{eq: strong conv}
\end{equation}
where $\partial \Phi(\theta^*)$ denotes the subgradient of $\Phi(\cdot)$ on $\theta^*$, and $\langle \partial \Phi(\theta^*), \widehat{\theta}_{B}-\theta^*\rangle\geq 0$ holds since $\theta^*$ is the minimum value of the convex function $\Phi(\cdot)$ on the set $\Theta_{B}.$ 
We next upper bound the left-hand side in the above inequality as follows:
\begin{equation}
\begin{aligned}
\Phi(\widehat{\theta}_{B})-\Phi(\theta^*)&=\Phi(\widehat{\theta}_{B})-\widehat{\Phi}(\widehat{\theta}_{B})+\widehat{\Phi}(\widehat{\theta}_{B})-\widehat{\Phi}(\theta^*)+\widehat{\Phi}(\theta^*)-\Phi(\theta^*) \\
&\leq \Phi(\widehat{\theta}_{B})-\widehat{\Phi}(\widehat{\theta}_{B})+\widehat{\Phi}(\theta^*)-{\Phi}(\theta^*)\\
&\leq 2 \max_{\theta\in \Theta_{B}}|\widehat{\Phi}(\theta)-\Phi(\theta)|,
\end{aligned}
\label{eq: key step}
\end{equation}
where the first inequality holds follows from the definition of $\widehat{\theta}_{B}$ being the minimizer of \eqref{eq: bound def}, implying $\widehat{\Phi}(\widehat{\theta}_{B})-\widehat{\Phi}(\theta^*)\leq 0$, and the second inequality holds as both $\widehat{\theta}_B,\theta^*\in \Theta_B$.

By the definitions of $\Phi(\theta)$ and $\widehat{\Phi}(\theta)$ in \eqref{eq: theta_star def appendix} and \eqref{eq: theta_init def appendix}, we have 
\begin{equation*}
\left|\widehat{\Phi}(\theta) - \Phi(\theta)\right|\leq \max_{1\leq l\leq L}\left|\theta^{\intercal}\left[\widehat{\mu}^{(l)}-\mu^{(l)}\right]+\widehat{S}(\theta)-{S}(\theta)\right|. 
\end{equation*}
Together with the results in \eqref{eq: strong conv} and \eqref{eq: key step}, we have 
\begin{equation}
\begin{aligned}
    \|\widehat{\theta}_B - \theta^*\|_2^2 &\lesssim \max_{\theta\in \Theta_{B}} \max_{1\leq l\leq L}\left|\theta^{\intercal}\left[\widehat{\mu}^{(l)}-\mu^{(l)}\right]+\widehat{S}(\theta)-{S}(\theta)\right|\\
    &\lesssim \max_{\theta\in \Theta_{B}} \max_{1\leq l\leq L}\left[\|\theta\|_2 \|\widehat{\mu}^{(l)} - \mu^{(l)}\|_2 + |\widehat{S}(\theta) - S(\theta)|\right].\\
\end{aligned}
    \label{eq: key step - 2}
\end{equation}
%Note that
%$$\widehat{h}^{(l)}(\theta)-h^{(l)}(\theta)= \theta^{\intercal}\left[\widehat{\mu}^{(l)}-\mu^{(l)}\right]+\widehat{S}(\theta)-{S}(\theta),$$
%which implies that
%\[
%\left|\widehat{h}^{(l)}(\theta)-h^{(l)}(\theta) \right| \leq \|\theta\|_2 \|\widehat{\mu}^{(l)} - \mu^{(l)}\|_2 + \|\widehat{S}(\theta) - S(\theta)\|_2.
%\]
On the event $\mathcal{G}$ defined in \eqref{eq: events}, we establish 
\begin{equation*}
\max_{\theta\in \Theta_{B}} \max_{1\leq l\leq L} \left[\|\theta\|_2 \|\widehat{\mu}^{(l)} - \mu^{(l)}\|_2 + |\widehat{S}(\theta) - S(\theta)|\right]\lesssim t\cdot (1+\|\theta_{*}\|_2) \sqrt{\frac{d}{n}}+\sqrt{\frac{d \log N}{N}}\lesssim t\sqrt{d/n},
\end{equation*}
where the last inequality holds since we assume $t\geq c_0$ in Condition {\rm (A1)} and $N\gg  n\log N$ and $\|\theta^*\|_2\leq C$ in Condition {\rm (A2)}.

Putting the preceding inequality back to \eqref{eq: key step - 2}, we establish that, on the event $\mathcal{G},$ %with probability larger than $1-\exp(-t^2)-N^{-cd},$
\begin{equation}
\|\widehat{\theta}_{B}-\theta^*\|_2\leq C\sqrt{t}\cdot (d/n)^{1/4},
    \label{eq: theta_B rate prop 1}
\end{equation}
for some constant $C>0$. Therefore, 
\begin{equation}
\|\widehat{\theta}_{B}-\theta^*\|_1\leq \sqrt{d}\|\widehat{\theta}_{B}-\theta^*\|_2 \leq C\sqrt{t}(d^3/n)^{1/4} \leq 1/2,
\label{eq: intermediate}
\end{equation}
where the last inequality holds once $t\leq \frac{1}{4 C^2}\sqrt{n/d^3}$.

In the following, we prove $\widehat{\theta}\in \Theta_{B}$, which further implies that $\widehat{\theta}_{B}=\widehat{\theta}$. We prove the result by a contradiction argument. We assume that there exists  $\widehat{\theta}$ outside $\Theta_{B}$ that minimizes $\widehat{\Phi}(\theta)$ over $\theta\in \RR^{dK}$, implying $\widehat{\Phi}(\thetainit)\leq \widehat{\Phi}(\thetainit_B)$.
Due to the convexity of $\widehat{\Phi}(\theta)$, for any $\nu\in [0,1]$, we have
\begin{equation*}
\widehat{\Phi}(\widehat{\theta}_{B}+t(\widehat{\theta}-\widehat{\theta}_{B}))\leq \nu \widehat{\Phi}(\widehat{\theta})+(1-\nu)\widehat{\Phi}(\widehat{\theta}_{B})\leq \widehat{\Phi}(\widehat{\theta}_B).
\end{equation*}
The pathway $\widehat{\theta}_{B}+\nu(\widehat{\theta}-\widehat{\theta}_{B})$ intersects $\Theta_{B}$ for some $\nu\in [0,1]$. We denote the intersection point as $\theta_{\rm inter}$, which satisfies $\widehat{\Phi}(\theta_{\rm inter})\leq \widehat{\Phi}(\widehat{\theta}_B)$ and $\theta_{\rm inter}\in \Theta_B$ with $\|\theta_{\rm inter}-\theta^*\|_1=1$.  Since  $\widehat{\theta}_B$ is the unique minimizer of \eqref{eq: bound def}, 
it implies 
\[
\widehat{\theta}_B = \theta_{\rm inter}\quad \text{and}\quad \|\widehat{\theta}_{B}-\theta^*\|_1= 1.
\]
% \begin{equation*}
% \theta_{\rm inter}\in \argmin_{\theta\in \Theta_{B}} \widehat{\Phi}(\theta) \quad \text{and}\quad \|\theta_{\rm inter}-\theta^*\|_1\geq 1.
% \end{equation*}
The above result contradicts \eqref{eq: intermediate} and hence $\widehat{\theta}\in \Theta_{B}$ and $\widehat{\theta}_B = \thetainit$. 
Then results in \eqref{eq: theta_B rate prop 1} and \eqref{eq: intermediate} imply the following results: on the event $\GG$,
\begin{equation}
    \|\widehat{\theta}-\theta^*\|_2\lesssim \sqrt{t}\cdot (d/n)^{1/4} \quad \textrm{and}\quad \|\widehat{\theta}-\theta^*\|_1 \leq 1/2.
    \label{eq: theta_hat is bounded}
\end{equation}
Together with Lemma \ref{lem: concentration lemma} which controls the probability of the event $\GG$, we establish that \eqref{eq: initial bound} holds with probability at least $1-N^{-c_1d}-e^{-c_1t^2}-\delta_n$.

%%%%%%%%%%%%%%%%%%%%%%%%%%%%%%%%%%%%%%%%%%%%%%%%%%%%%%%%%%%
\subsection{Proof of Theorem \ref{thm: quad convergence}}
\label{sec: quad convergence}
%%%%%%%%%%%%%%%%%%%%%%%%%%%%%%%%%%%%%%%%%%%%%%%%%%%%%%%%%%%

To facilitate our discussion, we denote 
$g(\theta) = \max_{\gamma\in \Delta^L} \theta^\intercal \U\gamma$ with $\U=(\mu^{(1)}, \mu^{(2)},..., \mu^{(L)})\in \RR^{dK\times L}$ as defined in \eqref{eq: U definition}, then $\theta^*$ and $\theta^*_{\rm ap}$, defined in \eqref{eq: minimax model} and \eqref{eq: approximate minimax}, are expressed as, 
\[
\theta^* = \argmin_{\theta\in \RR^{dK}} g(\theta) + S(\theta), \quad\textrm{and}\quad \theta^*_{\rm ap} = \argmin_{\theta\in \RR^{dK}} g(\theta) + Q(\theta).
\]

\subsubsection{Proof of \eqref{eq: quad convergence}} 

For the global minimizer $\theta^*$, its optimality condition implies 
$$
0\in \partial g(\theta^*)+\nabla S(\theta^*),
$$
where $\partial g(\theta)$ denotes the subgradient of the convex function $g(\theta)$.
Meanwhile, the optimality condition of ${\theta}^*_{\rm ap}$ implies 
$$
0\in \partial g({\theta}^*_{\rm ap})+\nabla S(\thetainit)+H(\thetainit) ({\theta}^*_{\rm ap}-\thetainit).
$$
That is, there exist $s_1\in \partial g(\theta^*_{\rm ap})$ and $s_2\in \partial g(\theta^*)$ such that 
\begin{equation}
s_1+\nabla S(\thetainit)+H(\thetainit) ({\theta}^*_{\rm ap}-\thetainit)=0 \quad \text{and}\quad s_2+\nabla S(\theta^*)=0.
\label{eq: subdiff exp}
\end{equation}
Since $g(\cdot)$ is convex with a monotone subdifferential operator,  $s_1$ and $s_2$ satisfy 
$
\langle s_1 - s_2,{\theta}^*_{\rm ap}-\theta^*\rangle\geq 0. $ We apply the expressions in \eqref{eq: subdiff exp} and obtain
\begin{equation*}
\langle\nabla S(\theta^*)-\nabla S(\thetainit)-H(\thetainit) ({\theta}^*_{\rm ap}-\thetainit), {\theta}^*_{\rm ap}-\theta^* \rangle \geq 0.
\end{equation*}
Rearranging the terms, we obtain that
\begin{equation*}
%\langle \partial g({\theta}^*_{\rm ap})-\partial g(\theta^*),{\theta}^*_{\rm ap}-\theta^*\rangle=
\left\langle \nabla S(\theta^*)-\nabla S(\thetainit)-H(\thetainit) ({\theta}^*-\thetainit),\;{\theta}^*_{\rm ap}-\theta^*\right\rangle\geq ({\theta}^*_{\rm ap}-{\theta}^*)^{\intercal}H(\thetainit)({\theta}^*_{\rm ap}-{\theta}^*).
\end{equation*}
Therefore, by the Cauchy-Schwarz inequality, we have
\begin{equation}
\begin{aligned}
    ({\theta}^*_{\rm ap}-{\theta}^*)^{\intercal}H(\thetainit)({\theta}^*_{\rm ap}-{\theta}^*)
    \leq \left\|\nabla S(\theta^*)-\nabla S(\thetainit)-H(\thetainit) ({\theta}^*-\thetainit)\right\|_2 \cdot \left\|{\theta}^*_{\rm ap}-\theta^*\right\|_2.
\end{aligned}
\label{eq: inequality 1}
\end{equation}
%Since $\nabla S(\theta)$ is continuous, we have 

To control the right-hand side in the above inequality, we observe that the following form holds:
\begin{equation*}
\nabla S(\theta^*)-\nabla S(\thetainit)-H(\thetainit) ({\theta}^*-\thetainit)=\int_{\nu=0}^{1} \left[H\left(\thetainit+\nu(\theta^*-\thetainit)\right) -H(\thetainit)\right]({\theta}^*-\thetainit) d\nu.
\end{equation*}
It implies that 
\begin{equation*}
    \begin{aligned}
    \left\|\nabla S(\theta^*)-\nabla S(\thetainit)-H(\thetainit) ({\theta}^*-\thetainit)\right\|_2 
    \leq \int_{\nu=0}^1 \left\|H\left(\thetainit+\nu(\theta^*-\thetainit)\right) -H(\thetainit)\right\|_2\left\|{\theta}^*-\thetainit\right\|_2 d\nu. 
\end{aligned}
\label{eq: thm1-interm1}
\end{equation*}
Note that we have established in \eqref{eq: Lip H} the Lipschitz continuity of the Hessian $H(\cdot)$, that is,
\begin{equation*}
    \left\|H\left(\thetainit+\nu(\theta^*-\thetainit)\right) -H(\thetainit)\right\|_2 \leq C \left(\nu\|\theta^* - \thetainit\|_2 + \nu^2\|\theta^* - \thetainit\|_2^2\right).
\end{equation*}
The combination of the preceding two inequalities yields
\begin{equation}
\begin{aligned}
    \left\|\nabla S(\theta^*)-\nabla S(\thetainit)-H(\thetainit) ({\theta}^*-\thetainit)\right\|_2 &\lesssim \int_0^1 \left(\nu \|\theta^* - \widehat{\theta}\|_2^2 + \nu^2\|\theta^* - \widehat{\theta}\|_2^3\right)  d\nu \\
    &\lesssim \|\theta^* - \widehat{\theta}\|_2^2,
\end{aligned}
\label{eq: thm1-interm2}
\end{equation}
where the last inequality holds as we have established in \eqref{eq: theta_hat is bounded} that on the event $\GG$,
\begin{equation*}
   \|\widehat{\theta}-\theta^*\|_2\leq \|\widehat{\theta}-\theta^*\|_1\leq 1/2 \quad \textrm{and}\quad \widehat{\theta}\in \Theta_B.
\end{equation*}

%The proof of  $\|\theta^* - \thetainit\|_2\leq 1$
%$\nu\in [0,1]$,
%We have established that on the event $\GG$, $\|\widehat{\theta}-\theta^*\|_2\lesssim \sqrt{t}\cdot (d/n)^{1/4}$, which implies that
%\[
%\|\widehat{\theta}-\theta^*\|_1 \leq \sqrt{d}\|\widehat{\theta}-\theta^*\|_2\leq C\sqrt{t}(d^3/n)^{1/4}\leq 1,
%\]
%where the last inequality holds when $t\leq C \sqrt{n/d^3}$ for some constant $C>0$. 
% In the proof of Proposition \ref{prop: initial bound}, we have established in \eqref{eq: theta_hat is bounded} that, on the event $\GG$, 
% \begin{equation*}
%    \|\widehat{\theta}-\theta^*\|_2\leq 1/2 \quad \textrm{and}\quad \widehat{\theta}\in \Theta_B.
% \end{equation*}
% Together with \eqref{eq: thm1-interm2}, we establish  
% $$
% \left\|\nabla S(\theta^*)-\nabla S(\thetainit)-H(\thetainit) ({\theta}^*-\thetainit)\right\|_2 \leq C \cdot \|\theta^* - \thetainit\|_2^2.$$
Putting \eqref{eq: thm1-interm2} back to \eqref{eq: inequality 1}, we have 
\begin{equation}
\lambda_{\min}(H(\thetainit))\cdot\|{\theta}^*_{\rm ap}-{\theta}^*\|_2^2\lesssim \|{\theta}^*-\thetainit\|_2^2 \cdot \|{\theta}^*_{\rm ap}-{\theta}^*\|_2.
\label{eq: immediate}
\end{equation}
Since $\widehat{\theta}\in \Theta_B$ on the event $\GG$, we apply Lemma \ref{lem: convexity of H} and establish that $\lambda_{\rm min}(H(\thetainit))\geq \kappa_1c_1^2$ for some positive constant $c_1>0$. Consequently, on the event $\GG$, \eqref{eq: immediate} implies 
\[
\|{\theta}^*_{\rm ap} - \theta^*\|_2 \lesssim \|\thetainit-\theta^*\|_2^2 \lesssim t\sqrt{d/n},
\]
where the last inequality holds due to \eqref{eq: theta_hat is bounded} established in the proof of Proposition \ref{prop: initial bound}.  Lastly, we leverage Lemma \ref{lem: concentration lemma} to control the probability of the event $
\GG$.

\subsubsection{Proof of \eqref{eq: stable optimizer}} To facilitate the analysis, we define $\widehat{g}(\theta) = \max_{\gamma\in \Delta^L}\theta^\intercal\widehat{\U}\gamma$, then $\thetainit$ and $\thetainit_{\rm ap}$, defined in \eqref{eq: sample optimizers} and \eqref{eq: inter estimator}, can be equivalently expressed as:
$$
\thetainit = \argmin_{\theta\in \RR^{dK}} \widehat{g}(\theta) + \widehat{S}(\theta)\quad \text{and} \quad \thetainit_{\rm ap} = \argmin_{\theta\in \RR^{dK}} \widehat{g}(\theta) + \widehat{Q}(\theta).$$ By the optimality conditions for $\thetainit$ and $\thetainit_{\rm ap}$, we obtain that there exist $s_1 \in \partial \widehat{g}(\widehat{\theta}_{\rm ap})$ and  $s_2\in \partial \widehat{g}(\widehat{\theta})$ such that 
\begin{equation}
s_1+\nabla \widehat{S}(\thetainit)+\widehat{H}(\thetainit) (\thetainit_{\rm ap}-\thetainit)=0,\quad \text{and}\quad s_2+ \nabla \widehat{S}(\widehat{\theta})=0.
\label{eq: sample expression}
\end{equation}
Since the function $\widehat{g}(\cdot)$ is convex towards $\theta$, the subdifferential operator is monotone in the sense that 
$
\langle s_1 - s_2, \; \widehat{\theta}_{\rm ap} - \widehat{\theta}\rangle \geq 0. 
$
Together with the corresponding expressions of $s_1,s_2$ in \eqref{eq: sample expression}, we establish that:
$$ 0\leq \langle \widehat{H}(\thetainit) (\thetainit-\thetainit_{\rm ap}),\thetainit_{\rm ap}-\widehat{\theta}\rangle = -(\widehat{\theta}-\thetainit_{\rm ap})^\intercal \widehat{H}(\widehat{\theta}) (\widehat{\theta}-\thetainit_{\rm ap}).$$ 
%\Zhenyu{check my rewriting about this subgradients. If it reads better, I can modify \eqref{eq: subdiff exp}.}
It follows from Lemma \ref{lem: strong convex of hatH} that, on the event $\GG$, $\widehat{H}(\thetainit)$ is positive definite and hence $\thetainit_{\rm ap}=\widehat{\theta}$. Together with Lemma \ref{lem: concentration lemma}, we establish that $\thetainit_{\rm ap}=\widehat{\theta}$, with a probability of at least $1-N^{-c_1d} - e^{-c_1t^2}-\delta_n$.

\subsection{Proof of Theorem \ref{thm: equi expressions}}
\label{proof of thm: equi expressions}
In the following, we mainly focus on the proof of \eqref{eq: closed form theta} and the proof of \eqref{eq: inter theta} can be established with a similar argument. The proof of \eqref{eq: closed form theta}, consists of two steps:
\begin{itemize}
    \item First, we show that the solution $\theta^*_{\rm ap}$ to the dual formulation in \eqref{eq: step 2} is well-defined and admits the closed-form expression given in \eqref{eq: closed form theta}.
\item Second, we demonstrate   that this $\theta^*_{\rm ap}$ also solves the original primal problem; that is, $\theta^*_{\rm ap}$ uniquely minimizes $\max_{\gamma \in \Delta^L} \phi_{\rm ap}(\theta, \gamma)$, where
\end{itemize}
\begin{equation}
    \phi_{\rm ap}(\theta,\gamma) = \theta^\intercal \U\gamma + Q(\theta),\quad \textrm{with}\quad Q(\theta) = \langle \nabla S(\thetainit),\theta-\thetainit\rangle+\frac{1}{2} (\theta-\thetainit)^{\intercal} H(\thetainit) (\theta-\thetainit).
    \label{eq: f_ap thm2}
\end{equation}

To show that $\theta^*_{\rm ap}$ for the dual formulation in \eqref{eq: step 2} is well defined, we start with justifying the following lemma. Note that Lemma \ref{lem: convexity of H} has established that $\lambda_{\rm min}H(\theta)>0$ for all $\theta\in \RR^{dK}$, thus $H(\thetainit)$ is positive definite.  
\begin{Lemma}
Suppose Condition {\rm (A3)} holds, and $H(\thetainit)$ is positive definite (implied by Lemma \ref{lem: convexity of H}). Then $\gamma^*_{\rm ap}$ in \eqref{eq: step 1} and $\theta^*_{\rm ap}$ in \eqref{eq: step 2} are uniquely defined. 
\end{Lemma}
Since $H(\thetainit)$ is positive definite, the function $\phi_{\rm ap}(\theta,\gamma)$ is strictly convex towards $\theta$ for any fixed $\gamma\in \Delta^L$.
By taking $\nabla_\theta \phi_{\rm ap}(\theta,\gamma) = 0$, we obtain the following expression of the minimizer
\begin{equation}
    \argmin_{\theta\in \RR^{dK}}\phi_{\rm ap}(\theta,\gamma) = \thetainit - [H(\thetainit)]^{-1}\left(\U\gamma + \nabla S(\thetainit)\right).
    \label{eq: argmin phi_ap}
\end{equation}
We further express the optimal value $F(\gamma)$ defined in \eqref{eq: step 1} as follows:
\begin{equation}
    F(\gamma) = \min_{\theta\in \RR^{dK}}\phi_{\rm ap}(\theta,\gamma)= -\frac{1}{2}\left(\U\gamma + \nabla S(\thetainit)\right)^\intercal [H(\thetainit)]^{-1}\left(\U\gamma + \nabla S(\thetainit)\right) + \thetainit^\intercal \U \gamma.
    \label{eq: expression of F(gamma)}
\end{equation}
Since $H(\thetainit)$ is positive definite, we apply  Condition (A3) to establish that $F(\gamma)$ is strictly concave towards $\gamma$, which justifies the uniqueness of the minimizer $\gamma^*_{\rm ap}=\argmax_{\gamma\in \Delta^L} F(\gamma)$. Moreover, by plugging the weight vector $\gamma^*_{\rm ap}$ into \eqref{eq: argmin phi_ap}, $\theta^*_{\rm ap}$ is uniquely defined as
\begin{equation}
    \theta^*_{\rm ap} = \argmin_{\theta\in \RR^{dK}} \phi_{\rm ap}(\theta,{\gamma}^*_{\rm ap})=\thetainit - [H(\thetainit)]^{-1} \left(\U {\gamma}^*_{\rm ap} + \nabla S(\thetainit)\right).
\label{eq: theta-star expression}
\end{equation}

Next, we turn to showing that $\theta^*_{\rm ap}$ uniquely minimizes $\max_{\gamma\in \Delta^L}\phi_{\rm ap}(\theta,\gamma).$
We introduce the following lemma establishing that  $({\theta}^*_{\rm ap}, {\gamma}^*_{\rm ap})$ is a  saddle point of $\phi_{\rm ap}(\theta, \gamma)$, whose proof is deferred to Section \ref{sec: saddle point}. 
\begin{Lemma} 
\label{lem: saddle point}
Suppose Conditions {\rm (A3)} holds, and $H(\widehat{\theta})$ is positive definite (implied by Lemma \ref{lem: convexity of H}). Then $({\theta}^*_{\rm ap}, {\gamma}^*_{\rm ap})$ is a saddle point for $\phi_{\rm ap}(\theta, \gamma)$ over $\theta\in \RR^{dK}$ and $\gamma\in \Delta^L$:
\begin{equation}
    \phi_{\rm ap}({\theta}^*_{\rm ap}, \gamma)\leq \phi_{\rm ap}({\theta}^*_{\rm ap}, {\gamma}^*_{\rm ap}) \leq \phi_{\rm ap}(\theta, {\gamma}^*_{\rm ap}), \quad \textrm{for all $\theta\in \RR^{dK}$ and $\gamma\in \Delta^L$}.
    \label{eq: saddle point thm2}
\end{equation}
\end{Lemma}
The saddle point
$({\theta}^*_{\rm ap}, {\gamma}^*_{\rm ap})$ implies that
{\small
\begin{equation}
    \min_{\theta\in \RR^{dK}}\max_{\gamma\in \Delta^L} \phi_{\rm ap}(\theta,\gamma) \leq \max_{\gamma\in \Delta^L} \phi_{\rm ap}({\theta}^*_{\rm ap},\gamma) = \phi_{\rm ap}({\theta}^*_{\rm ap}, {\gamma}^*_{\rm ap}) = \min_{\theta\in \RR^{dK}} \phi_{\rm ap}(\theta, {\gamma}^*_{\rm ap}) \leq \max_{\gamma\in \Delta^L}\min_{\theta\in \RR^{dK}}\phi_{\rm ap}(\theta,\gamma),
    \label{eq: long ineq thm2}
\end{equation}}
where the middle two equalities hold due to \eqref{eq: saddle point thm2}.
Together with the fact that
\[
\min_{\theta\in \RR^{dK}}\max_{\gamma\in \Delta^L} \phi_{\rm ap}(\theta,\gamma) \geq \max_{\gamma\in \Delta^L}\min_{\theta\in \RR^{dK}}\phi_{\rm ap}(\theta,\gamma),
\]
we obtain that the equalities hold throughout in \eqref{eq: long ineq thm2}, with 
$$\max_{\gamma\in \Delta^L} \phi_{\rm ap}({\theta}^*_{\rm ap},\gamma)=\min_{\theta\in \RR^{dK}}\max_{\gamma\in \Delta^L} \phi_{\rm ap}(\theta,\gamma).$$ That is, 
$$
{\theta}^*_{\rm ap}\in \argmin_{\theta\in \RR^{dK}}\max_{\gamma\in \Delta^L}\phi_{\rm ap}(\theta, \gamma).
$$
Since $\phi_{\rm ap}(\theta,\gamma)$ is strictly convex towards $\theta$ for each fixed $\gamma$, we have $\max_{\gamma\in \Delta^L}\phi_{\rm ap}(\theta,\gamma)$ is strictly convex towards $\theta$. Therefore, ${\theta}^*_{\rm ap}$ is the unique minimizer of  $\max_{\gamma\in \Delta^L}\phi_{\rm ap}(\theta, \gamma)$ such that
\[
{\theta}^*_{\rm ap}= \argmin_{\theta\in \RR^{dK}}\max_{\gamma\in \Delta^L}\phi_{\rm ap}(\theta, \gamma).
\]

Now, we move on to the proof of \eqref{eq: inter theta}.
Before that, we first provide the dual form of identifying $\widehat{\theta}_{\rm ap}$, analogous to \eqref{eq: step 1} and \eqref{eq: step 2} in the main text.
\begin{itemize}
    \item \textbf{Step 1.} We solve the maximization problem with respect to $\gamma$ and compute the optimal weight as
    \begin{equation}
        \widehat{\gamma}_{\rm ap} = \argmax_{\gamma\in \Delta^L} \widehat{F}(\gamma), \quad \textrm{with}\quad \widehat{F}(\gamma) = \min_{\theta\in \RR^{dK}}\widehat{\phi}_{\rm ap}(\theta,\gamma),
        \label{eq: step 1 hat}
    \end{equation}
    \item \textbf{Step 2.} With the optimal weight $\widehat{\gamma}_{\rm ap}$, we then solve the minimization problem:
    \begin{equation}
        \widehat{\theta}_{\rm ap} = \argmin_{\theta\in \RR^{dK}} \widehat{\phi}_{\rm ap}(\theta,\widehat{\gamma}_{\rm ap}).
        \label{eq: step 2 hat}
    \end{equation}
\end{itemize}
Similarly, the proof of \eqref{eq: inter theta} proceeds in two steps.
\begin{itemize}
    \item First, we show that the solution $\widehat\theta_{\rm ap}$ to the dual formulation in \eqref{eq: step 2 hat} is well-defined and admits the closed-form expression given in \eqref{eq: inter theta}, on the event $\GG.$
    \item Second, we demonstrate   that this $\widehat{\theta}_{\rm ap}$ also solves the original primal problem on the event $\GG$; that is, $\widehat\theta_{\rm ap}$ uniquely minimizes $\max_{\gamma \in \Delta^L} \widehat\phi_{\rm ap}(\theta, \gamma)$, where
\end{itemize}
\[
\widehat{\phi}_{\rm ap}(\theta,\gamma)=\theta^\intercal \widehat{\U}\gamma + \widehat{Q}(\theta),\quad \textrm{with}\quad \widehat{Q}(\theta) = \langle \nabla \widehat{S}(\thetainit),\theta-\thetainit\rangle+\frac{1}{2} (\theta-\thetainit)^{\intercal} \widehat{H}(\thetainit) (\theta-\thetainit).
\]
We emphasize that the main difference compared to the proof of \eqref{eq: closed form theta} is that the statements are now carried out on the high probability event $\GG$.

Since $\widehat{\theta}\in \Theta_B$ on the event $\GG$, Lemma \ref{lem: strong convex of hatH} ensures that $\lambda_{\rm min}(\widehat{H}(\widehat{\theta}))\geq c'$ for some constant $c'>0$. This implies that $\widehat{\phi}_{\rm ap}(\theta,\gamma)$ is strongly convex towards $\theta$ for any fixed $\gamma\in \Delta^L$. By taking $\nabla_\theta \widehat{\phi}_{\rm ap}(\theta,\gamma)=0$, we obtain the following expression of the minimizer:
\begin{equation}
    \argmin_{\theta\in \RR^{dK}}\widehat{\phi}_{\rm ap}(\theta,\gamma) = \widehat{\theta} - [\widehat{H}(\widehat{\theta})]^{-1}\left(\widehat{\U}\gamma + \nabla \widehat{S}(\widehat{\theta})\right).
    \label{eq: argmin phi_ap hat}
\end{equation}
Then the optimal value $\widehat{F}(\gamma)$ can be expressed as follows:
\begin{equation}
    \label{eq: expression of F(gamma) hat}
    \widehat{F}(\gamma) = \min_{\theta\in \RR^{dK}}\widehat{\phi}_{\rm ap}(\theta,\gamma) = -\frac{1}{2}\left(\widehat{\U}\gamma + \nabla \widehat{S}(\thetainit)\right)^\intercal [\widehat{H}(\thetainit)]^{-1}\left(\widehat{\U}\gamma + \nabla \widehat{S}(\thetainit)\right) + \thetainit^\intercal \widehat{\U} \gamma.
\end{equation}
On the event $\GG$, we shall establish that $\widehat{F}(\gamma)$ is strongly concave towards $\gamma$, which justifies the uniqueness of $\widehat{\gamma}_{\rm ap} = \argmax_{\gamma\in \Delta^L} \widehat{F}(\gamma)$. By plugging the weight vector $\widehat{\gamma}_{\rm ap}$ into \eqref{eq: argmin phi_ap hat}, $\widehat{\theta}_{\rm ap}$ is uniquely defined as:
\[
\widehat{\theta}_{\rm ap} = \argmin_{\theta\in \RR^{dK}}\widehat{\phi}_{\rm ap}(\theta,\widehat{\gamma}_{\rm ap}) = \widehat{\theta} - [\widehat{H}(\widehat{\theta})]^{-1}\left(\widehat{\U} \widehat{\gamma}_{\rm ap} + \nabla\widehat{S}(\widehat{\theta})\right).
\]

Similarly to Lemma \ref{lem: saddle point}, we can show that on the event $\GG$, $(\thetainit_{\rm ap}, \widehat{\gamma}_{\rm ap})$ forms a saddle point for $\widehat{\phi}_{\rm ap}(\theta,\gamma)$ over $\theta\in \RR^{dK}$ and $\gamma\in \Delta^L$. We omit the proof as it essentially follows the similar proof of Lemma \ref{lem: saddle point}. Then we have $\thetainit_{\rm ap}\in \argmin_{\theta\in \RR^{dK}}\max_{\gamma\in \Delta^L}\widehat{\phi}_{\rm ap}(\theta,\gamma)$. Since $\widehat{\phi}_{\rm ap}(\theta,\gamma)$ is strongly convex in $\theta$ for any fixed $\gamma\in \Delta^L$, after taking the maximization, $\max_{\gamma\in \Delta^L}\widehat{\phi}_{\rm ap}(\theta,\gamma)$ is still strongly convex. Therefore, we conclude that $\thetainit_{\rm ap}$ uniquely minimizes the function $\max_{\gamma\in \Delta^L}\widehat{\phi}_{\rm ap}(\theta,\gamma)$ on the event $\GG$.   Lastly, we apply Lemma \ref{lem: concentration lemma} which controls the probability of the event $\GG$ to complete the proof.

%%%%%%%%%%%%%%%%%%%%%%%%%%%%%%%%%%%%%%%%%%%%%%%%%%%%%%%%%%%%%%%%%%%%%%%%
\subsection{Proof of Theorem \ref{thm: gamma true}}
\label{proof of thm: gamma true}
%%%%%%%%%%%%%%%%%%%%%%%%%%%%%%%%%%%%
%%%%%%%%%%%%%%%%%%%%%%%%%%%%%%%%%%%
We recall the definition of $\Theta_{\rm loc}$ in \eqref{eq: Theta local}
\begin{equation*}
\Theta_{\rm loc}=\left\{\theta\in \RR^{dK}: \|\theta-\theta^*\|_2\leq 2\tau \sqrt{d/n}\right\} \quad \text{with} \quad  \tau= C \left(1+\frac{1}{\sigma_L^2(\U)}\right)t,
\end{equation*}
for some constant $C>0$.
Note that for $\theta\in \Theta_{\rm loc}$, on the event $\GG$, we have 
\begin{equation}
\begin{aligned}
\left|S(\theta)-Q(\theta)\right|&=\left|S(\theta)-\left[S(\thetainit)+\langle \nabla S(\thetainit),\theta-\thetainit\rangle+\frac{1}{2} (\theta-\thetainit)^{\intercal} H(\thetainit) (\theta-\thetainit)\right]\right|\\
&=\left|\frac{1}{2} \int_{\nu=0}^{1} (\theta-\widehat{\theta})^{\intercal} \left[H(\thetainit+\nu(\theta-\thetainit))-H(\thetainit)\right](\theta-\widehat{\theta}) d\nu\right|\\
&\lesssim  \|\theta-\widehat{\theta}\|_2^3\lesssim\left(\|\theta-\theta^*\|_2+\|\widehat{\theta}-\theta^*\|_2\right)^3\lesssim \tau^3 (d/n)^{3/2},
\end{aligned}
\label{eq: local bound thm5}
\end{equation}
where the first inequality follows from the continuity property of $H(\cdot)$ established in \eqref{eq: Lip H} of Lemma \ref{lem: key property}, the second inequality holds due to the triangle inequality and the last inequality follows from the definition of $\Theta_{\rm loc}$ and Theorem \ref{thm: final rate}.

Recall that 
\[
\phi(\theta,\gamma) = \theta^\intercal \U \gamma + S(\theta),\quad \phi_{\rm ap}(\theta,\gamma) = \theta^\intercal \U \gamma + Q(\theta).
\]
We introduce or recall the following notations to facilitate discussions,
\begin{equation}
    F^*(\gamma) = \min_{\theta\in \RR^{dK}}\phi(\theta,\gamma), \quad \textrm{and}\quad F(\gamma) = \min_{\theta\in \RR^{dK}}\phi_{\rm ap}(\theta,\gamma)
    \label{eq: def F}
\end{equation}
and 
\begin{equation}
    F^*_{\rm loc}(\gamma) = \min_{\theta\in \Theta_{\rm loc}}\phi(\theta,\gamma), \quad \textrm{and}\quad F_{\rm loc}(\gamma) = \min_{\theta\in \Theta_{\rm loc}}\phi_{\rm ap}(\theta,\gamma).
    \label{eq: def F loc}
\end{equation}
In the above notations, the superscript $^*$ indicates whether the function involves the original objective $\phi(\cdot, \cdot)$ or its approximation $\phi_{\rm ap}(\cdot, \cdot)$. The subscript ${\rm loc}$ indicates whether the minimization is taken over the local set $\Theta_{\rm loc}$ or the full space \(\RR^{dK}\).
Then, we recall the definitions of  $\gamma^*$ in \eqref{eq: one saddle} and $\gamma^*_{\rm ap}$ in \eqref{eq: step 1} as 
\begin{equation}
\gamma^*\in \argmax_{\gamma\in \Delta^L} F^*_{\rm loc}(\gamma) \quad \text{and} \quad \gamma^*_{\rm ap} = \argmax_{\gamma\in \Delta^L} F(\gamma).
\label{eq: recall gamma_star}
\end{equation}

In the following, we establish the upper bound of $\|\gamma^*_{\rm ap} - \gamma^*\|_2$, by controlling the objective gap $F(\gamma^*_{\rm ap})-F(\gamma^*)$. We leverage the function $F(\gamma)$ because $F(\gamma)$ is strongly concave in $\gamma\in \Delta^L$, as established in \eqref{eq: expression of F(gamma)}. This strong concavity ensures that the objective gap allows us to bound the distance between $\gamma^*_{\rm ap}$ and $\gamma^*$ pointwisely. The proof for bounding $\|\gamma^*_{\rm ap} - \gamma^*\|_2$
consists of two steps: 
\begin{enumerate}
\item[\bf Step 1.] We first establish that, on the event $\GG$,
\begin{equation}
    F_{\rm loc}(\gamma_{\rm ap}^*) - F_{\rm loc}(\gamma^*) \lesssim \tau^3(d/n)^{3/2}.
    \label{eq: min value diff thm5}
\end{equation}
\item[\bf Step 2.] We then show that, on the event $\GG$, 
$$F_{\rm loc}(\gamma^*_{\rm ap}) = F(\gamma^*_{\rm ap}) \quad \text{and} \quad F_{\rm loc}(\gamma^*) = F(\gamma^*).$$ 
Combining Step 1, we establish the upper bound for the objective gap $F(\gamma^*_{\rm ap}) - F(\gamma^*)$. Then we leverage the strong concavity of the function $F(\gamma)$ for $\gamma\in \Delta^L$, as established in \eqref{eq: expression of F(gamma)}, to obtain the convergence rate of $\|\gamma^*_{\rm ap} - \gamma^*\|_2$.
\end{enumerate}
%%%%%%%%%%%%%%%%%%%%%%%%%%%%%%%%%%%%%%%%%%%%%%%%%%%%%%%%%%%%%%%%%%%%%%%%%%%%%%%%%
\subsubsection{Proof for Step 1}
%%%%%%%%%%%%%%%%%%%%%%%%%%%%%%%%%%%%%%%%%%%%%%%%%%%%%%%%%%%%%%%%%%%%%%%%%%%%%%%%%
We now establish an upper bound for $F_{\rm loc}(\gamma_{\rm ap}^*) - F_{\rm loc}(\gamma^*).$
For any $\gamma\in \Delta^L$, on the event $\GG$, we have
\[
\begin{aligned}
    \left|F_{\rm loc}(\gamma) - F^*_{\rm loc}(\gamma)\right| %&= \left|\min_{\theta\in \Theta_{\rm loc}}\left\{\theta^\intercal \U \gamma + Q(\theta) \right\} - \min_{\theta\in \Theta_{\rm loc}}\left\{\theta^\intercal \U \gamma + S(\theta) \right\}\right|\\
    &= \left| \min_{\theta\in \Theta_{\rm loc}}\left\{\theta^\intercal \U \gamma+Q(\theta) \right\} -\min_{\theta\in \Theta_{\rm loc}}\left\{\theta^\intercal \U \gamma+ S(\theta) \right\} \right|\\
    &\leq \max_{\theta\in \Theta_{\rm loc}}\left|\theta^\intercal\U\gamma+ S(\theta)- \theta^\intercal \U\gamma- Q(\theta) \right| \\
    &= \max_{\theta\in \Theta_{\rm loc}}|S(\theta) - Q(\theta)| \leq C \tau^3(d/n)^{3/2},
\end{aligned}
\]
where the last inequality follows from \eqref{eq: local bound thm5}. Moreover, the following holds:
\[
\begin{aligned}
    F_{\rm loc}(\gamma_{\rm ap}^*) - F_{\rm loc}(\gamma^*) &= F_{\rm loc}(\gamma_{\rm ap}^*) - F_{\rm loc}^*(\gamma_{\rm ap}^*) + F_{\rm loc}^*(\gamma_{\rm ap}^*) - F_{\rm loc}^*(\gamma^*) + F^*_{\rm loc}(\gamma^*)- F_{\rm loc}(\gamma^*) \\
    &\leq F_{\rm loc}(\gamma_{\rm ap}^*) - F_{\rm loc}^*(\gamma_{\rm ap}^*) +  F_{\rm loc}^*(\gamma^*)- F_{\rm loc}(\gamma^*) \\
    &\leq 2\max_{\gamma\in \Delta^L}\left| F_{\rm loc}(\gamma) - F^*_{\rm loc}(\gamma)\right|,
\end{aligned}
\]
where the first inequality holds since the definition of $\gamma^*$ in \eqref{eq: recall gamma_star} implies $F^*_{\rm loc}(\gamma^*_{\rm ap})\leq F^*_{\rm loc}(\gamma^*)$.
Combining the above two inequalities, we establish \eqref{eq: min value diff thm5}.
%%%%%%%%%%%%%%%%%%%%%%%%%%%%%%%%%%%%%%%%%%%%%%%%%%%%%%%%%%%%%%%%%%%%%%%%%%%%%%%%%
\subsubsection{Proof for Step 2}
%%%%%%%%%%%%%%%%%%%%%%%%%%%%%%%%%%%%%%%%%%%%%%%%%%%%%%%%%%%%%%%%%%%%%%%%%%%%%%%%%
% In the following, we shall show that on the event $\GG$, $F_{\rm loc}(\gamma_{\rm ap}^*) = F(\gamma_{\rm ap}^*)$ and $F_{\rm loc}(\gamma^*)=F(\gamma^*)$. Then together with the strongly concavity of $F(\gamma)$ for $\gamma\in \Delta^L$, as established in \eqref{eq: expression of F(gamma)}, we obtain the convergence rate of $\|\gamma_{\rm ap}^* - \gamma^*\|_2$.
We begin with the proof of  $F_{\rm loc}(\gamma_{\rm ap}^*) = F(\gamma_{\rm ap}^*)$. 
On the event $\GG$, Lemma \ref{lem: saddle point} establishes that $\phi_{\rm ap}(\theta^*_{\rm ap},\gamma^*_{\rm ap})\leq \phi_{\rm ap}(\theta,\gamma^*_{\rm ap})$ for all $\theta\in \RR^{dK}$. Therefore, for a smaller parameter set $\Theta_{\rm loc}$, the inequality still holds such that:
\[
\phi_{\rm ap}({\theta}^*_{\rm ap}, {\gamma}^*_{\rm ap}) \leq \phi_{\rm ap}(\theta, {\gamma}^*_{\rm ap}) \quad \textrm{for all $\theta\in \Theta_{\rm loc}$}.
\]
In addition, by Theorem \ref{thm: quad convergence}, we have shown that, on the event $\GG$, 
%$\|{\theta}^*_{\rm ap}-\theta^*\|_2\leq $, for some constant $C>0$. \Zijian{Why the next sentence? We need to conditon on event $\mathcal{G}$?} Since  as given in \eqref{eq: final rate}, it holds that
\begin{equation}
\|{\theta}^*_{\rm ap}-\theta^*\|_2\leq Ct \sqrt{d/{n}}\leq\tau \sqrt{d/n},
\label{eq: bar theta bound thm5}
\end{equation}
where the last inequality holds since $\tau = Ct (1+\frac{1}{\sigma_L^2(\U)})\geq Ct.$
Hence, ${\theta}^*_{\rm ap} \in \Theta_{\rm loc}$. The preceding two inequalities imply that  ${\theta}^*_{\rm ap} \in \argmin_{\theta\in \Theta_{\rm loc}} \phi_{\rm ap}(\theta,{\gamma}^*_{\rm ap})$, with
\begin{equation}
    \phi_{\rm ap}(\theta_{\rm ap}^*,{\gamma}^*_{\rm ap})=\min_{\theta\in \Theta_{\rm loc}} \phi_{\rm ap}(\theta,{\gamma}^*_{\rm ap})= F_{\rm loc}(\gamma^*_{\rm ap}),
    \label{eq: thm5 step2 inter1}
\end{equation}
where the second equality follows from the definition of $F_{\rm loc}(\gamma)$ in \eqref{eq: def F loc}.
Note that the dual formulation of $\theta^*_{\rm ap}$ in \eqref{eq: step 2} implies 
\[
\phi_{\rm ap}(\theta_{\rm ap}^*,{\gamma}^*_{\rm ap})=\min_{\theta\in \RR^{dK}} \phi_{\rm ap}(\theta,{\gamma}^*_{\rm ap})= F(\gamma^*_{\rm ap}),
\]
where the last equality holds from the definition of $F(\gamma)$ in \eqref{eq: def F}.
Therefore, we establish that
\begin{equation}
\begin{aligned}
    F_{\rm loc}(\gamma_{\rm ap}^*) = F(\gamma_{\rm ap}^*).
\end{aligned}
    \label{eq: min f_ap gamma_bar thm5}
\end{equation}

Next, we establish that $F_{\rm loc}(\gamma^*)=F(\gamma^*)$. Together with \eqref{eq: min f_ap gamma_bar thm5} and \eqref{eq: min value diff thm5}, we obtain an upper bound for the objective gap $F(\gamma^*_{\rm ap})-F(\gamma^*)$, which in turn controls $\|\gamma^*_{\rm ap}-\gamma^*\|_2$. We divide the discussion into two cases for $\gamma^*$, based on whether the following condition holds or not: 
\begin{equation}
    \left\|[H(\thetainit)]^{-1}\U (\gamma^* - {\gamma}^*_{\rm ap})\right\|_2 \leq \tau \sqrt{d/n}.
    \label{eq: cond thm5}
\end{equation}

\noindent{\bf Case 1: $\gamma^*$ satisfies Condition \eqref{eq: cond thm5}.} 
We define
\begin{equation}
    \theta(\gamma^*) = \argmin_{\theta\in \RR^{dK}}\phi_{\rm ap}(\theta,\gamma^*),
    \label{eq: case1 eq1 thm5}
\end{equation}
and apply \eqref{eq: argmin phi_ap} to obtain 
\begin{equation*}
    \theta(\gamma^*)= \thetainit - [H(\thetainit)]^{-1}\left(\U\gamma^* + \nabla S(\thetainit)\right).
\end{equation*}
In \eqref{eq: closed form theta}, we show ${\theta}^*_{\rm ap}$ admits the closed-form expression ${\theta}^*_{\rm ap}=\thetainit-[H(\thetainit)]^{-1}\left(\U\gamma^*_{\rm ap}+\nabla S(\thetainit)\right)$. Therefore, we establish that
\begin{equation*}
\|\theta(\gamma^*) - {\theta}^*_{\rm ap}\|_2 = \left\|[H(\thetainit)]^{-1}\U (\gamma^* - {\gamma}^*_{\rm ap})\right\|_2 \leq \tau\sqrt{d/n},
    \label{eq: case1 eq2 thm5}
\end{equation*}
where the inequality holds due to Condition \eqref{eq: cond thm5}.
We now apply the above inequality together with \eqref{eq: bar theta bound thm5} and establish 
\begin{equation*}
\|\theta(\gamma^*) - \theta^*\|_2 \leq \|\theta(\gamma^*)-{\theta}^*_{\rm ap}\|_2 + \|{\theta}^*_{\rm ap} - \theta^*\|_2 \leq 2\tau\sqrt{d/n}.
    \label{eq: case1 eq3 thm5}
\end{equation*}
Therefore, we conclude that $\theta(\gamma^*)\in \Theta_{\rm loc}$. Since $\theta(\gamma^*)$ is the global minimizer of $\phi_{\rm ap}(\theta,\gamma^*)$ over the entire space $\RR^{dK}$ as defined in \eqref{eq: case1 eq1 thm5}, we obtain that $\theta(\gamma^*)$ is also the minimizer over the local parameter set $\Theta_{\rm loc}$. Formally, this yields
\[
\phi_{\rm ap}\left(\theta(\gamma^*),\gamma^*\right)=\min_{\theta\in \RR^{dK}}\phi_{\rm ap}(\theta,\gamma^*) = \min_{\theta\in \Theta_{\rm loc}}\phi_{\rm ap}(\theta,\gamma^*) .
\]
Applying the expressions of $F(\gamma)$ and $F_{\rm loc}(\gamma)$ in \eqref{eq: def F} and \eqref{eq: def F loc}, respectively, we establish that on the event $\GG$, when \eqref{eq: cond thm5} holds, 
\begin{equation}
    F_{\rm loc}(\gamma^*) = F(\gamma^*).
    \label{eq: min f_ap gamma_star thm5}
\end{equation}
% where the last inequality is due to the expression in \eqref{eq: F_gamma def thm2}.
Putting \eqref{eq: min f_ap gamma_bar thm5} and \eqref{eq: min f_ap gamma_star thm5} back to \eqref{eq: min value diff thm5}, we establish that
\begin{equation}
\begin{aligned}
F(\gamma_{\rm ap}^*) - F(\gamma^*) \lesssim \tau^3(d/n)^{3/2}.
\end{aligned}
    \label{eq: case1 eq4 thm5}
\end{equation}

The optimality condition of $\gamma^*_{\rm ap}$, which maximizes the concave function $F(\gamma)$ for $\gamma\in \Delta^L$, as defined in \eqref{eq: recall gamma_star}, implies that 
\begin{equation*}
    \langle \nabla F(\gamma_{\rm ap}^*), \gamma - \gamma_{\rm ap}^*\rangle \leq 0, \quad \textrm{for all $\gamma\in \Delta^L$.}
\end{equation*}
We substitute $\gamma$ with $\gamma^*$ in the above inequality to obtain that
\begin{equation}
    \langle \nabla F(\gamma_{\rm ap}^*), \gamma^* - \gamma_{\rm ap}^*\rangle \leq 0.
    \label{eq: optimality gamma_ap_star}
\end{equation}
Moreover, on the event $\GG$,
it follows from \eqref{eq: expression of F(gamma)} that $F(\gamma)$ is strongly concave for $\gamma\in \Delta^L$,
with its Hessian being expressed as
\[
\nabla^2 F(\gamma) = -\U^\intercal [H(\thetainit)]^{-1} \U \preceq - \lambda_{\rm min}([H(\thetainit)]^{-1})\sigma^2_L(\U)\cdot {\bf I}_L \preceq - \frac{\sigma_L^2(\U)}{\kappa_2(1-c_1)}\cdot {\bf I}_{L},
\]
where $\sigma_L(\U)>0$ is the $L$-th largest singular value of the matrix $\U$ as assumed in Condition {\rm (A3)}, and the last inequality follows from Lemma \ref{lem: convexity of H} with a constant $c_1\in (0,1/2)$.
 The strong concavity of $F(\cdot)$ implies 
\begin{equation*}
    \begin{aligned}
    F(\gamma^*) &\leq F(\gamma_{\rm ap}^*) + \langle \nabla F(\gamma_{\rm ap}^*), \gamma^* - \gamma_{\rm ap}^*\rangle - \frac{\sigma_L^2(\U)}{2\kappa_2(1-c_1)} \|\gamma^* - {\gamma}^*_{\rm ap}\|_2^2 \\
    &\leq F(\gamma_{\rm ap}^*) - \frac{\sigma_L^2(\U)}{2\kappa_2(1-c_1)} \|\gamma^* - {\gamma}^*_{\rm ap}\|_2^2,
\end{aligned}
\label{eq: case1 eq5 thm5}
\end{equation*}
% \begin{equation*}
%     \begin{aligned}
%     F(\gamma^*) &\leq F(\gamma_{\rm ap}^*) + \langle \nabla F(\gamma_{\rm ap}^*), \gamma^* - \gamma_{\rm ap}^*\rangle - \frac{\lambda_{\rm min}(\U^\intercal [H(\thetainit)]^{-1} \U)}{2} \|\gamma^* - {\gamma}^*_{\rm ap}\|_2^2 \\
%     &\leq F(\gamma_{\rm ap}^*) - \frac{\lambda_{\rm min}(\U^\intercal [H(\thetainit)]^{-1} \U)}{2} \|\gamma^* - {\gamma}^*_{\rm ap}\|_2^2,
% \end{aligned}
% \label{eq: case1 eq5 thm5}
% \end{equation*}
where the second inequality holds due to \eqref{eq: optimality gamma_ap_star}. 

Combining the above inequality with \eqref{eq: case1 eq4 thm5}, we further establish that on the event $\GG$,
\begin{equation*}
\|\gamma^* - {\gamma}^*_{\rm ap}\|_2 \lesssim \frac{1}{\sigma_L(\U)}\tau^{3/2}(d/n)^{3/4}.
\end{equation*}
Since $\tau^{1/2}= \sqrt{Ct(1+\frac{1}{\sigma_L^2(\U)})}\gtrsim \frac{1}{\sigma_L(\U)}$, where $t$ is any value larger than $c_0$ as specified in Condition {\rm (A1)}, we further establish that
\begin{equation}
\|\gamma^* - {\gamma}^*_{\rm ap}\|_2 \lesssim  \tau^2 (d/n)^{3/4}.
    \label{eq: case1 eq6 thm5}
\end{equation}

\noindent {\bf Case 2: $\gamma^*$ does not satisfy Condition \eqref{eq: cond thm5}.} In this case, $\gamma^*$ satisfies
\[
\left\|[H(\thetainit)]^{-1}\U (\gamma^* - {\gamma}^*_{\rm ap})\right\|_2 > \tau \sqrt{d/n}.
\]
We then define $\gamma_{\rm proj}$ as follows:
\[
\gamma_{\rm proj} = {\gamma}^*_{\rm ap} + \nu (\gamma^* -{\gamma}^*_{\rm ap}), \quad \textrm{with}\quad \nu = \frac{\tau\sqrt{d/n}}{\left\|[H(\widehat{\theta})]^{-1}\U (\gamma^* - {\gamma}^*_{\rm ap})\right\|_2} \in (0, 1).
\]
By the above definition, we have $\left\|[H(\widehat{\theta})]^{-1}\U (\gamma_{\rm proj} - {\gamma}^*_{\rm ap})\right\|_2 = \tau \sqrt{d/n}$, thus
\[
\tau^2\frac{d}{n} = \|[H(\widehat{\theta})]^{-1}\U(\gamma_{\rm proj} - {\gamma}^*_{\rm ap})\|_2^2 \leq \lambda_{\rm max}(\U^\intercal [H(\thetainit)]^{-2} \U) \|\gamma_{\rm proj} - {\gamma}^*_{\rm ap}\|_2^2.
\]
Therefore, we obtain that
\begin{equation}
    \|\gamma_{\rm proj} - {\gamma}^*_{\rm ap}\|_2 \geq \frac{\tau\sqrt{d/n}}{\sqrt{\lambda_{\rm max}(\U^\intercal [H(\thetainit)]^{-2} \U)}} \gtrsim \tau\sqrt{d/n},
\label{eq: gamma_proj diff thm5}
\end{equation}
where the second inequality holds from Lemma \ref{lem: convexity of H} that $\lambda_{\rm min}(H(\thetainit))\geq c'$ on the event $\GG$ {and $\|\U\|_2\leq C$ assumed in Condition (A4)}.

Recall that $F_{\rm loc}(\gamma)$ is concave towards $\gamma\in \Delta^L$, as defined in \eqref{eq: def F loc}. Therefore, we have
\[
\begin{aligned}
    F_{\rm loc}(\gamma_{\rm proj}) = F_{\rm loc}(\nu\gamma^* + (1-\nu){\gamma}^*_{\rm ap}) \geq \nu \cdot F_{\rm loc}(\gamma^*) + (1-\nu)\cdot F_{\rm loc}({\gamma}^*_{\rm ap}).
\end{aligned}
\]
Then it holds that
{
\begin{equation}
    F_{\rm loc}({\gamma}^*_{\rm ap}) -
F_{\rm loc}(\gamma_{\rm proj}) \leq \nu\cdot \left(F_{\rm loc}({\gamma}^*_{\rm ap}) - F_{\rm loc}(\gamma^*)\right)\lesssim\tau^3(d/n)^{3/2},
\label{eq: F_loc case 2 diff bound}
\end{equation}
}
where the last inequality follows from \eqref{eq: min value diff thm5} and $\nu\in (0,1)$. 

Since $\gamma_{\rm proj}$  satisfies $\left\|[H(\widehat{\theta})]^{-1}\U (\gamma_{\rm proj} - {\gamma}^*_{\rm ap})\right\|_2 \leq \tau \sqrt{d/n}$,  we shall apply the same argument as those from \eqref{eq: case1 eq1 thm5} to \eqref{eq: min f_ap gamma_star thm5} in Case 1, by replacing $\gamma^*$ by $\gamma_{\rm proj}$, and establish
\[
F_{\rm loc}(\gamma_{\rm proj}) = F(\gamma_{\rm proj}).
\]
Putting the above equality and \eqref{eq: min f_ap gamma_bar thm5} back to \eqref{eq: F_loc case 2 diff bound}, we establish that
\[
F(\gamma^*_{\rm ap}) - F(\gamma_{\rm proj}) \lesssim \tau^{3}(d/n)^{3/2}.
\]
Lastly, we essentially follow the argument as those from \eqref{eq: case1 eq4 thm5} to \eqref{eq: case1 eq6 thm5} in Case 1, by replacing $\gamma^*$ by $\gamma_{\rm proj}$, and  establish that
\begin{equation}
    \|\gamma_{\rm proj} - {\gamma}^*_{\rm ap}\|_2 \lesssim\tau^{2} (d/n)^{3/4}.
    \label{eq: gamma_proj diff thm5 contradict}
\end{equation}

As we consider the value $t$ that satisfies $t\ll \sigma_L^2(\U) (n/d)^{1/4}$, it holds that
$\tau \asymp (1+ \frac{1}{\sigma_L^2(\U)}) t \ll (d/n)^{-1/4}.$
Then, we have
\[
\tau^2 (d/n)^{3/4} \ll \tau (d/n)^{1/2}.
\]
Therefore, the inequality in \eqref{eq: gamma_proj diff thm5 contradict} contradicts \eqref{eq: gamma_proj diff thm5} on the event $\GG$. Then, Case 2 does not occur. To conclude, we have shown that only Case 1 occurs, for which we establish the convergence rate with $\|\gamma^* - \gamma^*_{\rm ap}\|_2 \lesssim \tau^2 (d/n)^{3/4}$, as in \eqref{eq: case1 eq6 thm5} on the event $\GG$. Lastly, we leverage Lemma \ref{lem: concentration lemma} to control the probability of the event $
\GG$.

\section{Additional Methods and Theories}
\label{appendix: additional methods}

This section is organized as follows. 
\begin{itemize}
    \item Section~\ref{sec: add literature} reviews additional literature related to algorithms, generalization bounds, and statistical inference for minimax optimization. 
    \item Section~\ref{sec: doubly robust} describes the complete procedure for solving the empirical \CGDRO\ model under the covariate shift setting, including (i) the construction of the DML estimators $\{\widehat{\mu}^{(l)}\}_{l\in [L]}$ in Section~\ref{subsec: construction of mu}, and (ii) solving the minimax optimization problem via the Mirror Prox algorithm in Section~\ref{subsec: MP}. 
    \item Section~\ref{subsec: properties of hatmu} investigates the statistical properties of the constructed DML estimators $\{\widehat{\mu}^{(l)}\}_{l\in [L]}$, including their convergence rates to the population counterparts and their asymptotic distributions. 
    \item Section~\ref{sec: entire infer} provides further technical details for the proposed perturbation-based statistical inference procedure, which were omitted from the main text. 
    \item Finally, Section~\ref{appedix: prior} discusses how prior information about the target distribution can be incorporated into the formulation of the \CGDRO\ model.
\end{itemize}

\subsection{Additional Literature about Optimization}
\label{sec: add literature}
This section reviews additional literature not covered in Section~\ref{subsec: literature} of the main text, focusing on algorithmic developments, generalization theory, and statistical inference related to minimax optimization problems.

\vspace{1mm}
\noindent\textbf{Algorithms for Minimax Optimization.}
Minimax optimization employs established algorithms like gradient descent maximization (updating the minimizer after finding a near-optimal maximizer) and gradient descent ascent (updating  both variables simultaneously or alternately) \cite{lin2020gradient}. These have been extended to stochastic settings \cite{lin2020gradient,yang2020catalyst}. For a summary of optimization convergence results, see \cite{zhang2025avoid}. We utilize the Mirror Prox algorithm  \cite{bubeck2015convex,mokhtari2020convergence,mokhtari2020unified} to solve the \CGDRO\ model. Our primary algorithmic contribution is integrating a DML estimator into the risk function. Our analysis focuses on the statistical convergence rate and uncertainty quantification of the minimax solution, rather than on optimization convergence guarantees.

\vspace{1mm}
\noindent\textbf{Generalization Bounds for Minimax Optimization.}
Research on stochastic minimax optimization also addresses generalization performance and statistical rates. For example, \cite{zhang2021generalization} examined the stability of empirical minimax optimization and established generalization bounds. \cite{zhang2022bring} extended this by combining these results with optimization convergence rates to derive bounds for complex algorithms. More recently, \cite{boob2024optimal} established stability and generalization bounds for the extra-gradient method. However, these existing results are not applicable to our setting due to their reliance on a strong concavity assumption for the inner maximization problem. When this assumption is violated, as in our case, the maximization problem may have multiple optimal solutions, deteriorating the convergence rate \cite{zhang2024generalization}. To overcome this, we develop surrogate optimization problems as theoretical bridges for our analysis. Additionally, our approach incorporates a DML estimator for the empirical objective, unlike prior work that assumes direct access to a simple sample-average empirical objective.

\vspace{1mm}
\noindent\textbf{Statistical Inference via Optimization Methods.}
Our work relates to statistical inference for optimization methods. Early work by \cite{chen2020statistical} established inference for stochastic gradient methods under strong convexity and smoothness, later generalized by \cite{davis2024asymptotic} for non-smooth objectives. Further works investigated statistical inference for covariance estimation using gradient methods \cite{jiang2025online} and constrained stochastic optimization \cite{duchi2021asymptotic,du2025online}. However, these analyses are not directly applicable to our minimax problem, and Figure \ref{fig: Inference Challenge} empirically shows that the limiting distribution of the empirical CG-DRO model is not normal. In contrast, we develop a novel inferential strategy for minimax optimization. We first quantify uncertainty from the inner maximization using a novel perturbation approach, then perform inference on the outer minimization via M-estimation. This decoupling overcomes the challenge of directly analyzing the full minimax solution, thereby enabling the construction of valid confidence intervals in the presence of a nonstandard limiting distribution. %

\subsection{Complete Procedure for Solving the Empirical \CGDRO\ \eqref{eq: sample optimizers}}
\label{sec: doubly robust}

In this subsection, we present the complete procedure for solving the empirical \CGDRO\ model \eqref{eq: sample optimizers}. This includes (i) the construction of the Double Machine Learning (DML) estimators $\{\widehat{\mu}^{(l)}\}_{l \in [L]}$ and (ii) solving the minimax problem using the Mirror Prox algorithm.

\subsubsection{Construction of DML estimators $\{\widehat{\mu}^{(l)}\}_{l \in [L]}$}
\label{subsec: construction of mu}
As shown in \eqref{eq: DR component}, the construction of $\{\widehat{\mu}^{(l)}\}_{l \in [L]}$ involves the implementation of the nuisance models $\{\widehat{f}^{(l)}(\cdot), \widehat{\omega}^{(l)}(\cdot)\}_{l \in [L]}$.

We start with the construction of the conditional probability models $\{\widehat{f}^{(l)}\}_{l\in [L]}$.
As discussed in \textbf{Step 2} of Section~\ref{sec: double robustness}, we require the fitted models $\{\widehat{f}^{(l)}\}_{l \in [L]}$ are independent of the data used for bias correction. To ensure such independence, we employ a cross-fitting strategy. Specifically, for each source $l\in[L]$, we randomly partition the samples $\{X^{(l)}_i, Y^{(l)}_i\}_{i\in [n_l]}$ into two disjoint subsets $\A_l$ and $\B_l$. The models fitted on one subset are thereby independent of the data in the other subset. Without loss of generality, we write $\A_l = \{1,2,..., \lfloor n_l/2\rfloor\}$ and $\B_l = [n_l]\setminus \A_l$. 
For each source $l\in [L]$, we fit the conditional probability models on each subset separately, yielding estimators $\hf{l}_\A$ and $\hf{l}_\B$, using data in $\A_l$ and $\B_l$, respectively.

Now, we describe the construction of $\hf{l}_\A$, and the same procedure applies to $\hf{l}_\B$. 
Given the data $\{\X{l}_i, \Y{l}_i\}_{i\in \A_l}$, where $\Y{l}_i\in \{0,1,...,K\}$, $\hf{l}_\A:\RR^d\to \Delta^{K}$ is learned by solving the following problem:
\[
\hf{l}_\A = \argmin_{f\in \mathcal{F}}\frac{1}{|\A_l|}\sum_{i\in \A_l}\ell(f(\X{l}_i), \Y{l}_i) + \lambda \cdot \Omega(f),
\]
where $\ell(f(x), y)$ is the cross-entropy loss with the predicted probability $f(x)$, $\Omega(f)$ is a regularization penalty, and $\lambda\geq 0$ is the parameter tuned by cross-validation. Here $\mathcal{F}$ is the function class associated with machine learning methods, say XGBoost, random forest classifiers, and neural networks.

Using the fitted models $\hf{l}_\A(\cdot)$ and  $\hf{l}_\B(\cdot)$, we define the following prediction for both source covariate $\X{l}_i$ and target covariate $\XQ_j$, 
\begin{equation}
    \hf{l}(\X{l}_i) = \begin{cases}
        \hf{l}_\A(\X{l}_i) & \textrm{if $i\in \B_l$} \\
        \hf{l}_\B(\X{l}_i) & \textrm{if $i\in \A_l$}
    \end{cases},
    \;\textrm{and}\quad
    \hf{l}(\XQ_j) = \frac{1}{2}\hf{l}_\A(\XQ_j) + \frac{1}{2}\hf{l}_\B(\XQ_j).
    \label{eq: fl sample-split}
\end{equation}
This construction ensures that the function $\hf{l}(\cdot)$ is independent of the sample to which the function is applied, and such a property is essential for conducting the bias correction for generic machine learning algorithms.

We apply a similar sample-splitting procedure to construct the density ratio estimators $\{\omega^{(l)}\}_{l\in [L]}$, which also require independence from the data used for bias correction. Specifically, for each source $l\in [L]$, we estimate the density ratio $\hw{l}_\A$ using the source covariates $\{\X{l}_i\}_{i\in \A_l}$ together with the target covariates $\{\XQ_j\}_{j\in [N]}$, and we estimate $\hw{l}_\B$ using the source covariates $\{\X{l}_i\}_{i\in \B_l}$ and the same target covariates.  Various approaches for estimating density ratios have been proposed in the literature \cite{sugiyama2007direct,gretton2009covariate, nguyen2010estimating, sugiyama2012density}, and we describe a concrete method based on Bayes’ formula in the following.

We detail the construction of the density ratio estimator $\hw{l}_\A$ by Bayes' formula; the similar procedure applies to $\hw{l}_\B$. To begin, we merge the source covariates $\{\X{l}_i\}_{i\in \A_l}$ with the target covariates $\{\XQ_j\}_{j\in [N]}$, and denote the combined dataset as $\{\widetilde{X}^{(l)}_i\}_{i\in [|\A_l|+N]}$, where the first $|\A_l|$ samples are from the $l$-th source domain (subset $\A_l$), and the remaining $N$ samples are from the target domain. We introduce an indicator random variable $G^{(l)}\in \{0,1\}$ to denote the domain label of $\widetilde{X}^{(l)}$, such that:
\[
G^{(l)} = \begin{cases}
    0 & \textrm{if $\widetilde{X}^{(l)}$ is drawn from the $l$-th source domain}\\
    1 & \textrm{if $\widetilde{X}^{(l)}$ is drawn from the target domain}
\end{cases}.
\]
Let $p(x | G^{(l)}=g)$ denote the conditional probability density of  $\widetilde{X}^{(l)}$ given the domain label $G^{(l)}=g$, for $g\in \{0,1\}$. Then we define the source and target covariate densities as:
\[
p_X^{(l)}(x) = p(x | G_i^{(l)}=0), \quad \textrm{and}\quad q_X(x) = p(x|G^{(l)}=1).
\]
The covariate density ratio between the target and the $l$-th source domain is given by
\[
\w{l}(x) = \frac{q_X(x)}{p_X^{(l)}(x)} = \frac{p(x|G^{(l)}=1)}{p(x | G_i^{(l)}=0)}.
\]
Let $p(G^{(l)}=g)$ denote the marginal probability of the label $G^{(l)}=g$, and $p(G^{(l)}=g| \widetilde{X}^{(l)}=x)$ denote the class posterior probability. Applying Bayes' formula yields:
\begin{equation}
    \w{l}(x) = \frac{p(G^{(l)}=0)}{p(G^{(l)}=1)}\cdot \frac{p(G^{(l)}=1| \widetilde{X}^{(l)}=x)}{p(G^{(l)}=0| \widetilde{X}^{(l)}=x)}.
    \label{eq: bayes density ratio}
\end{equation}
% \[
% \w{l}(x) = \frac{d \QQ_X(x)}{d \PP_X^{(l)}(x)} = \frac{p(G_i^{(l)} = 0)}{p(G_i^{(l)}=1)}\cdot \frac{p(G_i^{(l)}=1|\widetilde{X}^{(l)}_i=x)}{p(G_i^{(l)}=0|\widetilde{X}^{(l)}_i=x)}.
% \]

We now estimate this density ratio. First, the marginal class ratio $p(G^{(l)}=0)/p(G^{(l)}=1)$ is approximated by the empirical ratio of sample sizes, $|\mathcal{A}_l| / N$. Second, the class posterior probabilities $p(G^{(l)}_i=1| \widetilde{X}^{(l)}_i=x)$ is estimated using a probabilistic binary classifier, such as logistic regression or random forests, trained on the combined dataset $\{(\widetilde{X}^{(l)}_i, G^{(l)}_i)\}_{i \in n[|\mathcal{A}_l| + N]}$. Denoting this the fitted probability by $\hat{p}(G^{(l)}_i=1| \widetilde{X}^{(l)}_i=x)$, we construct $\hw{l}_\A(x)$ as follows:
\[
\hw{l}_\A(x) = \frac{|\A_l|}{N}\cdot \frac{\hat{p}(G^{(l)}_i=1| \widetilde{X}^{(l)}_i=x)}{1-\hat{p}(G^{(l)}_i=1| \widetilde{X}^{(l)}_i=x)}.
\]

Using these sample-split models, we then define the following prediction for source data:
\begin{equation}
    \hw{l}(\X{l}_i) = \begin{cases}
        \hw{l}_\A(\X{l}_i) & \textrm{if \;$i\in \B_l$}\\
        \hw{l}_\B(\X{l}_i) & \textrm{if \;$i\in \A_l$}\\
    \end{cases}.
    \label{eq: wl sample-split}
\end{equation}
This design ensures that $\hw{l}(\cdot)$ remains independent from the data on which the final estimator is computed, thereby enabling rigorous bias correction.

Provided the nuisance models $\hf{l},\hw{l}$ defined in \eqref{eq: fl sample-split} and \eqref{eq: wl sample-split}, we plug them into \eqref{eq: DR component} to complete the construction of the DML estimator $\widehat{\mu}^{(l)}$ under the general covariate shift regime.

\subsubsection{Mirror Prox Algorithm}
\label{subsec: MP}
We now present the Mirror Prox algorithm to solve the empirical \CGDRO\ problem in \eqref{eq: sample optimizers}. Mirror Prox performs gradient descent on $\theta$ and mirror ascent~\cite{nemirovski2009robust} on $\gamma$ due to the simplex constraint. Therefore, we first present the expressions of the gradients of $\widehat{\phi}(\theta,\gamma)$ in \eqref{eq: sample optimizers} with respect to $\theta$ and $\gamma$. Given the expressions $\widehat{\mu}^{(l)}, \widehat{S}(\theta)$ as established in Section \ref{sec: refined analysis}, it holds
\begin{equation*}
\nabla_{\theta} \widehat{\phi}(\theta,\gamma)=\left(\left[\sum_{l=1}^{L} \gamma_{l}\cdot \widehat{\mu}_1^{(l)}+\nabla_{\theta_1} \widehat{S}(\theta)\right]^{\intercal},\cdots, \left[\sum_{l=1}^{L} \gamma_{l}\cdot \widehat{\mu}^{(l)}_K+\nabla_{\theta_K} \widehat{S}(\theta)\right]^{\intercal}\right)^{\intercal},
\label{grad_theta}
\end{equation*}
\begin{equation*}
\nabla_{\gamma} \widehat{\phi}(\theta, \gamma)= \left(\theta^{\intercal}\widehat{\mu}^{(1)},\cdots, \theta^{\intercal}\widehat{\mu}^{(L)}\right).
\end{equation*}

Each iteration for Mirror Prox consists of two sequential steps: an \textbf{intermediate step} that computes an intermediate point to better estimate the gradient direction, followed by a \textbf{correction step} that updates from the current iterate using gradients evaluated at the intermediate point. The literature has shown that this two-step procedure  effectively reduces oscillations near saddle points and achieves faster convergence rates in smooth convex-concave settings \cite{nemirovski2004prox, bubeck2015convex, mokhtari2020unified}. As we shall show below, such a two-step procedure facilitates Mirror Prox yielding a fast optimization convergence rate of $\mathcal{O}(T^{-1})$. In contrast, without computing intermediate points for gradient evaluation as in the Mirror Prox, the single-step algorithm with gradient descent-mirror ascent yields a slower optimization convergence rate of  $\mathcal{O}(T^{-1/2})$, as discussed in Section 5.2 of \cite{bubeck2015convex}.

\begin{algorithm}[ht!]
\DontPrintSemicolon
\SetAlgoLined
\SetNoFillComment
\LinesNotNumbered 
\caption{Data-dependent Mirror Prox Algorithm}
\label{algo: mp}
\KwData{Estimated $\{\widehat{\mu}^{(l)}\}_{l\in [L]}$; target covariates $\{X_j^{\QQ}\}_{j \in [N]}$;  learning rates for Mirror Prox: $\eta_\theta,\eta_\gamma$; maximum iteration $T$, tolerance $\epsilon$.}

\KwResult{Mirror Prox estimators $\widehat{\theta}_T$, $\widehat{\gamma}_T$.}

    Set $\theta_{0}=\bar{\theta}_0=0$, $\gamma_{0}=\bar{\gamma}_0=(1/L,\cdots,1/L)$, and  ${\rm gap}= 1$;

\While{$0 \leq t \leq T$ and ${\rm diff} \geq \epsilon$}{
    \textbf{Intermediate Step}:
    {
    \[
    \begin{aligned}
\bar{\theta}_{t+1}&=\theta_{t}-\eta_\theta\cdot \nabla_{\theta} \widehat{\phi}(\bar{\theta}_{t},\bar{\gamma}_{t}),\;\;\textrm{and}\;\;\\
[\bar{\gamma}_{t+1}]_l&= \frac{\gamma_{t,l}\cdot \exp\left(\eta_\gamma\cdot \left[\nabla_{\gamma} \widehat{\phi}(\bar{\theta}_t, \bar{\gamma}_t)\right]_{l} \right)}{\sum_{j=1}^{L}\gamma_{t,j}\cdot \exp\left(\eta_\gamma \cdot \left[\nabla_{\gamma} \widehat{\phi}(\bar{\theta}_t, \bar{\gamma}_t)\right]_{j} \right)} , \quad \textrm{for $1\leq l\leq L$}.
\end{aligned}
    \]}

    \textbf{Correction Step}:
    {
    \[
    \begin{aligned}
    {\theta}_{t+1}&=\theta_{t}-\eta_\theta\cdot \nabla_{\theta} \widehat{\phi}(\bar{\theta}_{t+1},\bar{\gamma}_{t+1}),\;\;\textrm{and}\\
    [{\gamma}_{t+1}]_l&= \frac{\gamma_{t,l}\cdot \exp\left(\eta_\gamma \cdot \left[\nabla_{\gamma} \widehat{\phi}(\bar{\theta}_{t+1},\bar{\gamma}_{t+1})\right]_{l} \right)}{\sum_{j=1}^{L}\gamma_{t,j}\cdot \exp\left(\eta_\gamma \cdot \left[\nabla_{\gamma} \widehat{\phi}(\bar{\theta}_{t+1},\bar{\gamma}_{t+1})\right]_{j} \right)}, \quad \textrm{for $1\leq l\leq L$}.
    \end{aligned}
    \]
    }

    Set $\thetainit_t = \frac{1}{t}\sum_{s=1}^t\bar{\theta}_s$, and $\widehat{\gamma}_t = \frac{1}{t}\sum_{s=1}^t \bar{\gamma}_s$.

    Evaluate the duality gap: ${\rm gap} = \max_{\gamma\in \Delta^L}\widehat{\phi}(\widehat{\theta}_t,\gamma) - \min_{\theta\in \RR^{dK}}\widehat{\phi}(\theta,\widehat{\gamma}_t)$.
}

Denote $\widehat{\theta}_T = \widehat{\theta}_t$ and $\widehat{\gamma}_T = \widehat{\gamma}_t$, after convergence.

\end{algorithm}

Specifically, at iteration $t$, the \textbf{intermediate step} computes $(\bar{\theta}_{t+1}, \bar{\gamma}_{t+1})$ synchronously. We update $\bar{\theta}_{t+1}$ via the standard gradient descent, while $\bar{\gamma}_{t+1}$ is computed using a mirror ascent step. The mirror update generalizes the standard gradient update by replacing the Euclidean geometry with the non-Euclidean geometry that better respects the constraint structure. When the constraint is the simplex, as in our case, the mirror update admits a closed form, which avoids costly projection operations. In addition, the optimization convergence rate of gradient-based methods typically depends on the Lipschitz constant of the objective function measured in the $\ell_2$ norm, which grows with the dimension, the mirror update based on $\ell_\infty$ geometry exploits a dimension-independent Lipschitz constant, thus our method achieves faster optimization convergence compared to the standard gradient updates \cite{bubeck2015convex}. 

In the subsequent \textbf{correction step}, the iterate $(\theta_{t+1},\gamma_{t+1})$ is computed using gradients evaluated at the intermediate point $(\bar{\theta}_{t+1},\bar{\gamma}_{t+1})$. We emphasize that both steps at iteration $t$ are initialized from the same previous iterate $(\theta_t, \gamma_t)$. 
Note that the gradient computed for the correction step at the current iteration is reusable for the immediate step in the next iteration. It implies that at each iteration except for the first one, one only needs to compute the gradient once. This design contrasts with the extra-gradient Mirror Prox algorithm, which requires two gradient computations per iteration: one at $(\theta_t,\gamma_t)$ in the intermediate step, and another at $(\bar{\theta}_{t+1},\bar{\gamma}_{t+1})$ in the correction step  \cite{mokhtari2020unified, mokhtari2020convergence}.

Upon convergence at iteration $T$, we take the average of all the intermediate iterates $\frac{1}{T}\sum_{t=1}^T \bar{\theta}_t$ as the output in Algorithm \ref{algo: mp}. We denote $\widehat{\theta}_T =\frac{1}{T}\sum_{t=1}^T \bar{\theta}_t$ and $\widehat{\gamma}_T = \frac{1}{T}\sum_{t=1}^T \bar{\gamma}_t$. It has been established in the literature, like \cite{mokhtari2020unified} and Section 4 of \cite{bubeck2015convex}, that the optimistic-gradient Mirror Prox algorithm achieves the following duality gap, in smooth convex-concave settings, as in our case,
\[
\max_{\gamma\in \Delta^L}\widehat{\phi}(\widehat{\theta}_T,\gamma) - \min_{\theta\in \RR^{dK}}\widehat{\phi}(\theta,\widehat{\gamma}_T)\lesssim T^{-1}.
\]
Furthermore, we observe that
{\small
\[
\begin{aligned}
    &\max_{\gamma\in \Delta^L}\widehat{\phi}(\widehat{\theta}_T,\gamma) - \min_{\theta\in \RR^{dK}}\widehat{\phi}(\theta,\widehat{\gamma}_T) \\
    &\geq \max_{\gamma\in \Delta^L}\widehat{\phi}(\widehat{\theta}_T,\gamma) - \min_{\theta\in \RR^{dK}}\left[\max_{\gamma\in \Delta^L}\widehat{\phi}(\theta,\gamma)\right] + \max_{\gamma\in \Delta^L}\left[\min_{\theta\in \RR^{dK}}\widehat{\phi}(\theta,\gamma)\right] - \min_{\theta\in \RR^{dK}}\widehat{\phi}(\theta,\widehat{\gamma}_T) \\
    &\geq \max_{\gamma\in \Delta^L}\widehat{\phi}(\widehat{\theta}_T,\gamma) - \min_{\theta\in \RR^{dK}}\left[\max_{\gamma\in \Delta^L}\widehat{\phi}(\theta,\gamma)\right],
\end{aligned}
\]
}
where the first inequality holds as min-max larger than max-min.
The combination of the above two inequalities yields
\begin{equation}
    0\leq \max_{\gamma\in \Delta^L}\widehat{\phi}(\widehat{\theta}_T,\gamma) - \min_{\theta\in \RR^{Kd}}\left[\max_{\gamma\in \Delta^L}\widehat{\phi}(\theta,\gamma)\right] \lesssim T^{-1},
    \label{eq: conv MP}
\end{equation}
where $T$ denotes the number of iterations. Consequently, Algorithm~\ref{algo: mp} globally solves the empirical \CGDRO\ problem \eqref{eq: sample optimizers} with an error that decays at the rate $1/T$. That is, when $T$ is sufficiently large, Algorithm~\ref{algo: mp}'s  $\widehat{\theta}_T $ converges to the empirical \CGDRO\ estimator $\widehat{\theta}$ in \eqref{eq: sample optimizers}.

We now detail the selection of learning rates $\eta_\theta,\eta_\gamma$ used in Algorithm \ref{algo: mp}. We notice that the choice of these learning rates mainly affects the practical convergence speed of the optimization algorithm, i.e., $\|\widehat{\theta}_T - \widehat{\theta}\|_2$, while it does not directly affect the statistical convergence rate $\|\widehat{\theta} - \theta^*\|_2$, where $\widehat{\theta}_T$ denotes Algorithm \ref{algo: mp}'s output, $\widehat{\theta}$ denotes the empirical \CGDRO\ estimator defined in \eqref{eq: sample optimizers}, and $\theta^*$ is population \CGDRO\ model defined in \eqref{eq: minimax model}, respectively. This fundamentally differs from the selection of the tuning parameters in the statistical analysis, where the statistical convergence rate depends on the choice of the tuning parameters, such as the penalty level in the penalized regression. Moreover, in practice, we can monitor the algorithm's optimization convergence using the duality gap evaluated during the algorithm. For practical implementation, we set $\eta_\theta = \eta$ and $\eta_\gamma = {\eta}/{\log L}$ with $\eta\in [0.05, 0.2]$, which produced stable and sufficiently fast convergence throughout our simulated experiments in Section \ref{sec: simus}.
Furthermore, one could also further explore the adaptive learning rate scheme proposed in \cite{bach2019universal}, which in practice can accelerate convergence by adjusting to the local smoothness of the objective.

\subsubsection{DML Facilitated Mirror Prox}
\label{subsec: DML MP}
The following Algorithm \ref{algo:overall} summarizes the complete procedure to solve the empirical \CGDRO\ model \eqref{eq: sample optimizers}, under the general covariate shift setting.

\begin{algorithm}[H]
\DontPrintSemicolon
\SetAlgoLined
\SetNoFillComment
\LinesNotNumbered 
\caption{Mirror Prox Algorithm with DML}
\label{algo:overall}
\label{alg: main}
\KwData{Source data $\{(X_i^{(l)}, Y_i^{(l)})\}_{i \in [n_l]}$ for $l \in [L]$; target covariates $\{X_j^{\QQ}\}_{j \in [N]}$; learning rates for Mirror Prox: $\eta_\theta,\eta_\gamma$; maximum iteration $T$, tolerance ${\rm tol}$.}
\KwResult{Mirror Prox estimators $\widehat{\theta}_T$, $\widehat{\gamma}_T$.}

\For{$l=1,2,..., L$:}{

    Construct the cross-fitting conditional probability estimator $\widehat{f}^{(l)}$ and the cross-fitting density ratio estimator $\widehat{\omega}^{(l)}$, as detailed in Section \ref{subsec: construction of mu};

    Construct the DML estimator $\widehat{\mu}^{(l)}$ as in \eqref{eq: DR component};
}
Construct $\widehat{S}(\theta)$ as in \eqref{eq: hatS def};

Solve the empirical \CGDRO\ \eqref{eq: sample optimizers} with Mirror Prox Algorithm \ref{algo: mp} and return $\widehat{\theta}_T$, $\widehat{\gamma}_T$.
\end{algorithm}

% \begin{algorithm}[H]
% \DontPrintSemicolon
% \SetAlgoLined
% \SetNoFillComment
% \LinesNotNumbered 
% \caption{Doubly Robust Estimators $\{\widehat{\mu}^{(l)}\}_{l\in [L]}$}
% \label{algo:doubly robust}
% \KwData{Source data $\{(X_i^{(l)}, Y_i^{(l)})\}_{i \in [n_l]}$ for $l \in [L]$; target covariates $\{X_j^{\QQ}\}_{j \in [N]}$;}
% \KwResult{Doubly robust estimators $\{\widehat{\mu}^{(l)}\}_{l\in [L]}$.}

% \For{$l=1,2,..., L$:}{

%     Construct the sample-split conditional probability estimator $\widehat{f}^{(l)}$ as in \eqref{eq: fl sample-split};

%     Construct the sample-split density ratio estimator $\widehat{\omega}^{(l)}$ as in \eqref{eq: wl sample-split};

%     Construct the doubly robust estimators $\widehat{\mu}^{(l)}$ as in \eqref{eq: DR component} and \eqref{eq: mu estimator}.
% }

% \end{algorithm}
% We combine the estimation of $\mu^{(l)}$ and the Mirror Prox algorithm and summarize our proposal in Algorithm \ref{algo:overall}. We refer to our proposal as the Doubly Robust Mirror Prox algorithm.

\subsection{Theoretical Properties of DML Estimators $\{\widehat{\mu}^{(l)}\}_{l\in [L]}$}
\label{appendix: DR}
\label{subsec: properties of hatmu}
In this section, we first discuss the general covariate shift setting by studying the properties of DML estimators $\{\widehat{\mu}^{(l)}\}_{l\in [L]}$ constructed in \eqref{eq: DR component} of Section \ref{sec: double robustness}.  Then, we continue studying the setting without covariate shift, for which $\{\widehat{\mu}^{(l)}\}_{l\in [L]}$ is constructed following \eqref{eq: no shift est}.

We introduce additional notations. For a random vector $X\in \RR^d$ following the distribution $\mathbb{T}$, and a function $g:\RR^d\to \RR$, we define the $\ell_p$ norm for $g(\cdot)$ on $\mathbb{T}$ as $\|g\|_{\ell_p(\mathbb{T})} = \left(\E_{\TT}[(g(X))^p]\right)^{1/p}$.
For the random vectors $\{X_i\}_{i\in [n]}$ i.i.d. drawn from the distribution $\mathbb{T}$, given the function $g:\RR^d\to \RR^p$, we denote the variance matrix as:
\[
{\bf V}_{\mathbb{T}}(g(X)) = \E_{\TT}\left[\left(g(X) - \E_{\TT}[g(X)]\right)\left(g(X) - \E_{\TT}[g(X)]\right)^\intercal\right].
\]
and the estimated variance matrix as  
\[
\widehat{\bf V}_{\mathbb{T}}(g(X)) = \frac{1}{n}\sum_{i=1}^n \left(g(X_i) - \overline{g(X)}\right) (g(X_i) - \overline{g(X)})^\intercal, 
\quad \textrm{with $\overline{g(X)} = \frac{1}{n}\sum_{i=1}^n g(X_i)$}.
\]

% Given another function $h(X):\RR^d\to \RR$, the covariance matrix is denoted as
% \[
% {\bf V}_{\TT}(g(X), h(X)) = \E_{\TT}\left[\left(g(X) - \E_{\TT}[g(X)]\right)\left(h(X) - \E_{\TT}[h(X)]\right)^\intercal\right],
% \]
% and the estimated covariance matrix is denoted as,
% \[
% \widehat{\bf V}_{\mathbb{T}}(g(X),h(X)) = \frac{1}{n}\sum_{i=1}^n \left(g(X_i) - \overline{g(X)}\right) \left(h(X_i) - \overline{h(X)}\right)^\intercal,
% \]
% with $\overline{h(X)} = \frac{1}{n}\sum_{i=1}^n h(X_i)$.

We impose the following necessary conditions to establish the properties of $\{\widehat{\mu}^{(l)}\}_{l}$:
\begin{itemize}
    \item [(B1)] For each source $l\in [L]$ and each sample $1\leq i\leq n_l$, the covariate vector $\X{l}_i$ is subgaussian with $\kappa_1'\leq \lambda_{\rm min}\left( \E[\X{l}_i (\X{l}_i)^\intercal]\right)\leq \lambda_{\rm max}\left( \E[\X{l}_i (\X{l}_i)^\intercal]\right)\leq \kappa_2'$  for two constants $\kappa_1',\kappa_2'>0$. Additionally, there exists a constant $c_1>0$ such that $\E[(\varepsilon^{(l)}_{k,i})^2|\X{l}_i]>c_1$, where $\varepsilon^{(l)}_{k,i} = {\bf 1}(\Y{l}_i=k) - \f{l}_k(\X{l}_i)$, for each class label $1\leq k\leq K$, and each sample $1\leq i\leq n_l$. After applying the density ratio function to the covariate vector, there exists two  constants $c_2,c_3>0$ such that $\w{l}(\X{l}_i)\in [c_2,c_3]$ for each sample $1\leq i\leq n_l$. 
    \item [(B2)] There exists sequences  $\eta_\omega = \eta_\omega(n)$ and $\eta_f = \eta_f(n)$ such that, with probability at least $1-\delta_n$ for some sequence $\delta_n\rightarrow 0$, the estimated density ratio functions $\{\hw{l}\}_l$ and the estimated class probability functions $\{\hf{l}_k\}_{l,k}$ satisfy:
    \[
    \max_{l \in [L]} \left\| \frac{\widehat{\omega}^{(l)}}{\omega^{(l)}} - 1 \right\|_{\ell_2(\PP^{(l)})} \leq \eta_\omega, \quad \max_{l \in [L],\, k \in [K]} \left\| \widehat{f}^{(l)}_k - f^{(l)}_k \right\|_{\ell_2(\PP^{(l)})} \leq \eta_f,
    \]
    and there exists a positive constant $C>0$ such that 
    \[
    \left|\frac{\hw{l}(\X{l}_i)}{\w{l}(\X{l}_i)}-1\right| \leq C \quad \textrm{for all $1\leq i\leq n_l$ and for all $1\leq l\leq L$}.
    \]
    %where $C>0$ is some positive constant.
\end{itemize}
Condition {\rm (B1)} is a standard regularity assumption in high-dimensional statistics, ensuring that the source covariates are subgaussian, and the second moment of the noise conditional on the covariates is strictly larger than $0$. The second part imposes a boundedness condition on $\w{l}$, which is also standard in the density ratio estimation literature \cite{sugiyama2012density, cortes2010learning, ma2023optimally}. This condition is also consistent with the positivity (or overlap) assumption commonly invoked in causal inference \cite{rosenbaum1983central, imbens2015causal}. 

Condition {\rm (B2)} controls the estimation errors of the nuisance components, including the density ratio functions and conditional outcome functions. We also require the pointwise boundedness of the density ratio estimate to avoid instability in reweighting. Next, we give concrete examples of $\eta_f$ and $\eta_\omega$ established in the literature. We begin with the sequence $\eta_f$. In high-dimensional sparse logistic regression models, where the conditional distribution follows $\PP^{(l)}[Y=1|X=x]= \exp(x^\intercal \beta^{(l)})/(1+\exp(x^\intercal \beta^{(l)})$ for each source $l\in [L]$, we shall establish that $\eta_f= C\sqrt{s_\beta \log d/n}$, where $s_\beta = \max_{l\in [L]}\|\beta^{(l)}\|_0$ denotes the largest sparsity level across all source domains; see Section 4.4 of \cite{negahban2012unified} and Theorem 9 in \cite{huang2012estimation} for the exact statement. In the context of random forests, \cite{biau2012analysis} shows that $\eta_f^2 = C\cdot n^{-0.75/ (S\log 2+0.75)}$, under the assumption that $S$ out of the $d$ features influence the model. For certain neural networks, \cite{farrell2021deep} demonstrates that when the true model $f^{(l)}$ belongs to a Sobolev ball $\mathcal{W}^{\alpha, \infty}([-1,1]^p)$ with smoothness level $\alpha$, one may take $\eta_f^2 = C\cdot(n^{-\alpha/(\alpha+d)} \log^8 n + \log\log n/n)$.

We now turn to $\eta_\omega$, the error due to estimating the density ratio. In the special case of no covariate shift, i.e. $\w{l}(\cdot)\equiv 1$, for all $l\in [L]$, we simply set $\hw{l}(\cdot)\equiv 1$ in \eqref{eq: DR component}, yielding $\eta_\omega=0$.
In high-dimensional covariate shift settings, suppose the class posterior probabilities $p(G^{(l)}=1|\widetilde{X}^{(l)}=x)$, as defined in \eqref{eq: bayes density ratio}, admit the form $\exp(x^\intercal \gamma^{(l)})/(1+\exp(x^\intercal \gamma^{(l)}))$ for some sparse  vector $\gamma^{(l)}$, for each source $l\in [L]$. Then under standard high-dimensional logistic regression theory, we obtain the rate $\eta_\omega\lesssim \sqrt{{s_\gamma\log d}/{n}}$, where $s_\gamma=\max_{l\in [L]}\|\gamma^{(l)}\|_0$ denotes the maximum sparsity level of $\gamma^{(l)}$ for $l\in[L]$. Additional formulas for $\eta_\omega$
under alternative density ratio estimation methods can be found in Section 14 of \cite{sugiyama2012density}.

In the covariate shift regime, the following proposition establishes both the non-asymptotic and asymptotic results for $\widehat{\mu}^{(l)} - \mu^{(l)}$, with proof provided in Section \ref{proof of prop: doubly robust estimator}.

\begin{Proposition}
Suppose Conditions {\rm (A2), (B1), (B2)} hold. For the covariate shift setting, there exist positive constants $c_0, c_1>0$ such that the DML estimators $\{\widehat{\mu}^{(l)}\}_{l\in[L]}$ defined in \eqref{eq: DR component} satisfy: with probability at least $1-e^{-c_1t^2} - \delta_n$,
\begin{equation}
\max_{l=1,...,L}\|\widehat{\mu}^{(l)} - \mu^{(l)}\|_{2}\lesssim t\cdot \left(\left(1+\eta_\omega+\eta_f\right) \cdot\sqrt{\frac{d}{n}} + \eta_\omega\cdot \eta_f\cdot \sqrt{d}\right).
    \label{eq: rates of mu}
\end{equation}
where $t$ is any value larger than $c_0$, and $\delta_n\to 0$ is the sequence specified in Condition {\rm (B2)}. Moreover, with probability at least $1-n^{-c_1d} - \delta_n$,
\begin{equation}
    \max_{l=1,...,L}\|n_l\cdot(\widehat{\bf V}_\mu^{(l)} - {\bf V}_\mu^{(l)})\|_2 \lesssim\sqrt{\frac{d\log n}{n}} +d\left[(\eta_\omega+\eta_f+\eta_\omega\eta_f)+(\eta_\omega+\eta_f+\eta_\omega\eta_f)^2\right],
    \label{eq: rates of cov_mu}
\end{equation}
where $\widehat{\bf V}_\mu^{(l)}, {\bf V}_\mu^{(l)}\in \RR^{dK\times dK}$ are given by:
\begin{equation}
    \widehat{\bf V}_\mu^{(l)} = \frac{1}{n_l} \widehat{\bf V}_{\PP^{(l)}}\left(\left[\hw{l}(X)\cdot (\hf{l}(X) - {\bf y})\right] \otimes X\right) + \frac{1}{N} \widehat{\bf V}_{\QQ} \left(\hf{l}(X)\otimes X \right),
\label{eq: hat V_mu expression - cs}
\end{equation}
and
\begin{equation}
    {\bf V}_\mu^{(l)} = \frac{1}{n_l} {\bf V}_{\PP^{(l)}}\left(\left[\w{l}(X)\cdot (\f{l}(X) - {\bf y})\right] \otimes X\right) + \frac{1}{N} {\bf V}_{\QQ} \left(\f{l}(X)\otimes X \right),
    \label{eq: V_mu expression - cs}
\end{equation}
with the one-hot outcome vector ${\bf y} = \left({\bf 1}(Y = 1),..., {\bf 1}(Y=K)\right)^\intercal$.
Therefore, if $\eta_\omega=o(1),\eta_f=o(1)$ and $\eta_\omega\cdot \eta_f=o(n^{-1/2})$, then the following inequality holds with probability at least $1-e^{-c_1t^2} - \delta_n$, 
\[
\max_{l=1,...,L}\|\widehat{\mu}^{(l)} - \mu^{(l)}\|_{2}\leq t\cdot \sqrt{d/n}.
\]
In the fixed dimension regime, when $n\to \infty$, we establish that
\begin{equation}
\left[{\bf V}_\mu^{(l)}\right]^{-1/2}\left(\widehat{\mu}^{(l)}-\mu^{(l)}\right)\xlongrightarrow{d} \mathcal{N}(0,{\bf I}_{dK}), \quad \|n_l \cdot (\widehat{\bf V}_\mu^{(l)} - {\bf V}_\mu^{(l)})\|_2 \xlongrightarrow{p} 0.
\label{eq: limiting dist of mu}
\end{equation}

\label{prop: double robustness}
\end{Proposition}

We now investigate the non-shift regime for which $\{\widehat{\mu}^{(l)}\}_{l\in [L]}$ is constructed in the way described in Section \ref{sec: Mirror Prox}. The proof is provided in Section \ref{proof of prop: doubly robust estimator - nocs}. Note that in the non-shift regime, the construction of $\{\widehat{\mu}^{(l)}\}_{l\in [L]}$ does not involve the nuisance components $\{\hf{l},\hw{l}\}_l$. Therefore the following proposition holds without assuming Condition {\rm (B2)}.
\begin{Proposition}
\label{prop: double robustness - nocs}
    Suppose Condition ${\rm (B1)}$ holds. For the no covariate shift setting, there exist positive constants $c_0,c_1>0$ such that the estimators $\{\widehat{\mu}^{(l)}\}_{l\in [L]}$ defined in \eqref{eq: no shift est} satisfy: with probability at least $1-e^{-c_1t^2}$, 
    \begin{equation}
    \max_{l=1,...,L}\|\widehat{\mu}^{(l)} - \mu^{(l)}\|_{2}\leq t\cdot\sqrt{d/n},
    \label{eq: rates of mu - nocs}
    \end{equation}
    where $t$ is any value larger than $c_0$.
     Moreover, with probability at least $1-n^{-c_1d}$,
    \[
    \max_{l=1,...,L} \|n_l\cdot (\widehat{\bf V}_\mu^{(l)} - {\bf V}_\mu^{(l)})\|_2\lesssim \sqrt{\frac{d\log n}{n}},
    \]
    where $\widehat{\bf V}_\mu^{(l)}, {\bf V}_\mu^{(l)}\in \RR^{dK\times dK}$ are given by:
    \begin{equation}
    \widehat{\bf V}_\mu^{(l)} = \frac{1}{n_l} \widehat{\bf V}_{\PP^{(l)}}\left({\bf y} \otimes X\right),\quad {\bf V}_\mu^{(l)} = \frac{1}{n_l} {\bf V}_{\PP^{(l)}}\left({\bf y} \otimes X\right),
        \label{eq: V_mu expression - nocs}
    \end{equation}
    with the one-hot outcome vector ${\bf y} = \left({\bf 1}(Y = 1),..., {\bf 1}(Y=K)\right)^\intercal$.
    When $n\to\infty$, we establish that
    \begin{equation}
    \label{eq: limiting dist of mu - nocs}
    \left[{\bf V}_\mu^{(l)}\right]^{-1/2}\left(\widehat{\mu}^{(l)}-\mu^{(l)}\right)\xlongrightarrow{d} \mathcal{N}(0,{\bf I}_{dK}),\quad \|n_l\cdot (\widehat{\bf V}_\mu^{(l)} - {\bf V}_\mu^{(l)})\|_2 \xlongrightarrow{p} 0.
    \end{equation}

\end{Proposition}

The following lemma establishes the properties for the scaled covariance matrix $n_l \cdot {\bf V}_\mu^{(l)}$. The proof is provided in Section \ref{proof of lem: V_mu scale}.
\begin{Lemma}
    Suppose Conditions {\rm (A2)} and {\rm (B1)} hold. Then there exist constants $c_1,c_2>0$ such that ${\bf V}_\mu^{(l)}$ given by \eqref{eq: V_mu expression - cs} and \eqref{eq: V_mu expression - nocs} for the covariate shift and non-shift regimes, respectively, satisfy that
    $n_l \cdot [{\bf V}_\mu^{(l)}]_{j,j} \in [c_1,c_2]$ for all $j\in [dK]$ and for all $l\in [L]$.
    Moreover, when the dimension $d$ is fixed, there exists a constant $C>0$, such that $\|n{\bf V}_\mu^{(l)}\|_2\leq C$.
    \label{lem: V_mu scale}
\end{Lemma}

Combining Propositions \ref{prop: double robustness},  \ref{prop: double robustness - nocs}, and Lemma \ref{lem: V_mu scale}, Conditions {\rm (A1)}, {\rm (A5)} introduced in the main article hold, provided that Conditions {\rm (B1)} and {\rm (B2)} hold.

\subsection{Additional Details for Statistical Inferential Procedure}
\label{sec: entire infer}

In this subsection, we first summarize the full procedures of the statistical inference for the \CGDRO\ model $\theta^*$ in the following Algorithm \ref{algo: uq}.
Then we provide the remaining details of the implementation of the proposed inferential procedure in Section \ref{sec: procedure}. Specifically, 
\begin{itemize}
    \item Section \ref{subsec: hat_mu, hat_V_mu} presents the construction of inputs $\{\widehat{\mu}^{(l)}\}_l$ and $\{\widehat{\bf V}_\mu^{(l)}\}_l$;
    \item Section \ref{subsec: hat_V_m} provides the expression of $\widehat{\rm se}^{[m]}$ used in each individual interval ${\rm Int}^{[m]}$ defined in \eqref{uq: ci_m} for $m\in \mathcal{M}$;
    \item Moreover, in Section \ref{subsec: justify claim}, we justify the claim in \eqref{eq: limit dist decomp} to motivate our inferential procedure;
    \item Lastly, Section \ref{subsec: expression of c_star} presents the definition of $c^*(\alpha_0, d)$ that appears in the expression of $\err(M)$ in \eqref{eq: errM}. 
\end{itemize}

\begin{algorithm}[H]
\DontPrintSemicolon
\SetAlgoLined
\SetNoFillComment
\LinesNotNumbered 
\caption{Uncertainty Quantification Algorithm with Perturbation}
\label{algo: uq}

\KwData{Target covariates $\{X_j^{\QQ}\}_{j \in [N]}$; the DML estimator $\widehat{\mu}^{(l)}$ for $1 \leq l \leq L$ with estimated covariance matrix $\widehat{\bf V}^{(l)}_{\mu}$; the resampling size $M$.}
\KwResult{Confidence interval ${\rm CI}$.}

\For{$1 \leq m \leq M$}{
Generate perturbed $\widehat{\mu}^{(l,m)}\sim \Nc\left(\widehat{\mu}^{(l)}, \widehat{\bf V}^{(l)}_{\mu}\right)$ for $1 \leq l \leq L$;
}

Construct the filtered index set $\mathcal{M}$ in \eqref{eq: filtered set}.

\For{$m\in \mathcal{M}$}{

Compute $\widehat{\gamma}^{[m]}$ as in \eqref{eq: resampled gamma} and estimate $\widehat{\theta}^{[m]}$ as in \eqref{eq: m-opt};

Construct the interval ${\rm Int}^{[m]}$ as in \eqref{uq: ci_m};

}

Output the final CI as in \eqref{uq:union ci}.
%Take a union and obtain the final confidence interval ${\rm CI}^{[U]}$ \ref{uq:union ci}.

\end{algorithm}

%%%%%%%%%%%%%%%%%%%%%%%%%%%%%%%%%%%%%%%%%%%%%%%%%%%%%%%%%%%%%%%%%%%%%%%%%%%%%%%%%
\subsubsection{Construction of $\{\widehat{\mu}^{(l)},\widehat{\bf V}_\mu^{(l)}\}_{l\in [L]}$}
\label{subsec: hat_mu, hat_V_mu}
%%%%%%%%%%%%%%%%%%%%%%%%%%%%%%%%%%%%%%%%%%%%%%%%%%%%%%%%%%%%%%%%%%%%%%%%%%%%%%%%%
%\noindent\textbf{Inputs: $\{\widehat{\mu}^{(l)},\widehat{\bf V}_\mu^{(l)}\}_{l\in [L]}$}

We begin with the covariate shift setting. Given observed labeled data $\{\X{l}_i, \Y{l}_i\}_{i\in [n_l], l\in [L]}$ from $L$ source domains, and unlabeled data $\{\XQ_i\}_{i\in [N]}$ from the target domain, we construct the nuisance models $\{\hf{l},\hw{l}\}_{l\in [L]}$ as described in Section \ref{sec: doubly robust}. We then plug these nuisance models to construct the DML estimators $\{\widehat{\mu}^{(l)}\}_{l\in [L]}$, and their corresponding covariance matrices $\{\widehat{\bf V}_\mu^{(l)}\}_{l\in [L]}$, as defined in \eqref{eq: DR component} and \eqref{eq: hat V_mu expression - cs}, respectively.

In the no covariate shift setting, the construction of $\widehat{\mu}^{(l)}$ does not require nuisance models. Instead, we directly compute $\{\widehat{\mu}^{(l)}\}_l$ using the expression in \eqref{eq: no shift est} and their covariance matrices $\{\widehat{\bf V}_\mu^{(l)}\}_l$ using \eqref{eq: V_mu expression - nocs}, based solely on the labeled  data $\{\X{l}_i, \Y{l}_i\}_{i\in [n_l], l\in [L]}$ from $L$ source domains.

%\noindent\textbf{Expression of $\widehat{\bf V}^{[m]}$}
%%%%%%%%%%%%%%%%%%%%%%%%%%%%%%%%%%%%%%%%%%%%%%%%%%%%%%%%%%%%%%%%%%%%%%%%%%%%%%%%%
\subsubsection{Construction of $\widehat{\rm se}^{[m]}$}
\label{subsec: hat_V_m}
%%%%%%%%%%%%%%%%%%%%%%%%%%%%%%%%%%%%%%%%%%%%%%%%%%%%%%%%%%%%%%%%%%%%%%%%%%%%%%%%%

We use $\widehat{\rm se}^{[m]}$ to denote the estimated standard deviation of the first entry of  $-\left[H(\theta^*)\right]^{-1}\sum_{l=1}^{L}{\gamma}^*_l\left(\widehat{\mu}^{(l)}-{\mu}^{(l)}\right).$ Therefore, in the following, we construct $\widehat{\bf V}^{[m]}$, the estimated covariance matrix for  $-\left[H(\theta^*)\right]^{-1}\sum_{l=1}^{L}{\gamma}^*_l\left(\widehat{\mu}^{(l)}-{\mu}^{(l)}\right)$. Given $\widehat{\bf V}^{[m]}$, we obtain the expression of $\widehat{\rm se}^{[m]}$ as follows:
\[
\widehat{\rm se}^{[m]} = \sqrt{\widehat{\bf V}^{[m]}_{1,1}}.
\]

For each $m\in \mathcal{M}$, given the perturbed quantities $\widehat{\theta}^{[m]},\widehat{\gamma}^{[m]}$, we construct  $\widehat{\bf V}^{[m]}$ as 
\begin{equation}
    \widehat{\bf V}^{[m]}= \left[\widehat{H}(\widehat{\theta}^{[m]})\right]^{-1} \widehat{W}^{[m]} \left[\widehat{H}(\widehat{\theta}^{[m]})\right]^{-1},
    \label{eq: hat V_m}
\end{equation}
where $\widehat{H}(\widehat{\theta}^{[m]})$ denotes the Hessian of $\widehat{S}(\theta)$ evaluated at $\widehat{\theta}^{[m]}$ as defined in \eqref{eq: hatS def}. Given the perturbed quantities $\widehat{\gamma}^{[m]}, \widehat{\theta}^{[m]}$, the inner term $\widehat{W}^{[m]}$ is constructed as:
\[
\widehat{W}^{[m]}= \sum_{l=1}^L (\widehat{\gamma}_l^{[m]})^2\cdot  \widehat{\bf V}_\mu^{(l)} + \frac{1}{N}\widehat{\bf V}_{\QQ_X}\left(p(X,\widehat{\theta}^{[m]})\otimes X\right),
\]
where $p(X,\theta)\in \RR^K$ denotes the class probability vector defined in \eqref{eq: p def}, and $\widehat{\bf V}_\mu^{(l)}$ is given in \eqref{eq: V_mu expression - cs} and \eqref{eq: V_mu expression - nocs} for the covariate shift and non-shift settings.

\subsubsection{Justification of \eqref{eq: limit dist decomp}}
\label{subsec: justify claim}
The following proposition presents a simplified expression of $\widehat{\theta}^{[m]}-\theta^*$, for all $m\in \mathcal{M}$, which justifies the claim in \eqref{eq: limit dist decomp} to motivate our inferential procedure. The proof is provided in Section \ref{proof of prop: simplify theta_hat_m}. 
\begin{Proposition}
\label{prop: simplify theta_hat_m}
Suppose Conditions {\rm (A1)--(A5)} hold, and the value $t>0$ satisfies $c_0\leq t\ll \sigma_L^2(\U) (n/d)^{1/8}$ with the constant $c_0$ specified in Condition {\rm (A1)}. Then on the high probability event $\GG\cap \GG_4\cap \Ec_1\cap \Ec_2$, for all $m\in \mathcal{M}$,
    \[
    \widehat{\theta}^{[m]} - \theta^* = -[H(\theta^*)]^{-1} \left[{\U}(\widehat{\gamma}^{[m]}-\gamma^*)+ (\widehat{\U}-\U)\gamma^*\right] + o(\sqrt{d/n})
    \]
    where the event $\GG$ is defined in \eqref{eq: events}, the events $\GG_4$ and $\Ec_1\cap\Ec_2$ are defined in the following \eqref{eq: event G4} and \eqref{eq: E1E2}, respectively.
\end{Proposition}
This proposition holds on the high probability event $\GG\cap \GG_4\cap \Ec_1\cap \Ec_2$, where we shall apply Lemma \ref{lem: concentration lemma}, and the following Lemmas \ref{lemma: grad Shat limiting} and \ref{lem: E1E2} to establish that
\[
\liminf_{n\to\infty}\mathbf{P}(\GG\cap \GG_4\cap \Ec_1\cap \Ec_2)\geq 1-\alpha_0,
\]
where $\alpha_0\in (0, 0.01]$ is the pre-specified constant used to construct $\mathcal{M}$. As for the requirements of $\sigma_L(\U)$ and $t$,  we provide the detailed discussions following Theorem \ref{thm: inference-length}.

We emphasize that this proposition establishes a decomposition of $\widehat{\theta}^{[m]}-\theta^*$ with a remainder term of order $o(\sqrt{d/n})$ on the right-hand side. In contrast, for proving the inferential results in Theorems \ref{thm: inference-coverage} and \ref{thm: inference-length}, the later Proposition \ref{prop: theta_m nor} provides a more refined decomposition of $\widehat{\theta}^{[m]}-\theta^*$, which expresses the remainder term explicitly.

\subsubsection{Expression of $c^*(\alpha_0,d)$}
\label{subsec: expression of c_star}
We present the definition of the value $c^*(\alpha_0, d)$, which appears in the expression of $\err(M)$ as shown in \eqref{eq: errM}.
Given the pre-specified $\alpha_0\in (0,0.01]$ used to construct the filtered index set $\mathcal{M}$ as in \eqref{eq: filtered set}, we define
\begin{equation}
    c^*(\alpha_0,d)=\min_{l\in [L]}\left\{\frac{\exp\left\{-dK \log \left(\frac{dKL}{\alpha_0}\right)\left(3\|{n_l}{\bf V}_\mu^{(l)}\|_2 + 4\right)\right\}}{\sqrt{2\pi} \left[\|{n_l} {\bf V}_\mu^{(l)}\|_2 + \frac{4}{3}\right]^{dK/2}} \right\}.
    \label{eq: c star def}
\end{equation}
We verify the rate of $c^*(\alpha_0,d)$ and $\err(M)$ defined in \eqref{eq: errM} under the mild condition. Suppose that the rescaled covariance matrix $n_l{\bf V}_\mu^{(l)}$, as defined in \eqref{eq: V_mu expression - cs} and \eqref{eq: V_mu expression - nocs}, satisfies that $c_1\leq \|{n_l}{\bf V}_\mu^{(l)}\|_2 \leq c_2$, for all $l\in [L]$, with some positive constants $c_1,c_2>0$. Then $c^*(\alpha_0,d)$ defined in \eqref{eq: c star def} can be simplified as $c^*(\alpha_0, d)\asymp \left(\alpha_0/dKL\right)^{C dK}$.
Together with the expression 
${\rm Vol}(dK)=\frac{\pi^{dK/2}}{\Gamma\left(\frac{dK}{2}+1\right)},$
we show that $\err(M)$ admits the form
$${\rm err}_n(M) \asymp \left(d/{\alpha_0}\right)^C\left(\frac{\log n}{M}\right)^{\frac{1}{dKL}}.$$

\subsection{\CGDRO\ with Prior}
\label{appedix: prior}
The \CGDRO\ model $\theta^*$ defined in \eqref{eq:h_loss_opt} might become conservative for a particular target domain since the uncertainty class $\mathcal{C}$ may contain many plausible target distributions. We may improve the model's predictiveness by leveraging available prior information. In particular, if we had the prior knowledge of how the target conditional outcome distribution is mixed from source ones, we shall refine the uncertainty set $\mathcal{C}$ in \eqref{eq:uncertainty_class} by incorporating such prior $\mathcal{H}\subseteq \Delta^L$, such that
$$\mathcal{C}_{\mathcal{H}}=\left\{(\QQ_{X}, \TT_{Y|X}): \TT_{Y|X}=\sum_{l=1}^{L} \gamma_l \cdot \PP^{(l)}_{Y|X}, \; \gamma\in \mathcal{H}\right\}.$$
With this refined uncertainty class $\mathcal{C}_{\mathcal{H}}$, we define $\theta^*_{\mathcal{H}}$ as the solution to the following problem:
\[
\begin{aligned}
    \theta^*_\mathcal{H} &= \argmin_{\theta } \ \max_{\mathbb{T} \in \mathcal{C}_\mathcal{H}} \ \mathbb{E}_{(X, Y)\sim \mathbb{T}} \ \ell(X, Y, \theta)\\
    &=\argmin_{\theta } \ \max_{\gamma \in \mathcal{H}} \sum_{l=1}^L \gamma_l \cdot \mathbb{E}_{X \sim \mathbb{Q}_X} \mathbb{E}_{Y\sim \mathbb{P}^{(l)}_{Y|X}} \ell(X, Y, \theta).
\end{aligned}
\]
After applying the cross-entropy loss, we obtain the following simplified formulation analogous to \eqref{eq: minimax model}:
$$
\theta^*_{\mathcal{H}}= \argmin_{\theta} \ \max_{\gamma\in \mathcal{H}} 
\sum_{l=1}^{L} \gamma_{l} \cdot \theta^{\intercal}\mu^{(l)}+S(\theta).$$

\section{Proofs of Other Main Theorems}
\label{sec: main proof supp}
Apart from the proofs of Proposition \ref{prop: initial bound},  Theorems~\ref{thm: quad convergence}, \ref{thm: equi expressions} and \ref{thm: gamma true}, whose proofs are provided in Section~\ref{sec: proof of approx},  we organize the proofs of the remaining theorems and propositions in this section. The proofs of supporting lemmas are deferred to Section~\ref{sec: lemma proof}. This section is structured as follows:
\begin{itemize}
\item \textbf{Estimation.} This part corresponds to the results in Section~\ref{sec: conv rate}, where we establish the convergence rate of $\|\widehat{\theta} - \theta^*\|_2$. Specifically, we present the proofs of Theorems ~\ref{thm: refined rate} and \ref{thm: final rate} in Sections~\ref{proof of thm refined rate} and \ref{proof of thm final rate}, respectively.

\item \textbf{Inference.} This part contains the proofs of the results in Section~\ref{sec: UQ}, which address the inferential procedure for $\theta^*$. The proof of Proposition~\ref{prop: local theta} is provided in Section~\ref{proof of prop: local theta}. The proofs of Theorems \ref{thm: min_gamma_m}, \ref{thm: inference-coverage}, and \ref{thm: inference-length} are presented in Sections \ref{proof of thm min_gamma_m}, and \ref{proof of thm: inference}, respectively. Additionally, Propositions~\ref{prop: theta_m nor} and \ref{prop: simplify theta_hat_m} are introduced to support the main inferential results, with its proof given in Sections~\ref{proof of prop theta_m nor} and \ref{proof of prop: simplify theta_hat_m}, respectively.

\item \textbf{Double Machine Learning.} This part examines the properties of $\widehat{\mu}^{(l)}$ as discussed in Appendix~\ref{sec: doubly robust}. The proofs of Propositions~\ref{prop: double robustness} and \ref{prop: double robustness - nocs} are provided in Sections~\ref{proof of prop: doubly robust estimator} and \ref{proof of prop: doubly robust estimator - nocs}, respectively.

\end{itemize}

\subsection{Proof of Theorem \ref{thm: refined rate}} %
\label{proof of thm refined rate}
Given the optimal weight $\widehat{\gamma}_{\rm ap}$ is defined in \eqref{eq: inter theta}, the dual formulation of $\thetainit_{\rm ap}$ in \eqref{eq: step 2 hat} admits the following expression: 
\[
\thetainit_{\rm ap} = \argmin_{\theta\in \RR^{dK}} \theta^\intercal \widehat{\U} \widehat{\gamma}_{\rm ap} + \widehat{Q}(\theta),\quad \textrm{with}\; \widehat{Q}(\theta) = \widehat{S}(\theta) + \langle \nabla\widehat{S}(\thetainit), \theta-\thetainit\rangle + \frac{1}{2}(\theta-\thetainit)^\intercal \widehat{H}(\thetainit)(\theta-\thetainit).
\]
By the optimality condition of $\thetainit_{\rm ap}$, we have
\[
\widehat{\U}\widehat{\gamma}_{\rm ap} + \nabla \widehat{S}(\thetainit) + \widehat{H}(\thetainit)(\thetainit_{\rm ap} - \thetainit) = 0.
\]
As shown in \eqref{eq: stable optimizer} of Theorem \ref{thm: quad convergence}, $\thetainit_{\rm ap} = \thetainit$ on the high probability event $\GG$, where $\GG$ is defined in \eqref{eq: events}. Together with the preceding equality, it implies on the event $\GG$, 
\begin{equation}
\widehat{\U}\widehat{\gamma}_{\rm ap} + \nabla \widehat{S}(\thetainit)=0.
\label{eq: opt inter 0}
\end{equation}
Recall the closed-form expression of ${\theta}^*_{\rm ap}$ in \eqref{eq: closed form theta} such that ${\theta}^*_{\rm ap}=\thetainit-[H(\thetainit)]^{-1}\left(\U\gamma^*_{\rm ap}+\nabla S(\thetainit)\right)$. Together with the fact that $\thetainit_{\rm ap} = \thetainit$ on the event $\GG$, we have
\begin{equation}
\begin{aligned}
\thetainit_{\rm ap} -{\theta}^*_{\rm ap}&= [H(\thetainit)]^{-1}\left[\U{\gamma}^*_{\rm ap}+\nabla S(\thetainit)\right]\\
&=[H(\thetainit)]^{-1}\left[\left(\U\gamma^*_{\rm ap} - \widehat{\U}\widehat{\gamma}_{\rm ap}\right)+\left(\nabla S(\thetainit)-\nabla \widehat{S}(\thetainit)\right)\right]\\
&=[H(\thetainit)]^{-1}\left[\widehat{\U}(\gamma^*_{\rm ap} - \widehat{\gamma}_{\rm ap}) + (\U-\widehat{\U})\gamma^*_{\rm ap}+\left(\nabla S(\thetainit)-\nabla \widehat{S}(\thetainit)\right)\right],
\end{aligned}
\label{eq: key decomp}
\end{equation}
where the second equality follows from \eqref{eq: opt inter 0}. 
Therefore, we have
\[
\begin{aligned}
    \|\thetainit_{\rm ap} - {\theta}^*_{\rm ap}\|_2 \leq \|[H(\thetainit)]^{-1}\|_2 \cdot \left[\|\widehat{\U}\|_2 \|\widehat{\gamma}_{\rm ap} - {\gamma}^*_{\rm ap}\|_2 + \|\widehat{\U} - \U\|_2 + \|\nabla \widehat{S}(\thetainit) - \nabla S(\thetainit)\|_2\right].
\end{aligned}
\]

In the proof of Proposition \ref{prop: initial bound}, we have established in \eqref{eq: theta_hat is bounded} that $\thetainit\in \Theta_B$ on the event $\GG$. Hence, by the definition of $\GG$ in \eqref{eq: events}, we have $\|\widehat{\U} - \U\|_2\leq t\cdot \sqrt{d/n}$ and $\|\nabla \widehat{S}(\thetainit) - \nabla S(\thetainit)\|_2\lesssim \sqrt{d\log N/N}$, 
on the event $\GG$.
We apply these results and establish that
\begin{equation}
    \|\thetainit_{\rm ap} - {\theta}^*_{\rm ap}\|_2 \lesssim \lambda_{\rm max}\left([H(\thetainit)]^{-1}\right)\cdot \left(\|\U\|_2\|\widehat{\gamma}_{\rm ap} - {\gamma}^*_{\rm ap}\|_2 + t\sqrt{d/n} + \sqrt{d\log N/N}\right).
    \label{eq: theta diff - thm 3}
\end{equation}
We shall continue the proof with
the following lemma, which facilitates the analysis of $\|\widehat{\gamma}_{\rm ap}-{\gamma}^*_{\rm ap}\|_2$. The proof is provided in Section \ref{section: proof of lemma distance between gammas}.
\begin{Lemma}
\label{lemma: distance between gammas}
For a positive definite matrix $A\in \RR^{L\times L}$ and a vector $b\in \RR^L$, we consider the optimization problem 
\[
\gamma^* = \argmin_{\gamma\in \Delta^L} f(\gamma), \quad\textrm{with}\; f(\gamma) = \gamma^\intercal A\gamma + 2b^\intercal \gamma + c,
\] 
together with its perturbed problem,
\[
\hat{\gamma} = \argmin_{\gamma\in \Delta^L}\hat{f}(\gamma), \quad \textrm{with}\; \hat{f}(\gamma)= \gamma^\intercal \widehat{A}\gamma + 2\hat{b}^\intercal \gamma + \hat{c}.
\]
If $\widehat{A}\succ 0$, 
then
\[
\|\hat{\gamma}-\gamma^*\|_2 \leq \frac{\|\widehat{A} - A\|_{2} + \|\hat{b} - b\|_2}{\lambda_{\rm min}(A)}.
\]
\end{Lemma}

We now apply the above lemma to control $\|\widehat{\gamma}_{\rm ap}-{\gamma}^*_{\rm ap}\|_2$, noting that we have established in Theorem \ref{thm: equi expressions} that
{
\[
{\gamma}^*_{\rm ap} = \argmax_{\gamma\in \Delta^L} F(\gamma) = \argmin_{\gamma\in \Delta^L}\left\{\frac{1}{2}\left(\U\gamma+\nabla S(\thetainit)\right)^{\intercal}[H(\thetainit)]^{-1}\left(\U\gamma + \nabla S(\thetainit)\right)-\thetainit^\intercal \U\gamma\right\},
\]
}
and
{
\[
\widehat{\gamma}_{\rm ap} = \argmax_{\gamma\in \Delta^L} \widehat{F}(\gamma) = \argmin_{\gamma\in \Delta^L}\left\{\frac{1}{2}\left(\widehat{\U}\gamma+\nabla \widehat{S}(\thetainit)\right)^{\intercal}[\widehat{H}(\thetainit)]^{-1}\left(\widehat{\U}\gamma+\nabla \widehat{S}(\thetainit)\right)-\thetainit^\intercal \widehat{\U}\gamma\right\}.
\]
}
 By Lemma \ref{lemma: distance between gammas}, we establish that
{
\begin{equation}
\begin{aligned}
    &\lambda_{\rm min}(\U^\intercal [H(\thetainit)]^{-1} \U) \|{\widehat{\gamma}_{\rm ap}} - {\gamma}^*_{\rm ap}\|_2 \\
    &\leq 
    \left\|\widehat{\U}^{\intercal}[\widehat{H}(\thetainit)]^{-1} \widehat{\U} - \U^\intercal [H(\thetainit)]^{-1}\U\right\|_{2}\\
    &\quad +\left\|([\widehat{H}(\thetainit)]^{-1} \nabla \widehat{S}(\thetainit) - \thetainit)^\intercal \widehat{\U} - ([H(\thetainit)]^{-1} \nabla S(\thetainit) - \thetainit)^\intercal \U\right\|_2.
\end{aligned}
\label{eq: decompose gamma distance - thm 3}
\end{equation}
}

Next, we establish the upper bound for the components on the right-hand side of the above inequality.
To facilitate the discussion, we define
\[
\Delta_1 = \widehat{\U} - \U, \quad \Delta_2 = [\widehat{H}(\thetainit)]^{-1} - [H(\thetainit)]^{-1}, \quad \Delta_3 = \nabla \widehat{S}(\thetainit) - \nabla S(\thetainit).
\]
On the event $\GG$, we have $\|\Delta_1\|_2\leq t\sqrt{d/n}$, $\|\Delta_2\|_2\lesssim\sqrt{d\log N/N}$, and $\|\Delta_3\|_2\lesssim \sqrt{d\log N/N}$.
For the first term on the right-hand side of the above inequality \eqref{eq: decompose gamma distance - thm 3}, we have
{
\begin{equation}
\begin{aligned}
    &\left\|\widehat{\U}^{\intercal}[\widehat{H}(\thetainit)]^{-1} \widehat{\U} - \U^\intercal [H(\thetainit)]^{-1}\U\right\|_{2} \\
    &\leq \|\Delta_2\|_2\left(\|\U\|_2 + \|\Delta_1\|_2\right)^2 + \|[H(\thetainit)]^{-1}\|_2\|\Delta_1\|_2 \left(2 \|\U\|_2 + \|\Delta_1\|_2\right)\\
    &\lesssim t\cdot\|\U\|_2\sqrt{d/n} + \|\U\|_2^2\sqrt{d\log N/N}.
\end{aligned}
    \label{eq: term 1 result in decompose gamma distance - thm 3}
\end{equation}
}
For the second term in the right-hand side of \eqref{eq: decompose gamma distance - thm 3}, we have
\begin{equation}
    \begin{aligned}
        &\left\|([\widehat{H}(\thetainit)]^{-1} \nabla \widehat{S}(\thetainit) - \thetainit)^\intercal \widehat{\U} - ([H(\thetainit)]^{-1} \nabla S(\thetainit) - \thetainit)^\intercal \U\right\|_2 \\
        &\quad \leq \|[H(\thetainit)]^{-1} \nabla S(\thetainit) - \thetainit\|_2 \|\Delta_1\|_2 + (\|\U\|_2 + \|\Delta_1\|_2) \|\nabla S(\thetainit)\|_2 \|\Delta_2\|_2 \\
        &\quad \quad\quad + (\|\U\|_2 + \|\Delta_1\|_2)(\|[H(\thetainit)]^{-1}\|_2 + \|\Delta_2\|_2) \|\Delta_3\|_2\\
        &\quad \leq (1+\|\theta^*\|_2 + \|\nabla S(\theta^*)\|_2) t\cdot \sqrt{d/n} + (\|\U\|_2 + \|\nabla S(\theta^*)\|_2)\sqrt{d\log N/N}.
    \end{aligned}
    \label{eq: term 2 result in decompose gamma distance - thm 3}
\end{equation}
The combination of results in \eqref{eq: decompose gamma distance - thm 3},  \eqref{eq: term 1 result in decompose gamma distance - thm 3} and \eqref{eq: term 2 result in decompose gamma distance - thm 3} results in
{\small
\begin{equation*}
\begin{aligned}
    &\lambda_{\rm min}(\U^\intercal [H(\thetainit)]^{-1} \U)\|{\widehat{\gamma}_{\rm ap}} - {\gamma}^*_{\rm ap}\|_2\\ &\lesssim (1+\|\theta^*\|_2 + \|\nabla S(\theta^*)\|_2 + \|\U\|_2)t\cdot\sqrt{d/n}+  (\|\U\|_2^2 + \|\U\|_2 + \|\nabla S(\theta^*)\|_2)\sqrt{d\log N/N} \\
    &\lesssim (1+ \|\theta^*\|_2 + \|\nabla S(\theta^*)\|_2 + \|\U\|_2 + \|\U\|_2^2)\left(t\cdot\sqrt{d/n}+\sqrt{d\log N/N}\right).
\end{aligned}
\end{equation*}
}
Condition {\rm (A2)} supposes the bounded $\|\theta^*\|_1$, thus $\|\theta^*\|_2 \leq C$ for some constant, which further simplifies the preceding equation as follows:
{\small
\[
\lambda_{\rm min}(\U^\intercal [H(\thetainit)]^{-1} \U)\|{\widehat{\gamma}_{\rm ap}} - {\gamma}^*_{\rm ap}\|_2 \leq (1 + \|\nabla S(\theta^*)\|_2 + \|\U\|_2 + \|\U\|_2^2)\left(t\cdot\sqrt{d/n}+\sqrt{d\log N/N}\right).
\]}
Notice that
\[
\lambda_{\rm min}(\U^\intercal [H(\thetainit)]^{-1} \U) \geq \lambda_{\rm min}([H(\thetainit)]^{-1}) \cdot \sigma_{L}^2(\U).
\]
Therefore, we establish that 
\begin{equation*}
    \|{\widehat{\gamma}_{\rm ap}} - {\gamma}^*_{\rm ap}\|_2 \lesssim \frac{1+\|\nabla S(\theta^*)\|_2 + \|\U\|_2 + \|\U\|_2^2}{\lambda_{\rm min}\left([H(\thetainit)]^{-1}\right)\cdot \sigma_{L}^2(\U)}\left(t\sqrt{d/n}+\sqrt{d\log N/N}\right)
\end{equation*}

Putting the above result back to \eqref{eq: theta diff - thm 3}, on the event $\GG$, we have
{\small
\[
\begin{aligned}
    &\|\thetainit_{\rm ap} - {\theta}^*_{\rm ap}\|_2 \\&\lesssim \frac{\lambda_{\rm max}\left([H(\thetainit)]^{-1}\right)}{\lambda_{\rm min}\left([H(\thetainit)]^{-1}\right) \cdot \sigma_{L}^2 (\U)}\cdot \left(1+\|\nabla S(\theta^*)\|_2 + \|\U\|_2 + \|\U\|_2^2\right) \cdot \left(t\sqrt{d/n} + \sqrt{d\log N/N}\right) \\
    &\lesssim \frac{\kappa_2}{\kappa_1 \cdot \sigma_{L}^2(\U)}\cdot \left(1+\|\nabla S(\theta^*)\|_2 + \|\U\|_2 + \|\U\|_2^2\right) \cdot \left(t\sqrt{d/n} + \sqrt{d\log N/N}\right),
\end{aligned}
\]
}
where the second inequality follows from Lemma \ref{lem: convexity of H} and \eqref{eq: theta_hat is bounded}, which establishes the bounds for the eigenvalues of $H(\widehat{\theta})$, and $\kappa_1,\kappa_2$ are positive constants given in Condition {\rm (A2)}.
Lastly, we apply Lemma \ref{lem: concentration lemma} to control the probability of the event $\GG$ and complete the proof.

\subsection{Proof of Theorem \ref{thm: final rate}}
\label{proof of thm final rate}
It follows from Theorems \ref{thm: quad convergence} and \ref{thm: refined rate} that, under Conditions {\rm (A1)}, {\rm (A2)}, {\rm (A3)}, there exist positive constants $c_0,c_1>0$ such that with probability larger than $1-\exp(-c_1t^2) - N^{-c_1d}-\delta_n$,
\[
\begin{aligned}
    \|\widehat{\theta}-\theta^*\|_2 &\leq \|\widehat{\theta} - \widehat{\theta}_{\rm ap}\|_2 + \|\widehat{\theta}_{\rm ap} - \theta^*_{\rm ap}\|_2 + \|\theta^*_{\rm ap} - \theta^*\|_2\\
    &\lesssim \left[\frac{t}{\kappa_1} + \frac{\kappa_2\cdot t}{\kappa_1\cdot \sigma_L^2(\U)} \left(1 + \|\nabla S(\theta^*)\|_2 + \|\U\|_2 + \|\U\|_2^2\right) \right]\sqrt{d/n},
\end{aligned}
\]
where $t$ is any value satisfying \eqref{eq: t condition}.
Additionally, provided Condition {\rm (A4)}, we shall further simplify the above inequality and complete the proof of Theorem \ref{thm: final rate}.

\subsection{Proof of Proposition \ref{prop: local theta}}
\label{proof of prop: local theta}
Due to Lemma \ref{lem: convexity of H}, $S(\theta)$ is a strictly convex function of $\theta$. Then the continuous function $\phi(\theta,\gamma)$ is convex in $\theta$ and concave in $\gamma$. 
Therefore, given the convex compact sets $\Theta_{\rm loc}$ and $\Delta^L$, Sion's Minimax Theorem \cite{sion1958general} implies that:
\[
\min_{\theta \in \Theta_{\rm loc}}\max_{\gamma\in \Delta^L} \phi(\theta,\gamma) = \max_{\gamma\in \Delta^L}\min_{\theta\in \Theta_{\rm loc}} \phi(\theta,\gamma).
\]
Since $\theta^* = \argmin_{\theta\in \RR^{dK}}\left[\max_{\gamma\in \Delta^L}\phi(\theta,\gamma)\right]$ and $\Theta_{\rm loc}\subset \RR^{dK}$, it further implies that
\[
\theta^* = \argmin_{\theta \in \Theta_{\rm loc}}\max_{\gamma\in \Delta^L} \phi(\theta,\gamma).
\]
Recall that we have defined $\gamma^* \in \argmax_{\gamma\in \Delta^L}\min_{\theta\in \Theta_{\rm loc}} \phi(\gamma,\theta)$ as in \eqref{eq: one saddle}. Then it follows from Proposition 3.4.1 of \cite{bertsekas2009convex}, $(\theta^*,\gamma^*)$ is a saddle point satisfying 
\begin{equation*}
    \phi(\theta^*,\gamma) \leq \phi(\theta^*,\gamma^*) \leq \phi(\theta, \gamma^*), \quad \textrm{for all } \theta\in \Theta_{\rm loc}, \gamma\in \Delta^L.
    \label{eq: saddle point local}
\end{equation*}
The second inequality of the above equation implies that
\begin{equation}
    \theta^* = \argmin_{\theta\in \Theta_{\rm loc}}\phi(\theta,\gamma^*). 
    \label{eq: identification theta_star loc - eq0}
\end{equation}
The above result implies that $\theta^*$ is a local minimum of $\phi(\theta,\gamma^*)$ in the neighborhood of $\Theta_{\rm loc}$.

In the following, we show that $\theta^*$ is also a global minimum over the entire space $ \RR^{dK}$, by the argument of contradiction. Suppose that $\theta^*$ is not a  global minimizer of $\phi(\theta,\gamma^*)$ over the entire space $\theta\in \RR^{dK}$. Together with \eqref{eq: identification theta_star loc - eq0}, we obtain that there exists a point $\Tilde{\theta}$ such that 
\[
\phi(\Tilde{\theta},\gamma^*)<  \phi(\theta^*,\gamma^*), \quad \textrm{and}\quad \Tilde{\theta}\notin \Theta_{\rm loc}.
\]
Since $\phi(\theta,\gamma^*)$ is convex towards $\theta$, for any $\nu\in [0,1)$, it holds that
\begin{equation}
    \phi\left(\Tilde{\theta} + \nu(\theta^* - \Tilde{\theta}),\;\gamma^*\right) \leq \nu \phi(\theta^*,\gamma^*) + (1-\nu)\phi(\Tilde{\theta},\gamma^*)< \phi(\theta^*,\gamma^*).
    \label{eq: identification theta_star loc - eq1}
\end{equation}

Since $\theta^*$ is an interior point of $\Theta_{\rm loc}$, and $\Tilde{\theta}\notin \Theta_{\rm loc}$, the pathway $\nu \mapsto \nu\theta^* + (1-\nu)\Tilde{\theta}$ intersects $\Theta_{\rm loc}$ at the point $\theta_{\rm inter}$, which admits the following expression:
\[
\theta_{\rm inter} = \nu\theta^* + (1-\nu)\Tilde{\theta}, \quad \textrm{with}\quad 
\nu = 1-\frac{2\tau \sqrt{d/n}}{\|\Tilde{\theta} - \theta^*\|_2} \in (0,1).
\]
And it follows from \eqref{eq: identification theta_star loc - eq1} that $\theta_{\rm inter}$ satisfies:
\[
\phi(\theta_{\rm inter}, \gamma^*) < \phi(\theta^*,\gamma^*), \quad \textrm{and}\quad \theta_{\rm inter}\in \Theta_{\rm loc},
\]
which contradicts \eqref{eq: identification theta_star loc - eq0}. Therefore, we establish 
$
\theta^* = \argmin_{\theta\in \RR^{dK}}\phi(\theta,\gamma^*).$ Together with \eqref{eq: identification theta_star loc - eq0}, we complete the proof.

\subsection{Proof of Theorem \ref{thm: min_gamma_m}}
\label{proof of thm min_gamma_m}
We start with the following lemma, which provides an upper bound for $\|\widehat{\gamma}^{[m]} - {\gamma}^*_{\rm ap}\|_2$ via $\|\widehat{\U}^{[m]} - \U\|_2$. The proof is provided in Section \ref{section: proof of equation gamma distance and mu distance.}.
% \Zhenyu{Check if it is ok to condition on $\GG$.}
\begin{Lemma} Suppose Conditions {\rm (A2)}--{\rm (A4)} hold. On the event $\GG$ defined in \eqref{eq: events}, the following inequality holds for all $1\leq m\leq M$, 
    \begin{equation}
    \|\widehat{\gamma}^{[m]} - {\gamma}^*_{\rm ap}\|_2 \lesssim \frac{\kappa_2}{\sigma_L^2(\U)}\cdot \left(\|\widehat{\U}^{[m]} - \U\|_2 + \sqrt{d\log N/ N}\right),
    \label{eq: gamma distance and mu distance}
\end{equation}
where $\kappa_2>0$ is given in Condition {\rm (A2)}.
\label{prop: gamma_m and gamma_ap dist}
\end{Lemma}
By taking the minimum over all $m\in \mathcal{M}$ with $\mathcal{M}\subseteq [M]$, we establish that, on the event $\GG$
% $\GG_2\cap \GG_3$,
\begin{equation}
    \min_{m\in \mathcal{M}}\|\widehat{\gamma}^{[m]} - {\gamma}^*_{\rm ap}\|_2 \lesssim \frac{\kappa_2}{\sigma_L^2(\U)}\cdot \left(\min_{m\in \mathcal{M}}\|\widehat{\U}^{[m]} - \U\|_2 + \sqrt{d\log N/ N}\right).
    \label{eq: gamma distance and mu distance - min M}
\end{equation}
Thus, to control $\min_{m\in \mathcal{M}}\|\widehat{\gamma}^{[m]} - {\gamma}^*_{\rm ap}\|_2$, it suffices to upper bound $\min_{m\in \mathcal{M}} \|\widehat{\U}^{[m]} - \U^{[m]}\|_2$.

To facilitate the theoretical analysis, we introduce additional high probability events.
We denote the observed data by $\mathcal{O}$, consisting of the labeled source data and the unlabeled target data, that is $\Oc = \{\X{l}_i, \Y{l}_i\}_{i\in [n_l], l\in [L]}\bigcup \{\XQ_i\}_{i\in [N]}$.
We consider the following events for the data $\Oc$.
\begin{equation}
\begin{aligned}
    \Ec_1 &= \left\{\max_{l\in [L]} \left\|n_l\cdot( \widehat{\bf V}_\mu^{(l)} - {\bf V}_\mu^{(l)})\right\|_2 \leq \frac{1}{3} \right\},\\
    \Ec_2 &= \left\{\max_{l\in [L]} \max_{j\in [dK]} \frac{\left|\widehat{\mu}^{(l)}_j - \mu^{(l)}_j\right|}{\sqrt{[\widehat{\bf V}^{(l)}_\mu]_{j,j}+ 1/n_l}}\leq 1.05\cdot z_{\alpha_0/(dKL)}\right\},
    \label{eq: E1E2}
\end{aligned}
\end{equation}
where $\alpha_0\in (0,0.01]$ is a pre-specified constant used to construct $\mathcal{M}$ in \eqref{eq: filtered set}.
We provide explanations on the above events: on the event $\Ec_1$, the scaled estimated variance matrix $n_l\widehat{\bf V}_\mu^{(l)}$ is not too far away from the true $n_l{\bf V}_\mu^{(l)}$; on the event $\Ec_2$, the estimator $\widehat{\mu}^{(l)}_j$ does not deviate significantly from $\mu^{(l)}_j$ for each $j\in [dK]$ and $l\in [L]$.
The following lemma shows that $\Ec_1\cap \Ec_2$ happens with high probability, whose proof is provided in Section \ref{proof of lemma E1E2}.
\begin{Lemma}
    Suppose Condition {\rm (A5)} holds. Then
    \label{lem: E1E2}
    \[
    \liminf_{n\to\infty}\mathbf{P}(\Ec_1\cap \Ec_2)\geq 1-\alpha_0,
    \]
    where $\alpha_0$ is a pre-specified constant used to construct $\mathcal{M}$ in \eqref{eq: filtered set}.
\end{Lemma}
    
Leveraging the high probability event $\Ec_1\cap\Ec_2$, the following Lemma \ref{prop: sampling prob} establishes the upper bound for $\min_{m\in \mathcal{M}}\|\widehat{\U}^{[m]} - \U\|_2$, whose proof is provided in \ref{section: proof of eq: mu resampling high prob}. 
\begin{Lemma} 
Suppose that the resampling size $M$ satisfies that
\begin{equation}
M\gtrsim \left(\frac{1}{\alpha_0}\right)^{dKL}\left(\frac{1}{2}c^*(\alpha_0,d){\rm Vol}(dK)\right)^{-L}\cdot\log n,
\label{eq: M condition}
\end{equation}
where the value $c^*(\alpha_0, d)$ is defined in \eqref{eq: c star def} within Section \ref{subsec: expression of c_star}, and ${\rm Vol}(dK)$ denotes the volume of the unit ball in $\RR^{dK}$.
Then the following inequality holds:
    \begin{equation}
    \mathbf{P}\left(\min_{m\in \mathcal{M}}\|\widehat{\U}^{[m]} - \U\|_2 \lesssim {\rm err}_n(M)/\sqrt{n}\right) \geq (1-n^{-1})\mathbf{P}(\Ec_1\cap \Ec_2),
    \label{eq: mu resampling high prob}
\end{equation}
where events $\Ec_1,\Ec_2$ are specified in \eqref{eq: E1E2}.
\label{prop: sampling prob}
\end{Lemma}

We continue investigating the left-hand side of the above equality \eqref{eq: mu resampling high prob}.
% \eqref{eq: gamma distance and mu distance - min M}, which controls $\min_{m\in \mathcal{M}}\|\widehat{\gamma}^{[m]} - {\gamma}^*_{\rm ap}\|_2$ via $\min_{m\in \mathcal{M}}\|\widehat{\U}^{[m]} - \U\|_2$, where $\min_{m\in \mathcal{M}}\|\widehat{\U}^{[m]} - \U\|_2$ itself can be controlled via Lemma \ref{prop: sampling prob}.
Note that 
{
\[
\begin{aligned}
    &\mathbf{P}\left(\min_{m\in \mathcal{M}}\|\widehat{\U}^{[m]} - \U\|_2 \lesssim {\rm err}_n(M)/\sqrt{n}\right)\\
    &=\mathbf{P}\left(\min_{m\in \mathcal{M}}\|\widehat{\U}^{[m]} - \U\|_2 \lesssim {\rm err}_n(M)/\sqrt{n}\; \textrm{and} \; \GG\right) \\
    &\quad\quad+ \mathbf{P}\left(\min_{m\in \mathcal{M}}\|\widehat{\U}^{[m]} - \U\|_2 \lesssim {\rm err}_n(M)/\sqrt{n}\; \textrm{and} \; \GG^c\right) \\
    &\leq \mathbf{P}\left(\min_{m\in \mathcal{M}}\|\widehat{\U}^{[m]} - \U\|_2 \lesssim {\rm err}_n(M)/\sqrt{n}\; \textrm{and} \; \GG\right) + \mathbf{P}(\GG^c) \\
    &\leq \mathbf{P}\left(\min_{m\in \mathcal{M}}\|\widehat{\gamma}^{[m]} - \gamma^*_{\rm ap}\|_2\lesssim \frac{\kappa_2}{\sigma_L^2(\U)}\left({\rm err}_n(M)/\sqrt{n} + \sqrt{d\log N/N}\right)\right) + \mathbf{P}(\GG^c),
\end{aligned}
\]
% \[
% \begin{aligned}
%     &\mathbf{P}\left(\min_{m\in \mathcal{M}}\|\widehat{\U}^{[m]} - \U\|_2 \lesssim {\rm err}_n(M)/\sqrt{n}\right)\\
%     &=\mathbf{P}\left(\min_{m\in \mathcal{M}}\|\widehat{\U}^{[m]} - \U\|_2 \lesssim {\rm err}_n(M)/\sqrt{n}\; \textrm{and} \; \GG_2\cap \GG_3\right) \\
%     &\quad\quad+ \mathbf{P}\left(\min_{m\in \mathcal{M}}\|\widehat{\U}^{[m]} - \U\|_2 \lesssim {\rm err}_n(M)/\sqrt{n}\; \textrm{and} \; (\GG_2\cap \GG_3)^c\right) \\
%     &\leq \mathbf{P}\left(\min_{m\in \mathcal{M}}\|\widehat{\U}^{[m]} - \U\|_2 \lesssim {\rm err}_n(M)/\sqrt{n}\; \textrm{and} \; (\GG_2\cap \GG_3)\right) + \mathbf{P}((\GG_2\cap \GG_3)^c) \\
%     &\leq \mathbf{P}\left(\min_{m\in \mathcal{M}}\|\widehat{\gamma}^{[m]} - \gamma^*_{\rm ap}\|_2\lesssim \frac{\kappa_2}{\sigma_L^2(\U)}\left({\rm err}_n(M)/\sqrt{n} + \sqrt{d\log N/N}\right)\right) + \mathbf{P}((\GG_2\cap \GG_3)^c),
% \end{aligned}
% \]
}
where the last inequality follows from \eqref{eq: gamma distance and mu distance - min M}, which controls $\min_{m\in \mathcal{M}}\|\widehat{\gamma}^{[m]} - {\gamma}^*_{\rm ap}\|_2$ via $\min_{m\in \mathcal{M}}\|\widehat{\U}^{[m]} - \U\|_2$.
Combining the above inequality with \eqref{eq: mu resampling high prob}, we establish that
{
\[
\begin{aligned}
    &\mathbf{P}\left(\min_{m\in \mathcal{M}}\|\widehat{\gamma}^{[m]} - \gamma^*_{\rm ap}\|_2\lesssim \frac{\kappa_2}{\sigma_L^2(\U)}\left({\rm err}_n(M)/\sqrt{n} + \sqrt{d\log N/N}\right)\right)\\
&\geq (1-n^{-1})\mathbf{P}(\Ec_1\cap \Ec_2)-\mathbf{P}(\GG^c),
\end{aligned}
\]
% \[
% \begin{aligned}
%     &\mathbf{P}\left(\min_{m\in \mathcal{M}}\|\widehat{\gamma}^{[m]} - \gamma^*_{\rm ap}\|_2\lesssim \frac{\kappa_2}{\sigma_L^2(\U)}\left({\rm err}_n(M)/\sqrt{n} + \sqrt{d\log N/N}\right)\right)\\
% &\geq (1-n^{-1})\mathbf{P}(\Ec_1\cap \Ec_2)-\mathbf{P}((\GG_2\cap \GG_3)^c),
% \end{aligned}
% \]
}
provided that $M$ satisfies \eqref{eq: M condition}. Therefore, after we take the liminf operator with respect to $M\to\infty$, it becomes
\begin{equation*}
\begin{aligned}
    &\liminf_{M\to\infty}\mathbf{P}\left(\min_{m\in \mathcal{M}}\|\widehat{\gamma}^{[m]} - \gamma^*_{\rm ap}\|_2\lesssim \frac{\kappa_2}{\sigma_L^2(\U)}\left({\rm err}_n(M)/\sqrt{n} + \sqrt{d\log N/N}\right)\right)\\
    &\geq (1-n^{-1})\mathbf{P}(\Ec_1\cap \Ec_2)-\mathbf{P}(\GG^c)
\end{aligned}
\end{equation*}
Then we further take the liminf operator with respect to $n\to\infty$ on both sides to obtain that:
\begin{equation*}
    \begin{aligned}
        &\liminf_{n\to\infty}\liminf_{M\to\infty}\mathbf{P}\left(\min_{m\in \mathcal{M}}\|\widehat{\gamma}^{[m]} - \gamma^*_{\rm ap}\|_2\lesssim \frac{\kappa_2}{\sigma_L^2(\U)}\left({\rm err}_n(M)/\sqrt{n} + \sqrt{d\log N/N}\right)\right)\\
        &\geq \liminf_{n\to\infty}\left[(1-n^{-1})\mathbf{P}(\Ec_1\cap \Ec_2) - \mathbf{P}(\GG^c)\right],
    \end{aligned}
\end{equation*}
For two sequences $a_n,b_n$, we apply the fact that $\liminf_{n} (a_n b_n)\geq \liminf_{n}a_n\cdot \liminf_{n} b_n$ to obtain that
\[
 \liminf_{n\to\infty}\left[(1-n^{-1})\mathbf{P}(\Ec_1\cap \Ec_2)\right] \geq \liminf_{n\to\infty} (1-n^{-1})\cdot \liminf_{n\to\infty}\mathbf{P}(\Ec_1\cap \Ec_2)\geq 1-\alpha_0,
\]
where the last inequality follows from Lemma \ref{lem: E1E2}, which lower bounds $\mathbf{P}(\Ec_1\cap \Ec_2)$. For the event $\GG$ defined in \eqref{eq: events}, we pick the value $t=\sqrt{\log n}$, thus it follows from Lemma \ref{lem: concentration lemma} that
\[
\lim_{n\to\infty} \mathbf{P}(\GG) = 1.
\]
Combining the preceding three equations, we establish that:
% together with Lemma \ref{lem: concentration lemma} showing that $\mathbf{P}((\GG_2\cap \GG_3)^c)\leq N^{-c_1d}$, to establish that
\[
\begin{aligned}
    &\liminf_{n\to\infty}\liminf_{M\to\infty}\mathbf{P}\left(\min_{m\in \mathcal{M}}\|\widehat{\gamma}^{[m]} - \gamma^*_{\rm ap}\|_2\lesssim \frac{\kappa_2}{\sigma_L^2(\U)}\left({\rm err}_n(M)/\sqrt{n} + \sqrt{d\log N/N}\right)\right) \\
    % &\geq \liminf_{n\to\infty} (1-n^{-1})\cdot \liminf_{n\to\infty}\mathbf{P}(\Ec_1\cap \Ec_2)-\lim_{n\to\infty} N^{-c_1d}\\
    &\geq 1-\alpha_0.
\end{aligned}
\]
% where the last inequality follows from Lemma \ref{lem: E1E2}, which lower bounds $\mathbf{P}(\Ec_1\cap \Ec_2)$, and from the fact that $N^{-c_1d}\to 0$ as $n\to \infty$, due to $N\gg n\log N$ as specified Condition {\rm (A2)}.

\subsection{Proof of Theorems \ref{thm: inference-coverage} and \ref{thm: inference-length}}
\label{proof of thm: inference}

Before presenting the formal proof, we first introduce several supporting propositions and lemmas. With slight abuse of notation, we redefine the local neighborhood of $\theta^*$ as:
\begin{equation}
    \Theta_{\rm loc}=\left\{\theta\in \RR^{dK}: \|\theta-\theta^*\|_2\lesssim \tau \sqrt{d/n}\right\} \quad \text{with} \; \tau \; \text{defined in} \; \eqref{eq: final rate}.
    \label{eq: Theta local - new}
\end{equation}
In contrast to the earlier definition of $\Theta_{\rm loc}$ in \eqref{eq: Theta local}, we do not restrict the radius to be exactly $2\tau$. Such a definition of $\Theta_{\rm loc}$ is only used in the proof of Theorems \ref{thm: inference-coverage} and \ref{thm: inference-length}.

Next, we define the following event $\GG_4$, which controls the deviation of the empirical $\widehat{S}(\theta)$ from the population $S(\theta)$ uniformly over the redefined local set $\Theta_{\rm loc}$:
{\small
\begin{equation}
    \GG_4= \left\{\sup_{\theta \in \Theta_{\rm loc}}\left\|\left(\nabla\widehat{S}(\theta)-\nabla \widehat{S}(\theta^*)\right) -\left(\nabla S(\theta) - \nabla S(\theta^*)\right)\right\|_2 \lesssim \tau\sqrt{\frac{d^3 \log N }{n \cdot N}}\right\}.
    \label{eq: event G4}
\end{equation}
}The following lemma shows that $\GG_4$ occurs with a high probability, whose proof is provided in Section \ref{proof of prop grad Shat limiting}.
\begin{Lemma}
\label{lemma: grad Shat limiting}
Suppose Condition {\rm (A2)} holds. There exists some positive constant $c_1>0$ such that
\begin{equation*}
\mathbf{P}(\GG_4)\geq 1-N^{-c_1d}.
\end{equation*}
\end{Lemma}

For each index $m\in [M]$, the following proposition decomposes $\widehat{\theta}^{[m]} - \theta^*$ into a term that is independent of the perturbation procedure and another term that captures the variability induced by each perturbation. We establish the asymptotic normality of the first term and provide a high-probability bound for the second term when $\widehat{\theta}^{[m]} \in \Theta_{\rm loc}$.
The proof is given in Section~\ref{proof of prop theta_m nor}.
\begin{Proposition}
\label{prop: theta_m nor}
    Suppose Conditions {\rm (A1)}--{\rm (A5)} hold. Then for each $m\in \mathcal{M}$, we decompose the estimation error of $\widehat{\theta}^{[m]}$ as:
    \begin{equation}
        \widehat{\theta}^{[m]} - \theta^* = \mathcal{D}^* + {\rm Rem}^{[m]},
        \label{eq: error decompose of theta_m}
    \end{equation}
    where the term $\Dc^*$ satisfies the asymptotic normality:
    \begin{equation}
        \left[{\bf V}^*\right]^{-1/2} \Dc^*\xlongrightarrow{d} \Nc(0, {\bf I}_{dK}), \quad \textrm{with}\quad {\bf V}^* = [H(\theta^*)]^{-1} W^* [H(\theta^*)]^{-1},
        \label{eq: asymp D*}
    \end{equation}
and
    {
    \begin{equation}
    W^*= \sum_{l=1}^L (\gamma_l^*)^2{\bf V}_\mu^{(l)} + \frac{1}{N}{\bf V}_{\QQ_X}\left(p(X,\theta^*)\otimes X\right).
    \label{eq: W_star def}
    \end{equation}
    }In addition, we assume the quantity $\tau$ defined in \eqref{eq: final rate} satisfies $\tau\lesssim \sqrt{n/d^2}$, and the value $t$ satisfies \eqref{eq: t condition}. On the event $\GG\cap \GG_4$, if $\widehat{\theta}^{[m]}\in \Theta_{\rm loc}$, then we have
    {
    \begin{equation}
    \label{eq: remainder bound theta_m}
    \left\| {\rm Rem}^{[m]}\right\|_2 \lesssim \|\widehat{\gamma}^{[m]}-\gamma^*\|_2 + \tau\sqrt{\frac{d^3\log N}{n\cdot N}} + \tau^2\frac{d}{n},
    \end{equation}} where $\GG$ and $\GG_4$ are defined in \eqref{eq: events} and \eqref{eq: event G4}, respectively, and $\Theta_{\rm loc}$ is given in \eqref{eq: Theta local - new}.
\end{Proposition}
We provide additional clarification on the remainder bound in \eqref{eq: remainder bound theta_m}. On the event $\GG\cap \GG_4$, which holds with high probability as established in Lemmas~\ref{lem: concentration lemma} and \ref{lemma: grad Shat limiting}, the bound \eqref{eq: remainder bound theta_m} applies to any index $m$ whose corresponding $\widehat{\theta}^{[m]}$ lies in $\Theta_{\rm loc}$. In such cases, the remainder term ${\rm Rem}^{[m]}$ is guaranteed to satisfy the stated bound.

In the following, we present the proofs of Theorem \ref{thm: inference-coverage} and \ref{thm: inference-length} in sequence, where we focus on the fixed dimension setting, so the dependence on the dimension $d$ is often omitted in the statements. For those statements where $d$ explicitly appears, it indicates that the corresponding result remains valid even when the dimension $d$ grows with the sample size $n$.

\subsubsection{Proof of Theorem \ref{thm: inference-coverage}}
We take $m^*$ as any index such that
\[
\|\widehat{\gamma}^{[m^*]} - \gamma^*\|_2 = \min_{m\in \mathcal{M}}\|\widehat{\gamma}^{[m]} - \gamma^*\|_2.
\]
By construction of the confidence interval ${\rm CI}=\bigcup_{m\in \mathcal{M}} {\rm Int}^{[m]}$, we obtain that
\begin{equation}
    \mathbf{P}(\theta^*_1\in {\rm CI})\geq \mathbf{P}(\theta^*_1\in {\rm Int}^{[m^*]}).
    \label{eq: cov through m_star}
\end{equation}
Thus, to establish the coverage validity of ${\rm CI}$, it suffices to show that the individual interval ${\rm Int}^{[m^*]}$ contains $\theta^*_1$ with a probability larger than $1-\alpha$, where $\alpha$ is the pre-specified significance level. The remainder of the proof will focus on analyzing this specific index $m^*$.

It follows from \eqref{eq: cov through m_star} that
\begin{equation}
\begin{aligned}
    \mathbf{P}(\theta^*_1\notin {\rm CI})&\leq \mathbf{P}(\theta^*_1\notin {\rm Int}^{[m^*]}) = \mathbf{P}(|\widehat{\theta}_1^{[m^*]}-\theta_1^*|> z_{\alpha'/2} \widehat{\rm se}^{[m^*]}) \\
    &=\mathbf{P}\left(\frac{|\widehat{\theta}_1^{[m^*]}-\theta_1^*|}{\sqrt{{\bf V}^*_{1,1}}}> z_{\alpha'/2} \sqrt{\frac{\widehat{\bf V}^{[m^*]}_{1,1}}{{\bf V}^*_{1,1}}}\right),
\end{aligned}
    \label{eq: proof cov interm 0}
\end{equation}
where the first equality holds by the definition of ${\rm Int}^{[m]}$ in \eqref{uq: ci_m}, the second equality follows from the definition of $\widehat{\rm se}^{[m]}=\sqrt{\widehat{\bf V}^{[m]}_{1,1}}$ in Section \ref{subsec: hat_V_m}, and $\alpha'=\alpha-\alpha_0$ with $\alpha_0\in (0, 0.01]$ being the pre-specified constant used to construct $\mathcal{M}$

We next analyze the components of the target event on the most right-hand side of \eqref{eq: proof cov interm 0}.
According to \eqref{eq: error decompose of theta_m}, the estimation error $\widehat{\theta}^{[m^*]}-\theta^*$ can be decomposed as $\widehat{\theta}^{[m^*]}-\theta^*=\Dc^* + {\rm Rem}^{[m^*]}$. Then, for the first components for these terms, we have
\[
\widehat{\theta}^{[m^*]}_1 - \theta^*_1 = \Dc^*_1 + {\rm Rem}^{[m^*]}_1,
\]
where $\Dc^*_1$ and ${\rm Rem}^{[m^*]}_{1}$ denote the first entry of the terms $\Dc^*$ and ${\rm Rem}^{[m^*]}$, respectively.
It follows from the triangle inequality that
\[
\left|\widehat{\theta}^{[m^*]}_1-\theta^*_1\right|=
\left|\Dc^*_1 + {\rm Rem}^{[m^*]}_1\right|\leq \left|\Dc^*_1\right| +\left|{\rm Rem}^{[m^*]}_1\right|\leq 
\left|\Dc^*_1\right| +\left\|{\rm Rem}^{[m^*]}\right\|_2.
\]
Arranging the terms and dividing by $\sqrt{{\bf V}^*_{1,1}}$ on both sides, we obtain that
{\small
\begin{equation}
    \label{eq: proof cov interm 1}
    \frac{\left|\widehat{\theta}^{[m^*]}_1-\theta^*_1\right|}{\sqrt{{\bf V}^*_{1,1}}}\leq \frac{|\Dc^*_1|}{\sqrt{{\bf V}^*_{1,1}}} + \frac{\|\sqrt{n}\cdot {\rm Rem}^{[m^*]}\|_2}{\sqrt{n{\bf V}^*_{1,1}}}.
\end{equation}}
Moreover, we observe that
{\small
\[
\begin{aligned}
    \left|\sqrt{\frac{\widehat{\bf V}^{[m^*]}_{1,1}}{{\bf V}^*_{1,1}}} - 1\right| &= \frac{\left|\sqrt{\widehat{\bf V}^{[m^*]}_{1,1}}-\sqrt{{\bf V}^*_{1,1}}\right|}{\sqrt{{\bf V}^*_{1,1}}} = \frac{\left|\widehat{{\bf V}}^{[m^*]}_{1,1} - {\bf V}^*_{1,1}\right|}{\sqrt{{\bf V}^*_{1,1}}\left(\sqrt{\widehat{{\bf V}}^{[m^*]}_{1,1}} + \sqrt{{\bf V}^*_{1,1}}\right)}\\
    &\leq \frac{\left|\widehat{{\bf V}}^{[m^*]}_{1,1} - {\bf V}^*_{1,1}\right|}{{\bf V}^*_{1,1}}\leq \frac{\left\|\widehat{{\bf V}}^{[m^*]} - {\bf V}^*\right\|_2}{{\bf V}^*_{1,1}}.
\end{aligned}
\]
}
Therefore, we establish that
{\small
\begin{equation*}
    \label{eq: proof cov interm 2}
    \sqrt{\frac{\widehat{\bf V}^{[m^*]}_{1,1}}{{\bf V}^*_{1,1}}}\geq 1- \frac{\left\|n\cdot(\widehat{{\bf V}}^{[m^*]} - {\bf V}^*)\right\|_2}{n\cdot{\bf V}^*_{1,1}}.
\end{equation*}
}
Combining the above inequality and \eqref{eq: proof cov interm 1}, we conclude that:
{\small
\[
\mathbf{P}\left(\frac{|\widehat{\theta}_1^{[m^*]}-\theta_1^*|}{\sqrt{{\bf V}^*_{1,1}}}> z_{\alpha'/2} \sqrt{\frac{\widehat{\bf V}^{[m^*]}_{1,1}}{{\bf V}^*_{1,1}}}\right)\leq \mathbf{P}\left[\frac{|\Dc^*_1|}{\sqrt{{\bf V}^*_{1,1}}} + \frac{\|\sqrt{n} {\rm Rem}^{[m^*]}\|_2}{\sqrt{n{\bf V}^*_{1,1}}} > z_{\alpha'/2}\left(1- \frac{\left\|n(\widehat{{\bf V}}^{[m^*]} - {\bf V}^*)\right\|_2}{n{\bf V}^*_{1,1}}\right)\right].
\]
}
Together with \eqref{eq: proof cov interm 0}, we further establish that
\begin{equation}
    \mathbf{P}(\theta^*_1\notin {\rm CI}) \leq \mathbf{P}\left(\frac{|\Dc^*_1|}{\sqrt{{\bf V}^*_{1,1}}}  \geq z_{\alpha'/2} - z_{\alpha'/2}\cdot \frac{\left\|n\cdot(\widehat{{\bf V}}^{[m^*]} - {\bf V}^*)\right\|_2}{n\cdot{\bf V}^*_{1,1}}- \frac{\|\sqrt{n}\cdot {\rm Rem}^{[m^*]}\|_2}{\sqrt{n{\bf V}^*_{1,1}}}\right).
    \label{eq: proof cov - main step 1}
\end{equation}

To facilitate the proof, we further introduce the following event to control the components involving $m^*$ within \eqref{eq: proof cov - main step 1}:
\begin{equation*}
    \Ec_3 = \left\{\frac{\left\|n\cdot(\widehat{{\bf V}}^{[m^*]} - {\bf V}^*)\right\|_2}{n\cdot{\bf V}^*_{1,1}}\leq \kappa_{n,M},\;\textrm{and}\; \frac{\|\sqrt{n}\cdot {\rm Rem}^{[m^*]}\|_2}{\sqrt{n{\bf V}^*_{1,1}}}\leq \eta_{n,M}  \right\},
\end{equation*}
where the two sequences $\kappa_{n,M}$ and $\eta_{n,M}$ are defined as:
{
\[
\begin{aligned}
    \kappa_{n,M}&= \max_{l\in [L]}\|n_l(\widehat{\bf V}_\mu^{(l)} - {\bf V}_\mu^{(l)})\|_2+ \frac{1}{\sigma_L^2(\U)}\left(\frac{{\rm err}_n(M)}{\sqrt{n}} +\sqrt{\frac{\log N}{N}}\right)\\
    &\quad\quad +\left(1+\frac{1}{\sigma_L^2(\U)}\right)^2 \log n\cdot n^{-3/4}+ \frac{n}{N} + \frac{1}{\sqrt{N}},\\
    \eta_{n,M}&=\frac{1}{\sigma_L^2(\U)}\left({{\rm err}_n(M)} +\sqrt{\frac{ n\log N}{N}}\right)+ \left(1+\frac{1}{\sigma_L^2(\U)}\right)^2 \log n\cdot n^{-1/4}\\
    &\quad\quad +\left(1+\frac{1}{\sigma_L^2(\U)}\right) \sqrt{\log N/N}.
\end{aligned}
\]}Note that $\err(M)\to 0$ as $M$ grows, as defined in \eqref{eq: errM}, and $\max_{l\in [L]}\|n_l (\widehat{\bf V}_\mu^{(l)} - {\bf V}_\mu^{(l)})\|_2\xlongrightarrow{p}0$, as assumed in Condition {\rm (A5)}. Therefore, whenever $\sigma_L^2(\U)\gg \max\{\sqrt{n\log N/N},\sqrt{\log n}\cdot n^{-1/8}\}$, we shall verify that both $\kappa_{n,M}$ and $\eta_{n,M}$ satisfy that
\begin{equation}
    \lim_{n\to\infty}\lim_{M\to\infty} \kappa_{n,M} = 0, \quad \lim_{n\to\infty}\lim_{M\to\infty} \eta_{n,M} = 0.
    \label{eq: lim kappa, eta}
\end{equation}

The following lemma shows that the event $\Ec_3$ happens with high probability, whose proof is provided in Section \ref{proof of lem: derive event3}.
\begin{Lemma}
    \label{lem: derive event3}
    Suppose Conditions {\rm (A1)--(A5)} hold, $\sigma_L^2(\U)\gg \max\{\sqrt{n\log N/N}, \; \sqrt{\log n} \cdot n^{-1/8}\}$. Then it holds that
    \begin{equation*} \lim_{n\to\infty}\lim_{M\to\infty}\mathbf{P}(\Ec_3)\geq 1-\alpha_0.
    \end{equation*}
\end{Lemma}
We now continue studying the right-hand side of \eqref{eq: proof cov - main step 1}, by leveraging the high probability event $\Ec_3$. It holds that
{
\[
\begin{aligned}
    &\mathbf{P}\left(\frac{|\Dc^*_1|}{\sqrt{{\bf V}^*_{1,1}}}  \geq z_{\alpha'/2} - z_{\alpha'/2}\cdot \frac{\left\|n\cdot(\widehat{{\bf V}}^{[m^*]} - {\bf V}^*)\right\|_2}{n\cdot{\bf V}^*_{1,1}}- \frac{\|\sqrt{n}\cdot {\rm Rem}^{[m^*]}\|_2}{\sqrt{n{\bf V}^*_{1,1}}}\right) \\
    &= \mathbf{P}\left(\Ec_3 \bigcap\left\{\frac{|\Dc^*_1|}{\sqrt{{\bf V}^*_{1,1}}}  \geq z_{\alpha'/2} - z_{\alpha'/2}\cdot \frac{\left\|n\cdot(\widehat{{\bf V}}^{[m^*]} - {\bf V}^*)\right\|_2}{n\cdot{\bf V}^*_{1,1}}- \frac{\|\sqrt{n}\cdot {\rm Rem}^{[m^*]}\|_2}{\sqrt{n{\bf V}^*_{1,1}}}\right\}\right) \\
    &\quad + \mathbf{P}\left(\Ec_3^c \bigcap\left\{\frac{|\Dc^*_1|}{\sqrt{{\bf V}^*_{1,1}}}  \geq z_{\alpha'/2} - z_{\alpha'/2}\cdot \frac{\left\|n\cdot(\widehat{{\bf V}}^{[m^*]} - {\bf V}^*)\right\|_2}{n\cdot{\bf V}^*_{1,1}}- \frac{\|\sqrt{n}\cdot {\rm Rem}^{[m^*]}\|_2}{\sqrt{n{\bf V}^*_{1,1}}}\right\}\right) \\
    &\leq \mathbf{P}\left(\frac{|\Dc^*_1|}{\sqrt{{\bf V}^*_{1,1}}} \geq (1-\kappa_{n,M})z_{\alpha'2}- \eta_{n,M}\right) + \mathbf{P}(\Ec_3^c),
\end{aligned}
\]
}where the last inequality follows from the definition of $\Ec_3$. Together with \eqref{eq: proof cov - main step 1}, we obtain that:
\[
\mathbf{P}(\theta^*_1\notin {\rm CI})\leq \mathbf{P}\left(\frac{|\Dc^*_1|}{\sqrt{{\bf V}^*_{1,1}}} \geq (1-\kappa_{n,M})z_{\alpha'2}- \eta_{n,M}\right) - \mathbf{P}(\Ec_3)+1,
\]
which implies that
\begin{equation}
\label{eq: eq: proof cov key 2}
    \mathbf{P}(\theta^*_1\in {\rm CI}) = 1-\mathbf{P}(\theta^*_1\notin {\rm CI}) \geq \mathbf{P}(\Ec_3) - \mathbf{P}\left(\frac{|\Dc^*_1|}{\sqrt{{\bf V}^*_{1,1}}} \geq (1-\kappa_{n,M})z_{\alpha'2}- \eta_{n,M}\right).
\end{equation}
It follows from \eqref{eq: asymp D*} of Proposition \ref{prop: theta_m nor} that $\Dc^*_1/\sqrt{{\bf V}^*_{1,1}}\xlongrightarrow{d}\Nc(0,1)$. We apply the properties of $\kappa_{n,M},\eta_{n,M}$ in \eqref{eq: lim kappa, eta} and the bounded convergence theorem to establish that
\[
\lim_{n\to\infty}\lim_{M\to\infty}\mathbf{P}\left(\frac{|\Dc^*_1|}{\sqrt{{\bf V}^*_{1,1}}} \geq (1-\kappa_{n,M})z_{\alpha'2}- \eta_{n,M}\right) = \mathbf{P}(|Z|\geq z_{\alpha'/2}) = \alpha',
\]
where $Z$ denotes the standard gaussian random variable. 

Lastly, we take the liminf operators on both sides of \eqref{eq: eq: proof cov key 2} to establish that:
\[
\liminf_{n\to\infty}\liminf_{M\to\infty}\mathbf{P}(\theta_1^*\in {\rm CI})\geq \liminf_{n\to\infty}\liminf_{M\to\infty}\mathbf{P}(\Ec_3)-\alpha'.
\]
We then apply Lemma \ref{lem: derive event3}, which controls the probability of the event $\Ec_3$, to conclude that: if $\sigma_L^2(\U)\gg \max\{\sqrt{n\log N/N}, \; \sqrt{\log n} \cdot n^{-1/8}\}$,
\[
\liminf_{n\to\infty}\liminf_{M\to\infty}\mathbf{P}(\theta_1^*\in {\rm CI})\geq 1-\alpha_0 - \alpha' = 1-\alpha,
\]
where the last inequality holds as $\alpha'=\alpha-\alpha_0$.

\subsubsection{Proof of Theorem \ref{thm: inference-length}}
By the construction of ${\rm CI}$, we observe that
\begin{equation}
    {\bf Leng}({\rm CI}) \leq 2\max_{m\in \mathcal{M}}\left(\left|\widehat\theta_1^{[m]}
 - \theta_1^*\right| + z_{\alpha'/2}\cdot \sqrt{\widehat{{\bf V}}^{[m]}_{1,1}}\right),
 \label{eq: length decompose}
\end{equation}
where $\sqrt{\widehat{\bf V}^{[m]}_{1,1}}=\widehat{\rm se}^{[m]}$ as defined in Section \ref{subsec: hat_V_m}.
In the following discussion, we shall control $\max_{m\in \mathcal{M}}\left|\widehat\theta_1^{[m]}
 - \theta_1^*\right|$ and $\max_{m\in \mathcal{M}}\sqrt{\widehat{{\bf V}}^{[m]}_{1,1}}$ in order. 

\noindent\underline{\textit{Control of $\max_{m\in \mathcal{M}} \left|\widehat\theta_1^{[m]} - \theta_1^*\right|$.}}
We observe that 
\[
\max_{m\in \mathcal{M}}\left|\widehat\theta_1^{[m]}
 - \theta_1^*\right|\leq \max_{m\in \mathcal{M}}\|\widehat{\theta}^{[m]}-\theta^*\|_2.
\]
The following lemma controls the error  $\|\widehat{\theta}^{[m]}-\theta^*\|_2$ via the perturbed weight error $\|\widehat{\gamma}^{[m]}-\gamma^*\|_2$.
The proof is provided in Section \ref{proof of lemma resample theta}.
\begin{Lemma} 
\label{lem: resample theta convergence}
Suppose Conditions {\rm (A1)}--{\rm (A4)} hold. On the event $\GG$ defined in \eqref{eq: events},  there exists some constant $C>0$ such that  if $C\sqrt{d}  \|\widehat{\gamma}^{[m]}-\gamma^*\|_2 < 1$, then
\begin{equation*}
\|\widehat{\theta}^{[m]}-\theta^*\|_2\lesssim \|\widehat{\gamma}^{[m]}-\gamma^*\|_2+  t\sqrt{d/n},
\end{equation*}
where $t>0$ satisfies \eqref{eq: t condition}.
\end{Lemma}
The above lemma implies that we can control $\max_{m\in \mathcal{M}}\|\widehat{\theta}^{[m]}-\theta^*\|_2$ via $\max_{m\in \mathcal{M}}\|\widehat{\gamma}^{[m]} - \gamma^*\|_2$, if $C\sqrt{d} \max_{m\in \mathcal{M}}\|\widehat{\gamma}^{[m]} - \gamma^*\|_2<1$ for some constant $C>0$. We now investigate the perturbed weight error $\max_{m\in \mathcal{M}}\|\widehat{\gamma}^{[m]} - \gamma^*\|_2.$ The proof of the following lemma is provided in Section \ref{proof of lemma: max gamma_m diff}.
\begin{Lemma}
    Suppose Conditions {\rm (A2)}--{\rm (A5)} hold.
    On the event $\GG\cap \Ec_1\cap \Ec_2$, with the event $\GG$  defined in \eqref{eq: events}, and the events $\Ec_1,\Ec_2$ defined in \eqref{eq: E1E2}, 
    \begin{equation}
    \max_{m\in \mathcal{M}}\|\widehat{\gamma}^{[m]} -\gamma^*\|_2 \lesssim \frac{z_{\alpha_0/dKL}}{\sigma_L^2(\U)} \sqrt{d/n}  + \tau^2(d/n)^{3/4},
    \label{eq: max gamma_m distance}
\end{equation}
where the value $t$ satisfies $c_0\leq  t\ll \sigma_L^2(\U) (n/d)^{1/4}$, and $\alpha_0\in (0, 0.01]$ is the constant used to construct $\mathcal{M}$ in \eqref{eq: filtered set}.
\label{lemma: max gamma_m diff}
\end{Lemma}
As $\tau\asymp (1+\frac{1}{\sigma_L^2(\U)}) t$, when $t\ll \sigma_L^2(\U) (n/d)^{1/4}$, it holds that $\tau\ll (n/d)^{1/4}$. Moreover, provided that $\sigma_L(\U)\gg n^{-1/4}$, in the fixed dimension scenario, we have
\[
C\cdot \frac{z_{\alpha_0/dKL}}{\sigma_L^2(\U)}\sqrt{d/n} \leq \frac{1}{2},\quad \textrm{and}\quad \tau^2 (d/n)^{3/4}\ll (d/n)^{1/4}\ll 1,
\]
for some constant $C>0$.
Therefore, the above lemma indicates that on the event $\GG\cap \Ec_1\cap \Ec_2$, the perturbed weight vectors satisfy $C\sqrt{d}\max_{m\in \mathcal{M}}\|\widehat{\gamma}^{[m]}-\gamma^*\|_2 < 1$. Then together with Lemma \ref{lem: resample theta convergence}, we establish that: when $t\ll \sigma_L^2(\U) n^{1/4}$, on the event $\GG\cap \Ec_1\cap\Ec_2,$
\begin{equation}
\begin{aligned}
     \max_{m\in \mathcal{M}} \|\widehat{\theta}^{[m]}-\theta^*\|_2 &\lesssim \max_{m\in \mathcal{M}}\|\widehat{\gamma}^{[m]} - \gamma^*\|_2 + t\sqrt{d/n}\lesssim \left(t+ \frac{z_{\alpha_0}}{\sigma_L^2(\U)}\right)n^{-1/2} + \tau^2 n^{-3/4}.
\end{aligned}
\label{eq: max_m theta_m - theta_star}
\end{equation}
where the first inequality holds by taking the maximum over $m\in \mathcal{M}$ of Lemma \ref{lem: resample theta convergence}, and the second inequality follows from \eqref{eq: max gamma_m distance} and we omit the dependence on $d$ due to the fixed dimension regime. 

Furthermore, if $t\ll \sigma_L^2(\U) n^{1/8}$ such that $\tau\asymp (1+\frac{1}{\sigma_L^2(\U)})t\ll n^{1/8}$, we have $\tau^2 n^{-3/4}\ll n^{-1/2}$. Then, \eqref{eq: max_m theta_m - theta_star} further implies that, on the event $\GG\cap \Ec_1\cap \Ec_2$, 
\begin{equation}
    \max_{m\in \mathcal{M}}\left|\widehat\theta_1^{[m]}
 - \theta_1^*\right|\leq\max_{m\in \mathcal{M}} \|\widehat{\theta}^{[m]}-\theta^*\|_2 \lesssim \left(t+ \frac{z_{\alpha_0}}{\sigma_L^2(\U)}\right)\frac{1}{\sqrt{n}}.
    \label{eq: max theta_m diff}
\end{equation}

\noindent\underline{\textit{Control of $\max_{m\in \mathcal{M}}\sqrt{\widehat{{\bf V}}^{[m]}_{1,1}}$.}} Note that 
\begin{equation}
    \label{eq: proof length interm-1}
\max_{m\in \mathcal{M}} n\cdot \widehat{{\bf V}}^{[m]}_{1,1}\leq \max_{m\in \mathcal{M}}\|n\widehat{{\bf V}}^{[m]}\|_2 \leq \|n {\bf V}^*\|_2 + \max_{m\in \mathcal{M}}\|n(\widehat{\bf V}^{[m]} - {\bf V}^*)\|_2,
\end{equation}
where the last inequality holds by the triangle inequality. Recall the definition of ${\bf V}^*$ in \eqref{eq: asymp D*}, we have
{
\[
\begin{aligned}
    \|n{\bf V}^*\|_2 &= \|[H(\theta^*)]^{-1}\|_2^2 \|n W^*\|_2 \\
    &\leq \|[H(\theta^*)]^{-1}\|_2^2\left\{\sum_{l=1}^L (\gamma^*_l)^2 \|n_l{\bf V}_\mu^{(l)}\|_2 + \frac{n}{N}\|{\bf V}_{\QQ_X}(p(X,\theta^*)\otimes X)\|_2\right\}.
\end{aligned}
\]
}According to Condition {\rm (A5)} and the boundedness of $\XQ$ in Condition {\rm (A2)}, we establish that both $\|n_l{\bf V}_\mu^{(l)}\|_2$ and $\|{\bf V}_{\QQ_X}(p(X,\theta^*)\otimes X)\|_2$ are upper bounded by some positive constant. Therefore, we establish that
\begin{equation}
    \label{eq: proof length interm-2}
    \|n{\bf V}^*\|_2 \leq C,
\end{equation}
by some constant $C>0$.

The following lemma establishes the upper bound of $\max_{m\in \mathcal{M}}\|n(\widehat{\bf V}^{[m]} - {\bf V}^*)\|_2$, with the proof provided in Section \ref{proof of lem: max_m V_m dist}
\begin{Lemma}
\label{lem: max_m V_m dist}
    Suppose Conditions {\rm (A2)}--{\rm (A5)} hold, and the value $t$ satisfies $c_0\leq t\ll \sigma_L^2(\U) n^{1/8}$.
    On the event $\GG\cap \Ec_1\cap \Ec_2$, with the event $\GG$ defined in \eqref{eq: events} and the events $\Ec_1,\Ec_2$ defined in \eqref{eq: E1E2}, we have
    \begin{equation*}
    \max_{m\in \mathcal{M}}\|n\cdot (\widehat{\bf V}^{[m]}-{\bf V}^*)\|_2\leq C,
    \end{equation*}
    for some positive constant $C>0$.
\end{Lemma}
We plug the results in \eqref{eq: proof length interm-2} and the above lemma back to \eqref{eq: proof length interm-1}, and obtain that: on the event $\GG\cap \Ec_1\cap\Ec_2$, if $c_0\leq t\ll \sigma_L^2(\U) n^{1/8}$, then
\begin{equation}
\label{eq: proof length interm-4}
    \max_{m\in \mathcal{M}} \sqrt{\widehat{{\bf V}}^{[m]}_{1,1}}\lesssim \frac{1}{\sqrt{n}}.
\end{equation}

\noindent\underline{\textit{Proof of Length \eqref{eq: resampling length}.}} We plug \eqref{eq: proof length interm-4} and \eqref{eq: max theta_m diff} into the upper bound of ${\bf Leng}(\rm CI)$ in \eqref{eq: length decompose} to establish that: on the event $\GG\cap \Ec_1\cap\Ec_2$, if the value $t$ satisfies $c_0\leq t\ll \sigma_L^2(\U) n^{1/8}$, then
\[
{\bf Leng}({\rm CI})\lesssim \max_{m\in \mathcal{M}}\left|\widehat\theta_1^{[m]}
 - \theta_1^*\right| + \max_{m\in \mathcal{M}} \sqrt{\widehat{{\bf V}}^{[m]}_{1,1}}\lesssim \left(t+ \frac{z_{\alpha_0}}{\sigma_L^2(\U)}\right)\frac{1}{\sqrt{n}}.
\]
Therefore,
\[
\begin{aligned}
    &\mathbf{P}\left({\bf Leng}({\rm CI})\lesssim \left(t+ \frac{z_{\alpha_0}}{\sigma_L^2(\U)}\right)\frac{1}{\sqrt{n}}\right)\geq\mathbf{P}(\GG\cap \Ec_1\cap \Ec_2)  \\
    &= 1-\mathbf{P}(\GG^c \cup (\Ec_1\cap \Ec_2)^c) \geq 1 - \mathbf{P}(\GG^c)- \mathbf{P}((\Ec_1\cap \Ec_2)^c) 
    = \mathbf{P}(\GG) + \mathbf{P}(\Ec_1\cap \Ec_2) - 1,
\end{aligned}
\]
We apply the liminf operators first with respect to $M\to\infty$ and then with respect to $n\to\infty$ on both sides. Together with Lemmas \ref{lem: concentration lemma}, \ref{lem: E1E2} that lower bound $\mathbf{P}(\GG)$ and $\mathbf{P}(\Ec_1\cap\Ec_2)$, respectively, this yields
\[
\liminf_{n\to\infty}\liminf_{M\to\infty}\mathbf{P}\left({\bf Leng}({\rm CI})\lesssim \left(t+ \frac{z_{\alpha_0}}{\sigma_L^2(\U)}\right)\frac{1}{\sqrt{n}}\right) \geq 1-\alpha_0-e^{-c_1t^2}.
\]

\subsection{Proof of Proposition \ref{prop: theta_m nor}}
\label{proof of prop theta_m nor}
Recall that in \eqref{eq: m-opt} and in Proposition \ref{prop: local theta}, we have shown that
\[
\widehat{\theta}^{[m]} = \argmin_{\theta\in \RR^{dK}} \theta^\intercal \widehat{\U}\widehat{\gamma}^{[m]} + \widehat{S}(\widehat{\theta}^{[m]}), \quad \textrm{and}\quad \theta^*=\argmin_{\theta\in \RR^{dK}}\theta^\intercal \U \gamma^* + S(\theta^*),
\]
respectively.
The optimality conditions of $\thetainit^{[m]}$ and $\theta^*$ imply that
\begin{equation*}
\widehat{\U}\hgammam + \nabla\widehat{S}(\widehat{\theta}^{[m]})=0 \quad \text{and}\quad \U{\gamma}^*+\nabla {S}(\theta^*)=0.
\end{equation*}
We apply the above equations and obtain
\begin{equation*}
\begin{aligned}
    \nabla {S}(\widehat{\theta}^{[m]})-\nabla {S}(\theta^*) &= \nabla S(\thetainit^{[m]}) - \nabla \widehat{S}(\thetainit^{[m]}) + \nabla \widehat{S}(\thetainit^{[m]}) - \nabla S(\theta^*)
    \\
    &=-\left(\nabla \widehat{S}(\widehat{\theta}^{[m]})-\nabla {S}(\widehat{\theta}^{[m]})\right)-\left(\widehat{\U}\hgammam - \U \gamma^*\right).
\end{aligned}
\end{equation*}
Note that we can also express $\nabla {S}(\widehat{\theta}^{[m]})-\nabla {S}(\theta^*)$ in the integral form as follows:
\begin{equation}
    \nabla {S}(\widehat{\theta}^{[m]})-\nabla {S}(\theta^*) = \bar{H}^{[m]} (\widehat{\theta}^{[m]} - \theta^*), \quad \textrm{with}\quad \bar{H}^{[m]} = \int_0^1 H\left(\theta^* + \nu \cdot (\widehat{\theta}^{[m]}-\theta^*)\right) d\nu.
    \label{eq: barH_m}
\end{equation}
% By the intermediate value theorem, there exists $\nu\in [0,1]$ such that  
% $$\nabla {S}(\widehat{\theta}^{[m]})-\nabla {S}(\theta^*) = H(\widetilde{\theta}^{[m]})(\widehat{\theta}^{[m]}-\theta^*) \quad \text{with}\quad \widetilde{\theta}^{[m]}=(1-\nu)\theta^*+ \nu\widehat{\theta}^{[m]}.$$
We combine the above two equations and obtain
\begin{equation*}
\widehat{\theta}^{[m]}-\theta^*=\left[\bar{H}^{[m]}\right]^{-1}\left[-\left(\nabla \widehat{S}(\widehat{\theta}^{[m]})-\nabla {S}(\widehat{\theta}^{[m]})\right)-\left(\widehat{\U}\hgammam - \U \gamma^*\right)\right].
\label{eq: limit dist}
\end{equation*}
% \begin{equation*}
% \widehat{\theta}^{[m]}-\theta^*=\left[H\left(\widetilde{\theta}^{[m]}\right)\right]^{-1}\left[-\left(\nabla \widehat{S}(\widehat{\theta}^{[m]})-\nabla {S}(\widehat{\theta}^{[m]})\right)-\left(\widehat{\U}\hgammam - \U \gamma^*\right)\right].
% \label{eq: limit dist}
% \end{equation*}
In the following, we shall decompose the above expression as the perturbation-free term and the remainder terms
\begin{equation}
\widehat{\theta}^{[m]}-\theta^*=\mathcal{D}^*+ \text{Rem}^{[m]},
\label{eq: theta_m error decomposition}
\end{equation}
where 
\begin{equation}
\mathcal{D}^*=-\left[H(\theta^*)\right]^{-1} \left(\nabla \widehat{S}({\theta}^{*})-\nabla {S}({\theta}^{*})+(\widehat{\U}-\U)\gamma^*\right),
\label{eq: theta_m error decomposition - dominating}
\end{equation}
and ${\rm Rem}^{[m]} = \sum_{j=1}^{3}\text{Rem}^{[m]}_{(j)}$ with
\begin{equation}
\begin{aligned}
\text{Rem}_{(1)}^{[m]}&=-\left[\bar{H}^{[m]}\right]^{-1}\left[\widehat{\U}(\widehat{\gamma}^{[m]} - \gamma^*)\right],\\
\text{Rem}_{(2)}^{[m]}&=\left[\bar{H}^{[m]}\right]^{-1}\left[\left(\nabla \widehat{S}({\theta}^{*})-\nabla {S}(\theta^*)\right)-\left(\nabla \widehat{S}(\widehat{\theta}^{[m]})-\nabla {S}(\widehat{\theta}^{[m]})\right)\right],\\
\text{Rem}_{(3)}^{[m]}&=-\left(\left[\bar{H}^{[m]}\right]^{-1}-\left[H(\theta^*)\right]^{-1}\right)\left(\nabla \widehat{S}({\theta}^{*})-\nabla {S}({\theta}^{*})+ (\widehat{\U}-\U)\gamma^*\right).
\end{aligned}
\label{eq: theta_m error decomposition - remainder}
\end{equation}
% \begin{equation}
% \begin{aligned}
% \text{Rem}_{(1)}^{[m]}&=-\left[H\left(\widetilde{\theta}^{[m]}\right)\right]^{-1}\left[\widehat{\U}(\widehat{\gamma}^{[m]} - \gamma^*)\right],\\
% \text{Rem}_{(2)}^{[m]}&=\left[H\left(\widetilde{\theta}^{[m]}\right)\right]^{-1}\left[\left(\nabla \widehat{S}({\theta}^{*})-\nabla {S}(\theta^*)\right)-\left(\nabla \widehat{S}(\widehat{\theta}^{[m]})-\nabla {S}(\widehat{\theta}^{[m]})\right)\right],\\
% \text{Rem}_{(3)}^{[m]}&=-\left(\left[H\left(\widetilde{\theta}^{[m]}\right)\right]^{-1}-\left[H(\theta^*)\right]^{-1}\right)\left(\nabla \widehat{S}({\theta}^{*})-\nabla {S}({\theta}^{*})+ (\widehat{\U}-\U)\gamma^*\right).
% \end{aligned}
% \label{eq: theta_m error decomposition - remainder}
% \end{equation}

The following lemma establishes the limiting distribution of the perturbation-free term $\Dc^*$ defined in \eqref{eq: theta_m error decomposition - dominating}. The proof is provided in Section \ref{proof of lemma: theta_m error decomposition - dominating}.
\begin{Lemma}
\label{lemma: theta_m error decomposition - dominating}
    Suppose Condition {\rm (A5)} holds. Then the term $\Dc^*$ defined in \eqref{eq: theta_m error decomposition - dominating} admits the following asymptotic distribution:
    \begin{equation}
\mathcal{D}^* \xlongrightarrow{d} \Nc\left(0, [H(\theta^*)]^{-1}W^* [H(\theta^*)]^{-1}\right),
\label{eq: D limit dist}
\end{equation}
where the matrix $W^*$ is defined in \eqref{eq: W_star def}.
\end{Lemma}

Next, we control the remainder term ${\rm Rem}^{[m]}$, with the proof provided in Section \ref{proof of lemma: theta_m error decomposition - remainder}.
\begin{Lemma}
\label{lemma: theta_m error decomposition - remainder}
    Suppose that Conditions {\rm (A2), (A3), (A4)} hold, the value $t>0$ satisfies \eqref{eq: t condition}, and $\tau$ defined in \eqref{eq: final rate} satisfies $\tau \lesssim\sqrt{n/d^2}$. Then for any index $m$ whose  $\widehat{\theta}^{[m]}\in \Theta_{\rm loc}$, with $\Theta_{\rm loc}$ defined in \eqref{eq: Theta local - new}, on the event $\GG\cap \GG_4$, the remainder terms satisfy:
    \[
    \begin{aligned}
        \|{\rm Rem}_{(1)}^{[m]}\|_2&\lesssim \|\widehat{\gamma}^{[m]} - \gamma^*\|_2 \\
        \|{\rm Rem}_{(2)}^{[m]}\|_2 &\lesssim \tau\sqrt{\frac{d^3 \log N }{n \cdot N}}\\
        \|{\rm Rem}_{(3)}^{[m]}\|_2 &\lesssim \tau\sqrt{\frac{d^2\log N}{n\cdot N}} + \tau^2 \frac{d}{n}.
    \end{aligned}
    \]
\end{Lemma}
We apply the preceding lemma to establish that: on the event $\GG\cap \GG_4$,
\[
\left\| {\rm Rem}^{[m]}\right\|_2\leq \sum_{j=1}^3 \|{\rm Rem}^{[m]}_{(j)}\|_2 \lesssim \|\widehat{\gamma}^{[m]}-\gamma^*\|_2 + \tau\sqrt{\frac{d^3\log N}{n\cdot N}} + \tau^2\frac{d}{n}.
\]

\subsection{Proof of Proposition \ref{prop: simplify theta_hat_m}}
\label{proof of prop: simplify theta_hat_m}

It follows from \eqref{eq: theta_m error decomposition} and \eqref{eq: theta_m error decomposition - remainder} that for each $m\in \mathcal{M}$, 
\[
\widehat{\theta}^{[m]}-\theta^* = \Dc^* + \sum_{j=1}^3{\rm Rem}^{[m]}_{(j)}.
\]

On the event $\GG_2$ defined in \eqref{eq: events}, we have 
$$\|\nabla \widehat{S}(\theta^*) - \nabla S(\theta^*)\|_2\lesssim \sqrt{d\log N/N} \ll \sqrt{d/n},$$ 
where the last inequality holds as $N\gg n\log N$ in Condition {\rm (A2)}. Then on the event $\GG_2$, the term $\Dc^*$ defined in \eqref{eq: theta_m error decomposition - dominating} can be expressed as:
\begin{equation}
    \Dc^* = -\left[H(\theta^*)\right]^{-1} \left((\widehat{\U}-\U)\gamma^*\right) + o(\sqrt{d/n}).
    \label{eq: larger N regime-1}
\end{equation}
And it follows from \eqref{eq: max theta_m diff} that
when $t\ll \sigma_L^2(\U) n^{1/8}$, on the event $\GG\cap \Ec_1\cap \Ec_2$,
\[
\widehat{\theta}^{[m]}\in \Theta_{\rm loc}, \quad \textrm{for all $m\in \mathcal{M}$}.
\]
And when $t\ll \sigma_L^2(\U) n^{1/8}$, it holds that $\tau\asymp (1+\frac{1}{\sigma_L^2(\U)})t \ll n^{1/8}$.
Then we apply Lemma \ref{lemma: theta_m error decomposition - remainder} to establish that: $\GG\cap \GG_4\cap \Ec_1\cap \Ec_2$,
\begin{equation}
\|{\rm Rem}_{(2)}^{[m]} + {\rm Rem}^{[m]}_{(3)}\|_2 = o(\sqrt{d/n}).
    \label{eq: larger N regime-2}
\end{equation}
And the term ${\rm Rem}^{[m]}_{(1)}$, it can be further expressed as:
\begin{equation}
    \begin{aligned}
   {\rm Rem}_{(1)}^{[m]} &= -[H(\theta^*)]^{-1}\U(\widehat{\gamma}^{[m]}-\gamma^*) - [H(\theta^*)]^{-1}\left(\widehat{\U}-\U\right)(\widehat{\gamma}^{[m]}-\gamma^*) \\
   &\quad\quad- \left\{[\bar{H}^{[m]}]^{-1} -[H(\theta^*)]^{-1} \right\} \widehat{\U}(\widehat{\gamma}^{[m]}-\gamma^*) 
\end{aligned}
\label{eq: rem_1 decompose}
\end{equation}

We now study the second term on the right-hand side of \eqref{eq: rem_1 decompose}.
On the event $\GG\cap\Ec_1\cap\Ec_2$, Lemma \ref{lemma: max gamma_m diff} implies that, 
\[
\begin{aligned}
    \left\|[H(\theta^*)]^{-1}\left(\widehat{\U}-\U\right)(\widehat{\gamma}^{[m]}-\gamma^*)\right\|_2 &\lesssim \|\widehat{\U}-\U\|_2\|\widehat{\gamma}^{[m]}-\gamma^*\|_2 \\
    &\lesssim t\sqrt{d/n} \left[\frac{1}{\sigma_L^2(\U)} \sqrt{d/n} + \tau^2(d/n)^{3/4}\right]\\
    &\lesssim \frac{t}{\sigma_L^2(\U)}\frac{d}{n},
\end{aligned}
\]
where the second inequality follows from the definition of $\GG$ and Lemma \ref{lemma: max gamma_m diff}, the third inequality holds by $\tau\ll n^{1/8}$.

We continue studying the third term on the right-hand side of \eqref{eq: rem_1 decompose}. It follows from Lemma \ref{lem: key property}, which establishes the continuity of $H(\theta)$, that
\[
\begin{aligned}
    \left\|\bar{H}^{[m]} - H(\theta^*) \right\|_2 &= \left\|\int_0^1 \left(H(\theta^* + \nu(\widehat{\theta}^{[m]}-\theta^*)) - H(\theta^*)\right) d\nu \right\|_2 \\
    &\leq \int_0^1 \nu \left(\|\widehat{\theta}^{[m]} - \theta^*\|_2 + \|\widehat{\theta}^{[m]} - \theta^*\|_2^2 \right) d\nu \\
    &= \|\widehat{\theta}^{[m]} - \theta^*\|_2 + \|\widehat{\theta}^{[m]} - \theta^*\|_2^2 \lesssim \tau \sqrt{d/n} + \tau^2(d/n),
\end{aligned}
\]
where the last inequality holds by $\widehat{\theta}^{[m]}\in \Theta_{\rm loc}$.
% \[
% \begin{aligned}
%     \left\|[H(\widetilde{\theta}^{[m]})]^{-1} -[H(\theta^*)]^{-1} \right\|&\lesssim \|\widetilde{\theta}^{[m]}-\theta^*\|_2 + \|\widetilde{\theta}^{[m]}-\theta^*\|_2^2 \\
%     &\leq \|\widehat{\theta}^{[m]}-\theta^*\|_2 + \|\widehat{\theta}^{[m]}-\theta^*\|_2^2 \\
%     &\lesssim \tau \sqrt{d/n} + \tau^2(d/n),
% \end{aligned}
% \]
% where the first inequality holds by \eqref{eq: Lip H} in Lemma \ref{lem: key property}, the second inequality holds due to the definition of $\widetilde{\theta}^{[m]}$, the third inequality holds by $\widehat{\theta}^{[m]}\in \Theta_{\rm loc}$. Then we show that
Then we show that
\[
\left\|[\bar{H}^{[m]}]^{-1} - [H(\theta^*)]^{-1} \right\|_2\lesssim \tau \sqrt{d/n} + \tau^2(d/n),
\]
and thus
\[
\begin{aligned}
    &\left\|\left([\bar{H}^{[m]}]^{-1} -[H(\theta^*)]^{-1} \right) \widehat{\U}(\widehat{\gamma}^{[m]}-\gamma^*)\right\|_2\\
    &\leq \left\|[\bar{H}^{[m]}]^{-1} -[H(\theta^*)]^{-1} \right\|_2 \|\widehat{\U}\|_2 \left\|\widehat{\gamma}^{[m]}-\gamma^*\right\|_2\\
    &\lesssim\left(\tau\sqrt{d/n} + \tau^2(d/n)\right)\left(\|\U\|+t\sqrt{d/n}\right) \left[\frac{1}{\sigma_L^2(\U)} \sqrt{d/n} + \tau^2(d/n)^{3/4}\right],
\end{aligned}
\]
where the last inequality holds due to event $\GG$ and Lemma \ref{lemma: max gamma_m diff}.

As we assume that $c_0\leq t\ll \sigma_L^2(\U) n^{1/8}$, we have $t\ll \sqrt{d/n}$ and $t\ll n^{1/8}$ and $\sigma_L^2(\U) \gg n^{-1/8}$.
Together with $\|\U\|\leq C_0$ in Condition {\rm (A4)}, we shall further simplify the preceding inequality such that:
\[
\left\|\left([\bar{H}^{[m]}]^{-1} -[H(\theta^*)]^{-1} \right) \widehat{\U}(\widehat{\gamma}^{[m]}-\gamma^*)\right\|_2 =o(\sqrt{d/n}).
\]

Putting the above results back to \eqref{eq: rem_1 decompose}, we have
\[
\begin{aligned}
    {\rm Rem}_{(1)}^{[m]} &= -[H(\theta^*)]^{-1}{\U}(\widehat{\gamma}^{[m]}-\gamma^*) + o(\sqrt{d/n}).
\end{aligned}
\]
We put the above equation, \eqref{eq: larger N regime-1} and \eqref{eq: larger N regime-2} into \eqref{eq: theta_m error decomposition} to establish that: on the event $\GG\cap \GG_4\cap \Ec_1\cap \Ec_2$, 
\[
\widehat{\theta}^{[m]}-\theta^*=  -[H(\theta^*)]^{-1}{\U}(\widehat{\gamma}^{[m]}-\gamma^*) -\left[H(\theta^*)\right]^{-1} \left((\widehat{\U}-\U)\gamma^*\right) + o(\sqrt{d/n}),
\]
provided that $c_0\leq t \ll \sigma_L^2(\U) n^{1/8}$.

\subsection{Proof of Proposition \ref{prop: double robustness}}
\label{proof of prop: doubly robust estimator}
For each source $l\in [L]$, the estimation error of the proposed estimator $\widehat{\mu}^{(l)}\in \RR^{dK}$ shall be decomposed as
{
\begin{equation}
    \widehat{\mu}^{(l)} - \mu^{(l)} = \Dc_1^{(l)} + \Dc_2^{(l)} + \sum_{j=1}^4 {\rm Rem}_j^{(l)},
    \label{eq: estimation error decomposition}
\end{equation}
}
where the dominant terms $\Dc_1^{(l)}, \Dc_2^{(l)}$ are:
\begin{equation}
    \begin{aligned}
        \Dc_1^{(l)} &= \frac{1}{N}\sum_{j=1}^N\left[-\f{l}(\XQ_j)\otimes \XQ_j\right] - \E\left[-\f{l}(\XQ)\otimes \XQ\right],\\
        \Dc_2^{(l)} &= \frac{1}{n_l}\sum_{i=1}^{n_l}\left[\left(\w{l}(\X{l}_i)\cdot(\f{l}(\X{l}_i) - {\bf y}_i^{(l)})\right)\otimes \X{l}_i\right]
    \end{aligned}
    \label{eq: D1D2}
\end{equation}
with the one-hot outcome vector ${\bf y}^{(l)}_i = \left({\bf 1}(Y^{(l)}_i=1),..., {\bf 1}(\Y{l}_i=K)\right)^\intercal$. The remainder terms ${\rm Rem}_j^{(l)}$ for $j=1,2,3,4$ are defined as:
\begin{equation}
    \begin{aligned}
        {\rm Rem}_1^{(l)} &= \frac{1}{n_l}\sum_{i=1}^{n_l}\left[\left(\w{l}(\X{l}_i)\cdot (\hf{l}(\X{l}_i) - \f{l}(\X{l}_i)) \right)\otimes \X{l}_i\right] \\
        &\quad\quad\quad - \E_{\PP^{(l)}}\left[\left(\w{l}(X)\cdot (\hf{l}(X) - \f{l}(X)) \right)\otimes X\right] \\
        {\rm Rem}_2^{(l)} &= -\frac{1}{N}\sum_{j=1}^N\left[ \left(\hf{l}(\XQ_j)-\f{l}(\XQ_j)\right)\otimes \XQ_j\right] + \E_{\QQ_X}\left[\left(\hf{l}(X)-\f{l}(X)\right)\otimes X\right]\\
        {\rm Rem}_3^{(l)} &= \frac{1}{n_l}\sum_{i=1}^{n_l}\left[\left(\hw{l}(\X{l}_i) - \w{l}(\X{l}_i)\right)\cdot \left(\f{l}(\X{l}_i) - \mathbf{y}^{(l)}_i\right)\right] \otimes \X{l}_i \\
        {\rm Rem}_4^{(l)} &= \frac{1}{n_l}\sum_{i=1}^{n_l}\left[\left(\hw{l}(\X{l}_i) - \w{l}(\X{l}_i)\right)\cdot \left(\hf{l}(\X{l}_i) - \f{l}(\X{l}_i)\right)\right]\otimes \X{l}_i.
    \end{aligned}
    \label{eq: Rem1:4}
\end{equation}

\subsubsection{Proof of \eqref{eq: rates of mu}}

We first establish the convergence rates of $\Dc_1^{(l)}$ and $\Dc_2^{(l)}$, and then discuss the remainder terms ${\rm Rem}_{j=1,...4}^{(l)}$ afterwards. To facilitate discussion, we introduce the following lemma, with the proof provided in Section \ref{sec: proof of lemma - subgaus f(X)X}.
\begin{Lemma}
    Let $X\in \mathbb{R}^d$ be a subgaussian random vector, and let $h:\mathbb{R}^d\to\mathbb{R}^K$ be a deterministic function independent of $X$. Suppose that for each entry $k=1,..., K$, $h_k(X)$ is a bounded random variable with $\ell_2$ norm $\|h_k\|_{\ell_2}= (\E[h_k(X)]^2)^{1/2}$.  Let $X_1, X_2, \dots, X_n$ be i.i.d. samples of $X$. Then, there exist positive constants $c_0,c_1>0$ such that with probability at least $1-e^{-c_1t^2}$,
    \[
    \left\|\frac{1}{n}\sum_{i=1}^n h(X_i)\otimes X_i - \E[h(X)\otimes X]\right\|_2 < t\cdot \max_{k=1...,K}\|h_k\|_{\ell_2} \cdot \sqrt{d/n},
    \]
    where $t$ is any value larger than $c_0$.
\label{lemma: subgaus of f(X)X}
\end{Lemma}

\noindent{\bf Control of $\Dc_1^{(l)}$.} 
For each category $k=1,...,K$, we have $\f{l}_k(X)\in [-1,1]$, with $\|\f{l}_k\|_{\ell_2}\leq 1$. Then we directly apply Lemma \ref{lemma: subgaus of f(X)X} to establish that: with probability at least $1-e^{-c_1t^2}$,
\begin{equation}
    \|\Dc_1^{(l)}\|_2 \leq t\cdot \sqrt{d/N},
    \label{eq: rate D1}
\end{equation}
for any $t$ larger than $c_0$.

\noindent{\bf Control of $\Dc_2^{(l)}$.} We observe that the expectation of $\Dc_2^{(l)}$ is zero such that
\[
\E_{\PP^{(l)}}\left(\omega^{(l)}(X) \cdot( \f{l}(X)-{\bf y})\right)\otimes X = 0.
\]
This holds because the following equations hold for each category $k\in [K]$
\[
\begin{aligned}
    &\E_{(X,Y)\sim \PP^{(l)}}[\omega^{(l)}(X) \cdot {\bf 1}(Y=k) \cdot X] \\
    &= \E_{X\sim \PP^{(l)}_X}[\omega^{(l)}(X) \cdot \E_{Y\sim \PP^{(l)}_{Y|X}}\left[{\bf 1}(Y=k)|X\right] \cdot X] =\E_{X\sim \PP^{(l)}_X}[\omega^{(l)}(X)\cdot f^{(l)}_k(X) \cdot X].
\end{aligned}
\]
Therefore, $\Dc_2^{(l)}$ admits the following expression:
{\small
\[
\Dc_2^{(l)}=\frac{1}{n_l}\sum_{i=1}^{n_l}\left(\w{l}(\X{l}_i)\cdot(\f{l}(\X{l}_i) - {\bf y}_i^{(l)})\right)\otimes \X{l}_i - \E_{\PP^{(l)}}\left(\omega^{(l)}(X) \cdot( \f{l}(X)-{\bf y})\right)\otimes X,
\]}
Considering that $\f{l}_k(X) - {\bf 1}(Y=k) \in [-1,1]$, and $\omega^{(l)}(\cdot)$ is bounded in Condition {\rm (B1)}, we have
\[
\max_{k=1,...,K}\E_{\PP^{(l)}}\left[\omega^{(l)}(X)\cdot (\f{l}_k(X) - {\bf 1}(Y=k))\right]^2\leq C,
\]
for some constant $C>0$. Then we apply Lemma \ref{lemma: subgaus of f(X)X} to establish that: with a probability of at least $1-e^{-c_1t^2}$,
\begin{equation}
    \|\Dc_2^{(l)}\|_2 \lesssim t\sqrt{d/n_l},
    \label{eq: rate D2}
\end{equation}
where $t$ is any value larger than $c_0$.

The following lemma controls the remainder terms ${\rm Rem}_{j=1,...,4}^{(l)}$, and its proof is postponed in Section \ref{Proof of Lemma remainder terms of mu}.
\begin{Lemma}
    Suppose Conditions {\rm (A2), (B1), (B2)} hold. Then there exist positive constants $c_0,c_1>0$ such that, for any $t \geq c_0$, the following inequality holds with probability at least $1-e^{-c_1t^2}-\delta_n$:
    \begin{equation}
        \|\sum_{j=1}^4 {\rm Rem}_{j}^{(l)}\|_2 \lesssim t\cdot (\eta_\omega + \eta_f)\sqrt{d/n_l} + t\cdot\eta_\omega \eta_f \sqrt{d},
        \label{eq: rate remainder terms of mu}
    \end{equation}
    where $\delta_n\to0$ is the sequence specified in Condition {\rm (B2)}.
    \label{lemma: remainder terms of mu}
\end{Lemma}

Combining the error decomposition \eqref{eq: estimation error decomposition}, the control of $\Dc_1^{(l)}$, $\Dc_2^{(l)}$ in \eqref{eq: rate D1},\eqref{eq: rate D2}, respectively, and Lemma \ref{lemma: remainder terms of mu}, we conclude that: with probability at least $1-e^{-c_1t^2}-\delta_n$,
\[
\begin{aligned}
    \|\widehat{\mu}^{(l)}-\mu^{(l)}\|_2
    &\leq \|\Dc^{(l)}_1\|_2 + \|\Dc^{(l)}\|_2 + \|\sum_{j=1}^4 {\rm Rem}_j^{(l)}\|_2 \\
    &\lesssim t\left((1+\eta_\omega +\eta_f)\sqrt{d/n_l} + \eta_\omega\eta_f \sqrt{d}\right),
\end{aligned}
\]
where $t$ is any value larger than $c_0$.
By taking the maximum over $l=1,...,L$,
we complete the proof of equation \eqref{eq: rates of mu}.

\subsubsection{Proof of \eqref{eq: rates of cov_mu}}
It follows from the expressions of $\widehat{\bf V}_\mu^{(l)}$ and ${\bf V}_\mu^{(l)}$ that
{\small
\begin{equation}
    \begin{aligned}
    &\widehat{\bf V}_\mu^{(l)} - {\bf V}_\mu^{(l)} = {\rm Term}_1 + {\rm Term}_2,\quad \textrm{with}\\ 
    &{\rm Term}_1=  \frac{1}{n_l}\left\{\widehat{\bf V}_{\PP^{(l)}}\left(\left[\widehat{\omega}^{(l)}(X)\cdot(\widehat{f}^{(l)}(X)-{\bf y})\right]\otimes X\right)-{\bf V}_{\PP^{(l)}}\left(\left[{\omega}^{(l)}(X)\cdot({f}^{(l)}(X)-{\bf y})\right]\otimes X\right) \right\} \\
    &{\rm Term}_2= \frac{1}{N}\left\{\widehat{\bf V}_{\QQ}\left[\widehat{f}^{(l)}(X) \otimes X\right] - {\bf V}_{\QQ}\left[{f}^{(l)}(X) \otimes X\right]\right\} 
\end{aligned}
\label{eq: V_mu variance diff decompose}
\end{equation}
}
For the ${\rm Term}_1$ in \eqref{eq: V_mu variance diff decompose}, it can be further decomposed into
{\small
\[
\begin{aligned}
    &{\rm Term}_1 = {\rm Term}_{1,a} + {\rm Term}_{1, b},\quad \textrm{with} \\
    &{\rm Term}_{1,a}= \frac{1}{n_l}\left\{\widehat{\bf V}_{\PP^{(l)}}\left(\left[\widehat{\omega}^{(l)}(X)\cdot(\widehat{f}^{(l)}(X)-{\bf y})\right]\otimes X\right)-{\bf V}_{\PP^{(l)}}\left(\left[\widehat{\omega}^{(l)}(X)\cdot(\widehat{f}^{(l)}(X)-{\bf y})\right]\otimes X\right) \right\} \\
    &{\rm Term}_{1,b}= \frac{1}{n_l}\left\{{\bf V}_{\PP^{(l)}}\left(\left[\widehat{\omega}^{(l)}(X)\cdot(\widehat{f}^{(l)}(X)-{\bf y})\right]\otimes X\right)-{\bf V}_{\PP^{(l)}}\left(\left[{\omega}^{(l)}(X)\cdot({f}^{(l)}(X)-{\bf y})\right]\otimes X\right) \right\}
\end{aligned}
\]
}
For the ${\rm Term}_{1,a}$, as we construct $\hw{l},\hf{l}$ in the cross-fitting manner, which is independent from the data,
we leverage the standard concentration inequality of the empirical covariance matrix to show that: with probability at least $1-n_l^{-c_1d}$
\[
\|n_l \cdot {\rm Term}_{1,a}\|_2 \leq C\sqrt{\frac{d\log n_l}{n_l}}.
\]

We now investigate the term ${\rm Term}_{1,b}$. For convenience, we denote the inner vectors as:
\[
\widehat{Z} = \left[\widehat{\omega}^{(l)}(X)\cdot(\widehat{f}^{(l)}(X)-{\bf y})\right]\otimes X \in \RR^{dK},\quad Z=\left[{\omega}^{(l)}(X)\cdot({f}^{(l)}(X)-{\bf y})\right]\otimes X \in \RR^{dK},
\]
and write their $k$-th block, for $1\leq k\leq K$, as follows:
\[
\widehat{Z}_{(k)} = \left[\widehat{\omega}^{(l)}(X)\cdot(\widehat{f}_k^{(l)}(X)-{\bf 1}(Y=k))\right] X \in \RR^{d}, 
\]
and
\[
Z_{(k)}=\left[{\omega}^{(l)}(X)\cdot({f}^{(l)}_k(X)-{\bf 1}(Y=k))\right] X \in \RR^{d}.
\]
For their difference, we write
\[
\Delta Z = \widehat{Z} - Z \in \RR^{dK}, \quad \Delta Z_{(k)} = \widehat{Z}_{(k)} - Z_{(k)}.
\]

For each $k$, we have
\[
\Delta Z_{(k)} = \w{l}\cdot \left[\left(\frac{\hw{l}}{\w{l}}-1\right)\left(\f{l}_k - {\bf 1}(Y=k)\right) + \left(\hf{l}_k - \f{l}_k\right) + \left(\frac{\hw{l} }{\w{l}}-1\right)\left(\hf{l}_k-\f{l}_k\right)\right]\cdot X.
\]
As $\f{l}_k(\cdot), \w{l}(\cdot)$ are bounded, we have
\[
\|\Delta Z_{(k)}\|_2 \lesssim \|X\|_2 \left[|\frac{\hw{l}}{\w{l}}-1| + |\hf{l}_k-\f{l}_k| + |\frac{\hw{l}}{\w{l}}-1|\cdot |\hf{l}_k-\f{l}_k|\right].
\]
Taking the expectation on the distribution $\PP^{(l)}_X$, we obtain that:
\[
\E[\|\Delta Z_{(k)}\|_2^2] \lesssim  \E\left\{\|X\|_2^2 \left[|\frac{\hw{l}}{\w{l}}-1| + |\hf{l}_k-\f{l}_k| + |\frac{\hw{l}}{\w{l}}-1|\cdot |\hf{l}_k-\f{l}_k|\right]^2\right\}.
\]
By the Cauchy-Schwartz inequality, we further establish that
\[
\E[\|\Delta Z_{(k)}\|_2^2] \lesssim \sqrt{\E[\|X\|_2^4]}\cdot \sqrt{\E\left[|\frac{\hw{l}}{\w{l}}-1| + |\hf{l}_k-\f{l}_k| + |\frac{\hw{l}}{\w{l}}-1|\cdot |\hf{l}_k-\f{l}_k|\right]^4}
\]
Therefore, it follows that
{\small
\[
\begin{aligned}
   \E\left[\|\Delta Z_{(k)}\|_2^2\right]^{1/2} &\lesssim \E[\|X\|_2^4]^{1/4}\left[\|\frac{\hw{l}}{\w{l}}-1\|_{\ell_4} + \|\hf{l}_k-\f{l}_k\|_{\ell_4} + \|\frac{\hw{l}}{\w{l}}-1\|_{\ell_4}\cdot \|\hf{l}_k-\f{l}_k\|_{\ell_4}\right] \\
   &\lesssim \E[\|X\|_2^4]^{1/4}\left[\|\frac{\hw{l}}{\w{l}}-1\|_{\ell_2} + \|\hf{l}_k-\f{l}_k\|_{\ell_2} + \|\frac{\hw{l}}{\w{l}}-1\|_{\ell_2}\cdot \|\hf{l}_k-\f{l}_k\|_{\ell_2}\right],
\end{aligned}
\]
}
where the second inequality holds since the $\ell_p$ norms differ only by constants for subguassian random variables. 
Moreover, as the source covariate is subgaussian with $\E[\|X\|_2^4]^{1/4}\asymp \sqrt{d}$, and together with Condition {\rm (B2)}, the following holds: with probability $1-\delta_n$,
\begin{equation}
    \E\left[\|\Delta Z\|_2^2\right]^{1/2}\lesssim \sqrt{d}(\eta_\omega+\eta_f+\eta_\omega\eta_f).
    \label{eq: term1b}
\end{equation}
For the term ${\rm Term}_{1,b}$, by the bilinearity property of covariance operator, we have
{
\[
\begin{aligned}
    n_l {\rm Term}_{1,b} &= {\bf V}_{\PP^{(l)}}(\widehat{Z}) - {\bf V}_{\PP^{(l)}}(Z)= {\bf V}_{\PP^{(l)}}(\Delta Z) + 2{\bf V}_{\PP^{(l)}}(Z, \Delta Z).
\end{aligned}
\]}
Therefore, we have
\[
\|n_l {\rm Term}_{1,b}\|_2 \leq \E\left[\|\Delta Z\|_2^2\right] + 2\E\left[\|Z\|_2^2\right]^{1/2} \E\left[\|\Delta Z\|_2^2\right]^{1/2}.
\]
Note that $\E \|Z\|_2^2\asymp d$ due to the boundedness of $\w{l},\f{l}$ and subgaussianity of $\X{l}$. Together with \eqref{eq: term1b}, we show that with probability at least $1-\delta_n$,
\[
\|n_l {\rm Term}_{1,b}\|_2 \lesssim  d\cdot \left[(\eta_\omega+\eta_f + \eta_\omega\eta_f)+ (\eta_\omega+\eta_f + \eta_\omega\eta_f)^2\right].
\]

For the ${\rm Term}_2$ of \eqref{eq: V_mu variance diff decompose}, we shall decompose
{\small
\[
\begin{aligned}
    {\rm Term_2} = {\rm Term}_{2,a} + {\rm Term}_{2,b} &= \frac{1}{N}\left\{\widehat{\bf V}_{\QQ}\left[\widehat{f}^{(l)}(X) \otimes X\right] - {\bf V}_{\QQ}\left[\widehat{f}^{(l)}(X) \otimes X\right]\right\} \\
    &\quad\quad + \frac{1}{N}\left\{{\bf V}_{\QQ}\left[\widehat{f}^{(l)}(X) \otimes X\right] - {\bf V}_{\QQ}\left[{f}^{(l)}(X) \otimes X\right]\right\}.
\end{aligned}
\]}
Similarly, by the standard concentration inequality of the covariance matrix and Condition {\rm (B2)}, we have with probability at least $1-N^{-c_1d}-\delta_n$:
\[
\|N\cdot {\rm Term}_{2,a}\|_2 \lesssim\frac{\sqrt{d\log N}}{N} ,\quad\textrm{and}\quad
\|N\cdot {\rm Term}_{2,b}\|_2 \lesssim d(\eta_f+\eta_f^2).
\]

We put the above results back to \eqref{eq: V_mu variance diff decompose} and establish that: with probability at least $1-n^{-c_1d}-N^{-c_1d}-\delta_n:$
{\small
\[
\begin{aligned}
   &\|n_l (\widehat{\bf V}_\mu^{(l)} - {\bf V}_\mu^{(l)})\|_2\\
   &\leq \|n_l\cdot {\rm Term}_{1,a}\|_2+\|n_l\cdot {\rm Term}_{1,b}\|_2+\frac{n_l}{N}\|N\cdot {\rm Term}_{2,a}\|_2+\frac{n_l}{N}\|N\cdot {\rm Term}_{2,a}\|_2\\
   &\lesssim\sqrt{\frac{d\log n_l}{n_l}} + d\left[(\eta_\omega+\eta_f + \eta_\omega\eta_f)+ (\eta_\omega+\eta_f + \eta_\omega\eta_f)^2\right]+ \frac{n_l}{N}\sqrt{\frac{d\log N}{N}} + \frac{n_l}{N} d(\eta_f + \eta_f^2). 
\end{aligned}
\]
}
Considering that $N> n_l$ and $n=\min_l n_l$,  we have with probability at least $1-n^{-c_1d}-\delta_n:$
\begin{equation}
    \|n_l (\widehat{\bf V}_\mu^{(l)} - {\bf V}_\mu^{(l)})\|_2\lesssim \sqrt{\frac{d\log n}{n}} + d\left[(\eta_\omega+\eta_f + \eta_\omega\eta_f)+ (\eta_\omega+\eta_f + \eta_\omega\eta_f)^2\right].
    \label{eq: V_mu hat distance}
\end{equation}
Lastly, taking the maximum over $l\in [L]$, we complete the proof of \eqref{eq: rates of cov_mu}.

\subsubsection{Proof of \eqref{eq: limiting dist of mu}}
We directly apply \eqref{eq: rates of cov_mu} to show that: as $n\to\infty$,
\[
\|n_l \cdot (\widehat{\bf V}^{(l)}_\mu -{\bf V}^{(l)}_\mu)\|_2 \xlongrightarrow{p} 0.
\]
For the limiting distribution of $\widehat{\mu}^{(l)} - \mu^{(l)}$, 
it follows its decomposition in \eqref{eq: estimation error decomposition} that
\begin{equation}
    \sqrt{n_l}(\widehat{\mu}^{(l)}-\mu^{(l)}) = \sqrt{n_l} \Dc_1^{(l)} + \sqrt{n_l}\Dc_2^{(l)} + \sqrt{n_l} \cdot \sum_{j=1}^4 {\rm Rem}_j^{(l)}.
    \label{eq: limiting dist of mu - eq 1}
\end{equation}
In the following, we establish the limiting distribution for the terms on the right-hand side of the above equation.

In the fixed dimension regime, by classic CLT, we have
\begin{equation*}
    \sqrt{N}\Dc_1^{(l)} \xlongrightarrow{d} \Nc \left(0, {\bf V}_{\QQ}[\f{l}(X)\otimes X]\right),
    \label{eq: limit_dist D1}
\end{equation*}
and
\begin{equation*}
\sqrt{n_l}\Dc_2^{(l)} \xlongrightarrow{d} \mathcal{N}\left(0, {\bf V}_{\PP^{(l)}}\left(\left[{\omega}^{(l)}(X)\cdot ({f}^{(l)}(X)-{\bf y})\right]\otimes X\right)\right).
    \label{eq: limit_dist D2}
\end{equation*}
Note that $\Dc_1^{(l)},\Dc_2^{(l)}$ are independent, as defined in \eqref{eq: D1D2}. By Slutsky' theorem, we have
\begin{equation}
    \sqrt{n_l} \Dc_1 + \sqrt{n_l}\Dc_2 \xlongrightarrow{d} \mathcal{N}\left(0, {\bf V}_{\PP^{(l)}}\left(\left[{\omega}^{(l)}(X)\cdot ({f}^{(l)}(X)-{\bf y})\right]\otimes X\right) + \frac{n_l}{N} {\bf V}_{\QQ}[\f{l}(X)\otimes X]\right).
    \label{eq: limiting dist of mu - eq 2}
\end{equation}

Next, we utilize Lemma \ref{lemma: remainder terms of mu} to establish the asymptotic properties of remainder terms.
It follows from the conditions $\eta_f=o(1)$, $\eta_\omega=o(1)$ and $\eta_f\cdot \eta_\omega=o(n^{-1/2})$, that with probability at least $1-e^{-c_1t^2} - \delta_n$,
\[
\|\sqrt{n_l} \cdot \sum_{j=1}^4 {\rm Rem}_j^{(l)}\|_2 =o(1) \cdot t,
\]
where $\delta_n\to 0$ is a vanishing sequence with growing $n$. We take $t=\min(\eta_f, \eta_\omega)$ to obtain that: with $n\to\infty$,
\[
\sqrt{n_l}\cdot \sum_{j=1}^4 {\rm Rem}_j^{(l)} =  o_p(1).
\]
Together with \eqref{eq: limiting dist of mu - eq 1} and \eqref{eq: limiting dist of mu - eq 2}, it follows from Slutsky' theorem that
\[
\sqrt{n_l}(\widehat{\mu}^{(l)}-\mu^{(l)})\xlongrightarrow{d} \mathcal{N}\left(0, {\bf V}_{\PP^{(l)}}\left(\left[{\omega}^{(l)}(X)\cdot ({f}^{(l)}(X)-{\bf y})\right]\otimes X\right) + \frac{n_l}{N} {\bf V}_{\QQ}[\f{l}(X)\otimes X]\right).
\]
Applying the expression of ${\bf V}_\mu^{(l)}$ in \eqref{eq: V_mu expression - cs}, we show that:
\[
\left[n_l{\bf V}^{(l)}_\mu\right]^{-1} \left[\sqrt{n}_l(\widehat{\mu}^{(l)} - \mu^{(l)})\right]\xlongrightarrow{d} \mathcal{N}(0, {\bf I}_{dK}).
\]

\subsection{Proof of Proposition \ref{prop: double robustness - nocs}}
\label{proof of prop: doubly robust estimator - nocs}
In the no covariate shift setting, as $\PP^{(l)}_X = \QQ_X$, we have
\[
\mu^{(l)}_c = -\E_{X\sim \QQ_X}\E^{(l)}_{Y|X}({\bf 1}(Y=c)\cdot X) = -\E_{(X,Y)\sim \PP^{(l)}}({\bf 1}(Y=c)\cdot X).
\]
Then the estimation error of the proposed estimator $\widehat{\mu}^{(l)}$ is expressed as:
\[
\begin{aligned}
    \widehat{\mu}^{(l)} - \mu^{(l)} = -\frac{1}{n_l}\sum_{i=1}^{n_l} {\bf y}^{(l)}_i\otimes \X{l}_i + \E_{\PP^{(l)}}[{\bf y}\otimes X].
\end{aligned}
\]
We apply Lemma \ref{lemma: subgaus of f(X)X} to establish that with probability at least $1-e^{-ct^2}$ to complete the proof of equation \eqref{eq: rates of mu - nocs}.
Moreover, we leverage the standard concentration inequality of the empirical covariance matrix to show that: with probability at least $1-n_l^{-c_1d}$,
\[
\left\|n_l \cdot (\widehat{\bf V}_\mu^{(l)} - {\bf V}_\mu^{(l)})\right\|_2 \leq \sqrt{d\log n_l/n_l}.
\]
Taking the maximum over $l=1,...,L$ on both sides, we then establish that: with probability at least $1-n^{-c_1d},$
\[
\max_{l=1,...,L}\left\|n_l \cdot (\widehat{\bf V}_\mu^{(l)} - {\bf V}_\mu^{(l)})\right\|_2 \leq \sqrt{d\log n/n}.
\]
Consequently, we conclude that $n\to \infty$
\[
\left\|n_l \cdot (\widehat{\bf V}_\mu^{(l)} - {\bf V}_\mu^{(l)})\right\|_2\xlongrightarrow{p} 0.
\]
And classic CLT implies that
\[
\left[n{\bf V}^{(l)}_\mu\right]^{-1} [\sqrt{n_l}\cdot (\widehat{\mu}^{(l)} - \mu^{(l)})] \xlongrightarrow{d} \mathcal{N}(0, {\bf I}_{dK}).
\]

\section{Proof of Lemmas}
\label{sec: lemma proof}
This section contains the proofs of supporting lemmas.
We begin with the following lemma to facilitate the multiplication of the Kronecker product. For a matrix $V\in \RR^{d\times K}$, we define ${\rm vec}(V)$ as the column vector that stacks the columns of $V$. 
\begin{Lemma}
Consider the matrices $V\in \mathbb{R}^{d\times K}$, $A\in \RR^{K\times p}$ and the vectors $X\in \RR^d$ and $v={\rm vec}(V)\in \RR^{dK}$. We have
\begin{equation}
v^{\intercal}\left(A\otimes XX^{\intercal}\right)v= {\rm Tr}(V^{\intercal}XX^{\intercal} V A)
\label{eq: key 1},
\end{equation}
and
\begin{equation}
\left|v^{\intercal}\left(A\otimes XX^{\intercal}\right)v\right|\leq \sigma_{1}(A)\cdot{\rm Tr}(V^{\intercal}XX^{\intercal} V),
\label{eq: key 2}
\end{equation}
where $\sigma_1(A)$ denotes the largest singular value of $A$.
\label{lem: basic lemma}
\end{Lemma}

\subsection{Proof of Lemma \ref{lem: convexity of H}}
\label{proof of lem conveixty of H}
We first show that $H(\theta)\succ 0$ for any $\theta\in \RR^{dK}$ under Condition {\rm (A2)}. 
For any vector $v\in \RR^{dK}$, we partition it into $K$ subvectors $v = (v_1^\intercal, v_2^\intercal,..., v_K^\intercal)^\intercal$ with each $v_k\in \RR^{d}$. Then
\[
\begin{aligned}
    v^\intercal H(\theta) v &= \E_{\QQ_X}\left[\sum_{k=1}^K \left[p_k(X,\theta) \cdot (X^\intercal v_k)^2\right] - \left(\sum_{k=1}^K p_k(X,\theta) \cdot (X^\intercal v_k)\right)^2\right]\\
    &= \E_{\QQ_X}\left[{\rm Var}_{p(X,\theta)}(X^\intercal v_{k})\right].
\end{aligned}
\]
The variance is non-negative and equals zero if and only if $X^\intercal v_1 = X^\intercal v_2 = ... = X^\intercal v_K$ almost surely under $p(X,\theta).$
With $v\neq 0$, for $X^\intercal v_1 = X^\intercal v_2 = ... = X^\intercal v_K$ to hold almost surely, we consider two scenarios:
\begin{itemize}
    \item If $v_1=v_2=...=v_K=a$, then $v^\intercal H(\theta)v = \E_{\QQ_X}[(a^\intercal X) (1-p_0)p_0]$, with $p_0 = 1-\sum_{k=1}^K p_k(X,\theta)$. Since $(1-p_0)p_0>0$ for any $X\in \RR^{d}$ and $\theta\in \RR^{dK}$, and $\E_{\QQ_X}[XX^\intercal]\succ 0$ as assumed in Condition {\rm (A2)}, we have $v^\intercal H(\theta) v>0$.
    \item If $v_i\neq v_j$ for some $i,j$, then $(v_i - v_j)^\intercal X\neq 0$ with positive probability. This ensures that ${\rm Var}_{p(X,\theta)}(\cdot)>0$ with positive measure.
\end{itemize}
Therefore, combining the two scenarios, we find that $v^\intercal H(\theta)v>0$ for all $v\neq 0$, implying that $H(\theta)\succ 0$ for all $\theta\in \RR^{dK}$.

Next, we move forward considering $H(\theta)$'s property for $\theta\in \Theta_B$.
By substituting $A=D(X,\theta)$ in \eqref{eq: key 1}, we obtain that for any vector $v\in \RR^{dK}$,
\begin{equation}
    v^{\intercal}H(\theta)v=\E_{\QQ_X}{\rm Tr}(V^{\intercal}XX^{\intercal} V D(X,\theta)),
    \label{eq: vHv}
\end{equation}
where the matrix $V\in \RR^{d\times K}$ satisfies that ${\rm vec}(V) = v$. By the von Neumann trace inequality, we have
{\small
\begin{equation}
\sum_{k=1}^K \sigma_k(V^\intercal XX^\intercal V)\cdot \sigma_{K-k+1}(D(X,\theta))\leq {\rm Tr}(V^{\intercal}XX^{\intercal} V D(X,\theta))\leq \sum_{k=1}^{K}\sigma_k(V^\intercal XX^\intercal V)\cdot \sigma_k(D(X,\theta)),
    \label{eq: von trace ineq}
\end{equation}
}
where $\sigma_k(\cdot)$ denotes the $k$-th largest singular value of the matrix. 
In the following, we complete the proof in two steps:
\begin{enumerate}
    \item Show that under Condition {\rm (A2)}, the matrix $c_1\cdot {\bf 1}\preceq D(X,\theta)\preceq c_2\cdot {\bf I}$ for any $X\sim \QQ_X$ and $\theta\in \Theta_B$, with some positive constant $c_1,c_2>0$.
    \item Together with \eqref{eq: von trace ineq}, we prove that $H(\theta)$'s property for $\theta\in \Theta_B$.
\end{enumerate}

Under Condition {\rm (A2)}, for the target covariate $X\sim \QQ_X$, and parameter $\theta\in \Theta_B$, we have $$|X^\intercal \theta| \leq \|X\|_\infty \|\theta\|_1 \leq \|X\|_\infty (\|\theta^*\|+1),$$ which is bounded. Therefore, there exists some constant $c'>0$ satisfying both $p_k(X,\theta)\in [c', 1-c']$ for all $k=1,..., K$ and $1-\sum_{k=1}^K p_k(X,\theta)\in [c', 1-c']$.

For any vector $u\in \RR^{K}$, we have
\[
u^\intercal D(X,\theta) u = \sum_{k=1}^K u_k^2p_k(X,\theta) - \left(\sum_{k=1}^K u_k p_k(X,\theta)\right)^2.
\]
By Cauchy-Schwarz inequality,
$$\left(\sum_{k=1}^{K}u_k p_k(X,\theta)\right)^2\leq \left(\sum_{k=1}^{K}u^2_k p_k(X,\theta)\right)\left(\sum_{k=1}^{K} p_k(X,\theta)\right).$$
Then we apply $1-\sum_{k=1}^{K} p_k(X,\theta)\geq c'$ and obtain  
\begin{equation}
u^{\intercal}D(X,\theta)u\geq \left(\sum_{k=1}^{K}u^2_k p_k(X,\theta)\right) c' \geq c'^2 \|u\|_2^2,
\label{eq: part lower}
\end{equation}
where the second inequality holds as $p_k(X,\theta)\geq c'$ for all $k\in [K].$
Moreover, it is evident that
\begin{equation}
    u^\intercal D(X,\theta) u \leq \sum_{k=1}^K u_k^2 p_k(X,\theta)\leq (1-c')\|u\|_2^2.
    \label{eq: part upper}
\end{equation}
Combining \eqref{eq: part lower} and \eqref{eq: part upper}, we establish that $\sigma_{\min}\left(D(X,\theta)\right)\geq c'^2$ and $\sigma_{\max}(D(X,\theta))\leq 1-c'.$

Together with \eqref{eq: von trace ineq}, we further show that
\[
c'^2\cdot {\rm Tr}(V^\intercal X X^\intercal V) \leq {\rm Tr}(V^\intercal XX^\intercal V D(X,\theta)) \leq (1-c')\cdot {\rm Tr}(V^\intercal XX^\intercal V).
\]
Then it follows from \eqref{eq: vHv} that
\[
c'^2 \E_{\QQ_X}{\rm Tr}(V^\intercal X X^\intercal V) \leq v^\intercal H(\theta) v\leq (1-c')\cdot \E_{\QQ_X}{\rm Tr}(V^\intercal XX^\intercal V).
\]
Notice that 
\[
\kappa_1 \|v\|^2_2 \leq \E_{\QQ_X}{\rm Tr}(V^\intercal X X^\intercal V) = \sum_{k=1}^K V_k^\intercal \E_{\QQ_X}[XX^\intercal] V_k \leq \kappa_2 \|v\|_2^2,
\]
where $V_k$ is the $k$-th column of the matrix $V$ that satisfies that ${\rm vec}(V) = v$, and $\kappa_1,\kappa_2$ are positive constants specified in Condition {\rm (A2)}. Combining the above two results, we establish that
\[
c'^2 \kappa_1 \|v\|^2\leq v^\intercal H(\theta) v\leq (1-c') \kappa_2 \|v\|^2,
\]
for all $v\in \RR^{dK}$.

\subsection{Proof of Lemma \ref{lem: key property}}
\label{proof of lem key property}

\subsubsection{Proof of \eqref{eq: Lip D}} Note that 
{\small
\begin{equation*}
\begin{aligned}
D(X,\theta)-D(X,\theta^{\prime})&={\rm diag}(p(X,\theta))-{\rm diag}(p(X,\theta^{\prime}))+p(X,\theta^{\prime})p^{\intercal}(X,\theta^{\prime})-p(X,\theta)p^{\intercal}(X,\theta)\\
&={\rm diag}(p(X,\theta))-{\rm diag}(p(X,\theta^{\prime}))+2(p(X,\theta^{\prime})-p(X,\theta))p^{\intercal}(X,\theta)\\
&\quad\quad +(p(X,\theta^{\prime})-p(X,\theta))(p(X,\theta^{\prime})-p(X,\theta))^{\intercal}.
\end{aligned}
\end{equation*}
}
Due to the definition of $p(X,\theta)$ in \eqref{eq: p def}, $\|p(X,\theta)\|_2\leq 1$  for any $\theta\in \RR^{dK}$, thus,
\begin{equation}
\left\|D(X,\theta)-D(X,\theta^{\prime})\right\|_2\leq 3\|p(X,\theta)-p(X,\theta^{\prime})\|_2+\|p(X,\theta)-p(X,\theta^{\prime})\|_2^2.
\label{eq: diff bound}
\end{equation}
Note that the Jacobian of $p(X,\theta)$ with respect to $\theta$ is $D(X,\theta)({\bf I} \otimes X^\intercal) \in \RR^{K\times dK}$, then we can write  $p(X,\theta)-p(X,\theta^{\prime})$ in the integral form as follows: 
\[
p(X,\theta)-p(X,\theta^{\prime}) = \int_0^1 D(X,\theta' + \nu(\theta-\theta')) \begin{pmatrix} X^{\intercal}(\theta_1-\theta^{\prime}_1)\\\vdots \\ X^{\intercal}(\theta_K-\theta^{\prime}_K)\end{pmatrix} d\nu.
\]
% for some $t\in [0, 1]$,
% \begin{equation}
% p(X,\theta)-p(X,\theta^{\prime})=D(X,\theta+t(\theta^{\prime}-\theta))\begin{pmatrix} X^{\intercal}(\theta_1-\theta^{\prime}_1)\\\vdots \\ X^{\intercal}(\theta_K-\theta^{\prime}_K)\end{pmatrix}.
% \end{equation}
Since $D(X,\theta) = {\rm diag}(p(X,\theta)) - p(X,\theta) p^\intercal(X,\theta)$, we have $D(X,\theta) \preceq {\rm diag}(p(X,\theta))$, which implies that
\[
\|D(X,\theta)\|_2 = \lambda_{\rm max}(D)\leq \lambda_{\rm max}({\rm diag}(p(X,\theta))) = \max_{j} [p(X,\theta)]_j \leq 1.
\]
Combining the above two equations yields
\begin{equation}
    \|p(X,\theta)-p(X,\theta^{\prime})\|_2\leq  \Delta(X,\theta-\theta^{\prime}) \quad \text{with}\quad \Delta(X,\theta-\theta^{\prime})=\sqrt{\sum_{k=1}^{K}\left[X^{\intercal}(\theta_k-\theta^{\prime}_k)\right]^2}.
    \label{eq: p theta diff}
\end{equation}
We establish \eqref{eq: Lip D} by combining the above inequality and \eqref{eq: diff bound}. 

\subsubsection{Proof of \eqref{eq: Lip H}} Since $H(\theta^{\prime})-H(\theta)$ is a symmetric matrix, we have   
\begin{equation*}
\begin{aligned}
\|H(\theta^{\prime})-H(\theta)\|_2&=\max_{\|v\|_2\leq 1} \left|v^{\intercal} \left(H(\theta^{\prime})-H(\theta)\right)v\right|\\
&=\max_{\|v\|_2\leq 1} \left|v^{\intercal} \left[\E_{X\sim \QQ_{X}}\left(D(X,\theta^{\prime})-D(X,\theta)\right)\otimes XX^{\intercal}\right]v\right| \\
&= \max_{{\rm vec}(V)=v, \|v\|_2\leq 1}\left|\E_{X\sim \QQ_X}{\rm Tr}(V^\intercal XX^\intercal V ((D(X,\theta') - D(X,\theta))\right|.
\end{aligned}
\end{equation*}
By combining the above inequality and \eqref{eq: key 2}, we establish that, for ${\rm vec}(V)=v,$ 
\begin{equation}
\begin{aligned}
&\|H(\theta^{\prime})-H(\theta)\|_2\\
&\leq \E_{X\sim \QQ_{X}} \left[\sigma_1\left(D(X,\theta^{\prime})-D(X,\theta)\right)\cdot{\rm Tr}(V^{\intercal}XX^{\intercal} V)\right]\\
&\lesssim \E_{X\sim \QQ_{X}} \left\{\left[\Delta(X,\theta^{\prime}-\theta)+\Delta^2(X,\theta^{\prime}-\theta)\right]\cdot{\rm Tr}(V^{\intercal}XX^{\intercal} V)\right\}\\
&\lesssim \sqrt{\E_{X\sim \QQ_{X}} \left[\Delta^2(X,\theta^{\prime}-\theta)+\Delta^4(X,\theta^{\prime}-\theta)\right]} \cdot \sqrt{\E_{X\sim \QQ_{X}}[{\rm Tr}(V^{\intercal}XX^{\intercal} V)]^2},
\end{aligned}
\label{eq: decomposition}
\end{equation}
where the second inequality follows from \eqref{eq: Lip D} and the third inequality follows from the Cauchy-Schwarz inequality.
Next, we continue upper bounding the terms in the most right-hand side of the above inequality.

We apply the fact that $X\sim \QQ_X$ is a Sub-Gaussian random vector and establish
\begin{equation*}
\frac{1}{\|\theta^{\prime}-\theta\|_2^2}\E_{X\sim \QQ_{X}} \Delta^2(X,\theta^{\prime}-\theta)=\E_{X\sim \QQ_{X}} {\sum_{k=1}^{K}\left[X^{\intercal}\frac{\theta_k-\theta^{\prime}_k}{\|\theta^{\prime}-\theta\|_2}\right]^2}\leq C K,
\label{eq: temp 1}
\end{equation*}
\begin{equation*}
\frac{1}{\|\theta^{\prime}-\theta\|_2^4}\E_{X\sim \QQ_{X}} \Delta^4(X,\theta^{\prime}-\theta)\lesssim \E_{X\sim \QQ_{X}}  {\sum_{k=1}^{K}\left[X^{\intercal}\frac{\theta_k-\theta^{\prime}_k}{\|\theta^{\prime}-\theta\|_2}\right]^4}\leq C K.
\end{equation*}
Therefore,
\begin{equation}
    \sqrt{\E_{X\sim \QQ_{X}} \left[\Delta^2(X,\theta^{\prime}-\theta)+\Delta^4(X,\theta^{\prime}-\theta)\right]} \lesssim \|\theta^{\prime} - \theta\|_2 + \|\theta^{\prime} - \theta\|_2^2.
    \label{eq: temp 2}
\end{equation}
Since
$
{\rm Tr}(V^{\intercal}XX^{\intercal} V)=\sum_{k=1}^{K} (X^{\intercal} V_k)^2,
$
with $V_k$ denoting the $k$-th column of $V,$ we have  
\begin{equation}
\E_{X\sim \QQ_{X}}[{\rm Tr}(V^{\intercal}X_iX_i^{\intercal} V)]^2\lesssim \E_{X\sim \QQ_{X}}\sum_{k=1}^{K} (X^{\intercal} V_k)^4 \leq C,
\label{eq: moment bound}
\end{equation}
where $\E_{X\sim \QQ_{X}}(X^{\intercal} V_k)^4$ is bounded due to $X$ being sub-Gaussian vector and $\|V_k\|_2\leq 1$.
 
We establish \eqref{eq: Lip H} by plugging in the above \eqref{eq: temp 2} and \eqref{eq: moment bound} into the error bound \eqref{eq: decomposition}.

\subsection{Proof of Lemma \ref{lem: concentration lemma}}
\label{proof of lem concentration lemma}

We denote the $\epsilon$-net of the space $\Theta_{B}$ as $\mathcal{N}_{\epsilon}(\Theta_B)=\{\theta^{1},\cdots, \theta^{|\mathcal{N}_{\epsilon}|}\}$ such that, for any $\theta\in \Theta_{B}$, there exists $\theta^{\tau}\in \mathcal{N}_{\epsilon}(\Theta_B)$ such that $\|\theta-\theta^{\tau}\|_2\leq \epsilon.$ Since $
\Theta_{B}\subset \bar{\Theta}_{B}\coloneqq \{\theta\in \RR^{dK}: \|\theta-\theta^*\|_2\leq 1\},$ we have 
\begin{equation*}
|\mathcal{N}_{\epsilon}(\Theta_B)|\leq\left({C}/{\epsilon}\right)^{dK}.
\label{eq: epsilon-net 1}
\end{equation*}
Moreover, we denote the $\delta$-net of the unit ball $\mathbb{B}^{dK} = \{v\in \RR^{dK}: \|v\|_2\leq 1\}$ as $\mathcal{N}_\delta(\mathbb{B}^{dK})$, and the cardinality of the $\delta$-net is given by
\begin{equation*}
    |\mathcal{N}_\delta(\mathbb{B}^{dK})|\leq (C/\delta)^{dK}.
    \label{eq: unit ball net}
\end{equation*}

\noindent {\bf Proof of $\GG_1$ in \eqref{eq: events}.} For $X_i\sim \QQ_X$, recall that the expressions 
$S(\theta,X_i)=\log\left({1+\sum_{k=1}^{K}\exp(\theta_k^{\intercal}X_i)}\right)$, and
\begin{equation*}
\widehat{S}(\theta)=\frac{1}{N}\sum_{i=1}^{N}S(\theta,X_i) \quad \text{and}\quad S(\theta)=\E_{\QQ_X} S(\theta,X).
\end{equation*} Since $S(\theta,X_i)$ is bounded for $\theta\in \Theta_B$, by the union bound, we have 
\begin{equation}
\begin{aligned}
    \PP\left(\max_{\theta\in \mathcal{N}_{\epsilon}(\Theta_B)}|\widehat{S}(\theta)-S(\theta)|\leq t \right) &\geq 1-|\mathcal{N}_{\epsilon}(\Theta_B)|\cdot \exp(-cN t^2)\\
    &\geq 1- \exp\left(-c (Nt^2-dK\log (C/\epsilon))\right).
\end{aligned}
\label{eq: max inequality 1}
\end{equation}
For any $\theta\in \Theta_{B}$, there exists $\theta^{\tau}\in \mathcal{N}_{\epsilon}(\Theta_B)$ such that $\|\theta-\theta^{\tau}\|_2\leq \epsilon$ and 
\begin{equation*}
\begin{aligned}
&\left(\widehat{S}(\theta)-S(\theta)\right)-\left(\widehat{S}(\theta^{\tau})-S(\theta^{\tau})\right)\\
&=\frac{1}{N}\sum_{i=1}^{N}\left[S(\theta,X_i)-S(\theta^{\tau},X_i)-\E_{\QQ_X}\left(S(\theta,X)-S(\theta^{\tau},X)\right)\right].
\end{aligned}
\end{equation*}
By the mean value theorem, for some $\nu\in [0,1]$, we have 
$$\left|S(\theta,X_i)-S(\theta^{\tau},X_i)\right|=\left|\sum_{k=1}^{K} \frac{\exp\left(X_i^{\intercal}(\nu\theta+(1-\nu)\theta^{\tau})\right)}{1+\sum_{k=1}^{K} \exp\left(X_i^{\intercal}(\nu\theta+(1-\nu)\theta^{\tau})\right)}X_i^{\intercal}(\theta-\theta^{\tau})\right|\leq C \sqrt{d}\epsilon.$$
Therefore, we have
\begin{equation*}
\left|\left(\widehat{S}(\theta)-S(\theta)\right)-\left(\widehat{S}(\theta^{\tau})-S(\theta^{\tau})\right)\right|\leq C\sqrt{d} \epsilon.
\end{equation*}
Since the above inequality holds for any $\theta\in \Theta_B$, it further implies 
\begin{equation*}
\max_{\theta\in \Theta_{B}}\left|\left(\widehat{S}(\theta)-S(\theta)\right)\right|-\max_{\theta\in \mathcal{N}_{\epsilon}(\Theta_B)}\left|\left(\widehat{S}(\theta)-S(\theta)\right)\right|\leq C\sqrt{d} \epsilon.
\end{equation*}
Together with \eqref{eq: max inequality 1}, we choose $\epsilon=1/\sqrt{N}$ and $t=C\sqrt{d\log N/N}$ and obtain that
\begin{equation}
\PP\left(\max_{\theta\in \Theta_{B}}|\widehat{S}(\theta)-S(\theta)|\leq C\sqrt{d\log N/N} \right)\geq 1-N^{-c_1d}.
\label{eq: event-1 bound}
\end{equation}

\noindent {\bf Proof of $\GG_3$ in \eqref{eq: events}.} 
Given a $\delta$-net $\Nc_\delta(\mathbb{B}^{dK})$ of the unit ball $\mathbb{B}^{dK}$ in $\RR^{dK}$, for any $\theta\in \RR^{Kd}$, it is known that
\begin{equation*}
\|\widehat{H}({\theta})-{H}({\theta})\|_2 \leq\frac{1}{1-2\delta-\delta^2}  \max_{v\in \mathcal{N}_{\delta}(\mathbb{B}^{dK})}v^{\intercal}(\widehat{H}({\theta})-{H}({\theta}))v,
\label{eq: enet for spectral}
\end{equation*}
provided that $1-2\delta-\delta^2>0$.
Then, we establish that
\begin{equation}
 \max_{\theta\in \Theta_{B}}\|\widehat{H}({\theta})-{H}({\theta})\|_2 \leq\frac{1}{1-2\delta-\delta^2}  \max_{\theta\in \Theta_{B}}\max_{v\in \mathcal{N}_{\delta}}v^{\intercal}(\widehat{H}({\theta})-{H}({\theta}))v.
 \label{eq: max concentration-1}
\end{equation}

By the definitions of $\widehat{H}(\theta)$ and $H(\theta)$, we have
\begin{equation}
    v^{\intercal}(\widehat{H}({\theta})-{H}({\theta}))v= v^{\intercal}\frac{1}{N}\sum_{i=1}^{N}\left[D(X_i,\theta)\otimes X_iX_i^{\intercal}-\E_{X_i\sim \QQ_{X}}D(X_i,\theta)\otimes X_iX_i^{\intercal}\right]v.
    \label{eq: v (Hhat - H) v}
\end{equation}
For any $\theta\in \Theta_{B}$, there exists $\theta^{\tau}\in \mathcal{N}_{\epsilon}(\Theta_B)$ such that $\|\theta^{\tau}-\theta\|_2\leq \epsilon.$  For $v={\rm vec}(V)$ and $v\in \mathcal{N}_{\delta}$, we apply Lemma \ref{lem: basic lemma} and obtain
{\small
\begin{equation*}
\left|v^{\intercal}\left(D(X_i,\theta)\otimes X_iX_i^{\intercal}-D(X_i,\theta^{\tau})\otimes X_iX_i^{\intercal}\right)v\right|\leq \lambda_{\max}(\left(D(X_i,\theta)-D(X_i,\theta^{\tau})\right)\cdot {\rm Tr}(V^{\intercal}X_i X_i^{\intercal} V).
\end{equation*}
}
It follows from \eqref{eq: Lip D} that
{\small
\[
\begin{aligned}
    \|D(X_i,\theta) - D(X_i, \theta^\tau)\|_2&\leq 3 \Delta(X_i,{\theta} - \theta^\tau) + \Delta(X_i, \theta - \theta^\tau) \lesssim\|X_i\|_2 \|\theta - \theta^\tau\| \lesssim \sqrt{Kd}\epsilon,
\end{aligned}
\]}
provided that $\epsilon\leq 1.$ Moreover, as \eqref{eq: moment bound} implies that
${\rm Tr}(V^{\intercal}X_i X_i^{\intercal} V)\leq C$, we combine the preceding two equations to esablish that:
\[
\left|v^{\intercal}\left(D(X_i,\theta)\otimes X_iX_i^{\intercal}-D(X_i,\theta^{\tau})\otimes X_iX_i^{\intercal}\right)v\right| \lesssim d^{1/2}\epsilon,
\]
for any $v\in \mathcal{N}_\delta(\mathbb{B}^{dK})$. Together with \eqref{eq: v (Hhat - H) v}, we establish that
\begin{equation*}
\max_{v\in \mathcal{N}_{\delta}(\mathbb{B}^{dK})}\left|v^{\intercal}\left[\left(\widehat{H}({\theta})-{H}({\theta})\right)-\left(\widehat{H}({\theta}^{\tau})-{H}({\theta}^{\tau})\right)\right]v\right|\lesssim d^{1/2} \epsilon.
\end{equation*}
Therefore, for any $\theta\in \Theta_{B}$, there exists $\theta^{\tau}\in \mathcal{N}_{\epsilon}(\Theta_B)$ such that 
\begin{equation*}
\begin{aligned}
&\left|\max_{v\in \mathcal{N}_{\delta}(\mathbb{B}^{dK})} v^{\intercal}(\widehat{H}({\theta})-{H}({\theta}))v-\max_{v\in \mathcal{N}_{\delta}(\mathbb{B}^{dK})}v^{\intercal}(\widehat{H}({\theta^{\tau}})-{H}({\theta^{\tau}}))v\right|\\
&\leq \max_{v\in \mathcal{N}_{\delta}(\mathbb{B}^{dK})}\left|v^{\intercal}\left(\widehat{H}({\theta})-{H}({\theta})-\widehat{H}({\theta}^{\tau}) +{H}({\theta}^{\tau})\right)v\right|\lesssim d^{1/2} \epsilon.
\end{aligned}
\end{equation*}
Consequently, we obtain 
\begin{equation}
\max_{\theta\in \Theta_{B}}\max_{v\in \mathcal{N}_{\delta}(\mathbb{B}^{dK})}v^{\intercal}(\widehat{H}({\theta})-{H}({\theta}))v\leq \max_{\theta\in \mathcal{N}_{\epsilon}(\Theta_B)}\max_{v\in \mathcal{N}_{\delta}(\mathbb{B}^{dK})}v^{\intercal}(\widehat{H}({\theta})-{H}({\theta}))v+ C d^{1/2} \epsilon.
 \label{eq: max concentration-2}
\end{equation}
It remains to control
\begin{equation}
\begin{aligned}
    &\max_{\theta\in \mathcal{N}_{\epsilon}(\Theta_B)}\max_{v\in \mathcal{N}_{\delta}(\mathbb{B}^{dK})} v^{\intercal}(\widehat{H}({\theta})-{H}({\theta}))v\\
    &=\max_{\theta\in \mathcal{N}_{\epsilon}(\Theta_B)}\max_{v\in \mathcal{N}_{\delta}(\mathbb{B}^{dK})} v^{\intercal}\frac{1}{N}\sum_{i=1}^{N}\left[D(X_i,\theta)\otimes X_iX_i^{\intercal}-\E_{X_i\sim \QQ_{X}}D(X_i,\theta)\otimes X_iX_i^{\intercal}\right]v.
\end{aligned}
\label{eq: max concentration}
\end{equation}

For a given $v$, we have
\[
\begin{aligned}
    v^\intercal \left[D(X_i,\theta)\otimes X_i X_i^\intercal\right] v &= \sum_{k=1}^K p_k (v_k^\intercal X_i)^2 - \left(\sum_{k=1}^K p_k v_k^\intercal X_i\right)^2, 
\end{aligned}
\]
where $p = p(X_i,\theta)\in \RR^k$ and $v=[v_1^\intercal, v_2^\intercal,..., v_K^\intercal]^\intercal \in \RR^{dK}$. Since $X_i$ is a sub-gaussian random vector, we have each $(v_k^\intercal X_i)^2$ being a sub-exponential random variable. Together with the fact that each $p_k\geq c_0$ and $\sum_{k=1}^K p_k \leq 1-c_0$; therefore $\sum_{k=1}^K p_k (v_k^\intercal X_i)^2$ is a sub-exponential random variable. Moreover, we have $\sum_{k=1}^K p_k v_k^\intercal X_i$ being sub-gaussian, which implies that $(\sum_{k=1}^K p_k v_k^\intercal X_i)^2$ is a sub-exponential random variable. As a result, we establish that
$v^{\intercal} D(X_i,\theta)\otimes X_iX_i^{\intercal} v$ is a sub-exponential random variable, which further implies that 
\begin{equation*}
\mathbb{P}\left(\left|v^{\intercal}\frac{1}{N}\sum_{i=1}^{N}\left[D(X_i,\theta)\otimes X_iX_i^{\intercal}-\E_{X_i\sim \QQ_{X}}D(X_i,\theta)\otimes X_iX_i^{\intercal}\right]v\right|\geq {t}\right)\leq \exp(-cNt^2).
\end{equation*}
Together with \eqref{eq: max concentration}, the union bound leads to 
\begin{equation*}
\mathbb{P}\left(\max_{\theta\in \mathcal{N}_{\epsilon}(\Theta_B)}\max_{v\in \mathcal{N}_{\delta}(\mathbb{B}^{dK})} \left|v^{\intercal}(\widehat{H}({\theta})-{H}({\theta}))v\right|\geq t\right) \leq |\mathcal{N}_{\epsilon}(\Theta_B)|\cdot |\mathcal{N}_{\delta}(\mathbb{B}^{dK})|\cdot \exp(-cNt^2).
\end{equation*}
By choosing $t=\sqrt{d \left(\log (C/\epsilon)+\log (1/\delta)\right)/N}$ and setting $\delta=\epsilon= \frac{1}{\sqrt{N}}$, we combine the results in  \eqref{eq: max concentration-1} and  \eqref{eq: max concentration-2} and establish that: with probability larger than $1-N^{-c_1d}$,
\begin{equation}
\max_{\theta\in \Theta_{B}}\|\widehat{H}({\theta})-{H}({\theta})\|_2
\leq C \sqrt{d \log N/N}.
\label{eq: event-3 bound}
\end{equation}

\noindent {\bf Proof of $\GG_2$ in \eqref{eq: events}.}
  Note that 
\begin{equation*}
\nabla S(\theta)=\E_{X_i\sim \QQ_{X}}\left(p(X_i,\theta) \otimes X_i\right), \quad \nabla \widehat{S}(\theta)=\frac{1}{N}\sum_{i=1}^{N} p(X_i,\theta)\otimes X_i.
\end{equation*}
Given the $\delta$-net $\mathcal{N}_\delta(\mathbb{B}^{dK})$ of the unit ball $\mathbb{B}^{dK}$ in $\RR^{dK}$, for any $\theta\in \RR^{Kd}$, 
\begin{equation*}
\|\nabla S(\theta)-\nabla \widehat{S}(\theta)\|_2 \leq\frac{1}{1-\delta}  \max_{v\in \mathcal{N}_{\delta}(\mathbb{B}^{dK})}v^{\intercal}(\nabla S(\theta)-\nabla \widehat{S}(\theta)),
\end{equation*}
which further implies 
\begin{equation}
\max_{\theta\in \Theta_{B}}\|\nabla S(\theta)-\nabla \widehat{S}(\theta)\|_2 \leq\frac{1}{1-\delta}  \max_{\theta\in \Theta_{B}}\max_{v\in \mathcal{N}_{\delta}(\mathbb{B}^{dK})}v^{\intercal}(\nabla S(\theta)-\nabla \widehat{S}(\theta)).
\label{eq: enet for grad}
\end{equation}

For any $\theta\in \Theta_{B}$, there exists a $\theta^{\tau}\in \mathcal{N}_{\epsilon}(\Theta_B)$ such that $\|\theta^{\tau}-\theta\|_2\leq \epsilon.$  Then the following holds: for any $v\in \mathcal{N}_{\delta}(\mathbb{B}^{dK})$ with $v=(v_1^\intercal,..., v_K^\intercal)^\intercal$ for each $v_k\in \RR^d$, we have
\begin{equation*}
\begin{aligned}
    &\left|v^\intercal \left[p(X_i,\theta)\otimes X_i\right] - v^\intercal \left[p(X_i,\theta^\tau)\otimes X_i\right]\right| \\
    &= \left|\sum_{k=1}^K [p_k(X_i,\theta) - p_k(X_i,\theta^\tau)] X_i^\intercal v_k\right| \leq \left\|p(X_i,\theta) - p(X_i,\theta^\tau)\right\|_2 \left(\sum_{k=1}^K (X_i^\intercal v_k)^2\right)^{1/2}\\
    &\leq \left\|p(X_i,\theta) - p(X_i,\theta^\tau)\right\|_2 \|X_i\|_2,
\end{aligned}
\end{equation*}
where the last inequality holds as $\sum_{k=1}^K (X_i^\intercal v_k)^2 \leq \|X_i\|_2^2\sum_{k=1}^K \|v_k\|_2^2 = \|X_i\|_2^2$, with $v\in \mathcal{N}_\delta(\mathbb{B}^{dK}).$
Then it follows from \eqref{eq: p theta diff}, we have
\[
\begin{aligned}
    \left|v^\intercal \left[p(X_i,\theta)\otimes X_i\right] - v^\intercal \left[p(X_i,\theta^\tau)\otimes X_i\right]\right| &\leq K\|X_i\|_2^2 \|\theta - \theta^\tau\|_2 \lesssim d\epsilon.
\end{aligned}
\]
Since the above inequality holds for any $v\in \Nc_\delta(\mathbb{B}^{dK})$, we establish that
\[
\max_{v\in \Nc_\delta(\mathbb{B}^{dK})}\left|v^\intercal (\nabla S(\theta) - \nabla \widehat{S}(\theta)) - v^\intercal (\nabla S(\theta^\tau) - \nabla \widehat{S}(\theta^\tau))\right| \lesssim d \epsilon.
\]
It implies that
{\begin{equation}
\begin{aligned}
    &\max_{\theta \in \Theta_B}\max_{v\in \Nc_\delta(\mathbb{B}^{dK})}\left|v^\intercal (\nabla S(\theta) - \nabla \widehat{S}(\theta))\right| \\
    &\leq \max_{\theta \in \Nc_\epsilon(\Theta_B)}\max_{v\in \Nc_\delta(\mathbb{B}^{dK})}\left|v^\intercal (\nabla S(\theta) - \nabla \widehat{S}(\theta))\right| + Cd\epsilon,
\end{aligned}
    \label{eq: enet for grad-2}
\end{equation}}
for some constant $C>0$.

Notice that for any $\theta\in \Theta_\epsilon(\Theta_B), v\in \Nc_\delta(\mathbb{B}^{dK})$, we have $v^\intercal p(X_i,\theta) \otimes X_i = \sum_{k=1}^K p_k(X_i,\theta) X_i^\intercal v$ being a sub-gaussian random variable, which implies that
\[
\begin{aligned}
    \PP\left(\left|v^\intercal (\nabla \widehat{S}(\theta) - \nabla S(\theta) )\right| \geq t\right)\leq \exp(-cN t^2).
\end{aligned}
\]
By union bound, we have
\[
\begin{aligned}
    &\PP\left(\max_{\theta\in \Nc_\epsilon(\Theta_B)}\max_{v\in \Nc_\delta(\mathbb{B}^{dK})}\left|v^\intercal (\nabla \widehat{S}(\theta) - \nabla S(\theta) ) \right|\geq t\right)\\
    &\leq |\Nc_\epsilon(\Theta_B)|\cdot|\Nc_\delta(\mathbb{B}^{dK})|\cdot\exp(-cN t^2) \leq (C/\epsilon)^{dK}(C/\delta)^{dK} \exp(-cN t^2).
\end{aligned}
\]
We combine the above inequality with the results in  \eqref{eq: enet for grad} and  \eqref{eq: enet for grad-2} and establish that:
\[
\PP\left(\max_{\theta\in \Theta_B}\|\nabla S(\theta) - \nabla\widehat{S}(\theta)\|_2 \leq \frac{t+Cd\epsilon}{1-\delta}\right) \leq 1-(C/\epsilon)^{dK}(C/\delta)^{dK} \exp(-cN t^2).
\]
By choosing $\epsilon = \delta = 1/N$ and $t = A\sqrt{d\log N/N}$ with some sufficiently large $A>0$, we simplify the above inequality as follows: with probability larger than $1-N^{-c_1d}$,
\begin{equation}
\max_{\theta\in \Theta_{B}}\|\nabla S(\theta) - \nabla \widehat{S}(\theta)\|_2
\leq C \sqrt{d \log N/N}.
\label{eq: event-2 bound}
\end{equation}

\noindent {\bf Proof of \eqref{eq: ALL concen}}
Under Condition {\rm (A1)}, we have: with probability at least $1-e^{-c_1t^2}-\delta_n$,
\[
\max_{1\leq l\leq L}\|\widehat{\mu}^{(l)}-\mu^{(l)}\|_2 \leq t\sqrt{d/n}.
\]
We then have with probability at least $1-e^{-c_1t^2}-\delta_n$,
\[
\|\widehat{\U} - \U\|_2 \leq \sqrt{L}\max_{1\leq l\leq L}\|\widehat{\mu}^{(l)}-\mu^{(l)}\|_2 \leq t\sqrt{L d/n},
\]
thus $\GG_0$ holds.
Together with the results in \eqref{eq: event-1 bound}, \eqref{eq: event-2 bound}, and \eqref{eq: event-3 bound}, we have proved that
\[
\mathbf{P}(\GG_0\cap \GG_1\cap \GG_2\cap \GG_3) \geq 1-N^{-c_1d}-e^{-c_1t^2}-\delta_n.
\]

\subsection{Proof of Lemma \ref{lem: strong convex of hatH}}
\label{proof of lem: strong convex of hatH}
On the event $\GG_3$, for any $\theta\in \Theta_B$, if $N\geq C \frac{4d\log N}{\kappa_1c_1^2}$, we have
\[
\|H(\theta) - \widehat{H}(\theta)\|_2 \leq \frac{\kappa_1c_1^2}{2},
\]
It follows from Weyl's inequality for Hermitian matrices that
\[
\left|\lambda_{\rm min}(H(\theta)) - \lambda_{\rm min}(\widehat{H}(\theta))\right|\leq \|H(\theta) - \widehat{H}(\theta)\|_2 \leq  \frac{\kappa_1c_1^2}{2}.
\]
Then it follows from Lemma \ref{lem: convexity of H} that
\[
\lambda_{\rm min}(\widehat{H}(\theta)) \geq \frac{\kappa_1c_1^2}{2}, \quad\textrm{for all $\theta\in \Theta_B$}.
\]

\subsection{Proof of Lemma \ref{lem: saddle point}}
\label{sec: saddle point}
By definition, $\theta^*_{\rm ap}=\argmin_{\theta\in \RR^{dK}}\phi_{\rm ap}(\theta,\gamma^*_{\rm ap})$. It implies that
\[
\phi_{\rm ap}(\theta_{\rm ap}^*,\gamma^*_{\rm ap}) \leq \phi_{\rm ap}(\theta,\gamma^*_{\rm ap}), \quad \textrm{for all $\theta\in \RR^{dK}$},
\]
hence we obtain the second inequality in \eqref{eq: saddle point thm2}. We now move on to establishing the first inequality in \eqref{eq: saddle point thm2}.  We use the expression of $\phi_{\rm ap}(\theta,\gamma)$ in \eqref{eq: f_ap thm2} to show that:
\begin{equation}
\phi_{\rm ap}({\theta}^*_{\rm ap}, {\gamma}^*_{\rm ap}) - \phi_{\rm ap}({\theta}^*_{\rm ap}, \gamma) = \left[{\theta}^*_{\rm ap}\right]^\intercal \U ({\gamma}^*_{\rm ap} - \gamma).
\label{eq: diff expression}
\end{equation}
Hence, to establish the first inequality in \eqref{eq: saddle point thm2}, it is sufficient to establish $\left[{\theta}^*_{\rm ap}\right]^\intercal \U ({\gamma}^*_{\rm ap} - \gamma)\geq 0$ for all $\gamma\in \Delta^L,$ which will be established in the following. 

It follows from \eqref{eq: theta-star expression} that $\thetainit$ can be expressed as:
\[
\widehat{\theta} = \theta_{\rm ap}^* + [H(\thetainit)]^{-1}\left(\U \gamma_{\rm ap}^* + \nabla S(\thetainit)\right).
\]
We apply the preceding expression of $\widehat{\theta}$ to the expression of $F(\gamma)$ in \eqref{eq: expression of F(gamma)} to obtain that:
{\small
\[
\begin{aligned}
    F(\gamma) &=-\frac{1}{2}\left(\U\gamma + \nabla S(\thetainit)\right)^\intercal [H(\thetainit)]^{-1}\left(\U\gamma + \nabla S(\thetainit)\right) + \widehat{\theta}^\intercal \U \gamma 
    \\
    &=-\frac{1}{2}\left(\U\gamma + \nabla S(\thetainit)\right)^\intercal [H(\thetainit)]^{-1}\left(\U\gamma + \nabla S(\thetainit)\right) + \left[\theta_{\rm ap}^* + [H(\thetainit)]^{-1}\left(\U \gamma_{\rm ap}^* + \nabla S(\thetainit)\right)\right]^\intercal \U \gamma \\
    &=-\frac{1}{2}\gamma^\intercal \U^\intercal [H(\thetainit)]^{-1} \U \gamma + \left(\theta^*_{\rm ap} + [H(\thetainit)]^{-1}\U \gamma_{\rm ap}^*\right)^\intercal \U\gamma - \frac{1}{2}[\nabla S(\thetainit)]^\intercal [H(\thetainit)]^{-1} \nabla S(\thetainit).
\end{aligned}
\]}
And we compute the gradient of $F(\gamma)$, denoted as
$$\nabla F(\gamma)=\U^\intercal [H(\thetainit)]^{-1} \U \left(\gamma_{\rm ap}^*-\gamma\right)+\U^{\intercal} \theta^*_{\rm ap}.$$

Recall the definition of ${\gamma}^*_{\rm ap}$ in \eqref{eq: step 1}, with $\gamma^*_{\rm ap}=\argmax_{\gamma\in \Delta^L} F(\gamma)$.
Hence, for any fixed $\gamma\in \Delta^L$, we have $F({\gamma}^*_{\rm ap})\geq F({\gamma}^*_{\rm ap} + \nu(\gamma - {\gamma}^*_{\rm ap}))$ for all $\nu\in (0,1)$. 
Since $F(\cdot)$ is concave and $\gamma^*_{\rm ap}$ is the maximizer, we have 
\[
\begin{aligned}
    0&\leq F(\gamma^*_{\rm ap})-F(\gamma^*_{\rm ap} + \nu(\gamma - \gamma^*_{\rm ap})) \\
    &\leq \langle \nabla F(\gamma^*_{\rm ap} + \nu(\gamma - \gamma^*_{\rm ap})) , -\nu(\gamma - \gamma^*_{\rm ap})\rangle\\
    &=\nu\cdot \left[{\theta}^*_{\rm ap}\right]^\intercal \U ({\gamma}^*_{\rm ap}-\gamma) + {\nu^2}\cdot (\gamma - {\gamma}^*_{\rm ap})^\intercal \U^\intercal [H(\thetainit)]^{-1} \U (\gamma - {\gamma}^*_{\rm ap}),
\end{aligned}
\]
where the last equality follows from plugging in the expression of $\nabla F.$

Dividing both sides of the above inequality by $\nu>0$, we take $\nu\to 0$ and establish %
\[
\left[{\theta}^*_{\rm ap}\right]^\intercal \U ({\gamma}^*_{\rm ap}-\gamma) \geq 0.
\]
Since the above inequality holds for all $\gamma\in \Delta^L$, we apply \eqref{eq: diff expression} and complete the proof of \eqref{eq: saddle point thm2}.

\subsection{Proof of Lemma \ref{lem: V_mu scale}}
\label{proof of lem: V_mu scale}
We first establish the theoretical results in the covariate shift setting, then we move forward studying the non-shift regime.

\noindent\textit{Covariate Shift Setting.}
Without loss of generality, we consider $j=1$ for clarity.
    In the covariate shift setting, it follows from the expression of ${\bf V}_\mu^{(l)}$ in \eqref{eq: V_mu expression - cs} that:
    \begin{equation}
        \label{eq: proof V_mu scale - key 1}
        n_l\cdot [{\bf V}_\mu^{(l)}]_{1,1} = {\bf V}_{\PP^{(l)}}\left(\w{l}(X)\cdot \varepsilon_{1}^{(l)}\cdot X_1\right) + \frac{n_l}{N} {\bf V}_{\QQ} \left(\f{l}_1(X)\cdot X_1\right),
    \end{equation}
    where $\varepsilon_1^{(l)} = \f{l}_1(X)-{\bf 1}(Y=1)$, and $X_1$ denotes the first entry of the random vector $X$. 

    In the following, we establish the lower and upper bounds for the components on the right-hand side of \eqref{eq: proof V_mu scale - key 1}. We start with ${\bf V}_{\PP^{(l)}}\left(\w{l}(X)\cdot \varepsilon_{1}^{(l)}\cdot X_1\right)$.
    Since $\E[\varepsilon^{(l)}_i|\X{l}] =0$, it holds that
    \[
    {\bf V}_{\PP^{(l)}}\left(\w{l}(X)\cdot \varepsilon_{1}^{(l)}\cdot X_1\right) = \E\left[\w{l}(X)\cdot \varepsilon^{(l)}_i\cdot \X{l}_1\right]^2.
    \]
    In Condition {\rm (B1)}, we consider $\w{l}(X)\in [c_2,c_3]$ for two positive constants $c_2,c_3>0$. Therefore,
    \[
    c_2^2 \E[\varepsilon^{(l)}_1 \X{l}_1]^2\leq \E\left[\w{l}(X)\cdot \varepsilon^{(l)}_1\cdot \X{l}_1\right]^2 \leq c_3^2 \E[\varepsilon^{(l)}_1 \X{l}_1]^2.
    \]
    Moreover, we consider $\E[(\varepsilon^{(l)}_1)^2|\X{l}]\geq c_1$ in Condition {\rm (B1)}, which further yields
    \[
    \E[\varepsilon^{(l)}_1 \X{l}_1]^2 = \E[\E[(\varepsilon^{(l)}_1)^2|\X{l}] (\X{l}_1)^2] \geq c_1^2 \E[(\X{l}_1)^2] \geq c_1^2\kappa_1'.
    \]
    Due to the definition of $\varepsilon^{(l)}_1$, it holds that $\varepsilon^{(l)}_1\in [-1,1]$, therefore,
    \[
    \E[\varepsilon^{(l)}_1 \X{l}_1]^2 \leq \E[\X{l}_i]^2\leq \kappa_2'.
    \]
    We combine the preceding four equations to establish that 
    \begin{equation}
        \label{eq: proof V_mu scale - interm 1}
        {\bf V}_{\PP^{(l)}}\left(\w{l}(X)\cdot \varepsilon_{1}^{(l)}\cdot X_1\right) \in [c_1^2c_2^2\kappa_1', c_3^2 \kappa_2'].
    \end{equation}
    Next, we study the term ${\bf V}_{\QQ} \left(\f{l}_1(X)\cdot X_1\right)$. It holds that
    \begin{equation}
        \label{eq: proof V_mu scale - interm 2}
        0\leq {\bf V}_{\QQ} \left(\f{l}_1(X)\cdot X_1\right) \leq \E_\QQ\left(\f{l}_1(X)\cdot X_1\right)^2\leq \E_\QQ[X_1]^2 \leq \kappa_2,
    \end{equation}
    where the second equality holds since $\f{l}_1(X)\in [0,1]$, and the last equality holds due to Condition {\rm (A2)} where we consider $\E[\XQ(\XQ)^\intercal]\preceq \kappa_2\cdot {\bf I}_d$.

    We plug the inequalities \eqref{eq: proof V_mu scale - interm 1} and \eqref{eq: proof V_mu scale - interm 2} into \eqref{eq: proof V_mu scale - key 1} to establish that:
    \[
    n_l\cdot [{\bf V}_\mu^{(l)}]_{1,1} \in [c_1^2c_2^2\kappa_1', \; c_3^2\kappa_2'+\kappa_2].
    \]

    In the following, we study the term $\|n{\bf V}_\mu^{(l)}\|_2$. For clarity, we denote
    \[
    Z= \w{l}(\X{l})\cdot \left(\f{l}(\X{l})-{\bf y}^{(l)}\right)\otimes \X{l} \in \RR^{dK}.
    \]
    Notice that $\E[Z] = 0$, which yields ${\bf V}_{\PP^{(l)}}(Z) = \E[ZZ^\intercal]$. Therefore,
    \[
    \|{\bf V}_{\PP^{(l)}}(Z)\|_2 = \lambda_{\rm max}\left(\E[ZZ^\intercal]\right)\leq {\rm Tr}(\E[ZZ^\intercal]) = \E[{\rm Tr}(ZZ^\intercal)]= \E[\|Z\|_2^2].
    \]
    For the term $\E[\|Z\|_2^2]$, due to $\w{l}(\cdot)\in [c_1,c_2]$ and $\f{l}_k(\cdot)\in [0,1]$, we have
    \[
    \begin{aligned}
        \E[\|Z\|_2^2] = &\E\left[\sum_{k=1}^K\sum_{j=1}^d \left[\w{l}(\X{l})\right]^2\cdot \left[\f{l}_k(\X{l}) - {\bf y}^{(l)}_k\right]^2\cdot \left[\X{l}_j\right]^2 \right]\\
        &\leq \E\left[\sum_{k=1}^K\sum_{j=1}^d c_3^2 \left[\X{l}_j\right]^2 \right] = Kc_3^2 \E \left[\|\X{l}\|_2^2\right] \\
        &=K c_3^2\E\left[{\rm Tr}(\X{l}(\X{l})^\intercal)\right] = K c_3^2 {\rm Tr}\left(\E[\X{l}(\X{l})^\intercal]\right) \\
        &\leq dK c_3^2 \kappa_2'.
    \end{aligned}
    \]
    Therefore, we establish that
    \begin{equation}
        \label{eq: proof V_mu scale - interm 3}
         \left\|{\bf V}_{\PP^{(l)}}\left(\left[\w{l}(X)\cdot (\f{l}(X) - {\bf y})\right] \otimes X\right)\right\|_2 \leq dK c_3^2 \kappa_2'.
    \end{equation}

    We then study the term ${\bf V}_{\QQ} \left(\f{l}(X)\otimes X \right)$. It holds that
    \[
    \left\|{\bf V}_{\QQ} \left(\f{l}(X)\otimes X \right)\right\|_2\leq \E_\QQ\left[\|\f{l}(X)\otimes X\|_2^2\right]= \E_{\QQ}\left[\|\f{l}(X)\|_2^2 \|X\|_2^2\right].
    \]
    as $\|a\otimes b\|_2^2 = \|a\|_2^2\|b\|_2^2$.
    Since $\sum_{k=1}^K \f{l}_k(X)\in [0,1]$ and each $\f{l}_k(X)>0$, we obtain that $\|\f{l}(X)\|_2^2\leq 1$. Thus,
    \[
    \E_{\QQ}\left[\|\f{l}(X)\|_2^2 \|X\|_2^2\right] \leq \E_{\QQ}[\|X\|_2^2]={\rm Tr}(\E_{\QQ}[XX^\intercal])\leq d\kappa_2.
    \]
    We combine the preceding two (in)equalities to establish that
    \begin{equation}
        \label{eq: proof V_mu scale - interm 4}
        \left\|{\bf V}_{\QQ} \left(\f{l}(X)\otimes X \right)\right\|_2\leq d\kappa_2.
    \end{equation}

    We now combine \eqref{eq: proof V_mu scale - interm 3} and \eqref{eq: proof V_mu scale - interm 4} to obtain that:
    \[
    \begin{aligned}
        \|n_l{\bf V}_\mu^{(l)}\|_2 &\leq \left\| {\bf V}_{\PP^{(l)}}\left(\left[\w{l}(X)\cdot (\f{l}(X) - {\bf y})\right] \otimes X\right)\right\|_2 + \left\|\frac{n_l}{N} {\bf V}_{\QQ} \left(\f{l}(X)\otimes X \right)\right\|_2 \\
    &\leq dK c_3^2\kappa_2' + d\kappa_2.
    \end{aligned}
    \]

    \noindent\textit{Non-shift Regime.}
    We consider $j=1$ without loss of generality. Then it follows from the expression of ${\bf V}_\mu^{(l)}$ in \eqref{eq: V_mu expression - nocs} that
    \[
    n_l [{\bf V}_\mu^{(l)}]_{1,1}= {\bf V}_{\PP^{(l)}}({\bf 1}(Y=1)\cdot X_1) = \E[(\varepsilon^{(l)}_1)^2 (\X{l}_1)^2].
    \]
    By Condition {\rm (B1)}, it holds that
    \[
    c_1^2\kappa_1'\leq c_1^2\E[(\X{l}_1)^2]\leq n_l [{\bf V}_\mu^{(l)}]_{1,1}\leq \E[(\X{l}_1)^2]\leq \kappa_2'.
    \]
    For the term $\|n{\bf V}_\mu^{(l)}\|_2$, it holds that:
    \[
    \|n{\bf V}_\mu^{(l)}\|_2 \leq \E\left[\|{\bf y}^{(l)}\|_2^2\|\X{l}\|_2^2\right]\leq d\kappa_2'.
    \]

\subsection{Proof of Lemma \ref{lemma: distance between gammas}}
\label{section: proof of lemma distance between gammas}
For any given $\gamma \in \Delta^L$, we have
    \[
    \nabla \hat{f}(\gamma) - \nabla f(\gamma) = 2(\widehat{A} - A)\gamma + 2(\hat{b} - b)^\intercal \gamma.
    \]
    Therefore,
    \begin{equation}
        \|\nabla \hat{f}(\gamma) - \nabla f(\gamma)\|_2 \leq 2\|\widehat{A} - A\|_2 \|\gamma\|_2 + 2\|\hat{b} - b\|_2 \leq 2(\|\widehat{A} - A\|_2  + \|\hat{b} - b\|_2),
        \label{eq: perturb bound}
    \end{equation}
    where the last inequality holds as $\|\gamma\|_2\leq 1$ for $\gamma\in \Delta^L.$

    From the optimality condition for $\gamma^*$ and $\hat{\gamma}$, we have
    \[
    \nabla f(\gamma^*)^\intercal (\gamma - \gamma^*) \geq 0, \quad \textrm{for all $\gamma\in \Delta^L$},
    \]
    and
    \[
    \nabla \hat{f}(\hat{\gamma})^\intercal (\gamma - \hat{\gamma})\geq 0\quad \textrm{for all $\gamma\in \Delta^L$}.
    \]
    Choose $\gamma = \hat{\gamma}$ in the first inequality and $\gamma = \gamma^*$ in the second, we shall establish that
    \[
    \nabla f(\gamma^*)^\intercal (\hat\gamma - \gamma^*) \geq 0, \quad \textrm{and}\quad \nabla \hat{f}(\hat{\gamma})^\intercal (\gamma^* - \hat{\gamma})\geq 0.
    \]
    Adding the above two inequalities, we obtain that
    \[
    \left(\nabla f(\gamma^*) - \nabla\hat f(\hat{\gamma})\right)^\intercal (\hat{\gamma} - \gamma^*) \geq 0.
    \]
    Therefore,
    \[
    \left(\nabla f(\gamma^*) - \nabla f(\hat{\gamma})+ \nabla f(\hat{\gamma}) - \nabla\hat f(\hat{\gamma})\right)^\intercal (\hat{\gamma} - \gamma^*) \geq 0.
    \]
    It implies that
    \[
    \left(\nabla f(\hat{\gamma}) - \nabla\hat f(\hat{\gamma})\right)^\intercal (\hat{\gamma} - \gamma^*) \geq \left(\nabla f(\hat\gamma) - \nabla f(\gamma^*)\right)^\intercal(\hat{\gamma} - \gamma^*).
    \]

    By the strong convexity of $f(\gamma)$, or the strong monotonicity of $\nabla f(\gamma)$, we have
    \[
    \left(\nabla f(\hat\gamma) - \nabla f(\gamma^*)\right)^\intercal(\hat{\gamma} - \gamma^*) \geq 2\lambda_{\rm min}(A) \|\hat{\gamma} - \gamma^*\|_2^2.
    \]
    Combining the above two inequalities and \eqref{eq: perturb bound}, we establish that
    \[
    \begin{aligned}
        2\lambda_{\rm min}(A)\|\hat{\gamma} - \gamma^*\|_2^2 &\leq \left(\nabla f(\hat{\gamma}) - \nabla\hat f(\hat{\gamma})\right)^\intercal(\hat{\gamma} - \gamma^*) \\
        &\leq \left\|\nabla f(\hat{\gamma}) - \nabla\hat f(\hat{\gamma})\right\|_2 \|\hat{\gamma} - \gamma^*\|_2 \\
        &\leq 2(\|\widehat{A} - A\|_2  + \|\hat{b} - b\|_2) \|\hat{\gamma} - \gamma^*\|_2.
    \end{aligned}
    \]
    Dividing both sides by $\|\hat{\gamma}-\gamma^*\|_2$, we obtain
    \[
    \|\hat{\gamma} - \gamma^*\|_2 \leq \frac{\|\widehat{A} - A\|_2  + \|\hat{b} - b\|_2}{\lambda_{\rm min}(A)}.
    \]

\subsection{Proof of Lemma \ref{prop: gamma_m and gamma_ap dist}}
\label{section: proof of equation gamma distance and mu distance.}
Recall that $\widehat{\gamma}^{[m]}$ and $\gamma^*_{\rm ap}$, as defined in \eqref{eq: resampled gamma}, admit the following expressions:
{\small
\[
\widehat{\gamma}^{[m]}=\argmax_{\gamma\in \Delta^{L}} \widehat{F}^{[m]}(\gamma) = \argmin_{\gamma\in \Delta^L}\left\{\frac{1}{2}\left(\widehat{\U}^{[m]}\gamma+\nabla \widehat{S}(\thetainit)\right)^{\intercal}[\widehat{H}(\widehat{\theta})]^{-1}\left(\widehat{\U}^{[m]}\gamma+\nabla \widehat{S}(\thetainit)\right) -\widehat{\theta}^\intercal \widehat{\U}^{[m]}\gamma\right\},
\]
and
\[
\gamma^*_{\rm ap}=\argmax_{\gamma\in \Delta^L} F(\gamma)=\argmin_{\gamma\in \Delta^L}\left\{\frac{1}{2}\left(\U \gamma+\nabla S(\thetainit)\right)^{\intercal}[H(\thetainit)]^{-1}\left(\U \gamma+\nabla S(\thetainit)\right)- \widehat{\theta}^\intercal \U \gamma\right\}.
\]
}
We apply Lemma \ref{lemma: distance between gammas} upper bound the distance $\|\widehat{\gamma}^{[m]} - {\gamma}^*_{\rm ap}\|_2$, which admits the following upper bound: 
{\small
\begin{equation}
\begin{aligned}
    &\lambda_{\rm min}(\U^\intercal [H(\thetainit)]^{-1} \U) \|\widehat{\gamma}^{[m]} - {\gamma}^*_{\rm ap}\|_2 \\
    &\leq  
    \left\|\widehat{\U}^{[m]\intercal}[\widehat{H}(\thetainit)]^{-1} \widehat{\U}^{[m]} - \U^\intercal [H(\thetainit)]^{-1}\U\right\|_2\\
    &\quad +\left\|([\widehat{H}(\thetainit)]^{-1} \nabla \widehat{S}(\thetainit) - \thetainit)^\intercal \widehat{\U}^{[m]} - ([H(\thetainit)]^{-1} \nabla S(\thetainit) - \thetainit)^\intercal \U\right\|_2.
\end{aligned}
\label{eq: decompose gamma distance}
\end{equation}
}

We denote
\[
\Delta_1 = \bmmum - \bmmu, \quad \Delta_2 = \widehat{H}(\thetainit)^{-1} - H(\thetainit)^{-1}, \quad \Delta_3 = \nabla \widehat{S}(\thetainit) - \nabla S(\thetainit).
\]
On the event $\GG_2\cap \GG_3$, we have $\|\Delta_2\|\lesssim\sqrt{d\log N/N}$, and $\|\Delta_3\|_2\lesssim \sqrt{d\log N/N}$.
Then for the first term on the right-hand side of the above inequality \eqref{eq: decompose gamma distance} we establish that
{\small
\begin{equation}
\begin{aligned}
    &\left\|\widehat{\U}^{[m]\intercal}[\widehat{H}(\thetainit)]^{-1} \widehat{\U}^{[m]} - \U^\intercal [H(\thetainit)]^{-1}\U\right\|_{2} \\
    &\leq \|\Delta_2\|_2\left(\|\U\|_2 + \|\Delta_1\|_2\right)^2 + \|[H(\thetainit)]^{-1}\|_2\|\Delta_1\|_2 \left(2 \|\U\|_2 + \|\Delta_1\|_2\right)\\
    &\lesssim \|\U\|_2 \|\widehat{\U}^{[m]} - \U\|_2 + \|\U\|_2^2\sqrt{d\log N/N}.
\end{aligned}
    \label{eq: term 1 result in decompose gamma distance}
\end{equation}
}
For the second term of \eqref{eq: decompose gamma distance}, we have
\begin{equation}
    \begin{aligned}
        &\left\|([\widehat{H}(\thetainit)]^{-1} \nabla \widehat{S}(\thetainit) - \thetainit)^\intercal \widehat{\U} - ([H(\thetainit)]^{-1} \nabla S(\thetainit) - \thetainit)^\intercal \U\right\|_2 \\
        &\quad \leq \|[H(\thetainit)]^{-1} \nabla S(\thetainit) - \thetainit\|_2 \|\Delta_1\|_2 + (\|\U\|_2 + \|\Delta_1\|_2) \|\nabla S(\thetainit)\|_2 \|\Delta_2\|_2 \\
        &\quad \quad\quad + (\|\U\|_2 + \|\Delta_1\|_2)(\|[H(\thetainit)]^{-1}\|_2 + \|\Delta_2\|_2) \|\Delta_3\|_2\\
        &\quad \leq (1+\|\theta^*\|_2 + \|\nabla S(\theta^*)\|_2) \|\widehat{\U}^{[m]} - \U\|_2 + (\|\U\|_2 + \|\nabla S(\theta^*)\|_2)\sqrt{d\log N/N}.
    \end{aligned}
    \label{eq: term 2 result in decompose gamma distance}
\end{equation}
The combination of results in \eqref{eq: decompose gamma distance},  \eqref{eq: term 1 result in decompose gamma distance} and \eqref{eq: term 2 result in decompose gamma distance} results in
\begin{equation*}
\begin{aligned}
    \lambda_{\rm min}(\U^\intercal [H(\thetainit)]^{-1} \U)\|{\widehat{\gamma}_{\rm ap}} - {\gamma}^*_{\rm ap}\|_2 &\lesssim (1+\|\theta^*\|_2 + \|\nabla S(\theta^*)\|_2 + \|\U\|_2)\|\widehat{\U}^{[m]} - \U\|_2 \\
    &\quad\quad +  (\|\U\|_2^2 + \|\U\|_2 + \|\nabla S(\theta^*)\|_2)\sqrt{d\log N/N}.
\end{aligned}
\end{equation*}
Under Conditions \textrm{(A2),(A4)}, we shall further simplify the above inequality:
\[
\lambda_{\rm min}(\U^\intercal [H(\thetainit)]^{-1} \U)\|{\widehat{\gamma}_{\rm ap}} - {\gamma}^*_{\rm ap}\|_2 \lesssim \|\widehat{\U}^{[m]} - \U\|_2 + \sqrt{d\log N/ N}.
\]

Notice that
\[
\lambda_{\rm min}(\U^\intercal [H(\thetainit)]^{-1} \U) \geq \lambda_{\rm min}([H(\thetainit)]^{-1}) \cdot \sigma_{L}^2(\U) \gtrsim  \frac{\sigma_{L}^2(\U)}{\kappa_2},
\]
where the second inequality holds since Lemma \ref{lem: convexity of H} establishes that $H(\theta)\lesssim \kappa_2$ for all $\theta\in \Theta_B$, and \eqref{eq: theta_hat is bounded} establishes that $\widehat{\theta}\in \Theta_B$ on the event $\GG$.
% and $\kappa_2>0$ is a constant given in Condition {\rm (A2)}.
The combination of the above two results establishes \eqref{eq: gamma distance and mu distance}.

\subsection{Proof of Lemma \ref{lem: E1E2}}
\label{proof of lemma E1E2}

For the event $\Ec_1$, by Condition {\rm (A5)}, we have
\begin{equation}
\lim_{n\to\infty}\mathbf{P}\left(\Ec_1\right) = 1.
    \label{eq: E1 lim}
\end{equation}
In the following, we continue studying the event $\Ec_2$. 
According to Condition {\rm (A5)}, we have
$$\frac{\widehat{\mu}^{(l)}_j - \mu^{(l)}_j}{\sqrt{[{\bf V}_{\mu}^{(l)}]_{j,j}}}\xlongrightarrow{d}\Nc(0,1),\quad \textrm{for all $1\leq j\leq dK$, and all $1\leq l\leq L$}.$$
Applying the Bonferroni inequality over all $dK$ components and $L$ sources yields
\begin{equation}
\liminf_{n\to\infty}\mathbf{P}\left(\Ec_{2}'\right) \geq 1-\alpha_0, \;\; \textrm{where}\;\;\Ec_{2}'= \left\{\max_{1\leq l\leq L}\max_{1\leq j\leq dK}\frac{\left|\widehat{\mu}^{(l)}_j - \mu^{(l)}_j\right|}{\sqrt{[{\bf V}^{(l)}_\mu]_{j,j}}} \leq z_{\alpha_0/(dKL)}\right\}.
    \label{eq: E21 lim}
\end{equation}

Note that on the event $\Ec_1$, the following holds:
\[
\begin{aligned}
    \frac{\left|\widehat{\mu}^{(l)}_j - \mu^{(l)}_j\right|}{\sqrt{[\widehat{\bf V}^{(l)}_\mu]_{j,j} + \frac{1}{n_l}}} &\leq \frac{\left|\widehat{\mu}^{(l)}_j - \mu^{(l)}_j\right|}{\sqrt{[{\bf V}^{(l)}_\mu]_{j,j} + \frac{2}{3n_l}}}\leq \frac{\left|\widehat{\mu}^{(l)}_j - \mu^{(l)}_j\right|}{\sqrt{[{\bf V}^{(l)}_\mu]_{j,j}}}.
\end{aligned}
\]
We take the maximum over $l\in [L]$ and $j\in [dK]$ on both sides, and establish that: on the event $\Ec_1\cap \Ec_{2}'$,
\[
\begin{aligned}
    \max_{1\leq l\leq L}\max_{1\leq j\leq dK} \frac{\left|\widehat{\mu}^{(l)}_j - \mu^{(l)}_j\right|}{\sqrt{[\widehat{\bf V}^{(l)}_\mu]_{j,j} + \frac{1}{n_l}}} \leq \max_{1\leq l\leq L}\max_{1\leq j\leq dK}\frac{\left|\widehat{\mu}^{(l)}_j - \mu^{(l)}_j\right|}{\sqrt{[{\bf V}^{(l)}_\mu]_{j,j}}} \leq z_{\alpha_0/(dKL)}.
\end{aligned}
\]
Consequently, we have shown that $\Ec_{1}\cap \Ec_{2}'\subseteq \Ec_2$, which implies that
\[
\mathbf{P}(\Ec_2) \geq \mathbf{P}(\Ec_1 \cap \Ec_{2}') \geq \mathbf{P}(\Ec_1) + \mathbf{P}(\Ec_{2}') - 1.
\]
Together with \eqref{eq: E1 lim} and \eqref{eq: E21 lim}, we have
\[
\liminf_{n\to\infty}\mathbf{P}(\Ec_2) \geq 1-\alpha_0.
\]

For the event $\Ec_1\cap \Ec_2$, it holds that
\[
\mathbf{P}(\Ec_1\cap \Ec_2) \geq \mathbf{P}(\Ec_1) + \mathbf{P}(\Ec_2) - 1,
\]
which yields that
\[
\liminf_{n\to\infty}\mathbf{P}(\Ec_1\cap \Ec_2) \geq \liminf_{n\to\infty}\mathbf{P}(\Ec_1) + \liminf_{n\to\infty}\mathbf{P}(\Ec_2) - 1\geq 1-\alpha_0
\]

\subsection{Proof of Lemma \ref{prop: sampling prob}}
\label{section: proof of eq: mu resampling high prob}

We observe that
\[
\|\widehat{\U}^{[m]} - \U\|_2 \leq \sqrt{L}\max_{l\in [L]}\|\widehat{\mu}^{(l,m)} - \mu^{(l)}\|_2.
\]
As we set $n=\min_{l\in [L]} n_l$, it holds that
\[
\min_{m\in \mathcal{M}}\sqrt{n} \|\widehat{\U}^{[m]} - \U\|_2 \leq \sqrt{L} \min_{m\in \mathcal{M}} \max_{l\in [L]}\|\sqrt{n_l}\cdot (\widehat{\mu}^{(l,m)} - \mu^{(l)})\|_2.
\]
Therefore, to prove \eqref{eq: mu resampling high prob}, it suffices to show that
\begin{equation}
    \mathbf{P}\left(\min_{m\in \mathcal{M}}\max_{l\in [L]}\|\sqrt{n_l}\cdot (\widehat{\mu}^{(l,m)} - \mu^{(l)})\|_2 \leq {\rm err}_n(M)\right) \geq (1-n^{-1})\mathbf{P}(\Ec_1\cap \Ec_2).
    \label{eq: Z resampling high prob}
\end{equation}

We denote the observed data by $\mathcal{O}$, consisting of the labeled source data and the unlabeled target data, that is $\Oc = \{\X{l}_i, \Y{l}_i\}_{i\in [n_l], l\in [L]}\bigcup \{\XQ_j\}_{j\in [N]}$.
Now, we lower bound the targeted probability in \eqref{eq: Z resampling high prob} as:
\begin{equation}
    \begin{aligned}
        &\mathbf{P}\left(\min_{m\in \mathcal{M}}\max_{l\in [L]}\|\sqrt{n_l}\cdot (\widehat{\mu}^{(l,m)} - \mu^{(l)})\|_2 \leq {\rm err}_n(M)\right)\\
        &=\mathbf{E}_\Oc\left[\mathbf{P}\left(\min_{m\in \mathcal{M}}\max_{l\in [L]}\|\sqrt{n_l}\cdot (\widehat{\mu}^{(l,m)} - \mu^{(l)})\|_2 \leq {\rm err}_n(M) \mid \Oc\right)\right]\\
        &\geq \mathbf{E}_\Oc\left[\mathbf{P}\left(\min_{m\in \mathcal{M}}\max_{l\in [L]}\|\sqrt{n_l}\cdot (\widehat{\mu}^{(l,m)} - \mu^{(l)})\|_2 \leq {\rm err}_n(M) \mid \Oc\right)\cdot {\bf 1}_{\Oc\in \mathcal{E}_1\cap \mathcal{E}_2} \right],
    \end{aligned}
    \label{eq: Z resampling high prob - step 1}
\end{equation}
where $\mathbf{P}(\cdot \mid \Oc)$ denotes the conditional probability given the observed data $\Oc$, and $\mathbf{E}_\Oc$ denotes the expectation taken with respect to the observed data $\Oc$.

Next, we shall show that the rightmost term in \eqref{eq: Z resampling high prob - step 1} can be further simplified. 
For $m\notin \mathcal{M}$, the definition of $\mathcal{M}$ in \eqref{eq: filtered set} implies the existence of some $1\leq l\leq L$ and $1\leq j\leq dK$ satisfying:
\begin{equation}
    \frac{\left|\widehat{\mu}^{(l)}_j - \widehat{\mu}^{(l,m)}_j\right|}{\sqrt{[\widehat{\bf V}^{(l)}_\mu]_{j,j} + 1/n_l}} > (1+\eta_0) \cdot z_{\alpha_0/(dKL)}.
    \label{eq: implication of set M}
\end{equation}
For clarity, we take $\eta_0 = 0.1$ in the following discussions.
Then on the event $\Ec_2$ defined in \eqref{eq: E1E2}, we have
\[
\begin{aligned}
    \frac{\left| \widehat{\mu}^{(l,m)}_j- \widehat{\mu}^{(l)}_j + (\widehat{\mu}^{(l)}_j-\mu^{(l)}_j)\right|}{\sqrt{[\widehat{\bf V}^{(l)}_\mu]_{j,j} + 1/n_l}} \geq \frac{\left|\widehat{\mu}^{(l)}_j - \widehat{\mu}^{(l,m)}_j\right|}{\sqrt{[\widehat{\bf V}^{(l)}_\mu]_{j,j} + 1/n_l}} - \frac{\left|\widehat{\mu}^{(l)}_j-\mu^{(l)}_j\right|}{\sqrt{[\widehat{\bf V}^{(l)}_\mu]_{j,j} + 1/n_l}} 
    > 0.05\cdot z_{\alpha_0/(KdL)},
\end{aligned}
\]
where the last inequality follows from \eqref{eq: implication of set M} and the definition of $\Ec_2$ in \eqref{eq: E1E2}.
Therefore, we establish that on the event $\Ec_2$,
\[
{\left|\sqrt{n_l}\cdot (\widehat{\mu}^{(l,m)}_j - \mu^{(l)}_j)\right|} > 0.05\cdot z_{\alpha_0/(dKL)}\cdot {\sqrt{n_l\cdot [\widehat{\bf V}_\mu^{(l)}]_{j,j} + 1}}. 
\]
Moreover, conditional on the event $\Ec_1$, since $\max_{l\in [L]}\|\sqrt{n_l}\cdot (\widehat{\bf V}_\mu^{(l)} - {\bf V}_\mu^{(l)})\|_2 \leq 1/3,$ we further establish that
\[
\begin{aligned}
    {\left|\sqrt{n_l}\cdot (\widehat{\mu}^{(l,m)}_j - \mu^{(l)}_j)\right|}&> 0.05\cdot z_{\alpha_0/(dKL)} \cdot {\sqrt{n_l \cdot [{\bf V}_\mu^{(l)}]_{j,j} + \frac{2}{3}}}> 0.04\cdot z_{\alpha_0/(dKL)},
\end{aligned}
\]
where the last inequality holds since $\sqrt{\frac{2}{3}} \cdot 0.05 > 0.04$ and $[{\bf V}_\mu^{(l)}]_{j,j}\geq 0$.
Therefore, conditional on the event $\Ec_1\cap \Ec_2$, we have
\begin{equation}
    \max_{l\in [L]}\|\sqrt{n_l}\cdot (\widehat{\mu}^{(l,m)} - \mu^{(l)})\|_2 \geq \left|\sqrt{n_l}\cdot (\widehat{\mu}^{(l,m)}_j - \mu^{(l)}_j)\right| \geq 0.04 \cdot z_{\alpha_0/(KdL)}.
    \label{eq: lower bounf of m notin M}
\end{equation}

Recall the definition of $\err(M)$ in \eqref{eq: errM}, which goes to $0$ as $M$ grows. Thus, there exists a positive integer $M_0>0$ such that for $M\geq M_0$, $\err(M) < 0.04 \cdot z_{\alpha_0/(KdL)}$. Together with \eqref{eq: lower bounf of m notin M}, we show that on the event $\Oc\in \Ec_1\cap \Ec_2$, for any $M\geq M_0$, the following inequality holds:
$$\min_{m\notin \mathcal{M}}\max_{l\in [L]}\|\sqrt{n_l}\cdot (\widehat{\mu}^{(l,m)} - \mu^{(l)})\|_2> \err(M),$$ 
which further implies that: for any $M\geq M_0$,
{
\[
\begin{aligned}
    &\mathbf{P}\left(\min_{m\in \mathcal{M}} \max_{l\in [L]} \|\sqrt{n_l}\cdot (\widehat{\mu}^{(l,m)} - \mu^{(l)})\|_2 \leq \err(M)\mid \Oc\right)\cdot {\bf 1}_{\Oc\in \mathcal{E}_1\cap \mathcal{E}_2}  \\&
= \mathbf{P}\left(\min_{1\leq m\leq M} \max_{l\in [L]} \|\sqrt{n_l}\cdot (\widehat{\mu}^{(l,m)} - \mu^{(l)})\|_2 \leq \err(M)\mid \Oc\right)\cdot {\bf 1}_{\Oc\in \mathcal{E}_1\cap \mathcal{E}_2}.
\end{aligned}
\]
}
In the following analysis, we consider the resampling size $M$ that is larger than $M_0$.
Together with \eqref{eq: Z resampling high prob - step 1}, we have
\begin{equation}
    \begin{aligned}
        &\mathbf{P}\left(\min_{m\in \mathcal{M}}\max_{l\in [L]}\|\sqrt{n_l}\cdot (\widehat{\mu}^{(l,m)} - \mu^{(l)})\|_2 \leq {\rm err}_n(M)\right)\\
        &\geq \mathbf{E}_\Oc\left[\mathbf{P}\left(\min_{1\leq m\leq M}\max_{l\in [L]}\|\sqrt{n_l}\cdot (\widehat{\mu}^{(l,m)} - \mu^{(l)})\|_2 \leq {\rm err}_n(M) \mid \Oc\right)\cdot {\bf 1}_{\Oc\in \mathcal{E}_1\cap \mathcal{E}_2} \right],
    \end{aligned}
    \label{eq: Z resampling high prob - step 1 - all M}
\end{equation}

Now, we study the inner term of the right-hand side of \eqref{eq: Z resampling high prob - step 1 - all M}.
Note that
\[
\begin{aligned}
    &\mathbf{P}\left(\min_{1\leq m\leq M}\max_{l\in [L]}\|\sqrt{n_l}\cdot (\widehat{\mu}^{(l,m)} - \mu^{(l)})\|_2 \leq \err(M) \mid \Oc\right) \\
    &= 1 - \mathbf{P}\left(\min_{1\leq m\leq M}\max_{l\in [L]}\|\sqrt{n_l}\cdot (\widehat{\mu}^{(l,m)} - \mu^{(l)})\|_2 > \err(M) \mid \Oc\right) \\
    &= 1 - \prod_{1\leq m\leq M}\left[1 - \mathbf{P}\left(\max_{l\in [L]}\|\sqrt{n_l}\cdot (\widehat{\mu}^{(l,m)} - \mu^{(l)})\|_2 \leq \err(M) \mid \Oc\right)\right],
\end{aligned}
\]
where the second equality follows from the conditional independence of those perturbed $\{\widehat{\mu}^{(l,m)}\}_{1\leq m\leq M}$, given the source $l\in [L]$ fixed and the observed data $\Oc$. Since $1-x\leq e^{-x}$, we further lower bound the above expression as:
\[
\begin{aligned}
    &\mathbf{P}\left(\min_{1\leq m\leq M}\max_{l\in [L]}\|\sqrt{n_l}\cdot (\widehat{\mu}^{(l,m)} - \mu^{(l)})\|_2 \leq \err(M) \mid \Oc\right) \\
    &\geq 1 - \prod_{1\leq m\leq M}\exp\left[- \mathbf{P}\left(\max_{l\in [L]}\|\sqrt{n_l}\cdot (\widehat{\mu}^{(l,m)} - \mu^{(l)})\|_2 \leq \err(M) \mid \Oc\right)\right] \\
    &=1 - \exp\left[-M \cdot\mathbf{P}\left(\max_{l\in [L]}\|\sqrt{n_l}\cdot (\widehat{\mu}^{(l,m)} - \mu^{(l)})\|_2 \leq \err(M) \mid \Oc\right) \right].
\end{aligned}
\]
Therefore, after multiplying the factor ${\bf 1}_{\Oc \in \Ec_1\cap \Ec_2}$ on both sides, we obtain that
\begin{equation}
\begin{aligned}
    &\mathbf{P}\left(\min_{1\leq m\leq M}\max_{l\in [L]}\|\sqrt{n_l}\cdot (\widehat{\mu}^{(l,m)} - \mu^{(l)})\|_2 \leq \err(M) \mid \Oc\right) \cdot {\bf 1}_{\Oc \in \Ec_1\cap \Ec_2}\\
    &\geq \left( 1 - \exp\left[-M \cdot\mathbf{P}\left(\max_{l\in [L]}\|\sqrt{n_l}\cdot (\widehat{\mu}^{(l,m)} - \mu^{(l)})\|_2 \leq \err(M) \mid \Oc\right) \right]\right)\cdot {\bf 1}_{\Oc \in \Ec_1\cap \Ec_2}\\
    &= 1 - \exp\left[-M \cdot\mathbf{P}\left(\max_{l\in [L]}\|\sqrt{n_l}\cdot (\widehat{\mu}^{(l,m)} - \mu^{(l)})\|_2 \leq \err(M) \mid \Oc\right)\cdot {\bf 1}_{\Oc\in \Ec_1\cap \Ec_2} \right]\\
    &= 1 - \exp\left[-M \cdot\prod_{l\in [L]}\mathbf{P}\left(\|\sqrt{n_l}\cdot (\widehat{\mu}^{(l,m)} - \mu^{(l)})\|_2 \leq \err(M) \mid \Oc\right)\cdot {\bf 1}_{\Oc\in \Ec_1\cap \Ec_2} \right],
\end{aligned}
\label{eq: Z resampling high prob - step 2}
\end{equation}
where the last equality holds because $\widehat{\mu}^{(l,m)}$ and $\widehat{\mu}^{(l', m)}$ are independent conditional on the observed data $\Oc$ for any pair $(l,l')\in [L].$

We continue investigating the term that appears in the right-most of \eqref{eq: Z resampling high prob - step 2}:
$$\mathbf{P}\left(\|\sqrt{n_l}\cdot (\widehat{\mu}^{(l,m)} - \mu^{(l)})\|_2 \leq \err(M) \mid \Oc\right)\cdot {\bf 1}_{\Oc\in \Ec_1\cap \Ec_2}.$$
Notice that
\[
\sqrt{n_l}\cdot (\widehat{\mu}^{(l,m)} - \mu^{(l)}) = \sqrt{n_l}\cdot (\widehat{\mu}^{(l,m)} - \widehat{\mu}^{(l)}) + \sqrt{n_l}\cdot (\widehat{\mu}^{(l)} - \mu^{(l)}).
\]
For convenience, we denote 
\[
Z^{(l,m)} = \sqrt{n_l}\cdot (\widehat{\mu}^{(l,m)} - \widehat{\mu}^{(l)}),\quad \textrm{and}\quad {Z}^{(l)}= \sqrt{n_l}\cdot (\mu^{(l)}- \widehat{\mu}^{(l)}).
\]
Given the data $\Oc$, our perturbation strategy ensures that the random vector $\sqrt{n_l}\cdot(\widehat{\mu}^{(l,m)}-\widehat{\mu}^{(l)})$ follows the distribution of $\Nc(0, {n_l}\widehat{\bf V}_\mu^{(l)} + {\bf I})$. Therefore, the density of $Z^{(l,m)}$ is
\[
p\left(Z^{(l,m)}=z \mid \Oc\right) = \frac{\exp(-\frac{1}{2}z^\intercal (n_l\widehat{\bf V}_\mu^{(l)}+ {\bf I})^{-1} z)}{\sqrt{2\pi \cdot{\rm det}(n_l\widehat{\bf V}_\mu^{(l)} +{\bf I})}}.
\]
On the event $\mathcal{E}_1$, the following inequalities hold for all $l\in [L]$:
\begin{equation}
   {n_l}{\bf V}_\mu^{(l)}+ \frac{4}{3} {\bf I}\succeq {n_l}\widehat{\bf V}_\mu^{(l)}+ {\bf I} \succeq {n_l}{\bf V}_\mu^{(l)} + \frac{2}{3}{\bf I},
    \label{eq: implication of E1}
\end{equation}
which implies that
\begin{equation}
    p\left(Z^{(l,m)}=z \mid \Oc\right)   \geq \frac{\exp(-\frac{1}{2}z^\intercal ({n_l}{\bf V}_\mu^{(l)} + \frac{2}{3} {\bf I})^{-1} z)}{\sqrt{2\pi \cdot{\rm det}({n_l}{\bf V}_\mu^{(l)} + \frac{4}{3}{\bf I})}} =: g^{(l)}(z).
    \label{eq: lower bound by g(z) - interm}
\end{equation}
Therefore, we establish that for any $z\in \RR^{dK}$,
\begin{equation}
    p\left(Z^{(l,m)}=z \mid \Oc\right) \cdot {\bf 1}_{\Oc\in \mathcal{E}_1}\geq g^{(l)}(z) \cdot {\bf 1}_{\Oc\in \mathcal{E}_1}.
    \label{eq: lower bound by g(z)}
\end{equation}

We emphasize here that for the given data $\Oc$, $\{Z^{(l)}\}_{l\in [L]}$ are fixed as $\widehat{\mu}^{(l)}$ is computed upon the observed data, while $\{Z^{(l,m)}\}_{l\in [L],m\in [M]}$ is random due to the randomness of perturbation. Therefore,
\[
\begin{aligned}
    &\mathbf{P}(\|\sqrt{n_l}\cdot (\widehat{\mu}^{(l,m)} - \mu^{(l)})\|_2 \leq {\rm err}_n(M) \mid \Oc) \cdot {\bf 1}_{\Oc \in\mathcal{E}_1\cap \mathcal{E}_2} \\
    &= \mathbf{P}\left(\|Z^{(l,m)} -Z^{(l)}\|_2 \leq {\rm err}_n(M) \mid \Oc\right) \cdot {\bf 1}_{\Oc \in\mathcal{E}_1\cap \mathcal{E}_2}\\
    &= \left[\int p\left(Z^{(l,m)} = z\mid \Oc\right) \cdot {\bf 1}_{\{\|z -Z^{(l)}\|_2\leq {\rm err}_n(M)\}} d z\right] \cdot {\bf 1}_{\Oc \in\mathcal{E}_1\cap \mathcal{E}_2}.
\end{aligned}
\]
Together with \eqref{eq: lower bound by g(z)}, we further show that
\begin{equation}
    \begin{aligned}
        &\mathbf{P}(\|\sqrt{n_l}\cdot (\widehat{\mu}^{(l,m)} - \mu^{(l)})\|_2 \leq {\rm err}_n(M) \mid \Oc) \cdot {\bf 1}_{\Oc \in\mathcal{E}_1\cap \mathcal{E}_2}\\
        &\geq \left[\int g^{(l)}(z)\cdot {\bf 1}_{\{\|z-Z^{(l)}\|_2\leq {\rm err}_n(M)\}} d z\right] \cdot {\bf 1}_{\Oc \in\mathcal{E}_1\cap \mathcal{E}_2} \\
    &= \left[\int g^{(l)}\left( Z^{(l)}\right)\cdot {\bf 1}_{\{\|z -Z^{(l)}\|_2\leq {\rm err_n(M)\}}} d z\right] \cdot {\bf 1}_{\Oc \in\mathcal{E}_1\cap \mathcal{E}_2} \\
    &\quad\quad + \left[\int \left[g^{(l)}(z) - g^{(l)}\left(Z^{(l)}\right)\right]\cdot {\bf 1}_{\{\|z -Z^{(l)}\|_2\leq {\rm err}_n(M)\}} d z\right] \cdot {\bf 1}_{\Oc \in\mathcal{E}_1\cap \mathcal{E}_2} \\
    &=: {\bf (I_1)} + {\bf (I_2)}.
    \end{aligned}
    \label{eq: Zm - Zhat}
\end{equation}

Next, we further lower bound the two components ${\bf (I_1)}$ and ${\bf (I_2)}$ in the most right-hand side of the above inequality. 

\noindent \underline{\textit{Control of ${\bf (I_1)}$.}} Recall the definition of $g^{(l)}(z)$ in \eqref{eq: lower bound by g(z) - interm}, we have
\[
g^{(l)}(Z^{(l)}) = \frac{\exp(-\frac{1}{2}Z^{(l)\intercal}({n_l}{\bf V}_\mu^{(l)} + \frac{2}{3} {\bf I})^{-1} Z^{(l)\intercal})}{\sqrt{2\pi \cdot{\rm det}({n_l}{\bf V}_\mu^{(l)} + \frac{4}{3}{\bf I})}}.
\]
On the event $\mathcal{E}_1\cap \mathcal{E}_2$ defined in \eqref{eq: E1E2}, the component in its nominator satisfies that
\begin{equation}
    \begin{aligned}
    &\frac{1}{2} {Z}^{(l)\intercal}({n_l}{\bf V}_\mu^{(l)}+ \frac{2}{3} {\bf I})^{-1} {Z}^{(l)} \\
    &\leq \frac{1}{2} \frac{dK\cdot \max_{j\in [dK]} ({Z}^{(l)}_j)^2}{\lambda_{\rm min}({n_l}{\bf V}_\mu^{(l)}) + \frac{2}{3}}\\
    &\leq  \frac{dK}{2} \frac{\max_{j\in [dK]}\{{n_l}[\widehat{\bf V}^{(l)}_\mu]_{j,j} + 1\} \cdot (1.05 \cdot z_{\alpha_0/(dKL)})^2}{\lambda_{\rm min}({n_l}{\bf V}_\mu^{(l)}) + \frac{2}{3}} \\
    &\leq \frac{dK (1.05)^2 z_{\alpha_0/(dKL)}^2}{2} \cdot \frac{\lambda_{\rm max}(n_l{\bf V}_\mu^{(l)}) + \frac{4}{3}}{\lambda_{\rm min}(n_l{\bf V}_\mu^{(l)}) + \frac{2}{3}},
\end{aligned}
\label{eq: I1 - analysis 1}
\end{equation}
where the second inequality holds since on the event $\Ec_2$, 
$$\max_{j} |Z_j^{(l)}| = \max_j |\sqrt{n_l}(\mu^{(l)} - \widehat{\mu}^{(l)})|\leq 1.05 \cdot z_{\alpha_0/dKL}\cdot \max_j \sqrt{n_l[\widehat{\bf V}_\mu^{(l)}]_{j,j}+1},$$
and the third inequality holds due to \eqref{eq: implication of E1}.

We further simplify \eqref{eq: I1 - analysis 1} by controlling the term $z_{\alpha_0/(dKL)}$. Since the standard normally distributed variable $U$ satisfies $\mathbf{P}(U\geq t) \leq e^{-t^2/2}$ for any $t\geq 0$, we set $t=\sqrt{2\log (dKL/\alpha_0)}$ and obtain,
\[
\mathbf{P}\bigl(U \ge \sqrt{2\log (dKL/\alpha_0)}\bigr)
\;\le\;
\exp\bigl(-\,\tfrac12\cdot 2\log(dKL/\alpha_0)\bigr)
\;=\;
\frac{\alpha_0}{dKL}.
\]
Hence, by the definition of the $z_{\alpha_0/(dKL)}$, we have
\[
z_{\alpha_0/(dKL)} \leq \sqrt{2\log (dKL/\alpha_0)}.
\]
Therefore, it follows from \eqref{eq: I1 - analysis 1} that
\begin{equation}
    \begin{aligned}
    \frac{1}{2} {Z}^{(l)\intercal}({n_l}{\bf V}_\mu^{(l)}+ \frac{2}{3} {\bf I})^{-1} {Z}^{(l)} &\leq {dK}\cdot \log \left(\frac{dKL}{\alpha_0}\right)\cdot \left(3\cdot\lambda_{\rm max}(n_l{\bf V}^{(l)}) + 4\right).
\end{aligned}
\label{eq: I1 - analysis 2}
\end{equation}

Moreover, for the denominator of $g^{(l)}(Z^{(l)})$, we have
{\small
\[
\sqrt{2\pi\cdot {\rm det}({n_l}{\bf V}_\mu^{(l)} + \frac{4}{3}{\bf I})} = \sqrt{2\pi}\prod_{j=1}^{dK}[ \lambda_j(n_l{\bf V}_\mu^{(l)}) + 4/3 ]^{1/2} \leq  \sqrt{2\pi} \left[ \lambda_{\rm max}(n_l{\bf V}^{(l)}) + \frac{4}{3}\right]^{dK/2}.
\]}
Then together with \eqref{eq: I1 - analysis 2}, we obtain:
{\small
\begin{equation*}
    g^{(l)}\left({Z}^{(l)}\right) \geq \min_{l\in [L]}\left\{\frac{\exp\left\{-dK \log \left(\frac{dKL}{\alpha_0}\right)\left(3\|{n_l}\cdot{\bf V}_\mu^{(l)}\|_2 + 4\right)\right\}}{\sqrt{2\pi} \left[\lambda_{\rm max}({n_l}\cdot {\bf V}_\mu^{(l)}) + \frac{4}{3}\right]^{dK/2}} \right\}=:c^*(\alpha_0,d).
\end{equation*}
}

Therefore, we control ${\bf (I_1)}$ as follows:
\begin{equation}
\begin{aligned}
    {\bf (I_1)}&=\left[\int g^{(l)}\left({Z}^{(l)}\right)\cdot {\bf 1}_{\{\|z -{Z}^{(l)}\|_2\leq {\rm err}_n(M)\}} d z\right] \cdot {\bf 1}_{\Oc \in\mathcal{E}_1\cap \mathcal{E}_2} \\
    &\geq c^*(\alpha_0,d) \cdot \left[\int {\bf 1}_{\{\|z - {Z}^{(l)}\|_2\leq {\rm err}_n(M)\}} d z\right] \cdot {\bf 1}_{\Oc \in\mathcal{E}_1\cap \mathcal{E}_2} \\
    & = c^*(\alpha_0,d) \cdot {\rm Vol}(dK) \cdot [{\rm err}_n(M)]^{dK} \cdot {\bf 1}_{\Oc \in\mathcal{E}_1\cap \mathcal{E}_2},
\end{aligned}
\label{eq: Zm - Zhat, term 1}
\end{equation}
where ${\rm Vol}(dK)$ denotes the volume of the unit ball in $dK$-dimension. 

\noindent \underline{\textit{Control of ${\bf (I_2)}$.}} 
By the Mean Value Theorem, there exists some value $\nu\in (0, 1)$ such that
\[
g^{(l)}(z) - g^{(l)}\left({Z}^{(l)}\right) = [\nabla g^{(l)}({Z}^{(l)} + \nu(z - {Z}^{(l)}))]^\intercal (z - {Z}^{(l)}).
\]
We denote $w = {Z}^{(l)} + \nu(z-{Z}^{(l)})$ and $\nabla g^{(l)}(w)$ as expressed:
\[
\nabla g^{(l)}(w) = -\frac{\exp(-\frac{1}{2}w^\intercal ({n_l}{\bf V}^{(l)}_\mu + \frac{2}{3}{\bf I})^{-1} w)}{\sqrt{2\pi \cdot{\rm det}({n_l}{\bf V}^{(l)}_\mu + \frac{4}{3}{\bf I})}} ({n_l}{\bf V}_\mu^{(l)} + \frac{2}{3} {\bf I})^{-1} w.
\]
By Cauchy-Schwarz inequality, we have
\begin{equation}
    |g^{(l)}(z) - g^{(l)}({Z}^{(l)})| \leq \|\nabla g^{(l)}(w)\|_2 \|z - {Z}^{(l)}\|_2.
    \label{eq: g_l bound}
\end{equation}

Since $\lambda_{\rm min}({n_l}{\bf V}_\mu^{(l)} + \frac{2}{3}{\bf I})\geq \frac{2}{3}$, we have
\[
\begin{aligned}
    \|\nabla g^{(l)}(w)\|_2^2 &= \frac{\left[\exp(-\frac{1}{2}w^\intercal ({n_l}{\bf V}^{(l)}_\mu + \frac{2}{3}{\bf I})^{-1} w)\right]^2}{{2\pi \cdot{\rm det}({n_l}{\bf V}^{(l)}_\mu + \frac{4}{3}{\bf I})}} w^\intercal ({n_l}{\bf V}_\mu^{(l)} + \frac{2}{3} {\bf I})^{-2} w \\
    &\leq \frac{3}{{4\pi \cdot{\rm det}({n_l}{\bf V}^{(l)}_\mu + \frac{4}{3}{\bf I})}} \exp(-u)\cdot u,
\end{aligned}
\]
with $u = w^\intercal ({n_l}{\bf V}_\mu^{(l)} + \frac{2}{3} {\bf I})^{-1} w$. The function $u\mapsto \exp(-u) u$ for $u\geq 0$ has a global maximizer at $u=1$ with the maximum value $e^{-1}$. Therefore, we obtain that
\[
\|\nabla g^{(l)}(w)\|_2^2 \leq \frac{3}{{4\pi e \cdot{\rm det}({n_l}{\bf V}^{(l)}_\mu + \frac{4}{3}{\bf I})}} \leq \frac{3}{4\pi e}.
\]

Putting the above result back to \eqref{eq: g_l bound}, we obtain 
$|g^{(l)}(z) - g^{(l)}({Z}^{(l)})| \lesssim \|z-{Z}^{(l)}\|_2$. Then we establish that
\begin{equation}
\begin{aligned}
    &\left|\left[\int \left[g^{(l)}(z) - g^{(l)}({Z}^{(l)})\right]\cdot {\bf 1}_{\{\|z - {Z}^{(l)}\|_2\leq {\rm err}_n(M)\}} d z\right] \cdot {\bf 1}_{\Oc \in\mathcal{E}_1\cap \mathcal{E}_2}\right| \\
    &\leq C\cdot \left|\left[\int \|z-Z^{(l)}\|_2\cdot {\bf 1}_{\{\|z - {Z}^{(l)}\|_2\leq {\rm err}_n(M)\}} d z\right] \cdot {\bf 1}_{\Oc \in\mathcal{E}_1\cap \mathcal{E}_2}\right|\\
    &\leq C\cdot {\rm err}_n(M)\cdot \int {\bf 1}_{\{\|z - {Z}^{(l)}\|_2\leq {\rm err}_n(M)\}} dz \cdot {\bf 1}_{\Oc \in\mathcal{E}_1\cap \mathcal{E}_2} \\
    &= C\cdot {\rm err}_n(M)\cdot {\rm Vol}(dK) \cdot [{\rm err_n}(M)]^{dK} \cdot {\bf 1}_{\Oc \in\mathcal{E}_1\cap \mathcal{E}_2}, 
\end{aligned}
\label{eq: Zm - Zhat, term 2}
\end{equation}
for some constant $C>0$.
Since $\err(M)$ decreases with $M$ and $\lim_{M\to \infty}\err(M) = 0$, there exists a positive integer $M_1$ such that for $M\geq M_1$, we have $\err(M)\leq \frac{1}{2}c^*(\alpha_0,d)$. 
Therefore, for $M\geq M_1$, we control ${\bf (I_2)}$ as:
\[
\left|{\bf (I_2)}\right| < \frac{1}{2}c^*(\alpha_0, d) \cdot {\rm Vol}(dK) \cdot [{\rm err_n}(M)]^{dK} \cdot {\bf 1}_{\Oc \in\mathcal{E}_1\cap \mathcal{E}_2}. 
\]

Together with \eqref{eq: Zm - Zhat}, \eqref{eq: Zm - Zhat, term 1}, and \eqref{eq: Zm - Zhat, term 2}, we obtain that for any resampling time $M\geq \max\{M_0,M_1\}$,
\[
\begin{aligned}
    &\mathbf{P}(\|\sqrt{n_l}\cdot (\widehat{\mu}^{(l,m)} - \mu^{(l)})\|_2 \leq {\rm err}_n(M) \mid \Oc) \cdot {\bf 1}_{\Oc \in\mathcal{E}_1\cap \mathcal{E}_2} \\
    &\geq \frac{1}{2}c^*(\alpha_0, d) \cdot {\rm Vol}(dK) \cdot [{\rm err}_n(M)]^{dK} \cdot {\bf 1}_{\Oc \in\mathcal{E}_1\cap \mathcal{E}_2}.
\end{aligned}
\]
Putting the above inequality to \eqref{eq: Z resampling high prob - step 2}, we have
\begin{equation*}
\begin{aligned}
    &\mathbf{P}\left(\min_{1\leq m\leq M}\max_{l\in [L]}\|\sqrt{n_l}\cdot (\widehat{\mu}^{(l,m)} - \mu^{(l)})\|_2 \leq \err(M) \mid \Oc\right) \cdot {\bf 1}_{\Oc \in \Ec_1\cap \Ec_2} \\
    &\geq \left(1 - \exp\left[-M \cdot \left(\frac{1}{2}c^*(\alpha_0,d)\cdot {\rm Vol}(dK) \cdot [{\rm err}_n(M)]^{dK}\right)^L\right]\right)  \cdot {\bf 1}_{\Oc \in \Ec_1\cap \Ec_2}.
\end{aligned}
\end{equation*}
Together with \eqref{eq: Z resampling high prob - step 1 - all M},
\begin{equation*}
    \begin{aligned}
        &\mathbf{P}\left(\min_{m\in \mathcal{M}}\max_{l\in [L]}\|\sqrt{n_l}\cdot (\widehat{\mu}^{(l,m)} - \mu^{(l)})\|_2 \leq {\rm err}_n(M)\right)\\
        &\geq \left(1 - \exp\left[-M \cdot \left(\frac{1}{2}c^*(\alpha_0,d)\cdot {\rm Vol}(dK) \cdot [{\rm err}_n(M)]^{dK}\right)^L\right]\right)  \cdot \mathbf{P}(\mathcal{E}_1\cap \mathcal{E}_2).
    \end{aligned}
    \label{eq: Z resampling high prob - step 3}
\end{equation*}
Recall that we set
\[
\begin{aligned}
    \err(M) = \left(\frac{1}{2}c^*(\alpha_0,d) {\rm Vol}(dK)\right)^{-\frac{1}{dK}}\left(\frac{\log n}{M}\right)^{\frac{1}{dKL}},
\end{aligned}
\]
then for any $M\geq \max(M_0, M_1)$,
\[
\mathbf{P}\left(\min_{m\in \mathcal{M}}\max_{l\in [L]}\|\sqrt{n_l}\cdot (\widehat{\mu}^{(l,m)} - \mu^{(l)})\|_2 \leq {\rm err}_n(M)\right) \geq (1 - n^{-1})\cdot \mathbf{P}(\mathcal{E}_1\cap \mathcal{E}_2).
\]

Now, we verify when $M\geq \max(M_0, M_1)$. For $M_0$, we need $\err(M)< 0.04\cdot z_{\alpha_0/dKL}$ for any $M\geq M_0$. And for any $M_1$, we need $\err(M)\leq C\cdot c^*(\alpha_0,d)=C\cdot (\alpha_0/dKL)^{CdK}$, for some constant $C>0$. Since both the values of $z_{\alpha_0/dKL}$ and $(\alpha_0/dKL)^{CdK}$ are increasing with $d$, it suffices to let
\[
\err(M)\leq C\cdot \min\left\{0.04 z_{\alpha_0}, \alpha_0\right\}.
\]
As we set $\alpha_0\in (0,0.01)$, we must have $z_{\alpha_0}\geq 2$, and $\min\left\{0.04 z_{\alpha_0}, \alpha_0\right\}=\alpha_0$. Thus, the preceding inequality holds if
\[
M\geq C\cdot  \left(\frac{1}{\alpha_0}\right)^{dKL}\left(\frac{1}{2}c^*(\alpha_0,d){\rm Vol}(dK)\right)^{-L} \cdot \log n.
\]

\subsection{Proof of Lemma \ref{lemma: grad Shat limiting}}
\label{proof of prop grad Shat limiting}

It follows the expression of $\nabla\widehat{S}(\theta)$ and $\nabla S(\theta)$ that
\[
\nabla\widehat{S}(\theta)-\nabla \widehat{S}(\theta^*) = \frac{1}{N}\sum_{j=1}^N \left[\left(p(X_j^\QQ,\theta) - p(X_j^\QQ,\theta^*)\right)\otimes X_j^\QQ\right]
\]
and
\[
\nabla S(\theta) - \nabla S(\theta^*) = \E_{\QQ_X}[\left(p(X,\theta) - p(X,\theta^*)\right)\otimes X].
\]
For any $\theta \in \Theta_{\rm loc}$, we define
\[
Z_j(\theta) = \left[\left(p(X_j^\QQ,\theta) - p(X_j^\QQ,\theta^*)\right)\otimes X_j^\QQ\right] - \E_{\QQ_X}[\left(p(X,\theta) - p(X,\theta^*)\right)\otimes X],
\]
then we have
\begin{equation}
    \left(\nabla\widehat{S}(\theta)-\nabla \widehat{S}(\theta^*)\right) -\left(\nabla S(\theta) - \nabla S(\theta^*)\right) = \frac{1}{N}\sum_{j=1}^N Z_j(\theta),
    \label{eq: grad Shat - grad S}
\end{equation}
with $\E_{\QQ}[Z_j(\theta)] = 0$.

We denote the $\epsilon$-net of the space $\Theta_{\rm loc}$ as $\mathcal{N}_{\epsilon}=\{\theta^{1},\cdots,\theta^{\tau}, \cdots, \theta^{|\mathcal{N}_{\epsilon}|}\}$ such that, for any $\theta\in \Theta_{\rm loc}$, there exists $\theta^{\tau}\in \mathcal{N}_{\epsilon}$ such that $\|\theta-\theta^{\tau}\|_2\leq \epsilon$. By the standard result in high-dimensional geometry, there exists an $\epsilon$-net of $\Theta_{\rm loc}$ (which has a radius $C\cdot\tau \sqrt{d/n}$, for some constant $C>0$) satisfying that
\[
|\mathcal{N}_\epsilon| \leq \left(C\cdot \frac{\tau \sqrt{d/n}}{\epsilon}\right)^{dK}.
\]
By setting $\epsilon=\tau\cdot \sqrt{d/n}\cdot  N^{-1}$, then we have
\[
|\mathcal{N}_\epsilon|\leq CN^{dK},
\]
for some constant $C>0$.

Now, we study the property for each fixed $\theta^\tau\in \Nc_\epsilon$. 
It follows from \eqref{eq: p theta diff} that
\[
\|p(X,\theta^\tau)-p(X,\theta^*)\|_2\leq \Delta(X,\theta^{\tau}-\theta^*) \quad \text{with}\quad \Delta(X,\theta^{\tau}-\theta^*)=\sqrt{\sum_{k=1}^{K}\left[X^{\intercal}(\theta^\tau_k-\theta^{*}_k)\right]^2}.
\]
Therefore, for each category $k\in [K]$, the function $\theta\mapsto p_k(x,\theta)$ is Lipschitz with
\[
\left|p_k(X;\theta^\tau) - p_k(X; \theta^*)\right|\lesssim \|X\|_2 \|\theta^\tau - \theta^*\|_2 \lesssim\tau \sqrt{d/n}\|X\|_2,
\]
where the last inequality holds as $\theta_\tau \in \Theta_{\rm loc}$ with $\|\theta^\tau - \theta^*\|_2\lesssim \tau\sqrt{d/n}.$
Then
\[
\E_{\QQ_X}\left[\left|p_k(X,\theta^\tau) - p_k(X,\theta^*)\right| \right] \lesssim \tau \sqrt{d/n} \E_{\QQ_X}[\|X\|_2] \lesssim\tau d/\sqrt{n},
\]
where the last inequality holds due to the covariate $X$ being assumed to be subgaussian. Then we apply the result in Lemma \ref{lemma: subgaus of f(X)X}, and establish that with probability at least $1-N^{-c_1d}$:
\[
\left\|\frac{1}{N}\sum_{j=1}^N Z_j(\theta^\tau)\right\|_2 \lesssim(\tau d/\sqrt{n}) \cdot \sqrt{d\log N/N}= \tau \sqrt{\frac{d^3 \log N}{n\cdot N}}.
\]
Applying the union bound over all $\theta^\tau \in \mathcal{N}_\epsilon$, we have with probability at least $1-N^{-c_1d + dK}$,
\begin{equation}
    \max_{\theta^\tau \in \mathcal{N}_\epsilon}\left\|\frac{1}{N}\sum_{j=1}^N Z_j(\theta^\tau)\right\|_2 \lesssim\tau \sqrt{\frac{d^3 \log N}{n\cdot N}}.
    \label{eq: theta_tau Z max}
\end{equation}

Now, we continue exploring for each $\theta \in \Theta_{\rm loc}$. 
For any $\theta \in \Theta_{\rm loc}$, we pick the $\theta^\tau \in \mathcal{N}_\epsilon$ that is closest to $\theta$, which satisfies that $\|\theta - \theta^\tau\|_2 \leq \epsilon = \tau \sqrt{d/n} N^{-1}$.
We observe that:
\[
\begin{aligned}
    &\frac{1}{N}\sum_{j=1}^N \left(Z_j(\theta) - Z_j(\theta^\tau)\right) \\
    &= \frac{1}{N}\sum_{j=1}^N \left[\left(p(\XQ_j,\theta) - p(\XQ_j, \theta^\tau)\right)\otimes \XQ_j\right] - \E_{\QQ_X}\left[\left(p(X,\theta) - p(X,\theta^\tau)\right)\otimes X\right].
\end{aligned}
\]
It follows that:
{\small
\[
\begin{aligned}
    &\left\|\frac{1}{N}\sum_{j=1}^N \left(Z_j(\theta) - Z_j(\theta^\tau)\right) \right\|_2\\
    &\leq \frac{1}{N}\sum_{j=1}^N \left\|\left(p(\XQ_j,\theta) - p(\XQ_j, \theta^\tau)\right)\otimes \XQ_j\right\| + \E_{\QQ_X}\left[\left\|\left(p(X,\theta) - p(X,\theta^\tau)\right)\otimes X\right\|_2\right] \\
    &=\frac{1}{N}\sum_{j=1}^N \left\|p(\XQ_j,\theta) - p(\XQ_j, \theta^\tau)\right\|_2\left\|\XQ_j\right\|_2 + \E_{\QQ_X}\left[\left\|p(X,\theta) - p(X,\theta^\tau)\right\|_2\|X\|_2\right],
\end{aligned}
\]
}
where the last equality holds as $\|u\otimes v\|_2 = \|u\|_2 \|v\|_2$ for two vectors $u, v$. We now apply \eqref{eq: p theta diff} to obtain that:
\[
\|p(X,\theta) - p(X,\theta^\tau)\|_2\lesssim \|X\|_2 \|\theta-\theta^\tau\|_2 \leq \tau\sqrt{d/n} N^{-1} \|X\|_2.
\]
Combining the preceding two results, we establish that:
{\small
\[
\begin{aligned}
    \left\|\frac{1}{N}\sum_{j=1}^N \left(Z_j(\theta) - Z_j(\theta^\tau)\right) \right\|_2 &\leq \left[\frac{1}{N}\sum_{j=1}^N \|\XQ_j\|_2^2 + \E_{\QQ_X}\|X\|_2^2\right]\cdot \tau\sqrt{d/n} N^{-1}.
\end{aligned}
\]
}
In Condition {\rm (A2)}, we consider the target covariate to be bounded for each component, thus we obtain that:
\[
\left\|\frac{1}{N}\sum_{j=1}^N \left(Z_j(\theta) - Z_j(\theta^\tau)\right) \right\|_2 \lesssim\tau \sqrt{d^3/n} N^{-1}.
\]

Together with \eqref{eq: theta_tau Z max}, we obtain that, for any $\theta\in \Theta_{\rm loc}$, with probability at least $1-N^{-c_1d}$,
\[
\begin{aligned}
    \left\|\frac{1}{N}\sum_{j=1}^N Z_j(\theta)\right\|_2 &\leq \max_{\theta^\tau\in \mathcal{N}^\epsilon}\left(\left\|\frac{1}{N}\sum_{j=1}^N \left(Z_j(\theta) - Z_j(\theta^\tau)\right) \right\|_2+\left\|\frac{1}{N}\sum_{j=1}^N Z_j(\theta^\tau)\right\|_2\right)\\
    &\lesssim   \tau \left(\sqrt{\frac{d^3 \log N }{n \cdot N}} + \sqrt{\frac{d^3}{n \cdot N^2}}\right).
\end{aligned}
\]
Since $N\geq \log N$, it follows from \eqref{eq: grad Shat - grad S} that
\[
\left\|\left(\nabla\widehat{S}(\theta)-\nabla \widehat{S}(\theta^*)\right) -\left(\nabla S(\theta) - \nabla S(\theta^*)\right)\right\|_2 \lesssim \tau\sqrt{\frac{d^3 \log N }{n \cdot N}}.
\]
Since it holds for any $\theta \in \Theta_{\rm loc}$, we establish that
\[
\sup_{\theta \in \Theta_{\rm loc}}\left\|\left(\nabla\widehat{S}(\theta)-\nabla \widehat{S}(\theta^*)\right) -\left(\nabla S(\theta) - \nabla S(\theta^*)\right)\right\|_2 \lesssim \tau\sqrt{\frac{d^3 \log N }{n \cdot N}}.
\]

\subsection{Proof of Lemma \ref{lem: resample theta convergence}}
\label{proof of lemma resample theta}
Recall that $\widehat{\phi}(\theta,\gamma) = \theta^\intercal \widehat{\U}\gamma + \widehat{S}(\theta)$, then $\thetainit^{[m]}$, defined in \eqref{eq: m-opt}, shall be expressed as 
\[
\thetainit^{[m]} = \argmin_{\theta\in \RR^{dK}}\widehat{\phi}(\theta,\hgammam).
\]
To control the distance of $\|\thetainit^{[m]} - \theta^*\|_2$, we first define an intermediate estimator $\thetainit^{[m]}_B$ as follows
\begin{equation}
\thetainit^{[m]}_B=\argmin_{\theta\in \Theta_{B} }\widehat{\phi}(\theta,\hgammam).
\label{eq: m-opt constrained}
\end{equation}
where $\Theta_{B}$ is the bounded set defined in \eqref{eq: bound para}.
In the following, we will establish the convergence rate of $\|\thetainit^{[m]}_B - \theta^*\|_2$, and then we show that $\thetainit^{[m]}_B=\thetainit^{[m]}$ to complete the proof.

The optimality condition of $\thetainit^{[m]}_B$ in \eqref{eq: m-opt constrained} implies that 
\begin{equation*}
\left\langle \widehat{\U}\widehat{\gamma}^{[m]} +\nabla \widehat{S}({\thetainit}_{B}^{[m]}),\theta-{\thetainit}_{B}^{[m]}\right\rangle \geq 0, \quad \text{for all} \quad \theta\in \Theta_{B}.
\end{equation*}
We substitute $\theta$ with $\theta^*$ in the above inequality:
\begin{equation}
\left\langle \widehat{\U}\widehat{\gamma}^{[m]} +\nabla \widehat{S}({\thetainit}_{B}^{[m]}),\theta^*-{\thetainit}_{B}^{[m]}\right\rangle \geq 0.
\label{eq: opt inequ}
\end{equation}
The optimality condition of $\theta^*$ in Proposition \ref{prop: local theta} implies
\begin{equation*}
\U\gamma^*+\nabla {S}(\theta^*)=0.
\end{equation*}
Together with \eqref{eq: opt inequ}, we have
\begin{equation*}
\left\langle \widehat{\U}\hgammam +\nabla \widehat{S}({\thetainit}_{B}^{[m]})-\U\gamma^*-\nabla {S}(\theta^*),\theta^*-{\thetainit}_{B}^{[m]}\right\rangle \geq 0.
\end{equation*}
Moreover, since
$$\nabla \widehat{S}({\thetainit}_{B}^{[m]})-\nabla {S}(\theta^*)=\nabla\widehat{S}({\thetainit}_{B}^{[m]})-\nabla\widehat{S}({\theta}^{*})+\nabla\widehat{S}({\theta}^*)-\nabla {S}(\theta^*),$$
the above inequality further implies 
\begin{equation*}
\left\langle \nabla\widehat{S}({\thetainit}_{B}^{[m]})-\nabla\widehat{S}({\theta}^{*}),{\thetainit}_{B}^{[m]}-\theta^*\right\rangle\leq \left\langle \widehat{\U}\hgammam-\U\gamma^*+\nabla\widehat{S}({\theta}^*)-\nabla {S}(\theta^*) ,\theta^*-{\thetainit}_{B}^{[m]}\right\rangle
\end{equation*}

Note that on the event $\GG$, we have
\[
\left|\langle \widehat{\U}\hgammam - \U\gamma^*, \frac{\theta^* - {\thetainit}_{B}^{[m]}}{\|\theta^* - {\thetainit}_{B}^{[m]}\|_2} \rangle \right| \leq \|\U\|_2 \|\hgammam - \gamma^*\|_2 + \|\widehat{\U}-\U\|_2 \leq \|\U\|_2\|\hgammam - \gamma^*\|_2 + Ct\sqrt{d/n}, 
\]
and
\[
\left|\langle \nabla \widehat{S}(\theta^*) - \nabla S(\theta^*), \theta^* - {\thetainit}_{B}^{[m]}\rangle \right| \leq C\sqrt{d\log N/N} \|\theta^* - {\thetainit}_{B}^{[m]}\|_2.
\]
Therefore, we show that
\begin{equation}
\begin{aligned}
     &\left\langle \nabla\widehat{S}({\thetainit}_{B}^{[m]})-\nabla\widehat{S}({\theta}^{*}),{\thetainit}_{B}^{[m]}-\theta^*\right\rangle \\
     &\leq \left(\|\U\|_2\|\hgammam - \gamma^*\|_2 + Ct\sqrt{d/n} + C\sqrt{d\log N/N}\right)\cdot \|\theta^* - {\thetainit}_{B}^{[m]}\|_2.
\end{aligned}
    \label{eq: key inter inequalities}
\end{equation}
For the left-hand side of the above inequality, by the mean value theorem, there exists $\nu\in [0,1]$ such that  
$$\left\langle \nabla\widehat{S}({\thetainit}_{B}^{[m]})-\nabla\widehat{S}({\theta}^{*}),{\thetainit}_{B}^{[m]}-\theta^*\right\rangle= 
\left({\thetainit}_{B}^{[m]}-{\theta}^{*}\right)^{\intercal} \widehat{H}\left((1-\nu){\theta}^{*}+\nu {\thetainit}_{B}^{[m]}\right)\left({\thetainit}_{B}^{[m]}-{\theta}^{*}\right).$$
On the event $\mathcal{G}$, Lemma \ref{lem: strong convex of hatH} implies the positive definiteness of $\widehat{H}(\theta)$ for $\theta\in \Theta_B$, then we establish the following lower bound, 
$$\left\langle \nabla \widehat{S}({\thetainit}_{B}^{[m]})-\nabla \widehat{S}({\theta}^{*}),{\thetainit}_{B}^{[m]}-\theta^*\right\rangle\gtrsim \|{\thetainit}_{B}^{[m]}-{\theta}^{*}\|_2^2.$$

Together with \eqref{eq: key inter inequalities}, and the bounded assumption $\|\U\|_2$ in Condition {\rm (A4)}, we have shown that 
\begin{equation}
\begin{aligned}
    \|{\thetainit}_{B}^{[m]}-{\theta}^{*}\|_2&\lesssim \|\widehat{\gamma}^{[m]}-\gamma^*\|_2+ t\sqrt{d/n}+  \sqrt{d\log (N)/N} \\
    &\lesssim \|\widehat{\gamma}^{[m]}-\gamma^*\|_2+ t\sqrt{d/n},
\end{aligned}
\label{eq: inter bound}
\end{equation}
where the second inequality holds as $N\gg n\log n$ in Condition {\rm (A2)} and $t>c_0$ in Condition {\rm (A1)}.
It follows that
\begin{equation*}
\begin{aligned}
    \|{\thetainit}_{B}^{[m]}-{\theta}^{*}\|_1 &\leq \sqrt{d}\|{\thetainit}_{B}^{[m]}-{\theta}^{*}\|_2 \lesssim\sqrt{d}\left( \|\widehat{\gamma}^{[m]}-\gamma^*\|_2+ t\sqrt{d/n}\right).
\end{aligned}
\end{equation*}
When the value $t$ satisfies \eqref{eq: t condition}, we apply the conditions of $n,N$ in Condition {\rm (A2)} to obtain that: there exists a positive constant $C>0$, such that if $C\sqrt{d} \|\widehat{\gamma}^{[m]}-\gamma^*\|_2 < 1$, then
\begin{equation}
    \|\thetainit^{[m]}_B - \theta^*\|_1< 1.
    \label{eq: inter bound B}
\end{equation}
In the following, we show that $\thetainit^{[m]}_B = \thetainit^{[m]}$ by the argument of contradiction.

Suppose that ${\thetainit}_{B}^{[m]}\neq \widehat{\theta}^{[m]}$, for any $\theta\in \Theta_B$, Lemma \ref{lem: strong convex of hatH} implies the positive definiteness of the Hessian matrix for $\widehat{\phi}(\theta,\widehat{\gamma}^{[m]})$. Therefore, $\widehat{\phi}(\theta, \widehat{\gamma}^{[m]})$ is strongly convex towards $\theta\in \Theta_B$, and we must have $\thetainit^{[m]}\notin \Theta_B$. By the definition of $\thetainit^{[m]}$, it holds that $\widehat{\phi}(\thetainit^{[m]}, \hgammam) \leq \widehat{\phi}(\thetainit^{[m]}_B, \hgammam)$. 
Then, by the convexity of $\widehat{\phi}(\theta,\hgammam)$ for $\theta\in \RR^{dK}$, for any $\nu\in [0,1]$,
\[
\widehat{\phi}(
\nu\thetainit^{[m]} + (1-
\nu)\thetainit^{[m]}_B,\; \hgammam)\leq 
\nu\widehat{\phi}(\thetainit^{[m]}, \hgammam) + (1-
\nu)\widehat{\phi}(\thetainit^{[m]}_B,\hgammam) \leq \widehat{\phi}(\thetainit^{[m]}_B, \hgammam).
\]
Note that the pathway $
\nu\to 
\nu\thetainit^{[m]} + (1-\nu)\thetainit^{[m]}_B$ intersects $\Theta_B$ at a point, denoted as $\theta^{\rm inter}$. Together with the above inequality, we have $\widehat{\phi}(\theta^{\rm inter},\hgammam)\leq \widehat{\phi}(\thetainit^{[m]}_B, \hgammam)$ and $\theta^{\rm inter}\in \Theta_B$. Therefore, we establish that
\[
\theta^{\rm inter} \in \argmin_{\theta\in \Theta_B}\widehat{\phi}(\theta, \hgammam), \quad \textrm{and}\quad \|\theta^{\rm inter} - \theta^*\|_1 = 1.
\]
By the strong convexity of $\widehat{\phi}(\theta,\hgammam)$ for $\theta\in \Theta_B$, we further obtain that $\theta^{\rm inter} = \thetainit^{[m]}_B$, which implies that
\[
\|\theta^{[m]}_B - \theta^*\|_2 = 1.
\]
It contradicts \eqref{eq: inter bound B}, therefore we show that $\thetainit^{[m]}_B = \thetainit^{[m]}$. Then together with \eqref{eq: inter bound}, we complete the proof.

\subsection{Proof of Lemma \ref{lem: derive event3}}
\label{proof of lem: derive event3}

To control the probability of the event $\Ec_3$, we shall study the terms $\|{\rm Rem}^{[m^*]}\|_2$, and $\|n(\widehat{\bf V}^{[m^*]}-{\bf V}^*)\|_2$. For the term $\|{\rm Rem}^{[m^*]}\|_2$, if we verify that $\widehat{\theta}^{[m^*]}\in \Theta_{\rm loc}$, we can control $\|{\rm Rem}^{[m^*]}\|_2$ using the remainder's upper bound \eqref{eq: remainder bound theta_m} in Proposition \ref{prop: theta_m nor}, such that: on the event $\GG\cap \GG_4$,
\[
\left\| {\rm Rem}^{[m^*]}\right\|_2 \lesssim \|\widehat{\gamma}^{[m^*]}-\gamma^*\|_2 + \tau\sqrt{\frac{d^3\log N}{n\cdot N}} + \tau^2\frac{d}{n}.
\]
As for $\|n(\widehat{\bf V}^{[m^*]}-{\bf V}^*)\|_2$, we now introduce the following lemma to establish its upper bound.
The proof is provided in Section \ref{proof of lem: consistency of Vm}.
\begin{Lemma}
\label{lem: consistency of Vm}
    Suppose Conditions {\rm (A1)}--{\rm (A5)} hold, and the value $\tau$ defined in \eqref{eq: final rate} satisfies $\tau\lesssim \sqrt{n/d^2}$. Then for any index $m$ whose corresponding $\widehat{\theta}^{[m]}\in \Theta_{\rm loc}$ with $\Theta_{\rm loc}$ defined in \eqref{eq: Theta local - new}, we have that on the event $\GG_3$,
    \[
        \|n\cdot (\widehat{\bf V}^{[m]}-{\bf V}^*)\|_2\lesssim
    \max_{l\in [L]}\|n_l\cdot (\widehat{\bf V}_\mu^{(l)} - {\bf V}_\mu^{(l)})\|_2 + \|\widehat{\gamma}^{[m]}-\gamma^*\|_2 + \frac{n}{N} + \sqrt{d\log N/N},
    \]
    where $\GG_3$ is defined in \eqref{eq: events}.
\end{Lemma}
This lemma implies that, if we verify $\widehat{\theta}^{[m^*]}\in \Theta_{\rm loc}$, then we can control $\|n\cdot (\widehat{\bf V}^{[m^*]}-{\bf V}^*)\|_2$ via $\|\widehat{\gamma}^{[m^*]}-\gamma^*\|_2$.
Therefore, the following proof is outlined as follows:
\begin{itemize}
    \item[\bf Step 1.] Establish the upper bound of $\|\widehat{\gamma}^{[m^*]}-\gamma^*\|_2$, and justify that $\widehat{\theta}^{[m^*]}\in \Theta_{\rm loc}$, with high probability.
    \item[\bf Step 2.] Apply \eqref{eq: remainder bound theta_m} in Proposition \ref{prop: theta_m nor} and Lemma \ref{lem: consistency of Vm} to control $\left\| {\rm Rem}^{[m^*]}\right\|_2$ and  $\|n\cdot (\widehat{\bf V}^{[m]}-{\bf V}^*)\|_2$, and then finish the proof.
\end{itemize}

\subsubsection{Proof for Step 1}
We set the value $t$ as $t=\sqrt{\log n}$ in this proof. Then $\tau$ defined in \eqref{eq: final rate} admits the form:
\[
\tau = C\left(1+\frac{1}{\sigma_L^2(\U)}\right)t = C\left(1+\frac{1}{\sigma_L^2(\U)}\right)\sqrt{\log n},
\]
and
$\Theta_{\rm loc}$ defined in \eqref{eq: Theta local - new} shall be expressed as:
\begin{equation}
   \Theta_{\rm loc}=\left\{\theta\in \RR^{dK}: \|\theta-\theta^*\|_2\lesssim \left(1+\frac{1}{\sigma_L^2(\U)}\right) \sqrt{d\log n/n}\right\}.
   \label{eq: local set - logn}
\end{equation}

We now present the following high-probability event to control the rate of $\|\widehat{\gamma}^{[m^*]}-\gamma^*\|_2$.
{\small
\begin{equation}
    \Ec_4= \left\{\|\widehat{\gamma}^{[m^*]}-\gamma^*\|_2\lesssim \frac{1}{\sigma_L^2(\U)}\left(\frac{{\rm err}_n(M)}{\sqrt{n}} +\sqrt{\frac{d\log N}{N}}\right) + \left(1+\frac{1}{\sigma_L^2(\U)}\right)^2 \log n \left(\frac{d}{n}\right)^{3/4}\right\},
    \label{eq: event E4}
\end{equation}
}
On the right-hand side of the inequality of the event $\Ec_4$, the first term corresponds to the upper bound for $\|\widehat{\gamma}^{[m^*]}-\gamma^*_{\rm ap}\|_2$, which is controlled via Theorem \ref{thm: min_gamma_m}, while the second term corresponds to the upper bound for $\|\gamma^*_{\rm ap}-\gamma^*\|_2$ that is established in Theorem \ref{thm: gamma true}, with the value $t=\sqrt{\log n}$.

The following lemma formally justifies that $\Ec_4$ happens with high probability, whose proof is provided in Section \ref{proof of lemma E4}.
\begin{Lemma}
\label{lem: E4}
    Suppose Conditions {\rm (A1)}--{\rm (A5)} hold, and $\sigma_L^2(\U)\gg \sqrt{\log n}(d/n)^{1/4}$. Then 
    \[
    \liminf_{n\to\infty}\liminf_{M\to\infty} \mathbf{P}(\Ec_4)\geq 1-\alpha_0,
    \]
    where $\alpha_0\in (0, 0.01]$ is the pre-specified constant used to construct $\mathcal{M}$ in \eqref{eq: filtered set}.
\end{Lemma}

In the following discussions, we omit the dependence on the dimension $d$, as we consider the fixed dimension scenario for statistical inference for clarity.
On the event $\Ec_4$, with $\sigma_L^2(\U)\gg \max\{\sqrt{n\log N/N}, \; \sqrt{\log n} \cdot n^{-1/8}\}$ and a sufficiently large resampling size $M$ such that $\err(M)\ll \sigma_L^2(\U)$, it holds that:
\[
\|\widehat{\gamma}^{[m^*]}-\gamma^*\|_2\lesssim \frac{1}{\sigma_L^2(\U)}\left(\frac{{\rm err}_n(M)}{\sqrt{n}} +\sqrt{\frac{\log N}{N}}\right) + \left(1+\frac{1}{\sigma_L^2(\U)}\right)^2\log n \cdot n^{-3/4} \ll n^{-1/2}.
\]
Combined with Lemma \ref{lem: resample theta convergence}, which provides an upper bound on $\|\widehat{\theta}^{[m^*]}-\theta^*\|_2$ via $\|\widehat{\gamma}^{[m^*]}-\gamma^*\|_2$, we obtain that on the event $\GG\cap \Ec_4$,
\[
\|\widehat{\theta}^{[m^*]}-\theta^*\|_2 \lesssim \|\widehat{\gamma}^{[m^*]}-\gamma^*\|_2+\sqrt{\log n/n}\lesssim \sqrt{\log n/n}.
\]
Therefore, we conclude that on the event $\GG\cap \Ec_4$, with $\sigma_L^2(\U)\gg \max\{\sqrt{n\log N/N}, \; \sqrt{\log n} \cdot n^{-1/8}\}$ and a sufficiently large resampling size $M$ such that $\err(M)\ll \sigma_L^2(\U)$, it follows from \eqref{eq: local set - logn} that
$$\widehat{\theta}^{[m^*]}\in \Theta_{\rm loc}.$$

\subsubsection{Proof for Step 2}
Provided that $\widehat{\theta}^{[m^*]}\in \Theta_{\rm loc}$ as established in step 1, 
we now turn to establishing upper bounds for the two quantities: $\|n\cdot (\widehat{\bf V}^{[m^*]}-{\bf V}^*)\|_2$ and $\|{\rm Rem}^{[m^*]}\|_2$. 
Still, we omit the dependence on the dimension $d$ in the proof, as we consider the fixed dimension for the statistical inference part, for clarity.

We apply Lemma \ref{lem: consistency of Vm} to obtain that, whenever $\sigma_L^2(\U)\gg \max\{\sqrt{n\log N/N}, \; \sqrt{\log n} \cdot n^{-1/8}\}$ and the resampling size $M$ satisfies $\err(M)\ll \sigma_L^2(\U)$, on the event $\GG\cap \Ec_4$, 
\[
\begin{aligned}
        \|n\cdot (\widehat{\bf V}^{[m^*]}-{\bf V}^*)\|_2&\lesssim
    \max_{l\in [L]}\|n_l\cdot (\widehat{\bf V}_\mu^{(l)} - {\bf V}_\mu^{(l)})\|_2 +  \|\widehat{\gamma}^{[m^*]}-\gamma^*\|_2 + \frac{n}{N} + \sqrt{\log N/N}.
\end{aligned}
\]
Using the remainder's upper bound \eqref{eq: remainder bound theta_m} in Proposition \ref{prop: theta_m nor}, we control $\|{\rm Rem}^{[m^*]}\|_2$ in the decomposition of $\widehat{\theta}^{[m^*]}-\theta^*$. We obtain that on the event $\GG\cap \GG_4\cap \Ec_4$, when $t=\sqrt{\log n}$,
{
\begin{equation*}
    \left\|{\rm Rem}^{[m^*]}\right\|_2 \lesssim \|\widehat{\gamma}^{[m^*]}-\gamma^*\|_2 + \left(1+\frac{1}{\sigma_L^2(\U)}\right)\sqrt{\frac{\log n\log N}{n\cdot N}} + \left(1+\frac{1}{\sigma_L^2(\U)}\right)^2\frac{\log n}{n}.
\end{equation*}}
As we consider $n{\bf V}^*_{1,1}\in [c_1,c_2]$ in Condition {\rm (A5)}, the above results yield the following inequalities: on the event $\GG\cap \GG_4\cap \Ec_4$,
{\small
\begin{equation}
    \begin{aligned}
        \frac{\|n\cdot (\widehat{\bf V}^{[m^*]} - {\bf V}^*)\|_2}{n\cdot {\bf V}^*_{1,1}} &\lesssim \max_{l\in [L]}\|n_l\cdot (\widehat{\bf V}_\mu^{(l)} - {\bf V}_\mu^{(l)})\|_2 +  \|\widehat{\gamma}^{[m^*]}-\gamma^*\|_2 + \frac{n}{N} + \frac{1}{\sqrt{N}},\\
        \frac{\|\sqrt{n}\cdot{\rm Rem}^{[m^*]}\|_2}{\sqrt{n {\bf V}^*_{1,1}}} &\lesssim \sqrt{n}\|\widehat{\gamma}^{[m^*]} - \gamma^*\|_2 + \left(1+\frac{1}{\sigma_L^2(\U)}\right)\sqrt{\frac{\log n\log N}{N}} + \left(1+\frac{1}{\sigma_L^2(\U)}\right)^2\frac{\log n}{\sqrt{n}}.
    \end{aligned}
    \label{eq: E3 key}
\end{equation}}
Next, we plug the upper bound of $\|\widehat{\gamma}^{[m^*]}-\gamma^*\|_2$ in $\Ec_4$ into \eqref{eq: E3 key}. On the event $\GG\cap\GG_4\cap\Ec_4$, we establish that:
{\small
\begin{equation*}
    \begin{aligned}
        \frac{\|n\cdot (\widehat{\bf V}^{[m^*]} - {\bf V}^*)\|_2}{n\cdot {\bf V}^*_{1,1}}
        &\lesssim \max_{l\in [L]}\|n_l\cdot (\widehat{\bf V}_\mu^{(l)} - {\bf V}_\mu^{(l)})\|_2+ \frac{1}{\sigma_L^2(\U)}\left(\frac{{\rm err}_n(M)}{\sqrt{n}} +\sqrt{\frac{\log N}{N}}\right) \\
        &\quad\quad+ \left(1+\frac{1}{\sigma_L^2(\U)}\right)^2\log n \cdot n^{-3/4} + \frac{n}{N} + N^{-1/2} \\
        &= \kappa_{n,M}.
    \end{aligned}
\end{equation*}
}
and
{\small
\begin{equation*}
    \begin{aligned}
        \frac{\|\sqrt{n}\cdot{\rm Rem}^{[m^*]}\|_2}{\sqrt{n {\bf V}^*_{1,1}}} 
        &\lesssim \frac{1}{\sigma_L^2(\U)}\left({{\rm err}_n(M)} +\sqrt{\frac{ n\log N}{N}}\right)+ \left(1+\frac{1}{\sigma_L^2(\U)}\right)^2\log n\cdot n^{-1/4}\\
        &\quad+\left(1+\frac{1}{\sigma_L^2(\U)}\right)\sqrt{\frac{\log n\log N}{N}} \\
        &= \eta_{n,M}.
    \end{aligned}
\end{equation*}
}
Therefore, it follows from the preceding discussions that 
$$\GG\cap \GG_4\cap \Ec_4\subseteq \Ec_3,$$ 
provided that $\sigma_L^2(\U)\gg \max\{\sqrt{n\log N/N}, \; \sqrt{\log n} \cdot n^{-1/8}\}$, and a sufficiently large resampling size $M$ such that $\err(M)\ll \sigma_L^2(\U)$.
And it holds that
\[
\begin{aligned}
\mathbf{P}(\Ec_3)&\geq \mathbf{P}(\GG\cap \GG_4\cap \Ec_4) = 1- \mathbf{P}(\GG^c\cup \GG_4^c\cup \Ec_4^c) \\
&\geq 1 - \mathbf{P}(\GG^c) - \mathbf{P}(\GG_4^c) - \mathbf{P}(\Ec_4^c)= \mathbf{P}(\GG) + \mathbf{P}(\GG_4) + \mathbf{P}(\Ec_4)-2.  
\end{aligned}
\]
By taking the liminf operators w.r.t $M\to\infty$ then $n\to\infty$ on both sides, we have
\[
\liminf_{n\to\infty}\liminf_{M\to\infty}\mathbf{P}(\Ec_3) \geq \liminf_{n\to\infty}\liminf_{M\to\infty}\left\{\mathbf{P}(\GG) + \mathbf{P}(\GG_4) + \mathbf{P}(\Ec_4)\right\}-2.
\]
Combined with Lemmas \ref{lem: concentration lemma}, \ref{lemma: grad Shat limiting}, and \ref{lem: E4}, which lower bounds $\mathbf{P}(\GG)$, $\mathbf{P}(\GG_4)$, and $\mathbf{P}(\Ec_4)$, respectively, we complete the proof.

\subsection{Proof of Lemma \ref{lemma: max gamma_m diff}}
\label{proof of lemma: max gamma_m diff}
It follows Lemma \ref{prop: gamma_m and gamma_ap dist} and Theorem \ref{thm: gamma true} that on the event $\GG$, 
\begin{equation}
\begin{aligned}
    \max_{m\in \mathcal{M}}\|\widehat{\gamma}^{[m]} -\gamma^*\|_2 &\leq \max_{m\in \mathcal{M}}\|\widehat{\gamma}^{[m]} -\gamma^*_{\rm ap}\|_2 + \|\gamma_{\rm ap}^* - \gamma^*\|_2\\
    &\leq \frac{C\kappa_2}{\sigma_L^2(\U)} \left(\max_{m\in \mathcal{M}}\|\widehat{\U}^{[m]} - \U\|_2 + \sqrt{d\log N/N}\right) + \tau^2(d/n)^{3/4}.
\end{aligned}
    \label{eq: max gamma_m and gamma_ap in M}
\end{equation}
The next Lemma upper bounds $ \max_{m\in \mathcal{M}}\|\widehat{\U}^{[m]} - \U\|_2$, with proof provided in Section \ref{proof of lemma max Um distance}.
\begin{Lemma}
\label{lemma: max Um distance}
    Suppose Condition {\rm (A5)} holds. Then on the event $\Ec_1\cap \Ec_2$, defined in \eqref{eq: E1E2}, we have
    \[
    \max_{m\in \mathcal{M}}\|\widehat{\U}^{[m]} - \U\|_2 \lesssim z_{\alpha_0/dKL} \cdot\sqrt{d/n}.
    \]  
\end{Lemma}
Combining \eqref{eq: max gamma_m and gamma_ap in M} and the above Lemma \ref{lemma: max Um distance}, we have obtained that on the event $\GG\cap \Ec_1\cap \Ec_2$, 
\begin{equation*}
    \max_{m\in \mathcal{M}}\|\widehat{\gamma}^{[m]} -\gamma^*\|_2 \lesssim \frac{1}{\sigma_L^2(\U)} \left(z_{\alpha_0/dKL}\sqrt{d/n} + \sqrt{d\log N/N}\right) + \tau^2(d/n)^{3/4}.
\end{equation*}
Moreover, as we consider $N\gg n\log N$, we shall further simplify the preceding result and complete the proof.

\subsection{Proof of Lemma \ref{lem: max_m V_m dist}}
\label{proof of lem: max_m V_m dist}

It follows from \eqref{eq: max theta_m diff} that,
on the event $\GG\cap \Ec_1\cap \Ec_2$, when $t\ll \sigma_L^2(\U) (n/d)^{1/8}$,
\[
\widehat{\theta}^{[m]}\in \Theta_{\rm loc},\quad\textrm{for all $m\in \mathcal{M}$}.
\]
Then we apply Lemma \ref{lem: consistency of Vm} to establish that
{\small
\[
\begin{aligned}
\max_{m\in \mathcal{M}}\|n\cdot (\widehat{\bf V}^{[m]}-{\bf V}^*)\|_2
    &\lesssim \max_{l\in [L]}\|n_l\cdot (\widehat{\bf V}_\mu^{(l)} - {\bf V}_\mu^{(l)})\|_2 + \max_{m\in \mathcal{M}} \|\widehat{\gamma}^{[m]}-\gamma^*\|_2 + \frac{n}{N} + N^{-1/2} \\
    &\lesssim
1 + \max_{m\in \mathcal{M}}\|\widehat{\gamma}^{[m]}-\gamma^*\|_2,
\end{aligned}
\]}
where the second inequality holds since $N\geq n$ and $\max_{l\in [L]}\|n_l\cdot (\widehat{\bf V}_\mu^{(l)} - {\bf V}_\mu^{(l)})\|_2\leq 1/3$  on the event $\Ec_1$.
Together with Lemma \ref{lemma: max gamma_m diff} where we establish the upper bound for $\max_{m\in \mathcal{M}}\|\widehat{\gamma}^{[m]}-\gamma^*\|_2$, we further obtain that: on the event $\GG\cap \Ec_1\cap \Ec_2$,
{\small
\[
\begin{aligned}
    \max_{m\in \mathcal{M}}\|n\cdot (\widehat{\bf V}^{[m]}-{\bf V}^*)\|_2
    &\lesssim
1+ \frac{z_{\alpha_0/dKL}}{\sigma_L^2(\U)} \sqrt{d/n} + \tau^2(d/n)^{3/4}.
\end{aligned}
\]
}
Since $\tau\asymp (1+\frac{1}{\sigma_L^2(\U)})t \ll (n/d)^{1/8}$,
we further simplify the preceding inequality, such that
\begin{equation*}
\max_{m\in \mathcal{M}}\|n\cdot (\widehat{\bf V}^{[m]}-{\bf V}^*)\|_2\leq C,
\end{equation*}
for some positive constant $C>0$.

\subsection{Proof of Lemma \ref{lemma: theta_m error decomposition - dominating}}
\label{proof of lemma: theta_m error decomposition - dominating}
It follows from the definition of $\nabla \widehat{S}(\theta^*), \nabla S(\theta^*)$ that
\[
\nabla \widehat{S}(\theta^*) - \nabla S(\theta^*) = \frac{1}{N}\sum_{j=1}^N p(X^\QQ_j,\theta^*)\otimes X^\QQ_j - \E_{\QQ_X}\left[p(X,\theta^*)\otimes X\right].
\]
Since $\{X^\QQ_j\}_{j\in [N]}$ are i.i.d. drawn from $\QQ_X$, by the classic CLT, we have
\[
\sqrt{N}\left(\nabla \widehat{S}(\theta^*) - \nabla S(\theta^*)\right) \xlongrightarrow{d} \mathcal{N}\left(0_{dK},\; {\bf V}_{\QQ_X}\left(p(X,\theta^*)\otimes X\right)\right).
\]
Note that $(\widehat{\U}-\U)\gamma^* = \sum_{l=1}^L \gamma^*_l (\widehat{\mu}^{(l)} - \mu^{(l)})$.
According to the limiting distribution of $\widehat{\mu}^{(l)}$ in Condition {\rm (A5)}, we establish that
\[
\left[\sum_{l=1}^L \gamma_l^* {\bf V}_\mu^{(l)}\right]^{-1/2} (\widehat{\U}-\U)\gamma^* \xlongrightarrow{d}\Nc(0_{dK}, {\bf I}_{dK}).
\]
Combining the above two results, we show that
\[
(\widehat{\U}-\U)\gamma^* +\nabla \widehat{S}(\theta^*) - \nabla S(\theta^*) \xlongrightarrow{d} \Nc\left(0_{dK}, \; \sum_{l=1}^L \gamma_l^*{\bf V}_\mu^{(l)} + \frac{1}{N}{\bf V}_{\QQ}(p(X,\theta^*)\otimes X)\right).
\]
By the expression of $W^*$ in \eqref{eq: W_star def}, it implies that
\[
(\widehat{\U}-\U)\gamma^* +\nabla \widehat{S}(\theta^*) - \nabla S(\theta^*) \xlongrightarrow{d}\Nc(0_{dK}, W^*).
\]
By Slutsky' theorem, we further establish that
\[
\Dc^*= -[H(\theta^*)]^{-1} \left[(\widehat{\U}-\U)\gamma^* +\nabla \widehat{S}(\theta^*) - \nabla S(\theta^*)\right] \xlongrightarrow{d}\Nc(0, [H(\theta^*)]^{-1} W^* [H(\theta^*)]^{-1}).
\]

\subsection{Proof of Lemma \ref{lemma: theta_m error decomposition - remainder}}
\label{proof of lemma: theta_m error decomposition - remainder}

We consider that $\widehat{\theta}^{[m]}\in \Theta_{\rm loc}$ with $\Theta_{\rm loc}$ defined in \eqref{eq: Theta local - new}. That is,
\[
\|\widehat{\theta}^{[m]}-\theta^*\|_2 \lesssim \tau\sqrt{d/n}.
\]
Then for any $\nu\in [0,1]$, it holds that
\begin{equation}
    \|\theta^* + \nu(\widehat{\theta}^{[m]}-\theta^*)\|_2\leq \|\widehat{\theta}^{[m]}-\theta^*\|_2 \lesssim \tau\sqrt{d/n},
    \label{eq: theta_tilde_m bound} 
\end{equation}
as well, which implies that 
$$\|\theta^* + \nu(\widehat{\theta}^{[m]}-\theta^*)\|_1\lesssim \sqrt{d}\|\theta^* + \nu(\widehat{\theta}^{[m]}-\theta^*)\|_2\lesssim \tau\sqrt{d^2/n}\leq 1,$$
where the last inequality holds if $\tau \lesssim\sqrt{n/d^2}$.
Therefore, we have 
$$\theta^* + \nu(\widehat{\theta}^{[m]}-\theta^*)\in \Theta_B,\quad\forall \nu\in [0,1],$$ where the bounded parameter set $\Theta_{B}$ is defined in \eqref{eq: bound para}. Then Lemma \ref{lem: convexity of H} implies that there exists some constant $c_1\in (0,1/2)$ such that: for all $\nu\in [0,1]$,
$$c_1^2\kappa_1 \leq \lambda_{\min}(H(\theta^* + \nu(\widehat{\theta}^{[m]}-\theta^*)))\leq \lambda_{\max}(H(\theta^* + \nu(\widehat{\theta}^{[m]}-\theta^*)))\leq (1-c_1)\kappa_2,$$
where $\kappa_1,\kappa_2$ are positive constants defined in Condition {\rm (A2)}. The definition $\bar{H}^{[m]}$ in \eqref{eq: barH_m} yields that:
\[
c_1^2\kappa_1{\bf I}\preceq \bar{H}^{[m]} = \int_0^1 H(\theta^* + \nu(\widehat{\theta}^{[m]}-\theta^*)) d\nu \preceq (1-c_1)\kappa_2 {\bf I}.
\]
Thus, we obtain that
$\lambda_{\max}\left(\left[\bar{H}^{[m]}\right]^{-1}\right)\leq C$, for some constant $C>0$.

% Since $\widetilde{\theta}^{[m]} = \nu\theta^* + (1-\nu)\widehat{\theta}^{[m]}$ for some $\nu\in [0,1]$, it holds that
% \begin{equation}
%     \|\widetilde{\theta}^{[m]} - \theta^*\|_2\leq \|\widehat{\theta}^{[m]}-\theta^*\|_2 \lesssim \tau\sqrt{d/n},
%     \label{eq: theta_tilde_m bound}
% \end{equation}
% as well, which implies that $$\|\widetilde{\theta}^{[m]} - \theta^*\|_1\lesssim \sqrt{d}\|\widetilde{\theta}^{[m]} - \theta^*\|_2\lesssim \tau\sqrt{d^2/n}\leq 1,$$
% where the last inequality holds if $\tau \lesssim\sqrt{n/d^2}$. 
% Therefore, we have 
% $\widetilde{\theta}^{[m]}\in \Theta_B,$ where the bounded parameter set $\Theta_{B}$ is defined in \eqref{eq: bound para}. Then it follows from Lemma \ref{lem: convexity of H} that $\lambda_{\max}\left(\left[H\left(\widetilde{\theta}^{[m]}\right)\right]^{-1}\right)\leq C$, for some constant $C>0$.

Regarding the term $\text{Rem}_1^{[m]}$, on the event $\GG$, we have 
\begin{equation}
\begin{aligned}
    \left\|\text{Rem}_1^{[m]}\right\|_2&\leq \lambda_{\max}\left(\left[\bar{H}^{[m]}\right]^{-1}\right) \|\widehat{\U}\|_2 \|\hgammam - \gamma^*\|_2 \\
    &\lesssim (\|\U\|_2 + t\sqrt{d/n})\|\widehat{\gamma}^{[m]} - \gamma^*\|_2\\
    &\lesssim \|\widehat{\gamma}^{[m]} - \gamma^*\|_2,
\end{aligned}
\label{eq: remainder resample m - 1}
\end{equation}
where the last inequality holds as we consider the bounded $\|\U\|_2$ in Condition {\rm (A4)} and $t\lesssim\sqrt{n/d^3}$ as specified in \eqref{eq: t condition}.
Regarding the term $\text{Rem}_2^{[m]}$, we have 
\begin{equation*}
\begin{aligned}
    \left\|\text{Rem}_2^{[m]}\right\|_2&\leq \lambda_{\max}\left(\left[\bar{H}^{[m]}\right]^{-1}\right)\left\|\left(\nabla \widehat{S}({\theta}^{*})-\nabla {S}(\theta^*)\right)-\left(\nabla \widehat{S}(\widehat{\theta}^{[m]})-\nabla {S}(\widehat{\theta}^{[m]})\right)\right\|_2\\
    &\lesssim \left\|\left(\nabla \widehat{S}({\theta}^{*})-\nabla {S}(\theta^*)\right)-\left(\nabla \widehat{S}(\widehat{\theta}^{[m]})-\nabla {S}(\widehat{\theta}^{[m]})\right)\right\|_2
\end{aligned}
\end{equation*}
Since $\widehat{\theta}^{[m]}\in \Theta_{\rm loc}$, we apply Lemma \ref{lemma: grad Shat limiting} to establish that on the event $\GG_4$:
\begin{equation}
\left\|\text{Rem}_2^{[m]}\right\|_2\lesssim \sqrt{\frac{d^3 \log N }{n \cdot N}}\cdot \tau.
\label{eq: remainder resample m - 2}
\end{equation}
With respect to the term $\text{Rem}_3^{[m]}$, the following inequality holds:
\begin{equation*}
\left\|\text{Rem}_3^{[m]}\right\|_2\leq \left\|\left[\bar{H}^{[m]}\right]^{-1}-\left[H(\theta^*)\right]^{-1}\right\|_2 \left\|\nabla \widehat{S}({\theta}^{*})-\nabla {S}({\theta}^{*})+(\widehat{\U}-\U)\gamma^*\right\|_2.
\end{equation*}
By Lemma \ref{lem: key property} which establishes the continuity property of $H(\cdot)$, we have
\[
\begin{aligned}
    \|\bar{H}^{[m]} - H(\theta^*)\|_2 &= \left\|\int_0^1 H(\theta^* + \nu(\widehat{\theta}^{[m]} - \theta^*))d\nu - H(\theta^*)\right\|_2 \\
    &\lesssim \int_0^1 \nu \|\widehat{\theta}^{[m]}-\theta^*\|_2 d\nu = \|\widehat{\theta}^{[m]}-\theta^*\|_2 \lesssim \tau\sqrt{d/n}.
\end{aligned}
\]
where the last inequality holds as $\widehat{\theta}^{[m]}\in \Theta_{\rm loc}$.
Moreover, on the event $\GG$, by its definition in \eqref{eq: events}, we have
\[
\begin{aligned}
    \left\|\nabla \widehat{S}({\theta}^{*})-\nabla {S}({\theta}^{*})+(\widehat{\U}-\U)\gamma^*\right\|_2 &\lesssim \|\nabla\widehat{S}({\theta}^{*})-\nabla {S}({\theta}^{*})\|_2 + \|\widehat{\U} - \U\|_2\\
    &\lesssim \sqrt{d\log N/N}+t\sqrt{d/n}.
\end{aligned}
\]
The combination of the above three equations leads to
\begin{equation}
    \left\|\text{Rem}_3^{[m]}\right\|_2 \lesssim \tau\sqrt{\frac{d^2\log N}{n\cdot N}} + \tau \cdot t\cdot \frac{d}{n}\lesssim \tau\sqrt{\frac{d^2\log N}{n\cdot N}} + \tau^2 \frac{d}{n},
    \label{eq: remainder resample m - 3}
\end{equation}
where the second inequality holds as $\tau= C(1+\frac{1}{\sigma_L^2(\U)}) t\gtrsim t$.
Now, we combine \eqref{eq: remainder resample m - 1}, \eqref{eq: remainder resample m - 2}, and \eqref{eq: remainder resample m - 3} to complete the proof.

\subsection{Proof of Lemma \ref{lemma: subgaus of f(X)X}}
\label{sec: proof of lemma - subgaus f(X)X}
We consider that $h(X)\in \mathbb{R}^K$ is a subgaussian random vector, then its $k$-entry $h_k(X)\in \RR$ is a subgaussian random variable for any $k=1,..., K.$
In the following discussions, we first study the concentration inequality for $h_k(X)\cdot X$ for each $k=1,..., K$. Then we will move on studying $h(X)\otimes X$.

For any $k=1,...,K$, we have $|h_k(X)| \leq M$ for some constant $M\geq 0$, as $h_k(X)$ is bounded. Since $X$ is a subgaussian random vector, their product $Z = h_k(X)\cdot X\in \RR^d$ is still a subgaussian random vector with
\[
\|Z\|_{\psi_2} = \|h_k(X)\cdot X\|_{\psi_2} \leq M \|X\|_{\psi_2}.
\]
Here $\|\cdot\|_{\psi_2}$ denotes the subgaussian norm.
We denote $h_k(X)$'s $\ell_2$ norm as $\|h_k\|_{\ell_2}= (\E[h_k(X)]^2)^{1/2}$. Then for any fixed unit vector $u\in \RR^d$, it holds that
\[
\E[(u^\intercal Z)^2] = \E[(h_k(X))^2\cdot (u^\intercal X)^2]\leq \|h_k\|_{\ell_2}^2\cdot \E[(u^\intercal X)^2] \lesssim \|h_k\|_{\ell_2}^2 \|X\|_{\psi_2}^2,
\]
where the last equality holds since $X$ is a subgaussian random vector.

% For any $k=1,..., K$, we denote $h_k(X)$'s subgaussian norm as $\|h_k(X)\|_{\psi_2}$. Since $X\in \RR^d$ is a subgaussian random vector with subgaussian norm $\|X\|_{\psi_2}$, their product $Z = h_k(X)\cdot X\in \RR^d$ is a subexponential random vector with subexponential norm:
% \[
% \|Z\|_{\psi_1} = \|h_k(X)\cdot X\|_{\psi_1}\lesssim \|h_k(X)\|_{\psi_2} \|X\|_{\psi_2}.
% \]
% Notice that the $\ell_2$ norm and the subgaussian norm are equivalent up to constants; therefore, we establish that
% \begin{equation}
%     \|Z\|_{\psi_1} \leq c \|h_k\|_{\ell_2} \|X\|_{\psi_2} \leq C\|h_k\|_{\ell_2}.
%     \label{eq: subexp norm}
% \end{equation}
% The last equality holds since we consider $\|X\|_{\psi_2}$ is upper bounded by some constant, implied by Condition {\rm (A1)}.
% We have $\{X_i\}_{i\in [n]}$ i.i.d. samples of $X$, and for each $i$, we denote $Z_i = h_k(X_i)\cdot X_i$. Clearly, $\{Z_i\}_{i\in [n]}$ are i.i.d. samples of $Z$.
% Our goal is to bound $\left\|\frac{1}{n}\sum_{i=1}^n Z_i - \E[Z] \right\|_2$ with high probability.

For any fixed unit vector $u\in \mathbb{R}^d$, we define the projection of the error along $u$ as follows:
\begin{equation}
    S_u = u^\intercal \left(\frac{1}{n}\sum_{i=1}^n Z_i - \E[Z]\right) = \frac{1}{n}\sum_{i=1}^n \left(u^\intercal Z_i - \E [u^\intercal Z]\right).
    \label{eq: S_u def}
\end{equation}
% It follows from \eqref{eq: subexp norm} that $u^\intercal Z_i$ is a sub-exponential random scalar with subexponential norm with $\|u^\intercal Z_i\|_{\psi_1}= \|Z_i\|_{\psi_1} \leq C\|h_k\|_{\ell_2}$. Then the Bernstein inequality for subexponential random variables implies that for any $t>0$:
We apply the variance-based Bernstein inequality to establish that:
\begin{equation}
    \mathbb{P}\Biggl(\Bigl|S_u\Bigr| \ge t\Biggr) \le 2\exp\left[-c n \min\left\{\frac{t^2}{\|h_k\|_{\ell_2}^2\|X\|_{\psi_2}^2}, \frac{t}{M\|X\|_{\psi_2}}\right\}\right],
    \label{eq: S_u bound single}
\end{equation}
where  $c>0$ is an absolute constant.

We now wish to obtain a bound on the full $\ell_2$ error:
\[
\left\|\frac{1}{n}\sum_{i=1}^n Z_i - \E[Z]\right\|_2 = \sup_{\|u\|_2=1} \left| u^\intercal\Bigl(\frac{1}{n}\sum_{i=1}^n Z_i - \E[Z]\Bigr) \right| = \sup_{\|u\|_2=1} |S_u|.
\]
We define a $\epsilon$-net $\Nc_{\epsilon}$ of the unit sphere $\mathbb{S}^{d-1}$. It follows from Corollary 4.2.13 of \cite{vershynin2018high} that the cardinality $|\Nc_\epsilon|$ satisfies
\[
|\Nc_\epsilon|\leq (1+2/\epsilon)^d.
\]
For a fixed $u\in \mathcal{N}_\varepsilon$, we already have the inequality established in \eqref{eq: S_u bound single}.
A union bound over all $u\in \mathcal{N}_\epsilon$ yields that, for any $t>0$,
\begin{equation}
    \mathbb{P}\left\{ \max_{u\in \mathcal{N}_\epsilon}|S_u|< t\right\} \ge 1-2\left(1+\frac{2}{\epsilon}\right)^d\exp\left[-c n \min\left\{\frac{t^2}{\|h_k\|_{\ell_2}^2}, \frac{t}{M}\right\}\right],
    \label{eq: S_u bound union}
\end{equation}
where we omit the dependence on $\|X\|_{\psi_2}$ which is bounded by some constant.

Next we pass from maximum over the net to the supreme over the whole sphere. By the definition of $\epsilon$-net $\mathcal{N}_\epsilon$, we know that for every $u\in \mathbb{S}^{d-1}$, there must exist a $v\in \mathcal{N}$ such that $\|u - v\|_2\leq \epsilon$. Then, for any fixed vector $u\in \mathbb{S}^{d-1}$, we pick the net point $v\in \mathcal{N}_\epsilon$, then by the triangle inequality, we establish the upper bound of $|S_u|$, defined in \eqref{eq: S_u def}, as follows:
\[
|S_u|\leq |S_v| + |S_{u-v}|.
\]
Using the Cauchy-Schwarz inequality, we have
\[
|S_{u-v}| = \left|(u - v)^\intercal \left(\frac{1}{n}\sum_{i=1}^n Z_i - \E[Z]\right)\right| \leq \|u-v\|_2 \left|\frac{1}{n}\sum_{i=1}^n Z_i - \E[Z]\right|_2 \leq \epsilon \left\|\frac{1}{n}\sum_{i=1}^n Z_i - \E[Z]\right\|_2.
\]
Thus, we obtain that:
\[
|S_u|\leq |S_v| + \epsilon \left\|\frac{1}{n}\sum_{i=1}^n Z_i - \E[Z]\right\|_2.
\]
Since this is true for every $u\in \mathbb{S}^{d-1}$, we can take the supremum over $u$:
\[
\left\|\frac{1}{n}\sum_{i=1}^n Z_i - \E[Z]\right\|_2 =\sup_{u\in \mathbb{S}^{d-1}}|S_u| \leq \max_{v\in \mathcal{N}_\epsilon}|S_v| + \epsilon \left\|\frac{1}{n}\sum_{i=1}^n Z_i - \E[Z]\right\|_2.
\]
Rearranging the above inequality, we shall obtain that
\[
\left\|\frac{1}{n}\sum_{i=1}^n Z_i - \E[Z]\right\|_2 \leq \frac{1}{1-\epsilon}\max_{v\in \mathcal{N}_\epsilon}|S_v|.
\]
Together with the result in \eqref{eq: S_u bound union}, we establish that for any $t>0$:
\[
\mathbb{P}\left\{ \left\|\frac{1}{n}\sum_{i=1}^n Z_i - \E[Z]\right\|_2 < \frac{t}{1-\varepsilon}\right\} \geq 1-2\exp\left[-c\cdot n \min\left\{\frac{t^2}{\|h_k\|_{\ell_2}^2}, \frac{t}{M}\right\} + d\log \left(1+\frac{2}{\epsilon}\right)\right].
\]
We pick $t=a\cdot \|h_k\|_{\ell_2}\cdot \sqrt{d/n}$ for some constant $a>0$, and $\epsilon = 1/2$. Together with $n\geq d$, the above inequality becomes:
\[
\mathbb{P}\left\{ \left\|\frac{1}{n}\sum_{i=1}^n Z_i - \E[Z]\right\|_2 < 2a \|h_k\|_{\ell_2} \sqrt{d/n}\right\} \geq 1-2\exp\left[-c\cdot d a^2+d\log 5\right].
\]
Therefore, as long as the value of $a$ satisfies that $a^2 > c\cdot \log 5$ for some constant $c>0$, we establish that
\[
\mathbb{P}\left\{ \left\|\frac{1}{n}\sum_{i=1}^n Z_i - \E[Z]\right\|_2 \leq 2a \|h_k\|_{\ell_2}\sqrt{d/n}\right\} \geq 1-2\exp(-ca^2).
\]
Since
\[
\left\|\frac{1}{n}\sum_{i=1}^n h(X_i)\otimes X_i - \E[h(X)\otimes X]\right\|_2 \leq \sum_{k=1}^K \left\|\frac{1}{n}\sum_{i=1}^n h_k(X_i)\cdot X_i - \E[h_k(X)\cdot X]\right\|_2,
\]
we further establish that: for any $a>c_0$ with $c_0$ being some positive constant, 
\[
\mathbb{P}\left\{ \left\|\frac{1}{n}\sum_{i=1}^n h(X_i)\otimes X_i - \E[h(X)\otimes X]\right\|_2 < a\cdot \max_{k=1...,K}\|h_k\|_{\ell_2} \cdot \sqrt{d /n}\right\} \ge 1-2\exp(-ca^2).
\]

\subsection{Proof of Lemma \ref{lemma: remainder terms of mu}}
\label{Proof of Lemma remainder terms of mu}
We now control remainder terms in order.

\noindent{\bf Control of ${\rm Rem}_1^{(l)}$.} 
Since $\w{l}(\X{l})$ is bounded in Condition {\rm (B1)}, and $\hf{l}_k(\X{l}) - \f{l}_k(\X{l})\in [-1,1]$, for each category $k=1,..., K$, we have $\w{l}(X)\cdot (\hf{l}_k(X) - \f{l}_k(X))$ as a bounded random variable with
\[
\max_{k=1,...,K}\E_{\PP^{(l)}}\left[\w{l}(X)\cdot (\hf{l}_k(X) - \f{l}_k(X))\right]^2\lesssim \max_{k=1,...,K}\left\|\widehat{f}^{(l)}_k-f^{(l)}_k\right\|_{\ell_2(\PP^{(l)})}^2.
\]
Then we apply Lemma \ref{lemma: subgaus of f(X)X} to obtain that: for any $t \geq c_0$, with probability at least $1-e^{-c_1t^2}$:
\begin{equation}
    \|{\rm Rem}_1^{(l)}\|_2\lesssim \max_{k=1,...K} \left\|\widehat{f}^{(l)}_k-f^{(l)}_k\right\|_{\ell_2(\PP^{(l)})}t\cdot \sqrt{d/n}.
    \label{eq: R31}
\end{equation}

\noindent{\bf Control of ${\rm Rem}_2^{(l)}$.} 
Similarly, we shall control ${\rm Rem}_2^{(l)}$ as follows: for any value $t \geq c_0$, with probability at least $1-e^{-c_1t^2}$,
\begin{equation}
    \|{\rm Rem}_2^{(l)}\|_2\lesssim \max_{k=1,...,K} \left\|\widehat{f}^{(l)}_k-f^{(l)}_k\right\|_{\ell_2(\QQ)}t\cdot \sqrt{d/N}.
    \label{eq: R32 - interm}
\end{equation} 
Additionally, since
\[
\begin{aligned}
    \E_{\QQ}[(\widehat{f}^{(l)}_k(X)-f^{(l)}_k(X))]^2 &= \E_{\PP^{(l)}} \left[\omega^{(l)}(X)(\widehat{f}^{(l)}_k(X)-f^{(l)}_k(X))^2\right]\\
    &\leq \left(\E_{\PP^{(l)}}[(\omega^{(l)}(X))^2]\right)^{1/2}\left(\E_{\PP^{(l)}}[(\widehat{f}^{(l)}_k(X)-f^{(l)}_k(X))^4]\right)^{1/2}\\
    &\lesssim \left(\E_{\PP^{(l)}}[(\widehat{f}^{(l)}_k(X)-f^{(l)}_k(X))^4]\right)^{1/2},
\end{aligned}
\]
where the last inequality follows from the bounded assumption of $\w{l}(\cdot)$ in Condition {\rm (B1)}. Therefore, we establish that
\[
\left\|\widehat{f}^{(l)}_k-f^{(l)}_k\right\|_{\ell_2(\QQ)} \lesssim \left\|\widehat{f}^{(l)}_c-f^{(l)}_c\right\|_{\ell_4(\PP^{(l)})}.
\]
Note $\widehat{f}^{(l)}_k(X^{(l)})-f^{(l)}_k(X^{(l)})\in [-1,1]$ is also a subgaussian random variable with $\ell_2$ and $\ell_4$ norms differing only by a constant. Therefore,
\[
\left\|\widehat{f}^{(l)}_k-f^{(l)}_k\right\|_{\ell_2(\QQ)} \lesssim \left\|\widehat{f}^{(l)}_k-f^{(l)}_k\right\|_{\ell_2(\PP^{(l)})}.
\] 
Then putting the above result back to \eqref{eq: R32 - interm}, we establish that: for any $t \geq c_0$, with probability at least $1-e^{-c_1t^2}$,
\begin{equation}
    \|{\rm Rem}_2^{(l)}\|_2\leq t\cdot\max_{k=1,...,K} \left\|\widehat{f}^{(l)}_k-f^{(l)}_k\right\|_{\ell_2(\PP^{(l)})}\sqrt{d/N}.
    \label{eq: R32}
\end{equation}

\noindent\textbf{Control of ${\rm Rem}_3^{(l)}$.}
${\rm Rem}_3^{(l)}$ admits the following form
{\small
\[
\begin{aligned}
    {\rm Rem}_3^{(l)} &= \frac{1}{n_l}\sum_{i=1}^{n_l}\left[\omega(\X{l}_i)\cdot \left(\frac{\hw{l}(\X{l}_i)}{\omega^{(l)}(\X{l}_i)}-1\right)\cdot \left(\f{l}(\X{l}_i)-{\bf y}\right)\otimes \X{l}_i\right] \\
    &\quad - \E_{(X,Y)\sim \PP^{(l)}}\left[\w{l}(X)\cdot \left(\frac{\hw{l}(X)}{\omega^{(l)}(X)}-1\right)\cdot \left(\f{l}(X) - {\bf y}\right) \otimes X\right],
\end{aligned}
\]
}
where the second term on the right-hand side satisfies that
\[
\E_{(X,Y)\sim \PP^{(l)}}\left[\w{l}(X)\cdot \left(\frac{\hw{l}(X)}{\omega^{(l)}(X)}-1\right)\cdot \left(\f{l}(X) - {\bf y}\right) \otimes X\right] = 0,
\]
since for each category $k=1,...,K$,
{\small
\[
\begin{aligned}
    &\E_{(X,Y)\sim \PP^{(l)}}\left[\w{l}(X)\cdot\left(\frac{\hw{l}(X)}{\omega^{(l)}(X)}-1\right)\cdot {\bf 1}(Y=k)\cdot X\right] \\
    &= \E_{X\sim \PP^{(l)}_X}\left[\w{l}(X)\cdot\left(\frac{\hw{l}(X)}{\omega^{(l)}(X)}-1\right) \E_{Y\sim \PP^{(l)}_{Y|X}}\left({\bf 1}(Y=k)|X\right) \cdot X\right] \\
&=\E_{X\sim \PP^{(l)}_X}\left[\w{l}(X)\cdot\left(\frac{\hw{l}(X)}{\omega^{(l)}(X)}-1\right) \cdot f^{(l)}_k(X) \cdot X\right].
\end{aligned}
\]
}
It follows from the bounded assumption of $\w{l}(\cdot)$ in Condition {(B1)}, and $\f{l}_k\in [-1,1]$, that
\[
\E_{X\sim \PP^{(l)}_X}\left[\w{l}(X)\cdot\left(\frac{\hw{l}(X)}{\omega^{(l)}(X)}-1\right) \cdot f^{(l)}_k(X)\right]^2\leq C\cdot \E_{X\sim \PP^{(l)}_X}\left[\frac{\hw{l}(X)}{\omega^{(l)}(X)}-1\right]^2 = C\cdot \left\|\frac{\widehat{\omega}^{(l)}}{\omega^{(l)}}-1\right\|_{\ell_2(\PP^{(l)})}^2.
\]
Then we apply Lemma \ref{lemma: subgaus of f(X)X} to obtain that: for any $t \geq c_0$, with probability at least $1-e^{-c_1t^2}$,
\begin{equation}
    \|{\rm Rem}_3^{(l)}\|_2\leq t\cdot \left\|\frac{\widehat{\omega}^{(l)}}{\omega^{(l)}}-1\right\|_{\ell_2(\PP^{(l)})} \sqrt{d/n}.
   \label{eq: R4}
\end{equation}

\noindent\textbf{Control of ${\rm Rem}_4^{(l)}$.}
We shall decompose ${\rm Rem}_4^{(l)}$ as follows:
{\small
\[
\begin{aligned}
    {\rm Rem}_4^{(l)} &= {\rm Rem}_{41}^{(l)} + {\rm Rem}_{42}^{(l)} ,\quad \textrm{with}\\
    {\rm Rem}_{41}^{(l)}&= \frac{1}{n_l}\sum_{i=1}^{n_l}\left[\w{l}(\X{l}_i)\cdot\left(\frac{\hw{l}(\X{l}_i)}{\w{l}(\X{l}_i)}-1\right)\cdot \left(\hf{l}(\X{l}_i) - \f{l}(\X{l}_i)\right)\right]\otimes \X{l}_i \\
    &\quad \quad -\E_{\PP^{(l)}_X}\left[\w{l}(X)\cdot \left(\frac{\hw{l}(X)}{\w{l}(X)}-1\right)\cdot\left(\hf{l}(X)-\f{l}(X)\right)\otimes X\right], \\
    {\rm Rem}_{42}^{(l)}&=\E_{\PP^{(l)}_X}\left[\w{l}(X)\cdot \left(\frac{\hw{l}(X)}{\w{l}(X)}-1\right)\cdot\left(\hf{l}(X)-\f{l}(X)\right)\otimes X\right].
\end{aligned}
\]
}
We apply Lemma \ref{lemma: subgaus of f(X)X} to establish: for any $t \geq c_0$, with probability at least $1-e^{-c_1t^2}$:
\[
\|{\rm Rem}_{41}^{(l)}\|_2\leq t\cdot \left\|\frac{\widehat{\omega}^{(l)}}{ \omega^{(l)}}-1\right\|_{\ell_2(\PP^{(l)})} \cdot \max_{k=1...,K}\left\|\widehat{f}^{(l)}_k-f^{(l)}_k\right\|_{\ell_2(\PP^{(l)})} \cdot \sqrt{d/n}.
\]
For ${\rm Rem}_{42}^{(l)}$,
we observe that
{\small
\[
\begin{aligned}
&\left\|\E_{\PP^{(l)}}\left[\left(\widehat{\omega}^{(l)}(X) - \omega^{(l)}(X)\right)\left(\widehat{f}^{(l)}_k(X)-f^{(l)}_k(X)\right) X\right]\right\|_2 \\
    &\leq \E_{\PP^{(l)}}\left[\left|\left(\widehat{\omega}^{(l)}(X) - \omega^{(l)}(X)\right)\left(\widehat{f}^{(l)}_k(X)-f^{(l)}_k(X)\right)\right| \cdot \left\| X\right\|_2 \right]\\
    &\leq \left(\E_{\PP^{(l)}}\left|\left(\widehat{\omega}^{(l)}(X) - \omega^{(l)}(X)\right)\left(\widehat{f}^{(l)}_k(X)-f^{(l)}_k(X)\right)\right|^2\right)^{1/2} \left(\E_{\PP^{(l)}}[\|X\|_2^2]\right)^{1/2} && ({\rm Cauchy-Schwarz})\\
    &= C \sqrt{d} \left\|\left(\widehat{\omega}^{(l)} -\omega^{(l)}\right) \left(\widehat{f}^{(l)}_k-f^{(l)}_k\right)\right\|_{\ell_2(\PP^{(l)})} && (\textrm{Bounded $\ell_2$ norm of $\|X\|_2$}) \\
    &\leq C\sqrt{d}\left\|\widehat{\omega}^{(l)} - \omega^{(l)}\right\|_{\ell_2(\PP^{(l)})}\left\|\widehat{f}^{(l)}_k-f^{(l)}_k\right\|_{\ell_2(\PP^{(l)})} \\
    &\lesssim \sqrt{d}\left\|\frac{\widehat{\omega}^{(l)}}{\omega^{(l)}}-1\right\|_{\ell_2(\PP^{(l)})} \left\|\widehat{f}^{(l)}_k-f^{(l)}_k\right\|_{\ell_2(\PP^{(l)})} && (\textrm{as $\omega(\cdot)$ is bounded}). 
\end{aligned}
\]
}
We combine the above results to conclude that: for any $t \geq c_0$, with probability at least $1-e^{-c_1t^2}$:
{\small
\begin{equation}
\begin{aligned}
    \|{\rm Rem}_4^{(l)}\|_2 \lesssim & t\cdot \left\|\frac{\widehat{\omega}^{(l)}}{ \omega^{(l)}}-1\right\|_{\ell_2(\PP^{(l)})} \cdot \max_{k=1...,K}\left\|\widehat{f}^{(l)}_k-f^{(l)}_k\right\|_{\ell_2(\PP^{(l)})} \cdot (\sqrt{d} + \sqrt{d/n}).
    \label{eq: R5}
\end{aligned}
\end{equation}}

\noindent\textbf{Proof of equation \eqref{eq: rate remainder terms of mu}.} 
We combine \eqref{eq: R31},\eqref{eq: R32},\eqref{eq: R4}, and \eqref{eq: R5} to establish that: with probability at least $1-e^{-c_1t^2}$:
{\small
\[
\begin{aligned}
    \left\|\sum_{j=1}^4 {\rm Rem}_j^{(l)}\right\|_2 &\lesssim t\cdot \left(\left\|\frac{\hw{l}}{\w{l}}-1\right\|_{\ell_2(\PP^{(l)})} + \max_{k=1,...,K}\|\hf{l} - \f{l}\|_{\ell_2(\PP^{(l)})}\right)\sqrt{d/n} \\
    &\quad\quad + t\cdot \left\|\frac{\hw{l}}{\w{l}}-1\right\|_{\ell_2(\PP^{(l)})}\cdot \max_{k=1,...,K}\|\hf{l} - \f{l}\|_{\ell_2(\PP^{(l)})} \cdot \sqrt{d}.
\end{aligned}
\]
}
Next, we apply Condition {\rm (B2)} to further establish that: for any $t \geq c_0$, with probability at least $1-e^{-c_1t^2} - \delta_n$:
{\small
\[
\left\|\sum_{j=1}^4 {\rm Rem}_j^{(l)}\right\|_2 \leq t\cdot (\eta_\omega + \eta_f)\sqrt{d/n} + t\cdot\eta_\omega \eta_f \sqrt{d},
\]
}
where $\delta_n\to 0$ is a vanishing sequence.

\subsection{Proof of Lemma \ref{lem: basic lemma}}
\label{proof of lem: basic lemma}
The proof of the lemma relies on the following result of handling the Kronecker product: for a matrix $A\in \mathbb{R}^{K\times K}$, $B\in \mathbb{R}^{d\times d},$ and $V\in \mathbb{R}^{d\times K}$,
we have 
\begin{equation}
(A\otimes B){\rm vec}(V)={\rm vec}(BVA^{\intercal}).
\label{eq: basic KP}
\end{equation}
We apply \eqref{eq: basic KP} and obtain 
\begin{equation*}
\begin{aligned}
\left(A\otimes XX^{\intercal}\right){\rm vec}(V)={\rm vec}(XX^{\intercal} V A).
\end{aligned}
\end{equation*}
Then we have
\begin{equation*}
\begin{aligned}
    v^{\intercal}\left(A\otimes XX^{\intercal}\right)v
    &=[{\rm vec}(V)]^{\intercal}
\left(A\otimes XX^{\intercal}\right){\rm vec}(V)\\
&= [{\rm vec}(V)]^\intercal {\rm vec}(XX^\intercal VA) = {\rm Tr}(V^\intercal XX^\intercal VA),
\end{aligned}
\end{equation*}
which completes the proof of \eqref{eq: key 1}. By the von Neumann trace inequality, we have 
\begin{equation*}
\left| {\rm Tr}(V^{\intercal}XX^{\intercal} V A)\right|\leq \sum_{k=1}^{\min(K,p)}\sigma_{k}\left(V^{\intercal}XX^{\intercal} V\right)\cdot \sigma_{k}(A)\leq \sigma_{1}(A)\cdot{\rm Tr}(V^{\intercal}XX^{\intercal} V).
\end{equation*}

\subsection{Proof of Lemma \ref{lem: consistency of Vm}}
\label{proof of lem: consistency of Vm}
We first study the term $W^*$. For clarity, we consider the case $n_l=n$ for all $l\in [L]$, without loss of generality. Then
\[
n\cdot W^* = \sum_{l=1}^L (\gamma_l^*)^2 (n {\bf V}_\mu^{(l)}) + \frac{n}{N}{\bf V}_{\QQ_X}(p(X,\theta^*) \otimes X).
\]
It follows from the triangle inequality that:
\[
\|n\cdot W^*\|_2 \leq \max_{l\in [L]} \|n{\bf V}_\mu^{(l)}\|_2 + \frac{n}{N}\|{\bf V}_{\QQ_X}(p(X,\theta^*) \otimes X)\|_2.
\]
Since each entry of the probability vector $p(\XQ,\theta^*)$ is bounded, together with Condition {\rm (A2)} where $\XQ$ is a bounded random vector, it follows that 
$\|{\bf V}_{\QQ_X}(p(X,\theta^*) \otimes X)\|_2$ is bounded as well when we consider the fixed dimension setting. Together with Condition {\rm (A5)} that $\|n{\bf V}_\mu^{(l)}\|_2$ is upper bounded, we have
\[
\|n\cdot W^*\|_2 \leq C_1 + \frac{n}{N}C_2 \leq C,
\]
where $C_1,C_2,C>0$ denote some positive constants.

By the expressions of $\widehat{\bf V}^{[m]}$ and ${\bf V}^*$, the error of $\|n \cdot (\widehat{\bf V}^{[m]} - {\bf V}^*)\|_2$ admits the following upper bound:
    \begin{equation}
        \begin{aligned}
            \|n\cdot (\widehat{\bf V}^{[m]} - {\bf V}^*)\|_2 &\leq C_1\|n\cdot(\widehat{W}^{[m]} - W^*)\|_2 + C_2\left\|\widehat{H}(\widehat{\theta}^{[m]}) - H(\theta^*)\right\|_2 \\
            &\quad+ C_3\left\|\widehat{H}(\widehat{\theta}^{[m]}) - H(\theta^*)\right\|_2^2,
        \end{aligned}
        \label{eq: V distance}
    \end{equation}
    where $C_1,C_2,C_3>0$ are some positive constants depending on $\|[H(\theta^*)]^{-1}\|_2$, $\|nW^*\|_2$. 

    We first study the term $\|\widehat{H}(\widehat{\theta}^{[m]})-H(\theta^*)\|_2$. If $\widehat{\theta}^{[m]}\in \Theta_{\rm loc}$, i.e.,
    $\|\widehat{\theta}^{[m]} - \theta^*\|_2\lesssim \tau \sqrt{d/n}.$
    we have
    $\|\widehat{\theta}^{[m]}-\theta^*\|_1\leq \sqrt{d}\|\widehat{\theta}^{[m]}-\theta^*\|_2\lesssim \tau \sqrt{d^2/n} <1$, where the last inequality holds as $\tau\lesssim \sqrt{n/d^2}$. Therefore, $\widehat{\theta}^{[m]}\in \Theta_B$. We apply event $\GG_3$ to study the term $\left\|\widehat{H}(\widehat{\theta}^{[m]}) - H(\theta^*)\right\|_2$. 
    On the event $\GG_3$,
    \[
    \left\|\widehat{H}(\widehat{\theta}^{[m]}) - H(\theta^*)\right\|_2 \leq C\sqrt{d\log N/N}.
    \]
    
    Next, we study the term $\|n\cdot (\widehat{W}^{[m]} - W^*)\|_2$.  By the expressions of $\widehat{W}^{[m]}$ and $W^*$, we have
    {\small
    \[
    \begin{aligned}
        &\widehat{W}^{[m]} - W^* \\
        &= \left\{\sum_{l=1}^L (\widehat{\gamma}^{[m]}_l)^2\widehat{\bf V}_\mu^{(l)} - \sum_{l=1}^L (\gamma^*_l)^2 {\bf V}_\mu^{(l)}\right\} + \left\{\frac{1}{N}\widehat{\bf V}_\QQ\left(p(X,\widehat{\theta}^{[m]})\otimes X\right) - \frac{1}{N}{\bf V}_\QQ\left(p(X,\theta^*)\otimes X\right) \right\} \\
        &= T_1 + T_2.
    \end{aligned}
    \]
}
    For the term $T_1$, we have
    \[
    \begin{aligned}
        n\cdot T_1 = \sum_{l=1}^L (\widehat{\gamma}^{[m]}_l)^2\left[n\cdot (\widehat{\bf V}_\mu^{(l)} - {\bf V}_\mu^{(l)})\right] + \sum_{l=1}^L \left[(\widehat{\gamma}^{[m]}_l)^2-(\gamma^*_l)^2 \right]\cdot \left[n\cdot {\bf V}_\mu^{(l)}\right].
    \end{aligned}
    \]
    Since $\widehat{\gamma}^{[m]}, \gamma^*\in \Delta^L$, we have
    \[
    \begin{aligned}
        \sum_{l=1}^L \left[(\widehat{\gamma}^{[m]}_l)^2-(\gamma^*_l)^2 \right] &= \|\widehat{\gamma}^{[m]}\|_2^2 - \|\gamma^*\|_2^2 = \langle  \widehat{\gamma}^{[m]}+\gamma^*, \widehat{\gamma}^{[m]}-\gamma^*\rangle \\
        &\leq \|\widehat{\gamma}^{[m]}+\gamma^*\|_2\|\widehat{\gamma}^{[m]}-\gamma^*\|_2
    \leq 2\|\widehat{\gamma}^{[m]}-\gamma^*\|_2
    \end{aligned}
    \]
    Therefore, we establish that
    \[
    \|n\cdot T_1\|_2\leq \max_{l\in [L]}\|n\cdot (\widehat{\bf V}_\mu^{(l)} - {\bf V}_\mu^{(l)})\|_2 + 2\max_{l\in [L]}\|n\cdot {\bf V}_\mu^{(l)}\|_2\|\widehat{\gamma}^{[m]}-\gamma^*\|_2.
    \]
    For the term $T_2$,
    since $\XQ$ is a bounded random vector, there exists a positive constant $C>0$ such that
    \[
    \widehat{\bf V}_\QQ\left(p(X,\widehat{\theta}^{[m]})\otimes X\right)\leq C, \quad {\bf V}_\QQ\left(p(X,\theta^*)\otimes X\right)\leq C.
    \]
    Therefore, 
    \[
    \|N\cdot T_2\|_2\leq C.
    \]
    Consequently, we establish that
    \[
    \begin{aligned}
        &\|n\cdot (\widehat{W}^{[m]}-W^*)\|_2 \\
        &\leq \|n\cdot T_1\|_2 + \|n\cdot T_2\|_2\\
        &\leq \max_{l\in [L]}\|n\cdot (\widehat{\bf V}_\mu^{(l)} - {\bf V}_\mu^{(l)})\|_2 + 2\max_{l\in [L]}\|n\cdot {\bf V}_\mu^{(l)}\|_2\|\widehat{\gamma}^{[m]}-\gamma^*\|_2 + C\cdot\frac{n}{N}.
    \end{aligned}
    \]

    Putting the above results back to \eqref{eq: V distance}, we establish that:
    \[
    \begin{aligned}
        &\|n\cdot (\widehat{\bf V}^{[m]}-{\bf V}^*)\|_2\\
        &\lesssim
    \max_{l\in [L]}\|n\cdot (\widehat{\bf V}_\mu^{(l)} - {\bf V}_\mu^{(l)})\|_2 + \max_{l\in [L]}\|n\cdot {\bf V}_\mu^{(l)}\|_2\|\widehat{\gamma}^{[m]}-\gamma^*\|_2 + \frac{n}{N} + \sqrt{d\log N/N}\\
    &\lesssim 
    \max_{l\in [L]}\|n\cdot (\widehat{\bf V}_\mu^{(l)} - {\bf V}_\mu^{(l)})\|_2 +\|\widehat{\gamma}^{[m]}-\gamma^*\|_2 + \frac{n}{N} + \sqrt{d\log N/N},
    \end{aligned}
    \]
    where the last inequality holds as $\max_{l\in [L]}\|n\cdot {\bf V}_\mu^{(l)}\|_2\leq C$ for some constant $C>0$ as assumed in Condition {\rm (A5)}.

\subsection{Proof of Lemma \ref{lem: E4}}
\label{proof of lemma E4}

We denote the following events to facilitate discussions.
{\begin{equation}
    \begin{aligned}
        \Ec_{5} &= \left\{\|\widehat{\gamma}^{[m^*]}-\gamma^*_{\rm ap}\|_2 \lesssim \frac{1}{\sigma_L^2(\U)}\left(\frac{{\rm err}_n(M)}{\sqrt{n}} +\sqrt{\frac{d\log N}{N}}\right)\right\} \\
        \Ec_{6} &= \left\{\|\gamma_{\rm ap}^*-\gamma^*\|_2\lesssim \left(1+\frac{1}{\sigma_L^2(\U)}\right)^2 \log n {\left(\frac{d}{n}\right)^{3/4}}\right\}
    \end{aligned}
\end{equation}
}
For the event $\Ec_5$, it follows from Theorem \ref{thm: min_gamma_m} that
$
\liminf_{n\to\infty}\liminf_{M\to\infty} \mathbf{P}(\Ec_{5})\geq 1-\alpha_0.
$
Regarding the event $\Ec_6$, we apply Theorem \ref{thm: gamma true} with $t=\sqrt{\log n}$ to establish that:
if $\sigma_L^2(\U)\gg \sqrt{\log n}(d/n)^{1/4}$, then there exists a constant $c_1>0$ such that
$\mathbf{P}(\Ec_{6}) \geq 1-n^{-c_1d}-\delta_n,$
where the sequence $\delta_n\to 0$ is given in Condition {\rm (A1)}. Then with $n\to\infty$, we have
\[
\lim_{n\to\infty}\mathbf{P}(\Ec_{6}) = 1.
\]

By the triangle inequality, on the event $\Ec_5\cap \Ec_6$, we have
\[
\begin{aligned}
    \|\widehat{\gamma}^{[m^*]}-\gamma^*\|_2 &\leq \|\widehat{\gamma}^{[m^*]}-\gamma^*_{\rm ap}\|_2 + \|\gamma_{\rm ap}^*-\gamma^*\|_2\\
    &\lesssim \frac{1}{\sigma_L^2(\U)}\left(\frac{{\rm err}_n(M)}{\sqrt{n}} +\sqrt{\frac{d\log N}{N}}\right) + \left(1+\frac{1}{\sigma_L^2(\U)}\right)^2 \log n  \left(\frac{d}{n}\right)^{3/4}.
\end{aligned}
\]
That is, $\Ec_4\supseteq \Ec_5 \cap \Ec_6$.
Therefore, 
{
\[
\begin{aligned}
    \mathbf{P}(\Ec_4)&\geq \mathbf{P}(\Ec_{5}\cap \Ec_{6})=1-\mathbf{P}(\Ec_{5}^c\cup \Ec_{6}^c) \geq 1-\mathbf{P}(\Ec_{5}^c) - \mathbf{P}(\Ec_{6}^c)\\
    &= \mathbf{P}(\Ec_{5}) + \mathbf{P}(\Ec_{6}) - 1.
\end{aligned}
\]
}
We combine the above results to obtain that:
\begin{equation*}
    \liminf_{n\to\infty}\mathbf{P}(\Ec_{4})\geq \liminf_{n\to\infty}\left[\mathbf{P}(\Ec_{5}) + \mathbf{P}(\Ec_{6}) - 1\right]\geq 1-\alpha_0.
\end{equation*}

\subsection{Proof of Lemma \ref{lemma: max Um distance}}
\label{proof of lemma max Um distance}
For any $m\in \mathcal{M}$, by the definition of $\mathcal{M}$, we have
$$
\max_{1\leq l\leq L}\max_{1\leq j\leq dK}\frac{|\widehat{\mu}^{(l,m)}_{j} - \widehat{\mu}^{(l)}_j|}{\sqrt{[\widehat{\bf V}^{(l)}_\mu]_{j,j} + 1/n_l}}\leq 1.1\cdot z_{\alpha_0/(dKL)} .
$$
On the event $\Ec_2$, specified in \eqref{eq: E1E2}, we have
\[
\max_{1\leq l\leq L}\max_{1\leq j\leq dK}\frac{|\widehat{\mu}^{(l)}_j - \mu^{(l)}_j|}{\sqrt{[\widehat{\bf V}^{(l)}_\mu]_{j,j} + 1/n_l}} \leq 1.05 \cdot z_{\alpha_0/(dKL)}.
\]
By the triangle inequality, the above two inequalities imply that
\[
\begin{aligned}
    \max_{1\leq l\leq L}\max_{1\leq j\leq dK}\frac{\left|\widehat{\mu}^{(l,m)}_j - {\mu}^{(l)}_j\right|}{\sqrt{[\widehat{\bf V}^{(l)}_\mu]_{j,j} + 1/n_l}}
    \leq 2.15\cdot z_{\alpha_0/(dKL)}.
\end{aligned}
\]
Then on the event $\Ec_1$, we establish that
\[
\begin{aligned}
    \max_{1\leq l\leq L}\max_{1\leq j\leq dK}\frac{\left|\widehat{\mu}^{(l,m)}_j - {\mu}^{(l)}_j\right|}{\sqrt{[{\bf V}^{(l)}_\mu]_{j,j} + \frac{4}{3}n_l}}
    \leq 2.15\cdot z_{\alpha_0/(dKL)}.
\end{aligned}
\]
In Condition {\rm (A5)}, we suppose that $n_l\cdot [{\bf V}_\mu^{(l)}]_{j,j}\in [c_1,c_2]$ for all $l\in [L]$ and $j\in [dK]$, with two positive constants $c_1,c_2>0$. Thus, we obtain that
\[
\max_{1\leq l\leq L}\max_{1\leq j\leq dK}\left|\widehat{\mu}^{(l,m)}_j - {\mu}^{(l)}_j\right| \leq C\cdot z_{\alpha_0/dKL}\cdot n^{-1/2},
\]
which further implies that
\[
\begin{aligned}
    \|\widehat{\U}^{[m]} - \U\|_2 &\leq \|\widehat{\U}^{[m]} - \U\|_F = \left(\sum_{l=1}^L \sum_{j=1}^{dK}\left|\widehat{\mu}^{(l,m)}_j - {\mu}^{(l)}_j\right|^2\right)^{1/2} \\
    &\leq \sqrt{dKL} \max_{1\leq l\leq L}\max_{1\leq j\leq dK}\left|\widehat{\mu}^{(l,m)}_j - \widehat{\mu}^{(l)}_j\right| \leq C\cdot z_{\alpha_0/dKL}\cdot \sqrt{d/n}.
\end{aligned}
\]
The above inequality holds for any $m\in \mathcal{M}$, resulting in
$$
\max_{m\in \mathcal{M}}\|\widehat{\U}^{[m]} - \U\|_2 \leq C\cdot z_{\alpha_0/dKL}\cdot \sqrt{d/n}.
$$

\section{Experimental Setups and Additional Experiments}
\label{appendix: simus}

This section provides the experimental setups omitted from the main text as well as additional simulation results. Specifically, 
\begin{itemize}
    \item Section~\ref{appendix: setups} presents the detailed setups used to generate Figures~\ref{fig:worst-case} and~\ref{fig: Inference Challenge}.
    \item Section~\ref{appendix: groupdro} reports further comparisons between the proposed \CGDRO\ framework and the classical GDRO method under both covariate-shift and no-shift regimes. For completeness, we also provide the implementation details of the GDRO procedure in Algorithm~\ref{algo: GDRO}.
    \item Section~\ref{sec: setup simus} specifies the setups corresponding to the simulation study in Section~\ref{sec: simus} of the main text.
    \item Finally, Section~\ref{appendix: infer simus} presents additional statistical inference results in more complex and challenging settings than the settings \textbf{(S3)} and \textbf{(S4)} in the main text.
\end{itemize}

\subsection{Setups for Figures \ref{fig:worst-case} and \ref{fig: Inference Challenge}}
\label{appendix: setups}
In this subsection, we present the detailed setups for the experiments in  Figures \ref{fig:worst-case} and \ref{fig: Inference Challenge}. All experiments involve generating labeled data from $L$ source domains and unlabeled data from the target domain. For each source domain $l\in [L]$, the labeled data $\{\X{l}_i, \Y{l}_i\}_{i\in [n_l]}$ are i.i.d. generated as follows: 
\begin{itemize}
    \item Covariates $\X{l}_i\sim \PP^{(l)}_X$.
    \item Labels $\Y{l}_i$ follows a multinomial distribution with $K+1$ categories, where 
    \begin{equation}
         \PP^{(l)}(Y=c|X) = \frac{\exp(X^\intercal \beta^{(l)}_c)}{\sum_{k=0}^K\exp(X^\intercal \beta_k^{(l)})}, \quad \textrm{for $c=0,1,..., K$.}
         \label{eq: Y|X linear}
    \end{equation}
\end{itemize}
 For the target domain, the unlabeled data $\{X^\QQ_j\}_{j\in [N]}$ are i.i.d. drawn from $\QQ_X$. We will specify the parameters $L, K, \{n_l\}, N,\{\beta^{(l)}_c\}_{l,c}$ and distribution $\{\PP^{(l)}_X\}_L, \QQ_X$ across settings.

\subsubsection{Setup for Figure \ref{fig:worst-case}} We consider a binary classification task $(K = 1)$ with $L = 2$ source domains under the covariate shift regime. Specifically, we set the source covariate distributions as \(\PP^{(1)}_X = \PP^{(2)}_X =\mathcal{N}(0_4, \mathbf{I}_4)\), while the target covariate distribution is $\QQ_X = \Nc((-1,-1,1,1)^\intercal,\mathbf{I}_4)$. The class-conditional coefficients are defined as
\[
\beta^{(l=1)}_{c=0} = 0_4, \; \beta^{(l=1)}_{c=1} = (-0.4, -0.4, 0.2, 0.2)^\intercal,\; \beta^{(l=2)}_{c=0} = 0_4, \; \beta^{(l=2)}_{c=1} = (1, 1, 0.2, 0.2)^\intercal.
\]
The total number of labeled source samples is fixed at $n_1 + n_2 = 4,000$, while the proportion from the first source varies over $\{0.1, 0.2, \ldots, 0.9\}$, as depicted on the horizontal axis of Figure~\ref{fig:worst-case}. The number of unlabeled target samples is set to $N = 10{,}000$. 

Next, we detail the implementation of each method evaluated in Figure~\ref{fig:worst-case}. The proposed method is implemented via Algorithm~\ref{alg: main}, where we use logistic regression to estimate both nuisance components $\widehat{f}^{(l)}$ and $\widehat{\omega}^{(l)}$ for each source $l=1,2$. For the \texttt{ERM}  baseline, we pool the source data and fit a single logistic regression model across the combined dataset. The Group DRO baseline is implemented using the Mirror Prox algorithm as described in Algorithm~\ref{algo: GDRO}.

% To compute the \textbf{non-reducible loss} for each source domain, we proceed as follows. For the first source, given the target covariates $\{\XQ_i\}_{1\leq i\leq N}$, we simulate binary outcomes $\{Y_i^\QQ\}_{1\leq i\leq N}$ from the distribution $\PP^{(1)}(Y=1|X=\XQ_i)$. The corresponding non-reducible loss is then computed as:
% \[
% -\frac{1}{N}\sum_{i=1}^N \sum_{c=0}^1 \mathbf{1}(Y_i^\QQ = c)\log \mathbb{P}^{(1)}(Y_i^\QQ = c \mid \XQ_i).
% \]
% Similarly, to compute the non-reducible loss for the second source, we simulate $\{Y_i^\QQ\}_{1\leq i\leq N}$ using $\PP^{(2)}(Y=1|X=\XQ_i)$, and evaluate the loss via:
% \[
% -\frac{1}{N}\sum_{i=1}^N \sum_{c=0}^1 \mathbf{1}(Y_i^\QQ = c)\log \mathbb{P}^{(2)}(Y_i^\QQ = c \mid \XQ_i).
% \]

\subsubsection{Setup for Figure \ref{fig: Inference Challenge}} We describe the settings used to illustrate the \textbf{nonregular}, \textbf{unstable}, and \textbf{regular} regimes in Figure~\ref{fig: Inference Challenge}. All settings involve binary classification (K = 1) under covariate shift. In particular, we set the covariate distributions $\PP^{(l)}_X = \mathcal{N}(0_{d}, {\bf I}_{d})$, for $l\in [L]$, and $\QQ_X = \mathcal{N}(0.1_{d}, {\bf I}_{d})$. The source sample size is fixed at $n_l=400$, for $l\in [L]$, while the target sample size is fixed at $N=4000$.
\begin{itemize}
    \item \textbf{Nonregular.} We set $L = 2$ and $d = 20$. The class-conditional coefficients are:
    \[
    \beta^{(l=1)}_{c=0} = 0_{d}, \; \beta^{(l=1)}_{c=1} = (6, 0.5_3, 0_{d-4})^\intercal,\; \beta^{(l=2)}_{c=0} = 0_{d}, \; \beta^{(l=2)}_{c=1} = (0.5_4, 0_{d-4})^\intercal.
    \]
    \item \textbf{Unstable.} We set $L = 4$ and $d = 4$. The class-conditional coefficients are: for $l\in [L]$,
    \[
    \beta^{(l)}_{c=0} = 0_{d}, \; \beta^{(l)}_{c=1} = (1,1,0,0)^\intercal + U^{(l)},\;\textrm{with } U^{(l)}\stackrel{i.i.d.}{\sim}\mathcal{N}(0_d, 0.01{\bf I}_d).
    \]
    
    \item \textbf{Regular.} We set $L = 2$ and $d = 20$. The class-conditional coefficients are:
    \[
    \beta^{(l=1)}_{c=0} = 0_{d}, \; \beta^{(l=1)}_{c=1} = (0.5_2, 0_{d-2})^\intercal,\; \beta^{(l=2)}_{c=0} = 0_{d}, \; \beta^{(l=2)}_{c=1} = (0_2, 0.5_2, 0_{d-4})^\intercal.
    \]
\end{itemize}
For both ERM and our proposed method (Algorithm~\ref{alg: main}), we fit logistic regression models.

\subsection{Additional Comparisons Against Group DRO}
\label{appendix: groupdro}

In this section, we numerically compare our proposed method implemented in Algorithm \ref{alg: main}, against the Group DRO approach in both regimes with and without covariate shift. The implementation of Group DRO proceeds as follows: given labeled data $\{\X{l}_i, \Y{l}_i\}_{1\leq i\leq n_l}$ for each source $l=1,...,L$, Group DRO solves the following minimax optimization problem:
\begin{equation}
    \widehat{\theta}_{\rm gdro} = \argmin_{\theta\in \RR^{dK}} \max_{\gamma\in \Delta^L} \widehat{\phi}_{\rm gdro}(\theta,\gamma), \quad \textrm{with}\quad \widehat{\phi}_{\rm gdro}(\theta,\gamma)= \sum_{l=1}^L \gamma_l \cdot \frac{1}{n_l}\sum_{i=1}^{n_l} \ell(\X{l}_i, \Y{l}_i, \theta),
    \label{eq: GDRO finite}
\end{equation}
where $\ell(x,y,\theta)$ denotes the cross-entropy loss. To solve \eqref{eq: GDRO finite}, we apply the optimistic-gradient Mirror Prox algorithm in Algorithm \ref{algo: GDRO}. 
As discussed in Section \ref{sec: Mirror Prox} of the main paper, the two-step Mirror Prox reduces oscillations near saddle points and achieves faster convergence rates in
smooth convex-concave settings.
As a remark, we also implemented the single-step gradient-based method from \cite{sagawa2019distributionally}. However, we found that this method often exhibits unstable convergence behavior and does not fully converge within the maximum number of iterations, even with extensive learning rate tuning in our experiments.

We now describe the experimental setup. We consider a binary classification task ($K=1$) with $L=2$ source domains. 
The covariate distributions for both source domains are $\PP^{(1)}_X=\PP^{(2)}_X = \Nc(0_4, {\bf I}_4)$. In the no covariate shift regime, the target covariate distribution matches the source one, i.e. $\QQ_X = \Nc(0_4, {\bf I}_4)$. In the presence of covariate shift, the target distribution is shifted to $\Nc(0.5_4, {\bf I}_4)$.
The class-conditional coefficients in \eqref{eq: Y|X linear} are defined as
\[
\beta^{(l=1)}_{c=0} = 0_4, \; \beta^{(l=1)}_{c=1} = (-0.4, -0.4, 0.2, 0.2)^\intercal,\; \beta^{(l=2)}_{c=0} = 0_4, \; \beta^{(l=2)}_{c=1} = (1, 1, 0.2, 0.2)^\intercal.
\]
The sample size for each source domain is fixed at $n\in\{500, 1000,..., 5000\}$. The number of unlabeled target samples is fixed at $N = 10{,}000$. 

The proposed approach is implemented using Algorithm~\ref{alg: main}, where we apply logistic regression to estimate both nuisance components $\widehat{f}^{(l)}$ and $\widehat{\omega}^{(l)}$ for each $l = 1, 2$. The Group DRO method is implemented via Algorithm~\ref{algo: GDRO} with learning rates set as $\eta_\theta = 0.01$ and $\eta_\gamma = 0.1$.
We evaluate both methods in terms of the estimation error $\|\widehat{\theta} - \theta^*\|_2$, where $\widehat{\theta}$ denotes the estimated model returned by each method, and $\theta^*$ is the population \CGDRO\ model, computed using Algorithm~\ref{alg: main} on a much larger dataset with $n = 100{,}000$ and $N = 200{,}000$. In addition, we assess both approaches by their worst-case risk, as defined in \eqref{eq: worst-case loss}.

\begin{algorithm}[!ht]
\DontPrintSemicolon
\SetAlgoLined
\SetNoFillComment
\LinesNotNumbered 
\caption{Mirror Prox Algorithm for Group DRO}
\label{algo: GDRO}
\KwData{Source data $\{(\X{l}_i, \Y{l}_i)\}_{i\in [n_l]}$ for $l\in [L]$;  learning rates for Mirror Prox: $\eta_\theta,\eta_\gamma$; maximum iteration $T$, tolerance $\epsilon$.}

    Set $\theta_{0}=\bar{\theta}_0=0$, $\gamma_{0}=\bar\gamma_0=(1/L,\cdots,1/L)$, and  ${\rm gap}= 1$;

\While{$0 \leq t \leq T$ and ${\rm diff} \geq \epsilon$}{
    \textbf{Intermediate Step}:
    {\small
    \[
    \begin{aligned}
\bar{\theta}_{t+1}&=\theta_{t}-\eta_\theta\cdot \nabla_{\theta} \widehat{\phi}_{\rm gdro}(\bar{\theta}_{t},\bar{\gamma}_{t}),\;\;\textrm{and}\;\;\\
[\bar{\gamma}_{t+1}]_l&= \frac{\gamma_{t,l}\cdot \exp\left(\eta_\gamma\cdot \left[\nabla_{\gamma} \widehat{\phi}_{\rm gdro}(\bar{\theta}_t, \bar{\gamma}_t)\right]_{l} \right)}{\sum_{j=1}^{L}\gamma_{t,j}\cdot \exp\left(\eta_\gamma \cdot \left[\nabla_{\gamma} \widehat{\phi}_{\rm gdro}(\bar{\theta}_t, \bar{\gamma}_t)\right]_{j} \right)}\quad \textrm{for $1\leq l\leq L$};
\end{aligned}
    \]}

    \textbf{Correction Step}:
    {\small
    \[
    \begin{aligned}
    {\theta}_{t+1}&=\theta_{t}-\eta_\theta\cdot \nabla_{\theta} \widehat{\phi}_{\rm gdro}(\bar{\theta}_{t+1},\bar{\gamma}_{t+1}),\;\;\textrm{and}\;\;\\
    [{\gamma}_{t+1}]_l&= \frac{\gamma_{t,l}\cdot \exp\left(\eta_\gamma \cdot \left[\nabla_{\gamma} \widehat{\phi}_{\rm gdro}(\bar{\theta}_{t+1},\bar{\gamma}_{t+1})\right]_{l} \right)}{\sum_{j=1}^{L}\gamma_{t,j}\cdot \exp\left(\eta_\gamma \cdot \left[\nabla_{\gamma} \widehat{\phi}_{\rm gdro}(\bar{\theta}_{t+1},\bar{\gamma}_{t+1})\right]_{j} \right)} \quad \textrm{for $1\leq l\leq L$};
    \end{aligned}
    \]
    }

    Set $\thetainit = \frac{1}{t}\sum_{s=1}^t\bar{\theta}_s$, and $\widehat{\gamma} = \frac{1}{t}\sum_{s=1}^t \bar{\gamma}_s$.

    Evaluate the duality gap: ${\rm gap} = \max_{\gamma\in \Delta^L}\widehat{\phi}(\widehat{\theta},\gamma) - \min_{\theta\in \RR^{Kd}}\widehat{\phi}(\theta,\widehat{\gamma})$.
}
\KwResult{Mirror Prox estimators $\widehat{\theta}_{\rm gdro}$, $\widehat{\gamma}_{\rm gdro}$.}
\end{algorithm}

\begin{figure}[H]
    \centering
    \includegraphics[width=0.7\linewidth]{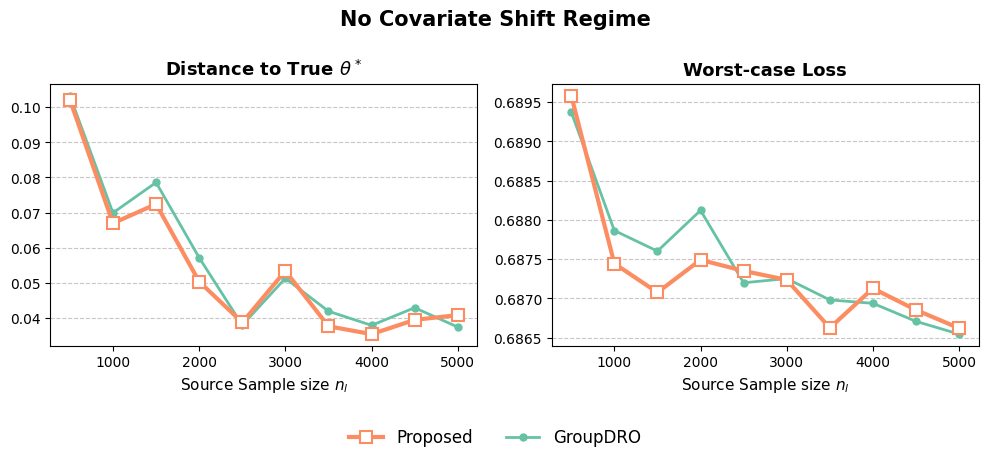}
    \caption{Comparison of the proposed method and Group DRO under the no covariate shift regime. The left panel displays the estimation error $\|\widehat{\theta} - \theta^*\|_2$, where $\widehat{\theta}$ is the estimator returned by each method, and $\theta^*$ is the population \CGDRO\ model. The right panel corresponds to the worst-case loss as defined in \eqref{eq: worst-case loss}. The sample size for each source domain is $n_1=n_2=n$, where $n$ varies over $\{500,1000,1500,...,5000\}$. The target sample size is fixed at $N=10000.$}
    \label{fig:groupdro-nocs}
\end{figure}

\begin{figure}[H]
    \centering
    \includegraphics[width=0.7\linewidth]{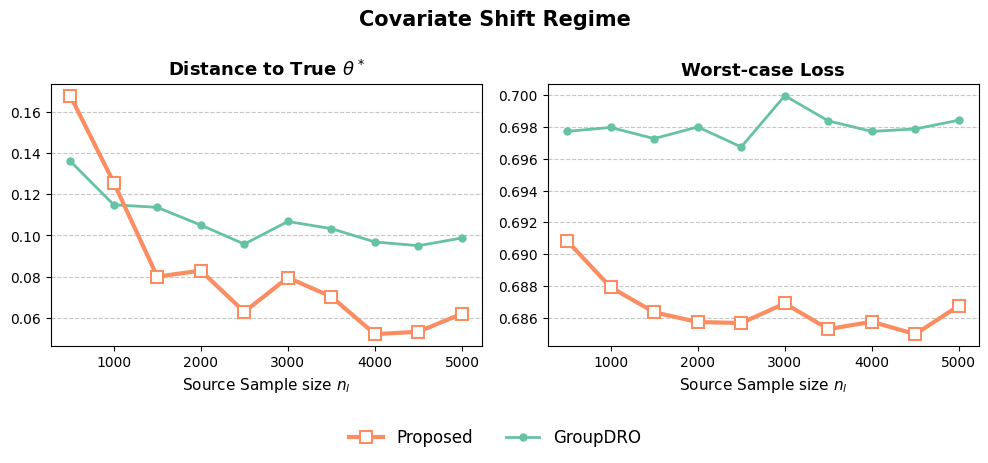}
    \caption{Comparison of the proposed method and Group DRO under the covariate shift regime. The left panel displays the estimation error $\|\widehat{\theta} - \theta^*\|_2$, where $\widehat{\theta}$ is the estimator returned by each method, and $\theta^*$ is the population \CGDRO\ model. The right panel corresponds to the worst-case loss as defined in \eqref{eq: worst-case loss}. The sample size for each source domain is $n_1=n_2=n$, where $n$ varies over $\{500,1000,1500,...,5000\}$. The target sample size is fixed at $N=10000.$}
    \label{fig:groupdro-cs}
\end{figure}

Figures~\ref{fig:groupdro-cs} and~\ref{fig:groupdro-nocs} compare the performance of the proposed method and Group DRO under the covariate shift and no covariate shift regimes, respectively. As shown in Figure~\ref{fig:groupdro-nocs}, both methods exhibit nearly identical estimation error and worst-case loss, and both achieve decreasing distance to $\theta^*$ as the sample size increases. This observation confirms that the two approaches align in the non-shift regime, consistent with the theoretical discussion in Section~\ref{sec: group dro}. In contrast, under covariate shift (Figure~\ref{fig:groupdro-cs}), the two methods differ from each other, and Group DRO has a higher worst-case risk compared to our proposed method.

\subsection{Setups of Simulations}
\label{sec: setup simus}
We now provide the setup details for the experiments in Section \ref{sec: simus}. Specifically, we present the values of the dimension $d$, the number of source domains $L$, the number of categories $K+1$, the specification of functions $\{\phi^{(l)}_c\}_{l,c}$ in \eqref{eq: phi_cl}, and the target covariate distribution $\QQ_X$ for each setting.
\begin{itemize}
    \item[\textbf{(S1)}] We consider a binary classification task with $K=1$ and $L=2$ source domains. We set $\QQ_X=\Nc(0.2_d, {\bf I}_d)$ with $d=20$. For  $l\in [L]$ and $k\in \{0,1\}$, we set
    $\phi^{(l)}_k(X) = X^\intercal \beta^{(l,k)},$ {with} $\beta^{(l,0)} = 0_d$ and $\beta^{(l,1)}\sim \Nc(0_d, 0.25 {\bf I}_d)$. 
    \item[\textbf{(S2)}] We consider a multi-class classification task with $K=3$ (i.e., four categories) and $L=4$ source domains. We set $\QQ_X = \Nc(0.2_d, 0.25{\bf I}_d)$ with dimension $d=5$. For each source $l\in [L]$ and category $k\in [K]$, the function $\phi_k^{(l)}(\cdot)$ in \eqref{eq: phi_cl} is generated as follows:
    {\small
    \[
    \phi_k^{(l)}(X) = a^{(l,k)} + \sum_{j=1}^d w^{(l)}_j\cdot \exp\left(-{(X_j - \frac{k}{4})^2}/{4}\right) + b^{(l,k)} X_1^{\intercal} X_2 + c^{(l,k)}\sin\left((X_3 - X_4) + {k}/{3}\right),
    \]
    }
    where the parameters $a^{(l,k)}, b^{(l,k)}, c^{(l,k)} \stackrel{i.i.d.}{\sim}{\rm Uniform}(-0.5, 0.5)$ and $w^{(l)}\sim {\rm Dirichlet}({\bf 1}_d)$.
\end{itemize}

\begin{itemize}
    \item[\textbf{(S3)}] This captures a spectrum of regimes ranging from \textbf{regular} to \textbf{nonregular}, controlled by the perturbation parameter {$\delta$}. We consider a binary classification task ($K=1$) on two source domains ($L=2$). We set $\QQ_X = \Nc(0.2_d, {\bf I}_d)$ with $d=20$ and  $\phi_k^{(l)}(X)=X^\intercal \beta^{(l,k)}$ with
    \[
    \beta^{(1,0)} = 0_d, \; \beta^{(1,1)} = (0.5+{\delta},  0.5_3, 0_{d-4})^\intercal, \; \beta^{(2,0)} = 0_d,\; \beta^{(2,1)} = (0_4, 0.5_2, 0_{d-6})^\intercal.
    \]
    The parameter $\delta\in \{0, 0.5, 1.0, ..., 4.0\}$ perturbs the first entry of the first source's class-1 coefficient vector. As $\delta$ increases, the discrepancy between the two source grows, increasing the nonregularity of the problem.
    \item[\textbf{(S4)}] This setting captures a spectrum of regimes ranging from \textbf{regular} to \textbf{unstable}, governed by the noise parameter {$\sigma$}. We consider a multi-class classification task with four categories ($K=3$) and $L=4$ source domains. We set $\QQ_X = \Nc(0.2_d, {\bf I}_d)$ with $d=20$ and $\phi_k^{(l)}(X)=X^\intercal \beta^{(l,k)}$ with
    \[
    \beta^{(l,0)} = 0_d, \; \beta^{(l,k)}_{1:5} = (1, 1, 0, 0, 0)^\intercal + \Nc(0_5, {\sigma^2}{\bf I}_5), \; \beta^{(l,k)}_{6:d} = 0_{d-5}\quad \textrm{for $k=1,2,3$.}
    \]
    The noise parameter $\sigma \in [0.1, 0.5]$ controls the level of instability across sources. When $\sigma$ is small, the coefficients across sources are nearly identical, inducing strong instability. As $\sigma$ increases, the variability between sources grows, alleviating the instability and leading to a more regular setting.
\end{itemize}

\subsection{Dependence on Resampling Size $M$}
\label{appendix: varyM}

\begin{figure}[H]
    \centering
    \includegraphics[width=0.7\linewidth]{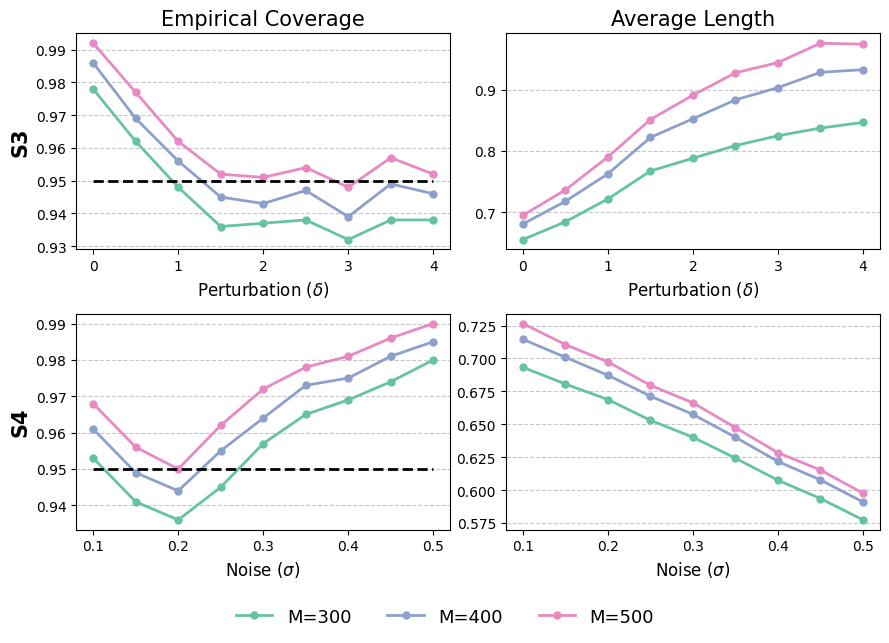}
    \caption{Empirical coverage and average interval lengths of the proposed method with varying resampling size $M\in \{300,400,500\}$ under two settings: \textbf{(S3)} (top panels) with sparse perturbation level $\delta \in \{0, 0.5, \ldots, 4.0\}$, and \textbf{(S4)} (bottom panels) with instability level $\sigma \in \{0.1, 0.15, \ldots, 0.50\}$. Each point is averaged over 500 simulation rounds.}
    \label{fig:varyM}
\end{figure}

We investigate how the performance of the proposed inference procedure depends on the resampling size $M$. Specifically, we examine the empirical coverage and average interval length for $M \in \{300, 400, 500\}$. The experiment is conducted under the settings \textbf{(S3)} and \textbf{(S4)} described in Section~\ref{sec: simus}.
As shown in Figure~\ref{fig:varyM}, increasing $M$ leads to a higher empirical coverage, while the corresponding average interval length also grows moderately but not substantially.

\subsection{Additional Inference Results}
\label{appendix: infer simus}
In this section, we introduce a more challenging setting to validate the proposed inferential procedure.
\begin{itemize}
    \item [\textbf{(S5)}] This setting captures a spectrum of regimes ranging from \textbf{regular} to \textbf{nonregular}, controlled by the parameter $\delta$. We consider a multi-class classification task with three categories ($K=3$) and $L=3$ source domains. The target distribution is $\QQ_X = \Nc(0.2_d, 0.25{\bf I}_d)$ with $d=5$. For each source $l\in [L]$ and category $k\in [K]$, the function $\phi_k^{(l)}(\cdot)$ in \eqref{eq: phi_cl} is generated as follows:
    {\small
    \[
    \begin{aligned}
        \phi_k^{(l)}(X) &= a^{(l,k)} + \sum_{j=1}^d w^{(l)}_j \exp\left(-\frac{(X_j - \frac{k}{4})^2}{4}\right)+ b^{(l,k)} X_1^{\intercal} X_2 + c^{(l,k)}\sin\left((X_3 - X_4) + \frac{k}{3}\right)\\
        &\quad + \delta X_1 {\bf 1}(l=1, k=1) ,
    \end{aligned}
    \]
    }
    where $a^{(l,k)}, b^{(l,k)}, c^{(l,k)} \stackrel{iid}{\sim}{\rm Uniform}(-0.5, 0.5)$ and $w^{(l)}\sim {\rm Dirichlet}({\bf 1}_d)$. The parameter $\delta\in \{0, 0.1, ..., 1.0\}$ perturbs the first entry of the first source's class-1 coefficient vector. Non-regularity gets pronounced when $\delta$ increases.
\end{itemize}
We implement the proposed approach using Algorithm~\ref{algo: uq}, which involves the estimation of $\widehat{\mu}^{(l)}$ and $\widehat{\bf V}_\mu^{(l)}$, both constructed based on the nuisance components $\widehat{f}^{(l)}$ and $\widehat{\omega}^{(l)}$ introduced in Section~\ref{sec: doubly robust}. In this experiment, we employ the XGBoost algorithm with cross-validation to estimate these quantities.

\begin{figure}[H]
    \centering
    \includegraphics[width=0.8\linewidth]{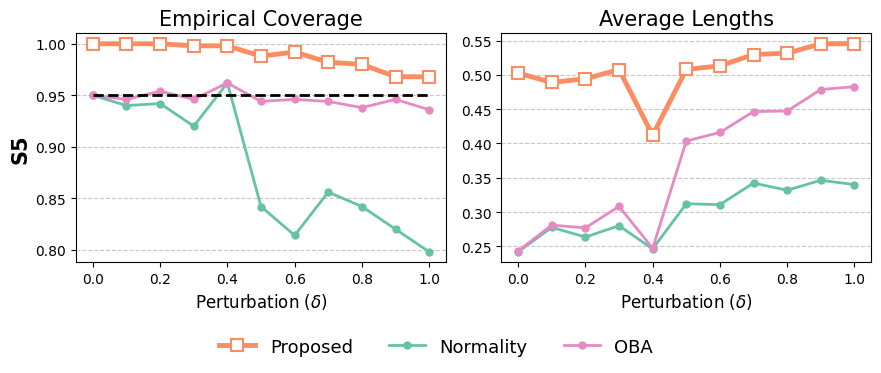}
    \caption{Empirical coverage and average interval lengths of different inference methods under \textbf{(S5)}. The \texttt{Bootstrap} method is omitted due to its high computational cost, which requires refitting 500 ML models per round. Results are averaged over 500 simulation rounds. }
    \label{fig:infer-nonparametric}
\end{figure}

Figure~\ref{fig:infer-nonparametric} presents the empirical coverage and average interval lengths for various methods for \textbf{(S5)}.
Due to the computational burden of the \texttt{Bootstrap} method, which requires fitting $B = 500$ ML models with cross-validation per simulation, we omit it from this experiment. Aligning with patterns for the parametric settings in Figure~\ref{fig:infer-parametric}, the \texttt{Normality} method achieves nominal $95\%$ coverage only in the regular regime when the perturbation level $\delta = 0$, but quickly deteriorates as $\delta$ increases and the problem becomes nonregular. In contrast, both the oracle \texttt{OBA} method and our \texttt{Proposed} approach maintain valid coverage uniformly across all values of $\delta$ in both dimensions. Even though our method does not require access to the true parameter $\theta^*_1$, it yields average interval lengths comparable to those of the oracle \texttt{OBA}, particularly in the more nonregular settings with larger $\delta$.

\end{document}